\documentclass[12pt,letterpaper]{article} 

\usepackage{epsfig}
\usepackage{cite}

\long\def\begincomment#1\endcomment{}

\begin{document}



\phantom{nothing}
\begin{center}
{\LARGE\bf Brillouin improvement for Wilson fermions}
\end{center}
\vspace{10pt}

\begin{center}
{\large\bf Stephan D\"urr$\,^{a,b}$}
\,\,\,and\,\,\,
{\large\bf Giannis Koutsou$\,^{a,b}$}
\\[10pt]
${}^a${\sl Bergische Universit\"at Wuppertal, Gau\ss$\,\!$stra\ss$\,\!$e\,20,
42119 Wuppertal, Germany}\\
${}^b${\sl J\"ulich Supercomputing Center, Forschungszentrum J\"ulich,
52425 J\"ulich, Germany}
\end{center}
\vspace{10pt}

\begin{abstract}
\noindent
We present a parameter-free Wilson-type lattice Dirac operator with an
81-point stencil for the covariant derivative and the Laplacian which
attempts to minimize the breaking of rotational symmetry near the
boundary of the Brillouin zone. The usefulness of this ``Brillouin
operator'' in practical applications is explored by studying the
scaling of pseudo-scalar decay constants in quenched QCD, with rather
good results in the physical charm region. We also investigate the
suitability of this operator as a kernel to the overlap procedure.
Here, the resulting overlap operator is found to be cheaper to
construct and significantly better localized than the variety with the
standard Wilson kernel.
\end{abstract}
\vspace{10pt}


\newcommand{\pad}{\partial}
\newcommand{\hqu}{\hbar}
\newcommand{\ovr}{\over}
\newcommand{\til}{\tilde}
\newcommand{\pri}{^\prime}
\renewcommand{\dag}{^\dagger}
\newcommand{\<}{\langle}
\renewcommand{\>}{\rangle}
\newcommand{\gaf}{\gamma_5}
\newcommand{\nab}{\nabla}
\newcommand{\lap}{\triangle}
\newcommand{\dal}{{\sqcap\!\!\!\!\sqcup}}
\newcommand{\trc}{\mathrm{tr}}
\newcommand{\Trc}{\mathrm{Tr}}
\newcommand{\Mpi}{M_\pi}
\newcommand{\Fpi}{F_\pi}
\newcommand{\Mka}{M_K}
\newcommand{\Fka}{F_K}
\newcommand{\Met}{M_\et}
\newcommand{\Fet}{F_\et}
\newcommand{\Mss}{M_{\bar{s}s}}
\newcommand{\Fss}{F_{\bar{s}s}}
\newcommand{\Mcc}{M_{\bar{c}c}}
\newcommand{\Fcc}{F_{\bar{c}c}}

\newcommand{\al}{\alpha}
\newcommand{\be}{\beta}
\newcommand{\ga}{\gamma}
\newcommand{\de}{\delta}
\newcommand{\ep}{\epsilon}
\newcommand{\ve}{\varepsilon}
\newcommand{\ze}{\zeta}
\newcommand{\et}{\eta}
\renewcommand{\th}{\theta}
\newcommand{\vt}{\vartheta}
\newcommand{\io}{\iota}
\newcommand{\ka}{\kappa}
\newcommand{\la}{\lambda}
\newcommand{\rh}{\rho}
\newcommand{\vr}{\varrho}
\newcommand{\si}{\sigma}
\newcommand{\ta}{\tau}
\newcommand{\ph}{\phi}
\newcommand{\vp}{\varphi}
\newcommand{\ch}{\chi}
\newcommand{\ps}{\psi}
\newcommand{\om}{\omega}

\newcommand{\psb}{\bar{\psi}}
\newcommand{\etb}{\bar{\eta}}
\newcommand{\psh}{\hat{\psi}}
\newcommand{\eth}{\hat{\eta}}
\newcommand{\psd}{\psi^{\dagger}}
\newcommand{\etd}{\eta^{\dagger}}
\newcommand{\qh}{\hat{q}}
\newcommand{\kh}{\hat{k}}

\newcommand{\bdm}{\begin{displaymath}}
\newcommand{\edm}{\end{displaymath}}
\newcommand{\bea}{\begin{eqnarray}}
\newcommand{\eea}{\end{eqnarray}}
\newcommand{\beq}{\begin{equation}}
\newcommand{\eeq}{\end{equation}}

\newcommand{\mr}{\mathrm}
\newcommand{\mb}{\mathbf}
\newcommand{\Nf}{N_{\!f}}
\newcommand{\Nc}{N_{ c }}
\newcommand{\Nt}{N_{ t }}
\newcommand{\ri}{\mr{i}}
\newcommand{\DW}{D_\mr{W}}
\newcommand{\Dsl}{D\!\!\!\!\slash\,}
\newcommand{\Dov}{D_\mr{ov}}
\newcommand{\Dst}{D_\mr{st}}
\newcommand{\Dke}{D_\mr{ke}}
\newcommand{\Dovm}{D_{\mr{ov},m}}
\newcommand{\Dstm}{D_{\mr{st},m}}
\newcommand{\Dkem}{D_{\mr{ke},m}}
\newcommand{\MeV}{\,\mr{MeV}}
\newcommand{\GeV}{\,\mr{GeV}}
\newcommand{\fm}{\,\mr{fm}}
\newcommand{\MSbar}{\overline{\mr{MS}}}

\hyphenation{topo-lo-gi-cal simu-la-tion theo-re-ti-cal mini-mum con-tinu-um}


\section{Introduction}


Apart from being formally correct, a good lattice action should satisfy several
requirements:
($i$) it should induce small cut-off effects in on-shell quantities,
($ii$) it should have a continuum-like dispersion relation, and
($iii$) it should be cheap to simulate.
Unfortunately, at least in the fermion sector these requirements tend to be in
conflict with each other.
For instance, the classic Wilson action \cite{Wilson:1974sk,Wilson:1975id} is
good on ($iii$), but not so much on the first two points.
By contrast, an overlap action \cite{Neuberger:1997fp,Neuberger:1998wv} with
this operator as a kernel is significantly better on ($i$), but worse on the
remaining two points.

In the literature, there are two main avenues for obtaining a better fermion
discretization.
The ``bottom-up'' approach is to expand physical quantities in powers of the
lattice spacing $a$, and to demand that the leading cut-off effects are
proportional to $\al^na$ (or even $a^2$), where $\al$ is the strong coupling
constant and $n$ some power.
This program of perturbative or non-perturbative $O(a)$-improvement has been
carried out successfully \cite{Sheikholeslami:1985ij,Heatlie:1990kg,
Martinelli:1990ny,Luscher:1996sc,Luscher:1996ug}.
The ``top-down'' approach starts from the concept of a perfect action with zero
cut-off effects, perfect chiral symmetry in the sense of the Ginsparg-Wilson
relation \cite{Ginsparg:1981bj}, and a continuum-like dispersion relation.
To realize these goals exactly, a Dirac operator $D(x,y)$ is needed with
non-zero entry for each $(x,y)$ pair, which is in strong conflict with
criterion $(iii)$ above.

In practice, one would like to maintain some degree of sparsity, that is to
have an operator which is zero whenever $x$ and $y$ are further apart than a
certain threshold.
In the literature the most prominent attempts to realize ultralocal approximate
derivatives of the ideal perfect action go by the name truncated perfect action
\cite{Hasenfratz:1993sp,Bietenholz:1995cy,Hasenfratz:2000xz}, hypercube action
\cite{Bietenholz:1996pf,DeGrand:1998pr,Bietenholz:1998ut}, and chirally
improved action \cite{Gattringer:2000js,Gattringer:2000qu}.
They differ by the extent through which they make use of the full
Dirac-Clifford algebra (the ``continuum'' operator uses only $\ga_\mu$ with
$\mu\!=\!1...4$ and the identity), and by the criteria used to pin down the
various coefficients.

In this article we pursue a similar approach, albeit with a different focus of
which properties should be optimized.
We do not attempt to reduce $O(a^2)$ cut-off effects or the amount of chiral
symmetry breaking, since it is known how one can get rid of these effects by
adding local improvement terms and/or using the overlap recipe.
By contrast, we strive for good overall appearance of the eigenvalue spectrum
of $D$, and for a continuum-like dispersion relation, because for these
properties no systematic improvement scheme is known.

To ease the discussion let us consider the improved Wilson (``clover'') Dirac
operator
\bdm
D(x,y)={1\ovr2}\sum_\mu
\Big\{
(\ga_\mu\!-\!I)U_\mu(x)\de_{x+\hat\mu,y}-
(\ga_\mu\!+\!I)U_\mu\dag(x\!-\!\hat\mu)\de_{x-\hat\mu,y}
\Big\}
+{1\ovr2\ka}\de_{x,y}
-{c_\mr{SW}\ovr2}\sum_{\mu<\nu}\si_{\mu\nu}F_{\mu\nu}\de_{x,y}
\edm
with $\si_{\mu\nu}\!=\!{\ri\ovr2}[\ga_\mu,\ga_\nu]$ and $F_{\mu\nu}$ the
hermitean clover-leaf field-strength tensor.
In the wavy brackets there is a discrete Laplacian whose job is to lift
15 out of the 16 species, such that the resulting operator is doubler-free.
In other words, the structure of the Wilson operator is
\beq
D(x,y)=\sum_\mu \ga_\mu \nab_\mu^\mr{std}(x,y)
-{a\ovr2}\lap^\mr{std}(x,y)+m_0\de_{x,y}+\mbox{improvement term}
\label{def_wils}
\eeq
where $\nab_\mu^\mr{std}$ denotes the forward-backward symmetric covariant
derivative with 2-point stencil, and $\lap^\mr{std}$ the standard covariant
Laplacian with 9-point stencil.
The mass parameters in the two representations above relate through
$1/(2\ka)=4\!+\!am_0$.

The idea explored in this paper is to start from (\ref{def_wils}), and to
replace the covariant derivative $\nab_\mu^\mr{std}$ and the Laplacian
$\lap^\mr{std}$ by similar discretizations with improved properties.
As a tribute to criterion $(iii)$ above, we shall include only one adjacent
layer in each positive or negative direction.
Accordingly, both stencils have support on at most $3^d$ points in $d$
space-time dimensions (which, in the following, will be referred to as
2D, 3D, 4D for $d=2,3,4$, respectively).
The choice of the final operator is based on a Darwinistic selection rule.
Both for $\nab_\mu$ and $\lap$, a few varieties with distinct properties are
considered, and for each combination the resulting Dirac operator is
implemented.
Based on the respective eigenvalue spectra and free field dispersion
relations, we select the most promising combination in 2D (Sec.\,2),
3D (Sec.\,3) and 4D (Sec.\,4).
Fortunately, it turns out that one choice fares best regarding either
criterion, and this choice is the same in any dimension.
The resulting operator has no tunable parameters, and maintains the property
of $\gaf$-hermiticity, i.e.\ $\gaf D\gaf=D\dag$.
Details of our implementation, including the overall link smearing strategy,
the gauge covariant derivatives (based on a summation over all shortest paths
with backprojection to the group) and tree-level clover improvement, are
specified in Sec.\,5.
Practical tests in quenched QCD, with a focus on scaling studies of simple
quantities, and in comparison to an analogously defined link-smeared tree-level
clover improved Wilson operator, are reported in Sec.\,6.
In Sec.\,7 we explore the suitability of our Brillouin operator as a kernel
to the overlap procedure, finding a noticeable reduction of the condition
number of the shifted hermitean kernel, and a significant improvement of the
locality of the resulting overlap operator.
A summary of our findings is given in Sec.\,8, and details of all stencils,
both in position and momentum space, are arranged in four appendices with the
hope that they might prove useful in applications beyond lattice QCD.


\section{Construction and main features in 2D}


\subsection{Summary of 2D Laplace stencils}

The ``standard'' stencil of the Laplacian in 2D and the ``tilted'' variety
(as defined in App.\,A) have the Fourier space representation (with $k_i=ap_i$
the dimensionless wave-number)
\bea
a^2\hat\lap^\mr{std}(k_1,k_2)&=&2\cos(k_1)+2\cos(k_2)-4
\nonumber
\\
&=&-4\sin^2(k_1/2)-4\sin^2(k_2/2)
\label{def_2Dstd}
\\
a^2\hat\lap^\mr{til}(k_1,k_2)&=&2\cos(k_1)\cos(k_2)-2
\nonumber
\\
&=&8\cos^2(k_1/2)\cos^2(k_2/2)-4\cos^2(k_1/2)-4\cos^2(k_2/2)
\label{def_2Dtil}
\eea
respectively.
From the stencil notation in App.\,A it is easy to see that in position space
the ``standard'' Laplacian (\ref{def_2Dstd}) has only 1-hop contributions
(apart from the center element), while the ``tilted'' Laplacian
(\ref{def_2Dtil}) has only support at the edge of the $3^2$-point area around
the center.
Both of them discretize the continuum Laplacian in the sense that they
deviate from the continuum behavior $\hat\lap^\mr{con}=-p_1^2-p_2^2$ through
$O(a^2)$-suppressed terms.
Note, however, that the ``tilted'' version (\ref{def_2Dtil}) differs from the
``standard'' variety (\ref{def_2Dstd}) by having a second zero at the edge of
the Brillouin zone, i.e.\ at $k_1=k_2=\pi$ (in the convention where the
Brillouin zone ranges from $-\pi/a$ to $\pi/a$ in every direction).

In 2D these two stencils form a basis of all Laplace filters with (at most) a
9-point stencil.
By taking a linear combination $\hat\lap(k_1,k_2)=
\al\hat\lap^\mr{std}(k_1,k_2)+(1\!-\!\al)\hat\lap^\mr{til}(k_1,k_2)$ one may
try to improve certain properties of the discretized Laplacian.
In particular, reducing the breaking of the rotational symmetry of the
continuum operator is important.
Two choices of $\al$ are popular in the literature.
First, $\al\!=\!1/2$ leads to (what we call) the ``Brillouin'' filter
\cite{McClellan:1973}
\bea
a^2\hat\lap^\mr{bri}(k_1,k_2)&=&\cos(k_1)\cos(k_2)+\cos(k_1)+\cos(k_2)-3
\nonumber
\\
&=&4\cos^2(k_1/2)\cos^2(k_2/2)-4
\label{def_2Dbri}
\eea
since $a^2\hat\lap^\mr{bri}(k_1,k_2)=-4$ whenever one of the momenta is
$\pm\pi/a$.
In other words, $a^2\hat\lap^\mr{bri}$ takes a constant value on the entire
boundary of the Brillouin zone.
Second, the choice $\al\!=\!2/3$ yields the ``isotropic'' filter or stencil
(see \cite{Kumar:2004} and Refs.\,6-7 therein)
\bea
a^2\hat\lap^\mr{iso}(k_1,k_2)&=&
[2\cos(k_1)\cos(k_2)+4\cos(k_1)+4\cos(k_2)-10]/3
\nonumber
\\
&=&
[8\cos^2(k_1/2)\cos^2(k_2/2)+4\cos^2(k_1/2)+4\cos^2(k_2/2)-16]/3
\label{def_2Diso}
\eea
since for small momenta $a^2\hat\lap^\mr{iso}(k_1,k_2)=
-a^2[p_1^2\!+\!p_2^2]+a^4[p_1^2\!+\!p_2^2]^2/12+O(a^6)$
has $O(a^4)$ terms which depend only on the combination $p_1^2\!+\!p_2^2$.
Put differently, the continuum relation $\hat\lap^\mr{con}=-p_1^2-p_2^2$ is
violated on-axis (to this order) in the same manner as off-axis.
Note that this improvement strategy differs from the usual one, where one tries
to remove $O(a^n)$ terms, for ever-larger $n$, along the axes only.

\begincomment
obj1:=2*cos(k1)+2*cos(k2)-4;
sort(expand(subs(k1=2*k1h,k2=2*k2h,obj1),trig));
series(subs({k1=a*p1,k2=a*p2},obj1),a,4);

obj2:=2*cos(k1)*cos(k2)-2;
sort(expand(subs(k1=2*k1h,k2=2*k2h,obj2),trig));
series(subs({k1=a*p1,k2=a*p2},obj2),a,4);

obj3:=4*cos(k1/2)^2*cos(k2/2)^2-4;
sort(expand(combine(obj3,trig),trig));
series(subs({k1=a*p1,k2=a*p2},obj3),a,4);

obj4:=(cos(k1)-1)*(cos(k2)+5)/3:
obj4:=sort(expand(obj4+ \
 subs({k1=k2,k2=k1},obj4) \
));
sort(expand(subs(k1=2*k1h,k2=2*k2h,obj4),trig));
factor(series(subs({k1=a*p1,k2=a*p2},obj4),a,6));
\endcomment

\begin{figure}[!p]
\centering
\epsfig{file=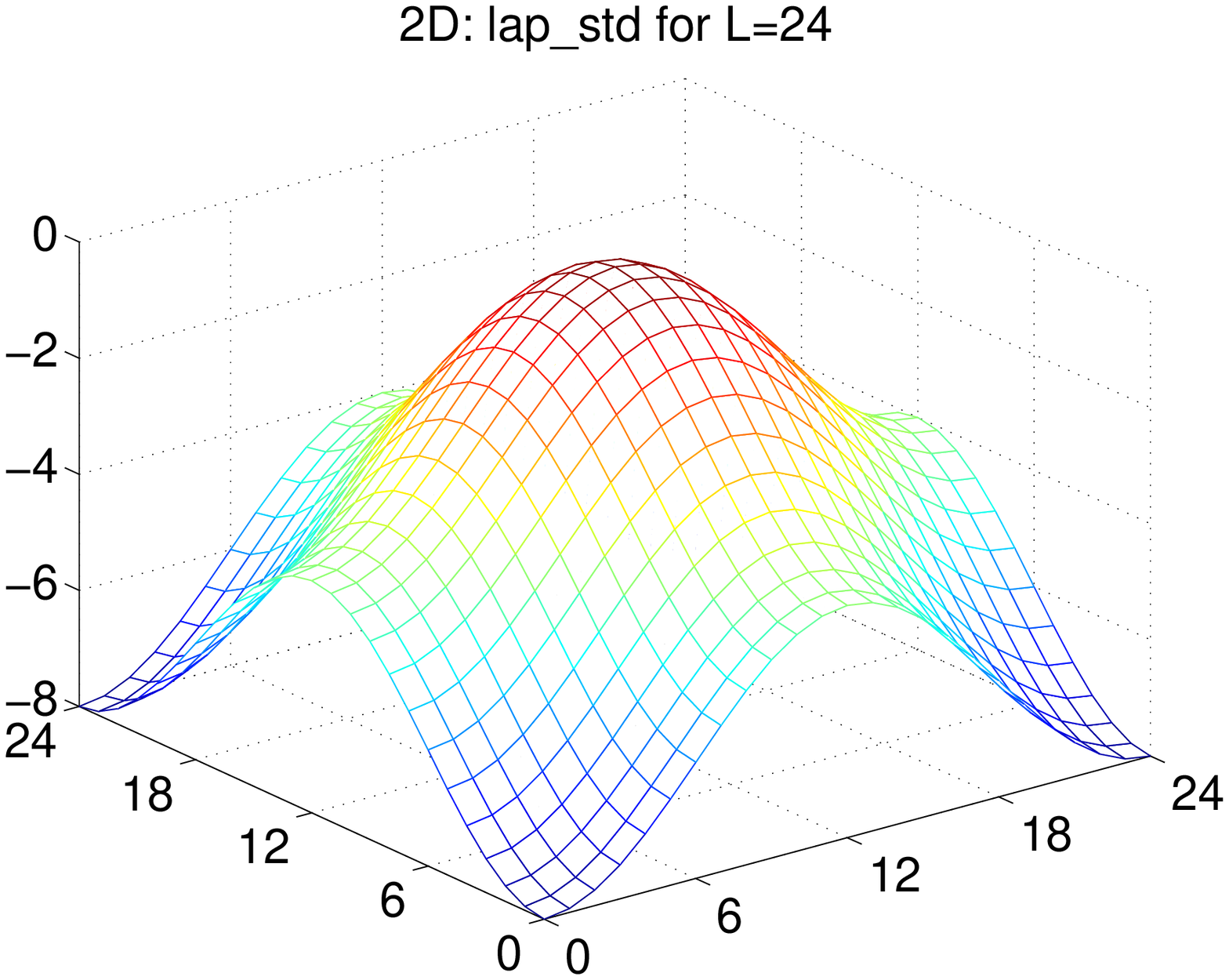,height=5.7cm}
\epsfig{file=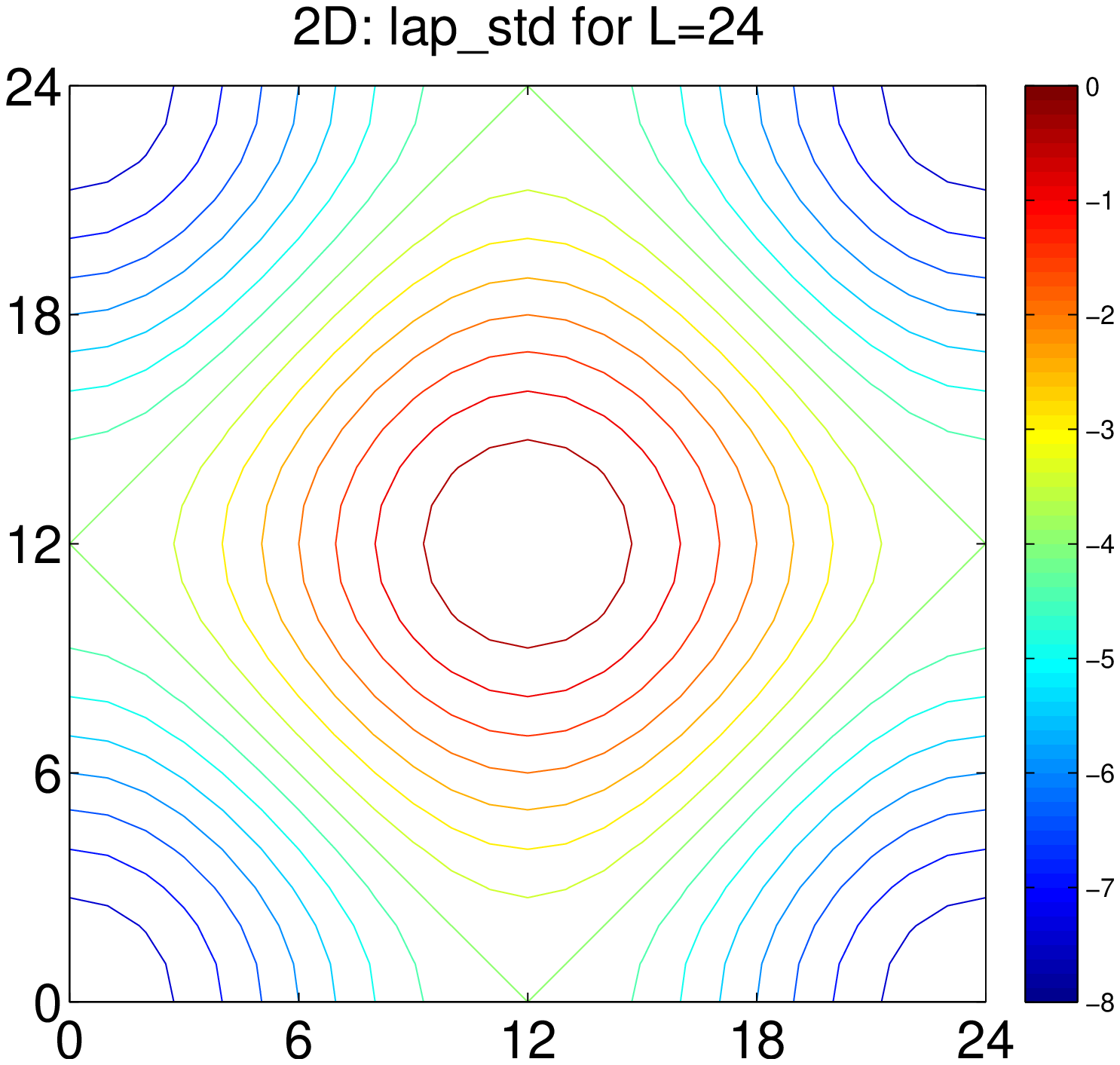,height=5.7cm}\\
\epsfig{file=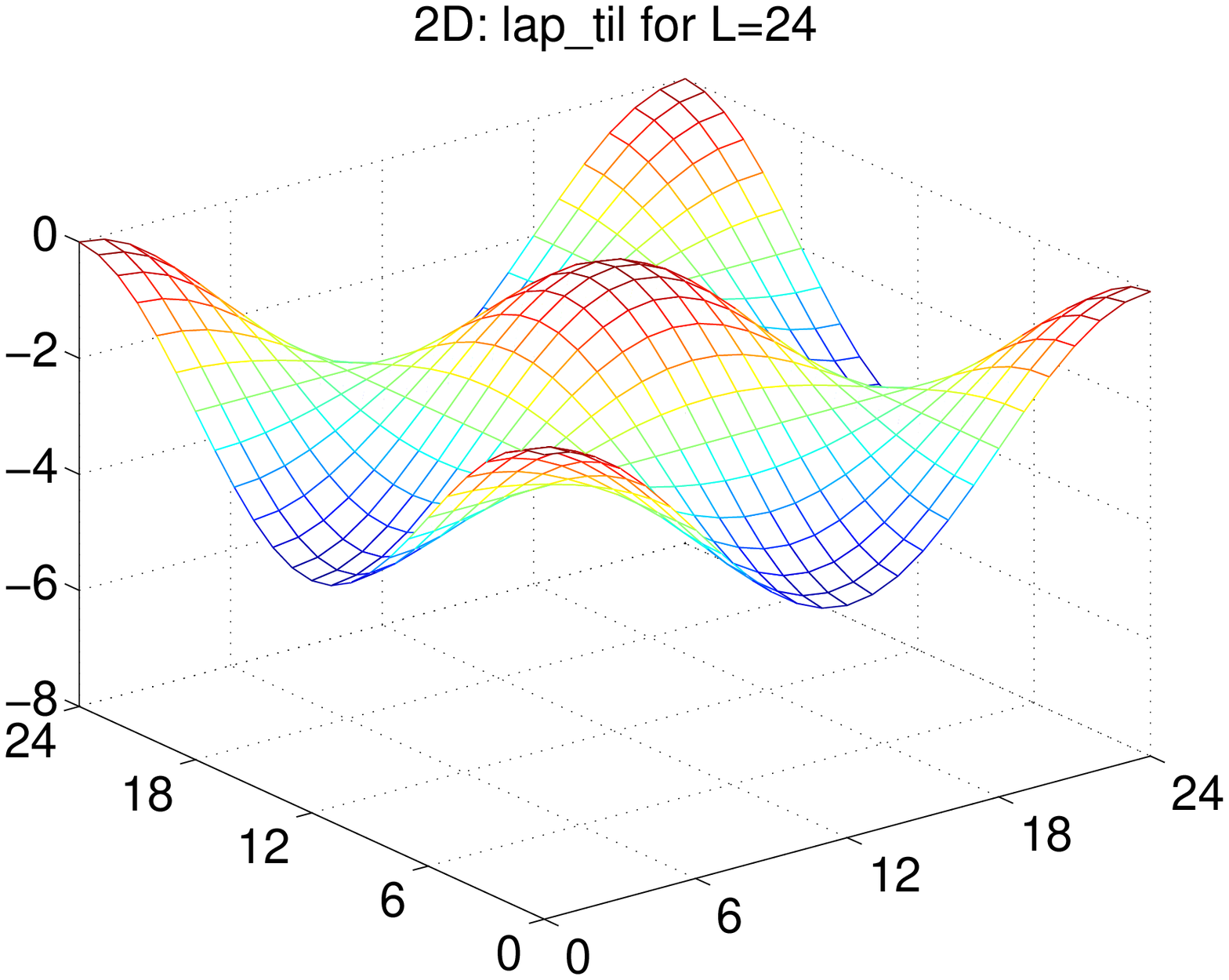,height=5.7cm}
\epsfig{file=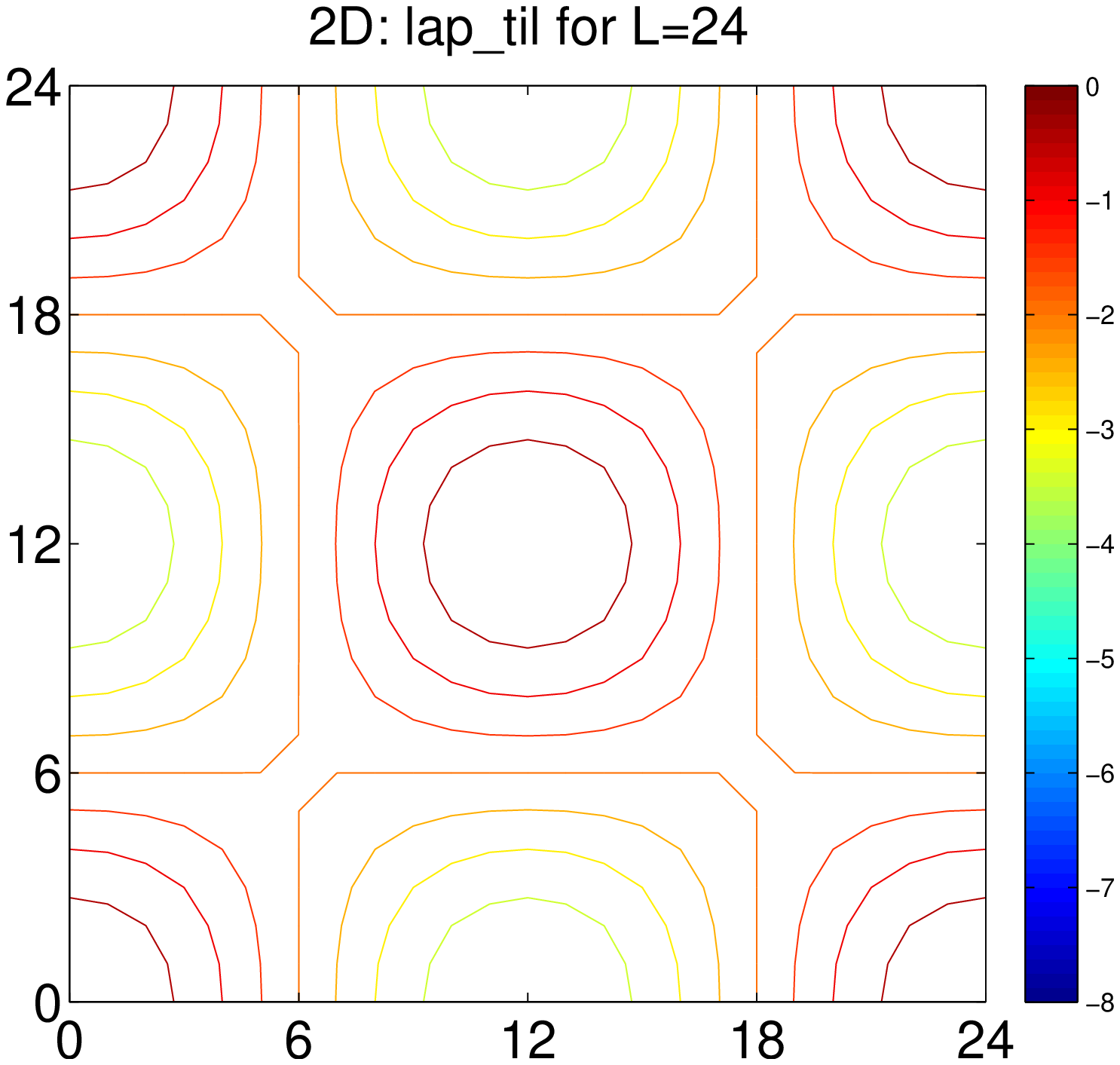,height=5.7cm}\\
\epsfig{file=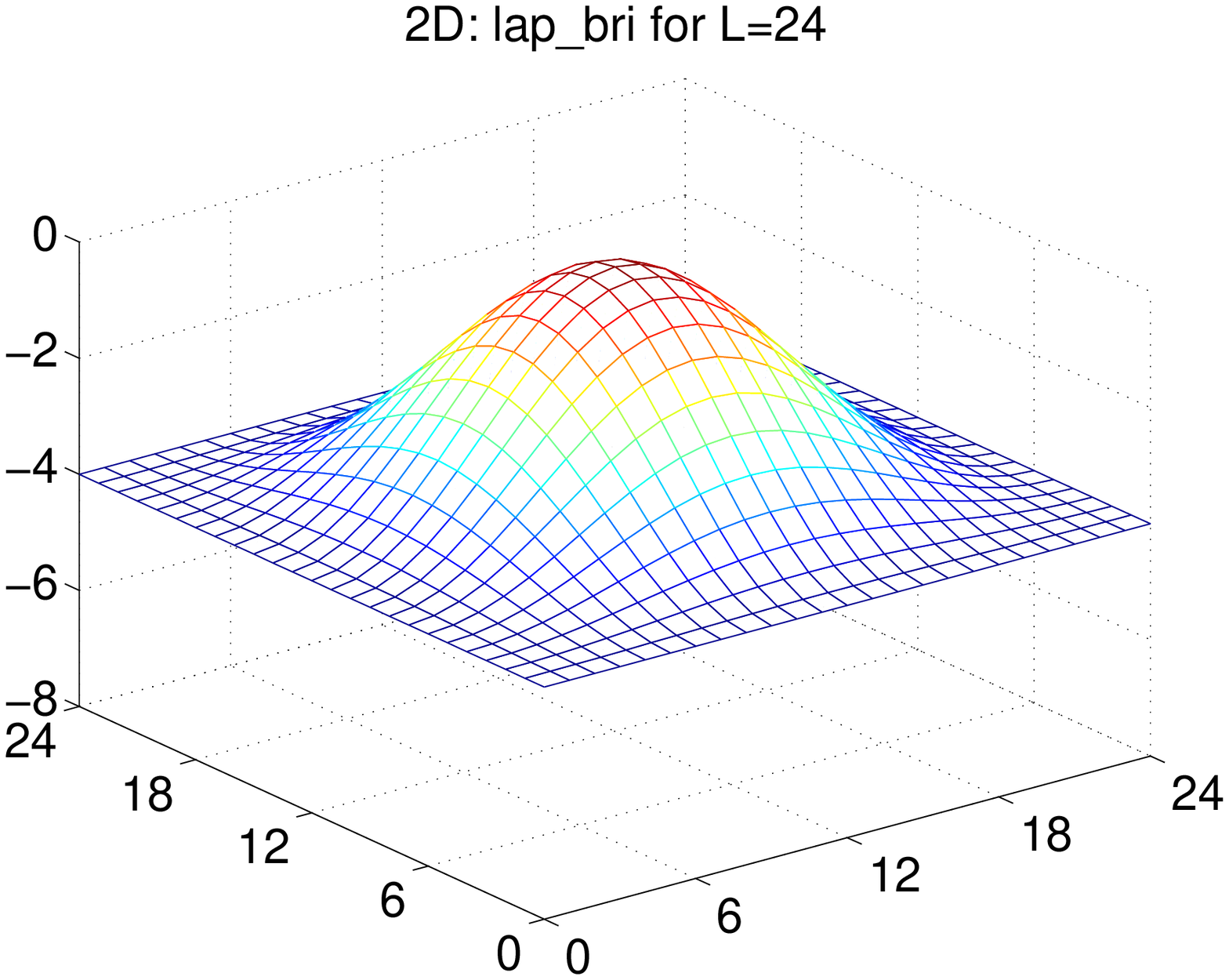,height=5.7cm}
\epsfig{file=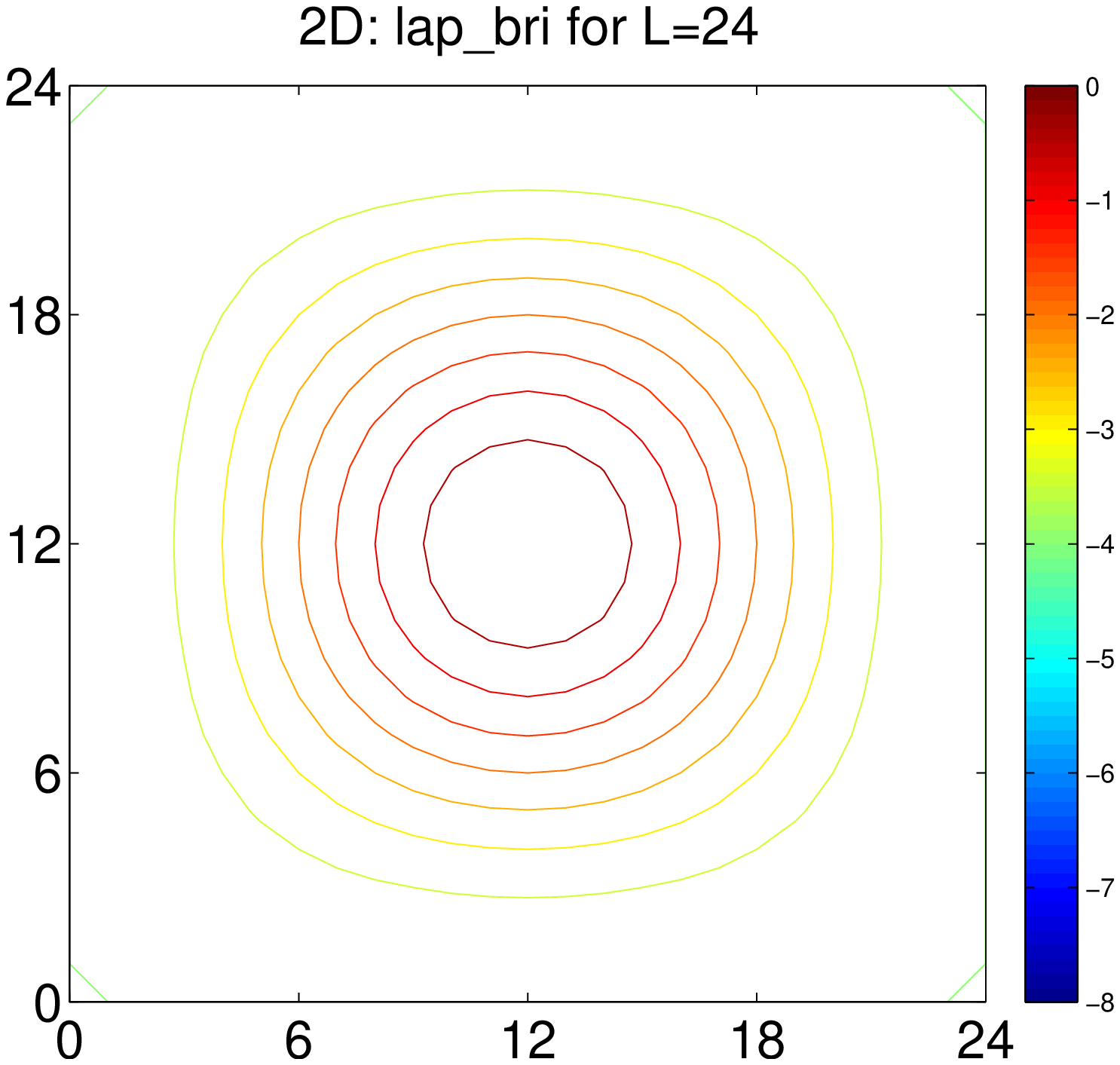,height=5.7cm}\\
\epsfig{file=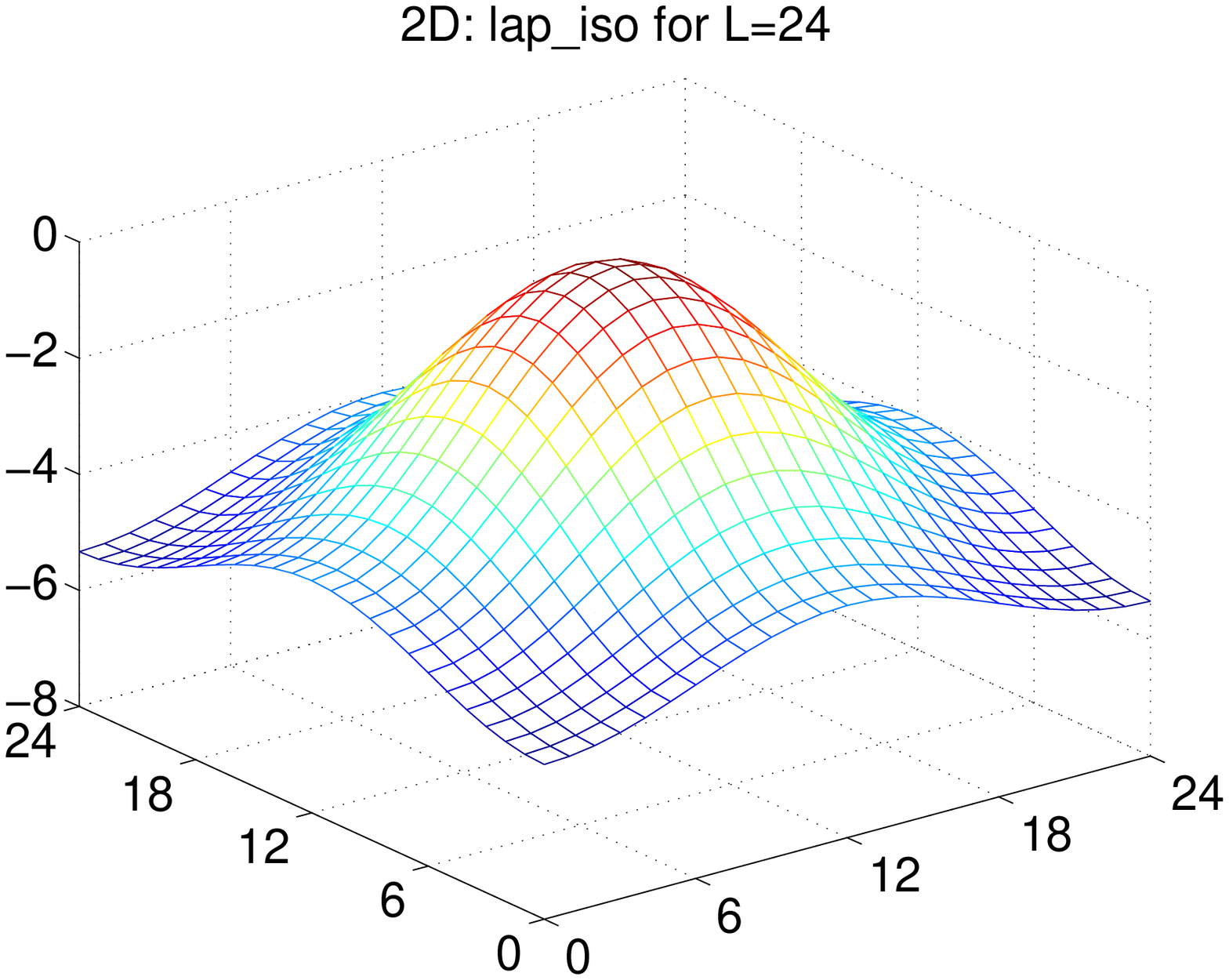,height=5.7cm}
\epsfig{file=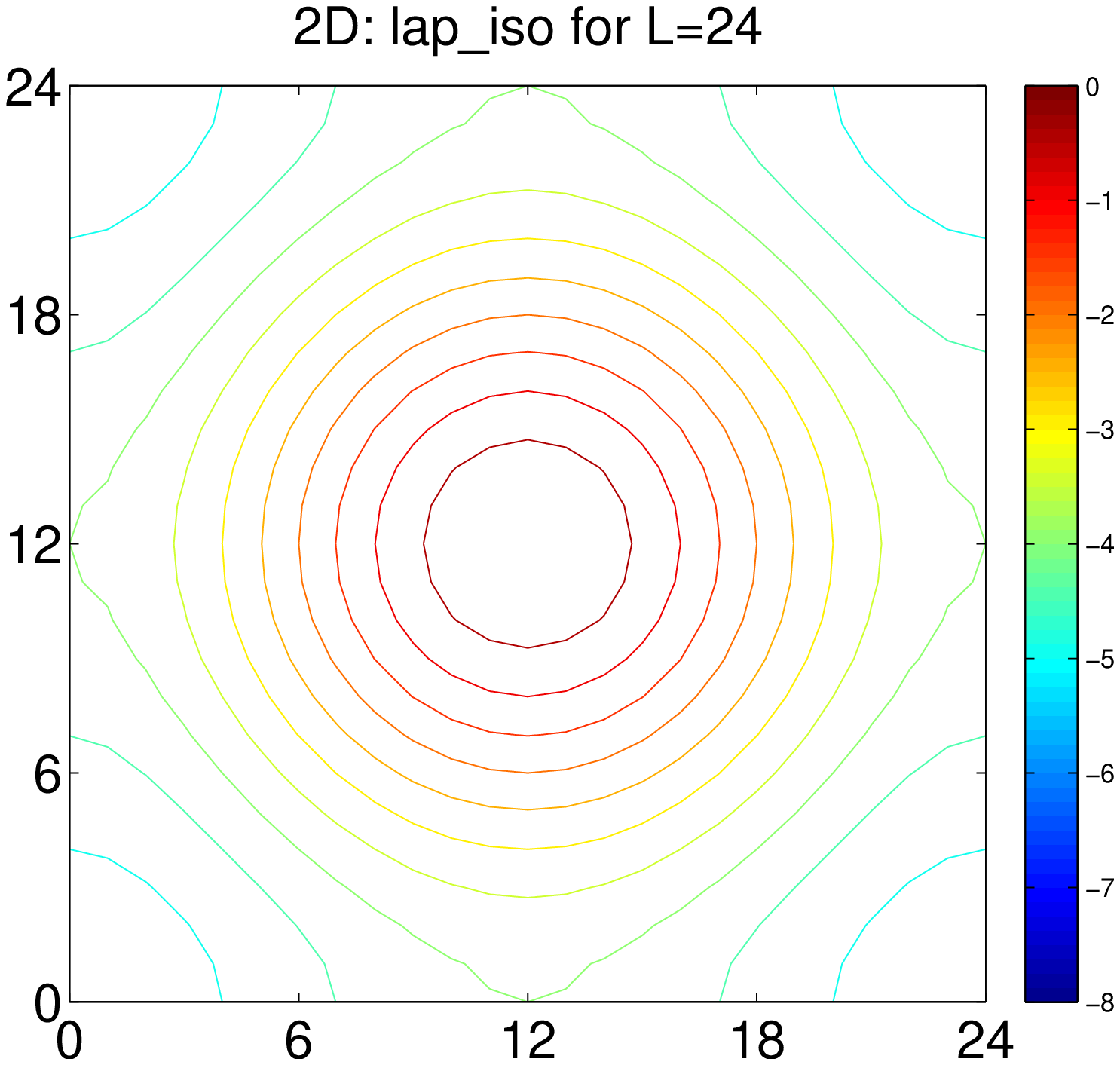,height=5.7cm}
\caption{\label{fig:fourier_lap}\sl
Fourier transformation of the four Laplace stencils considered in 2D.}
\end{figure}

\begin{figure}[!tb]
\centering
\epsfig{file=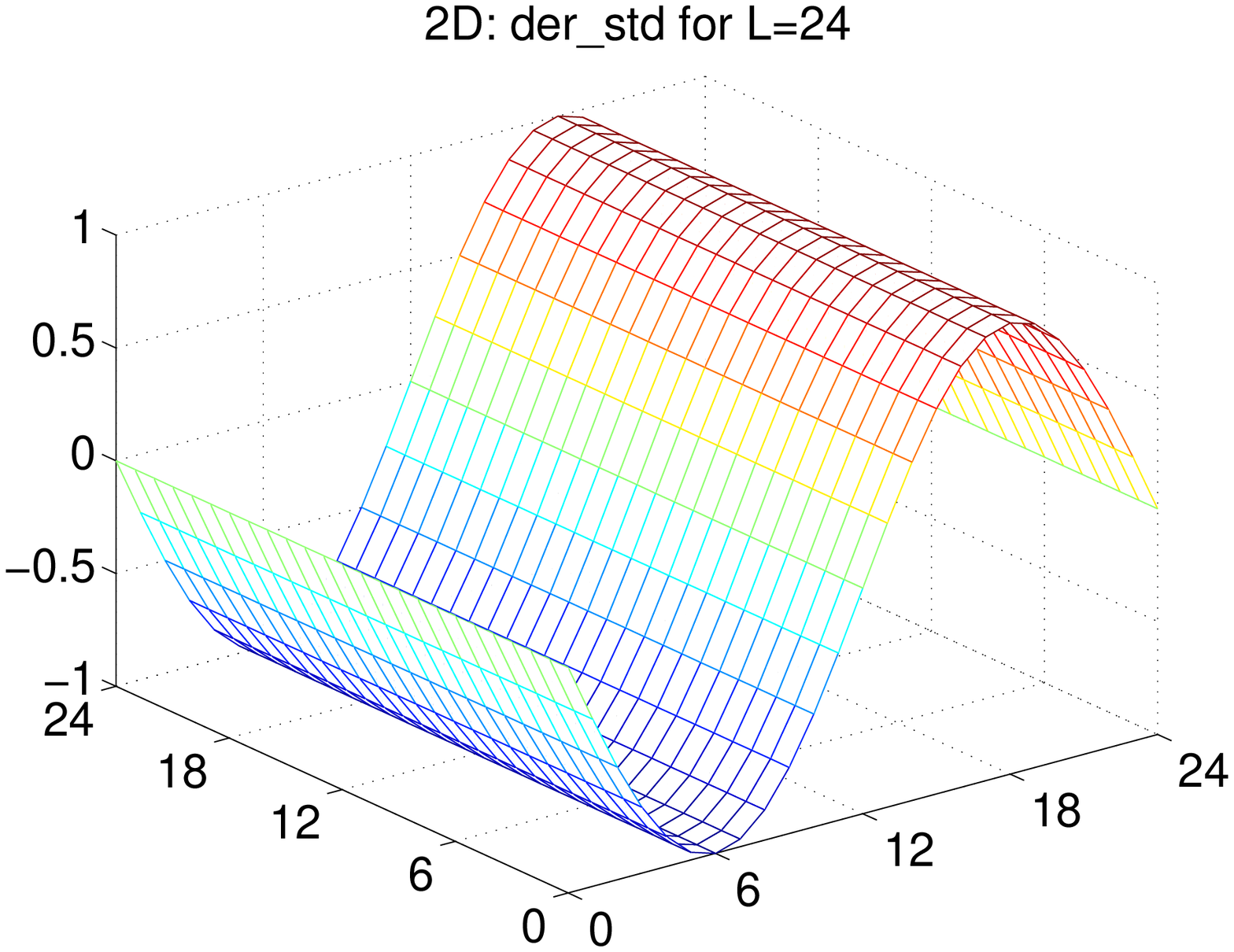,height=5.7cm}
\epsfig{file=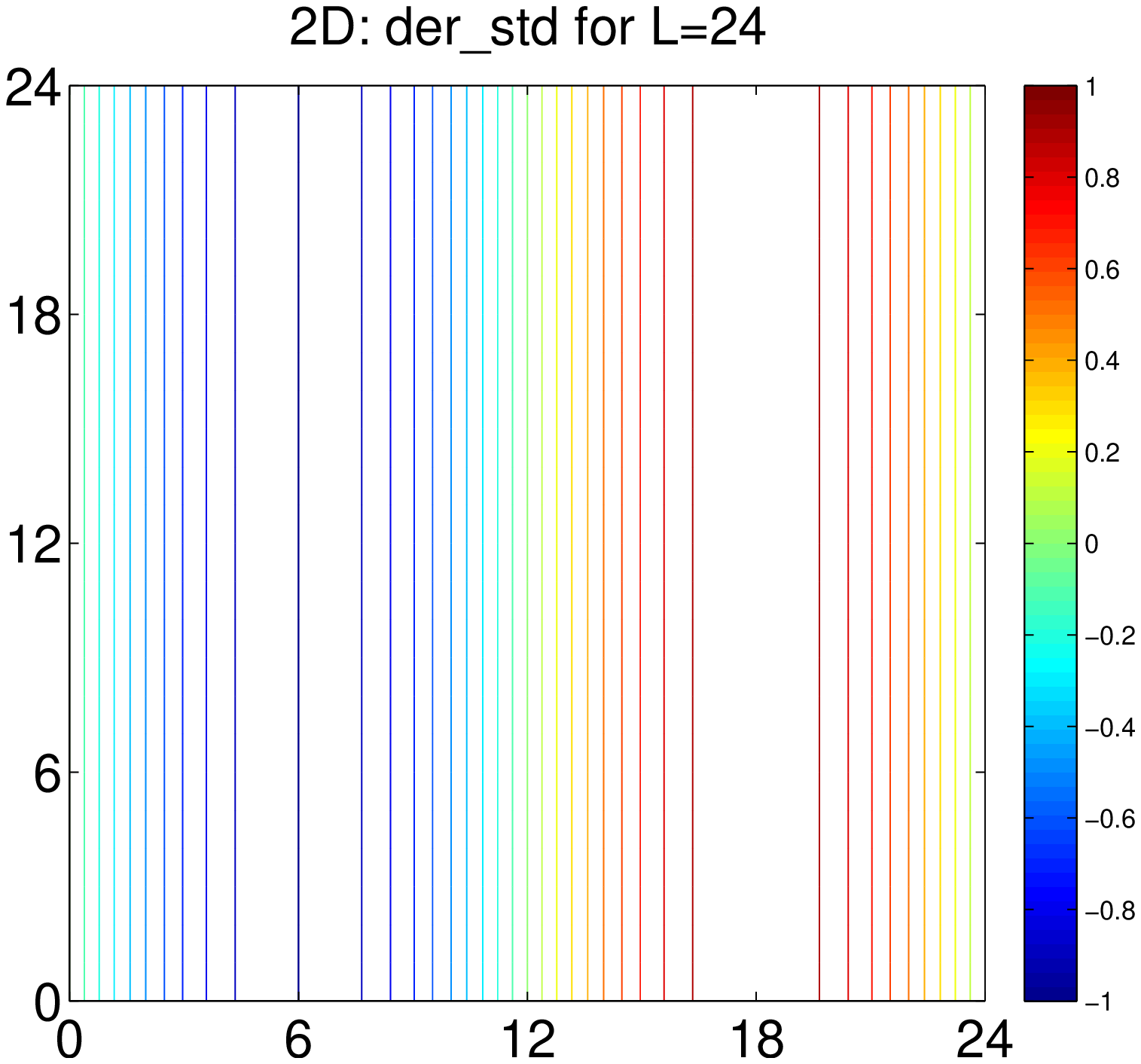,height=5.7cm}\\
\epsfig{file=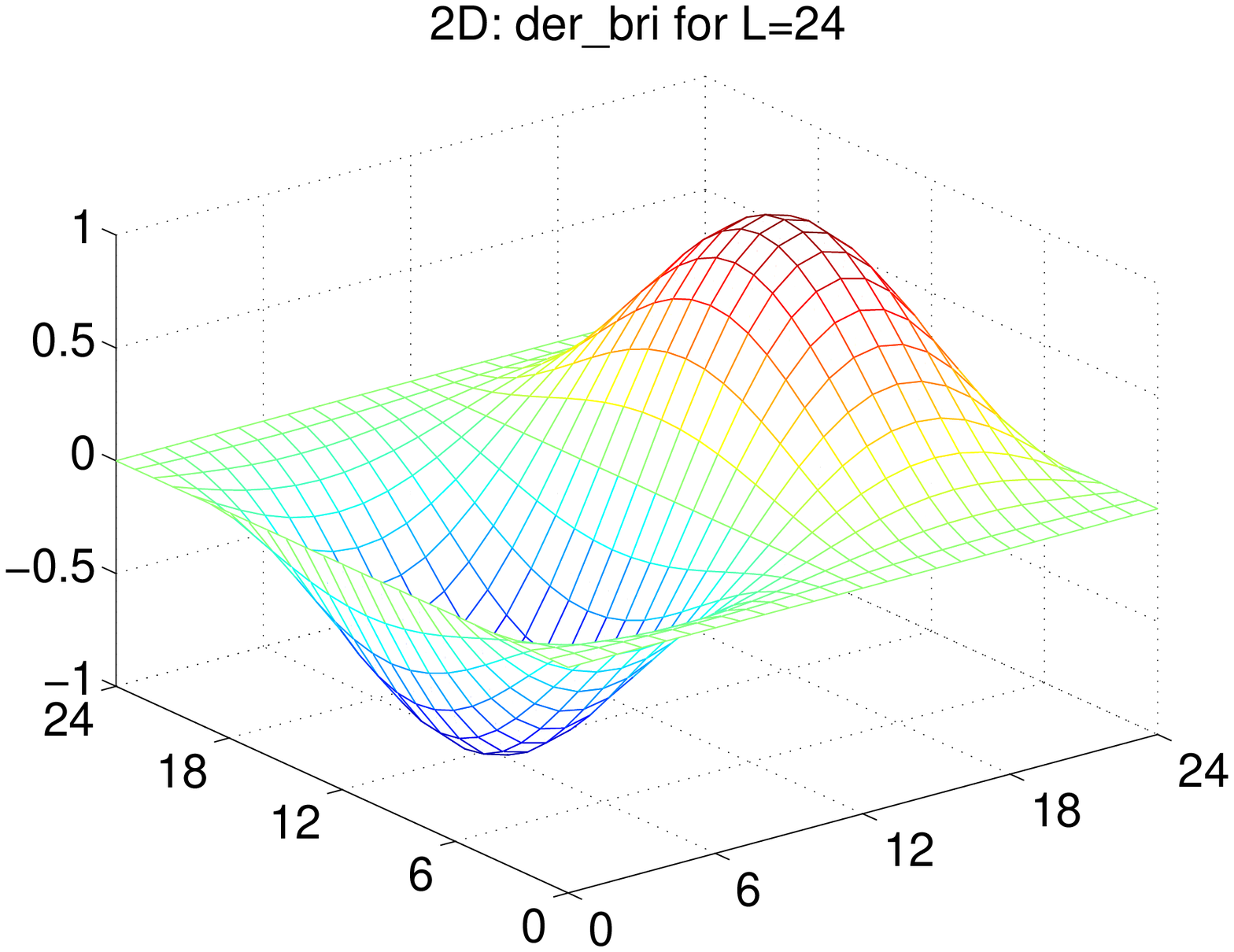,height=5.7cm}
\epsfig{file=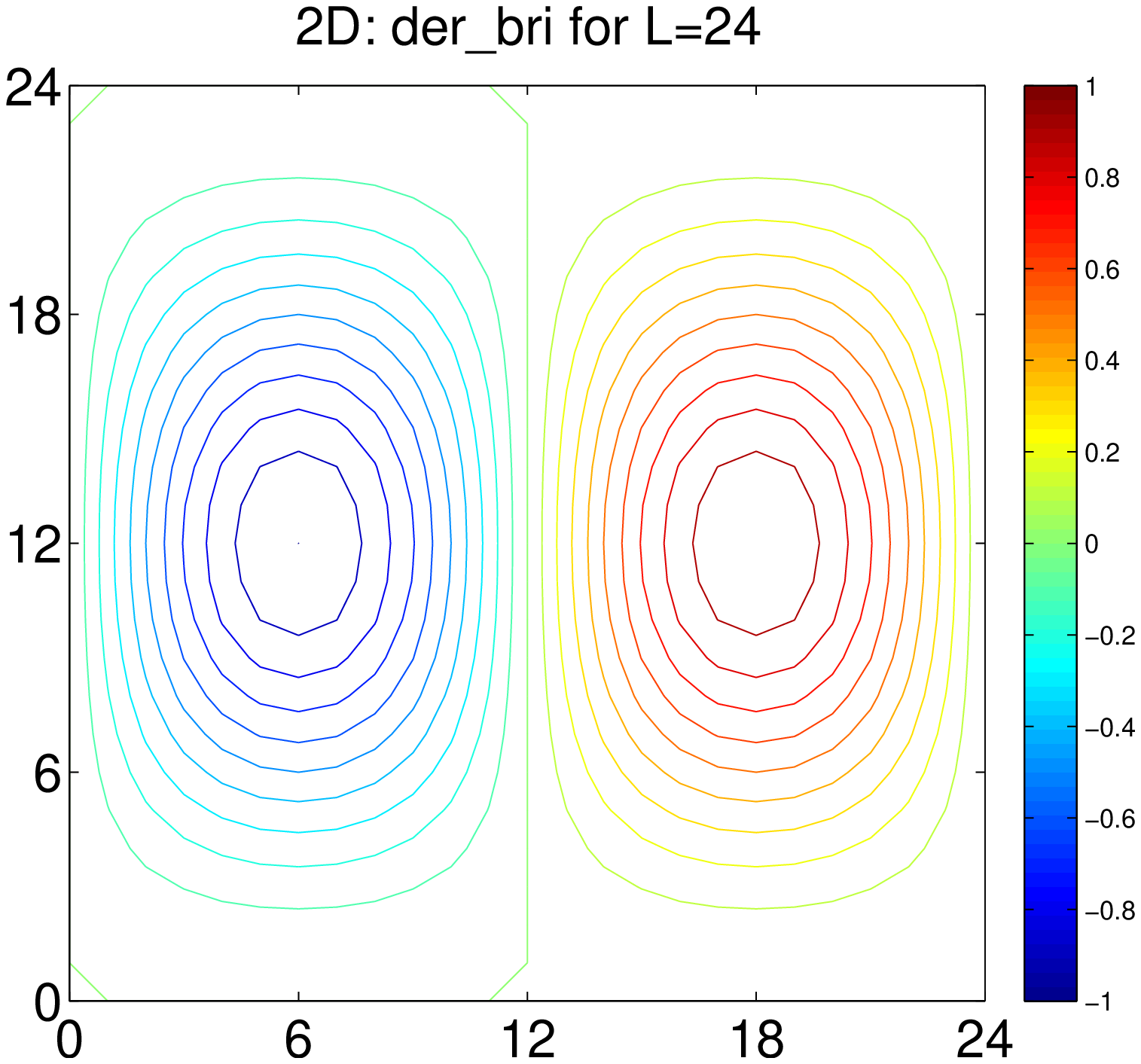,height=5.7cm}\\
\epsfig{file=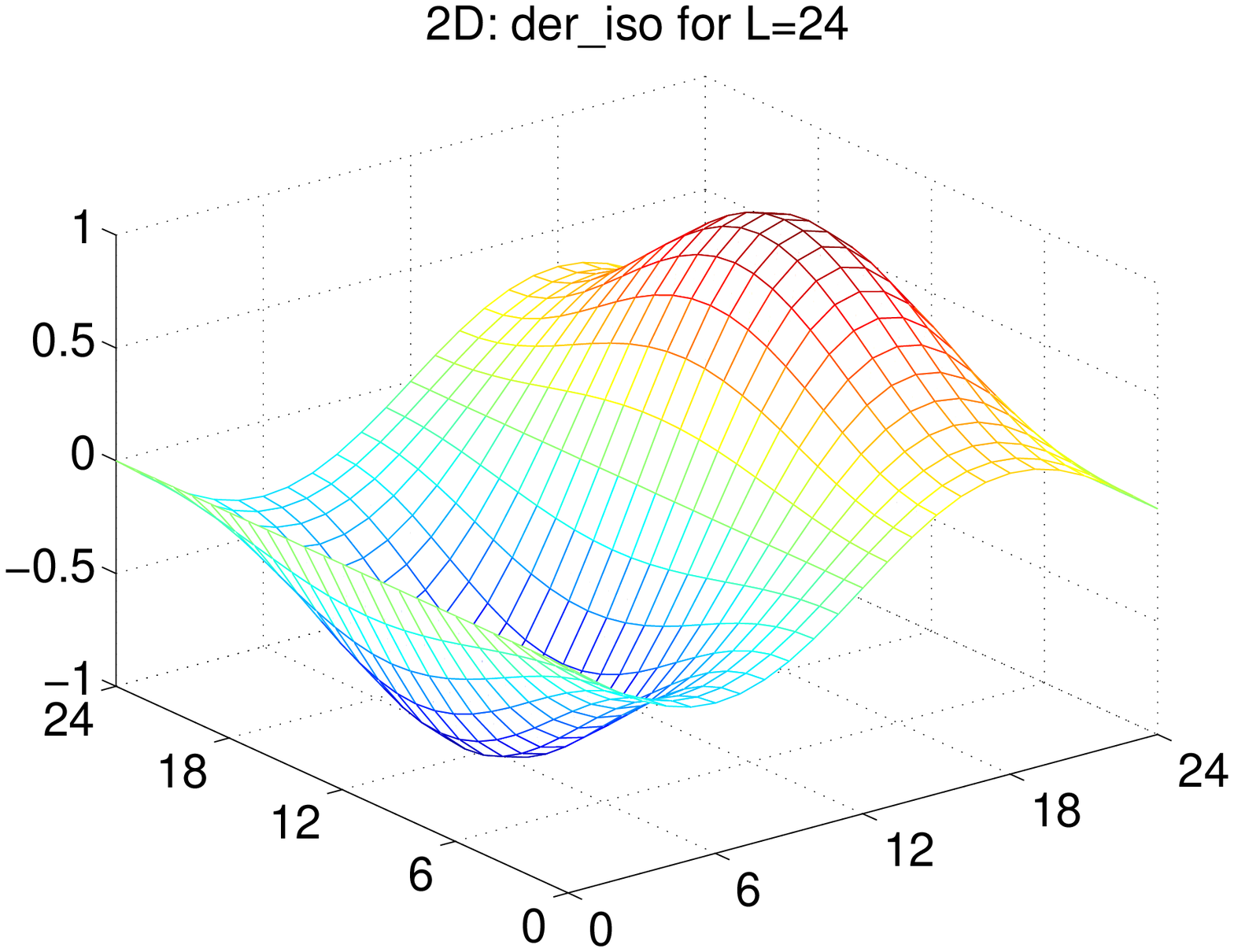,height=5.7cm}
\epsfig{file=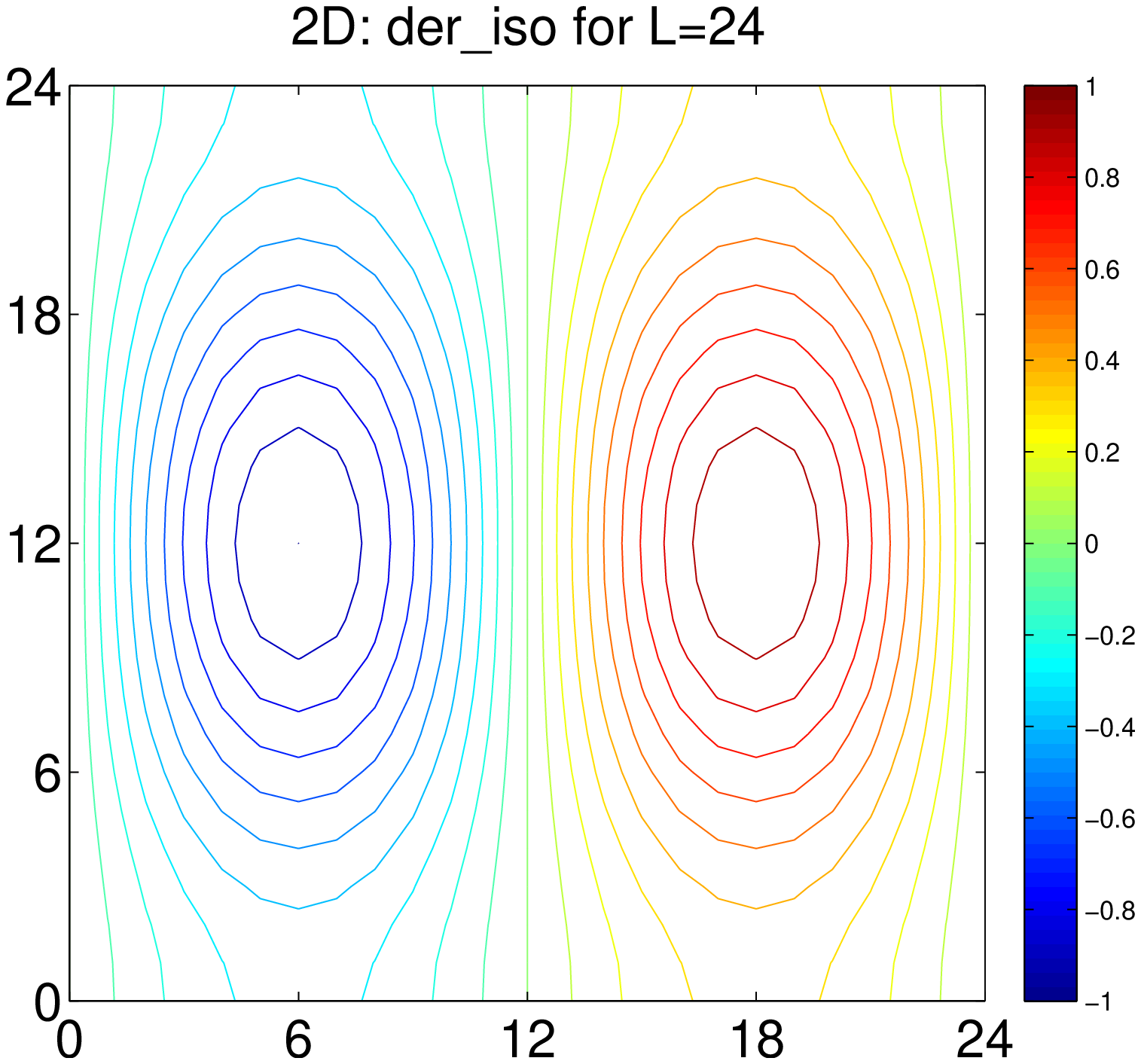,height=5.7cm}
\caption{\label{fig:fourier_der}\sl
Fourier transformation of the three derivative stencils considered in 2D.}
\end{figure}

In Fig.\,\ref{fig:fourier_lap} the momentum space representation of the four
Laplacians is shown as a mesh plot (left) and as a contour plot (right).
We choose a $24^2$ lattice, and arrange the center of the Brillouin zone
($p\!=\!0$) in the center of the frame, i.e.\ the boundaries correspond to
momenta $p\!=\!\pm\pi/a$.
The standard Laplacian that appears in the Wilson operator has a zero in the
center, and decreases quadratically as one moves away from this point.
As we follow the boundary of the Brillouin zone, it oscillates between $-4$
and $-8$.
The tilted Laplacian is rather different, since it shows a second zero at
$p\!=\!(\pi/a,\pi/a)$, a quarter of which is seen in each corner.
The Brillouin Laplacian has just one zero and achieves complete flatness at the
boundary of the Brillouin zone.
Finally, the isotropic Laplacian achieves best isotropy near the center of the
Brillouin zone, meaning that one can move relatively far out from the center
until its equipotential lines become noticeably non-circular.

In Fig.\,\ref{fig:fourier_der} the momentum space representations of the three
derivatives specified in App.\,A are shown as a mesh plot (left) and as a
contour plot (right).
The standard derivative that appears in the Wilson operator is a pure
$\sin(p_1/a)$, without any structure in the transverse direction.
The Brillouin derivative modulates the transverse direction to the point that a
strict zero is realized on the entire boundary.
The isotropic operator modulates the transverse direction in a less pronounced
manner.
For the reasons behind the name of this latter operator, which may sound a bit
paradoxical, see App.\,D.


\subsection{Eigenvalue spectra in 2D}

Given the four choices of $\lap$ discussed above and the three choices of
$\nab_\mu$, we can construct 12 Dirac operators and study their eigenvalue
spectra.
As the gauge group is irrelevant in this step, we prepare a thermalized
background in the $U(1)$ gauge theory with $L/a\!=\!24$ at $\be\!=\!3.3$.

\begin{figure}[!p]
\centering
\epsfig{file=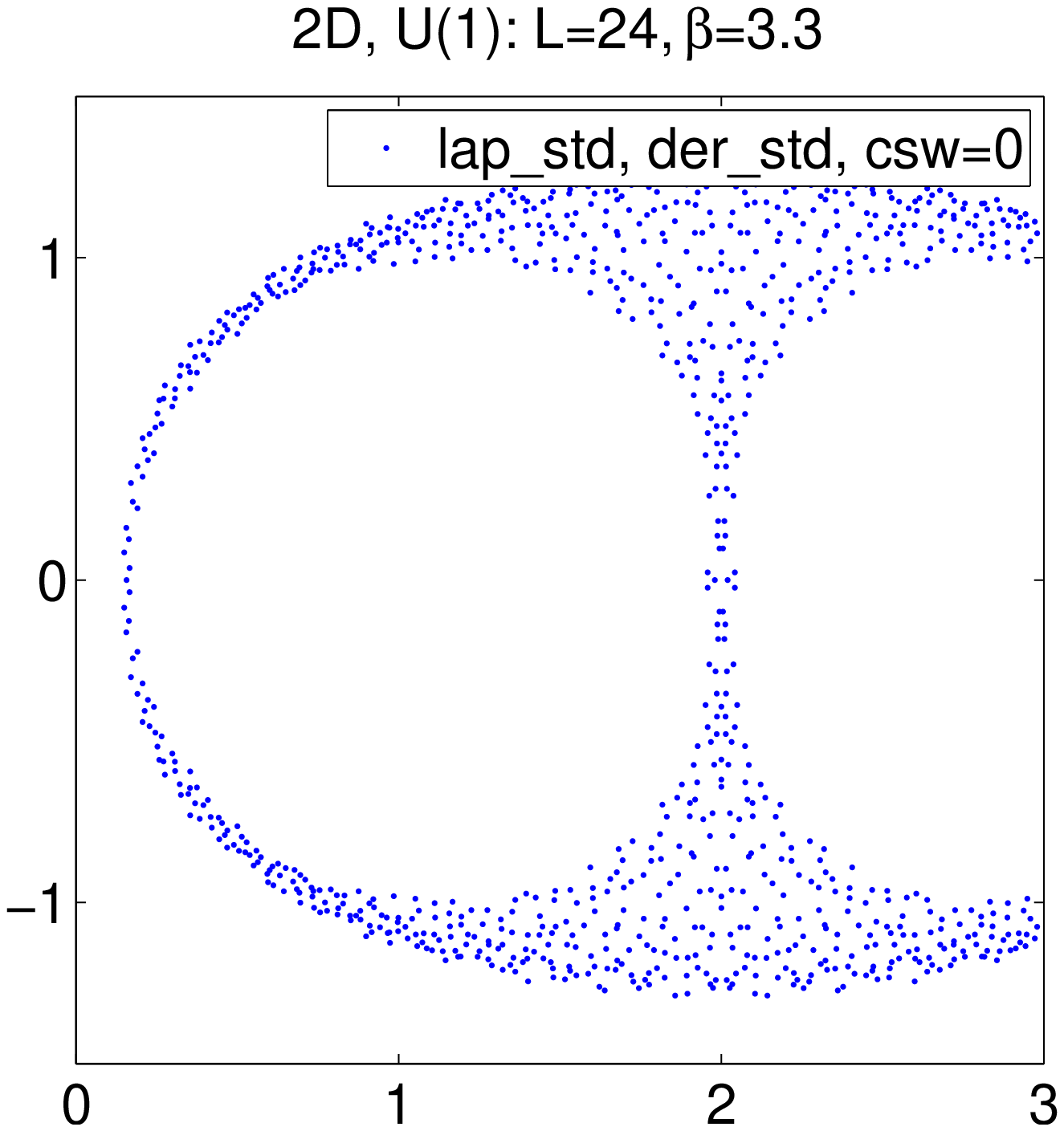,height=5.7cm}
\epsfig{file=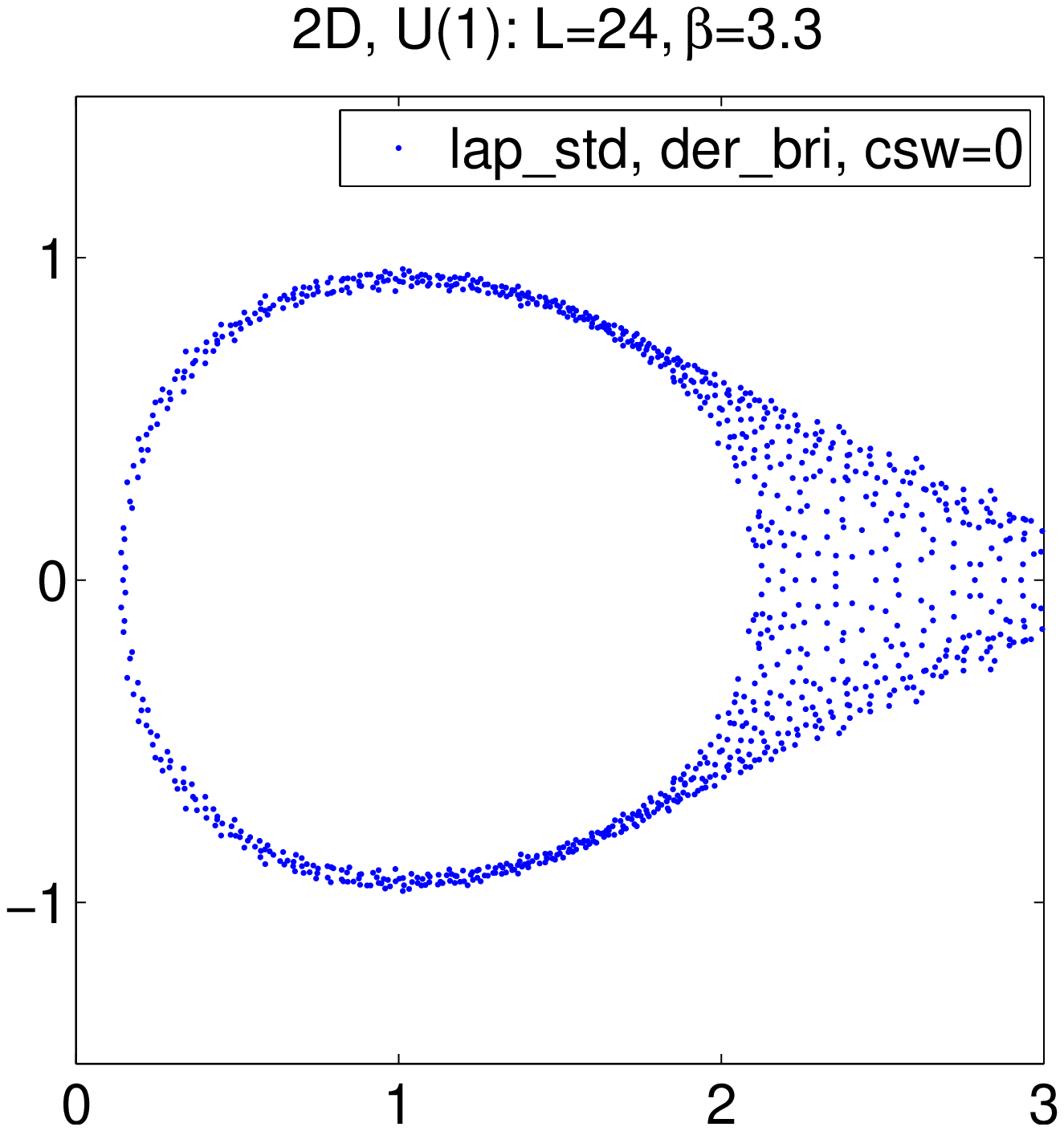,height=5.7cm}
\epsfig{file=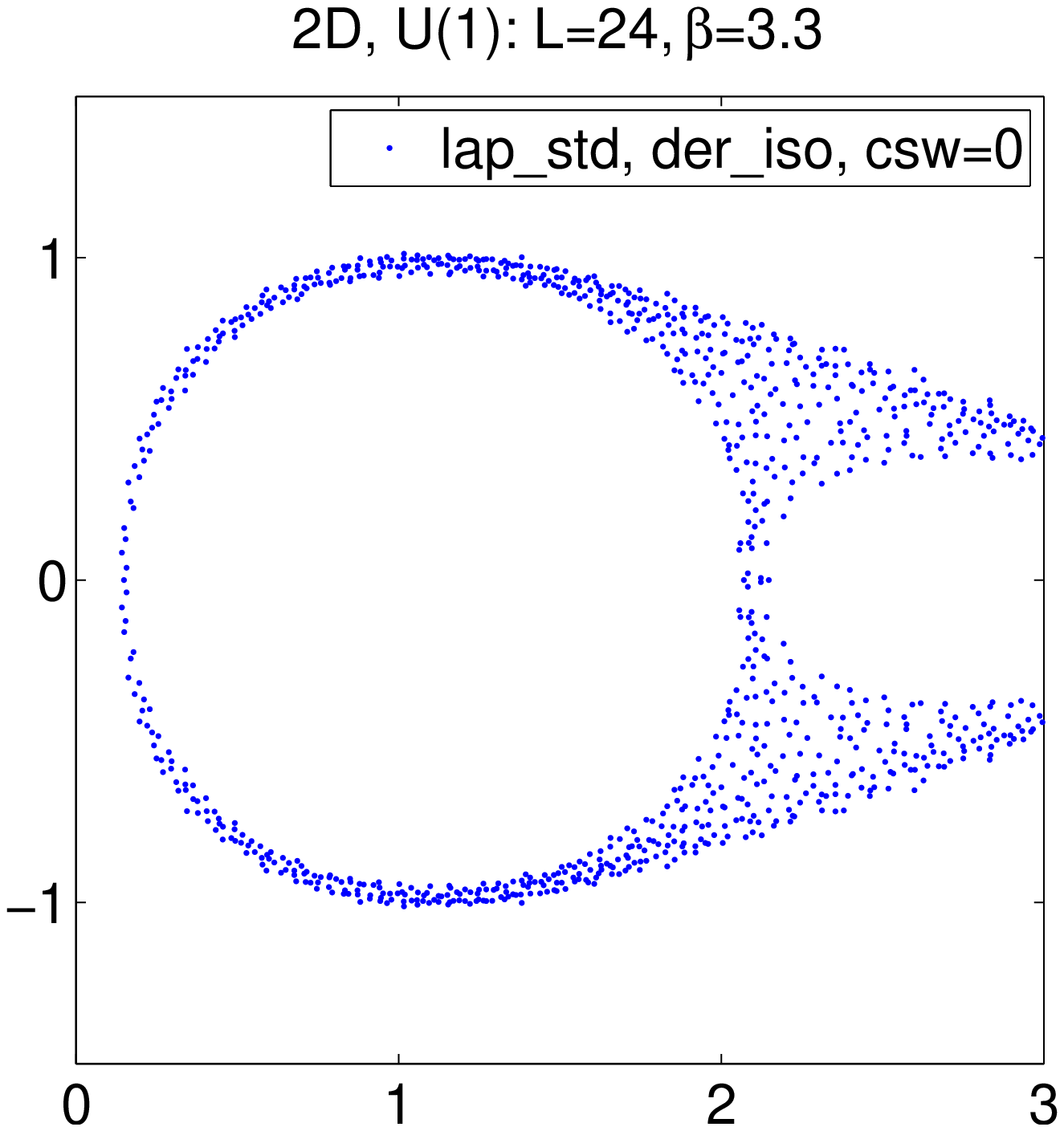,height=5.7cm}\\
\epsfig{file=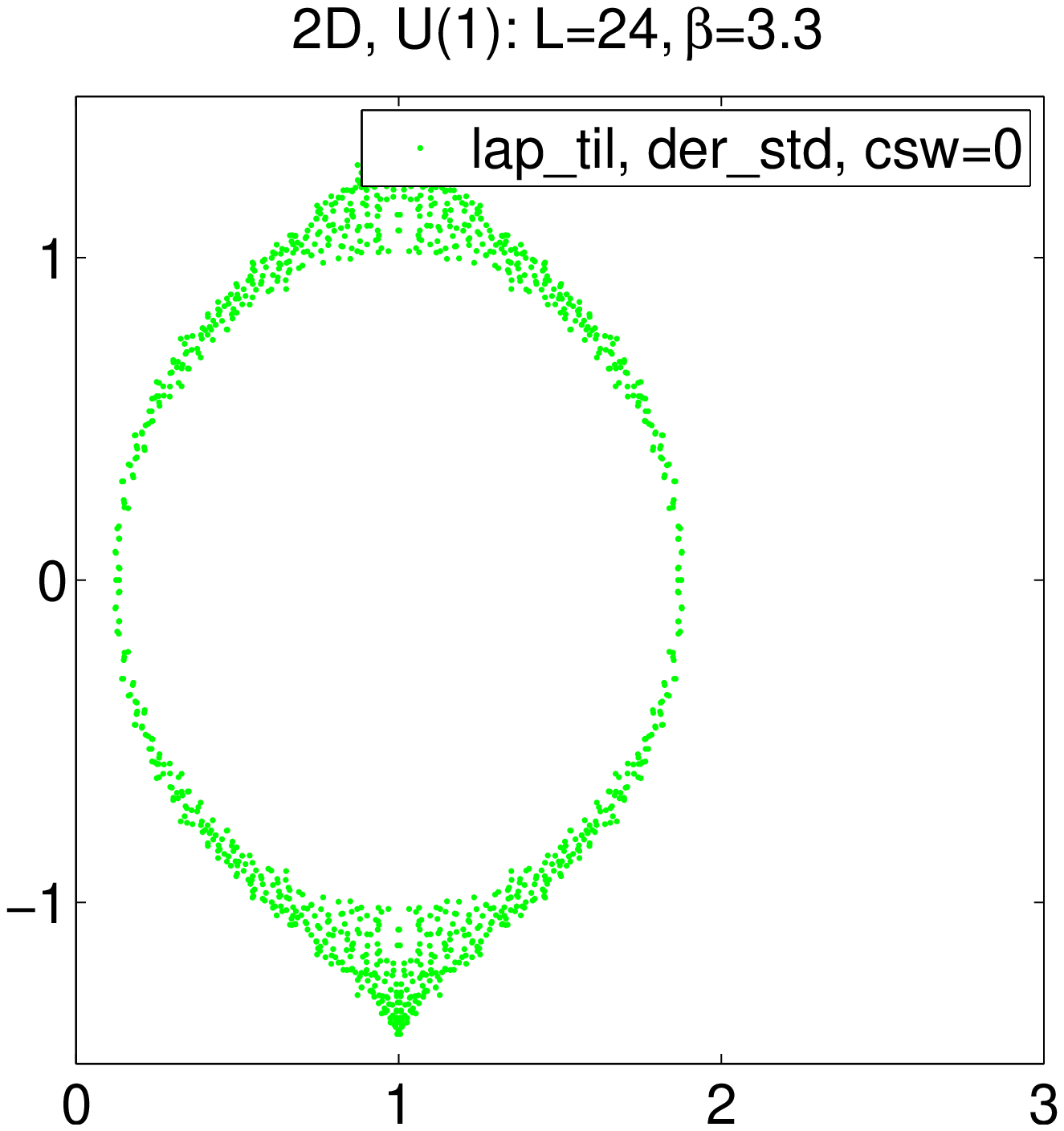,height=5.7cm}
\epsfig{file=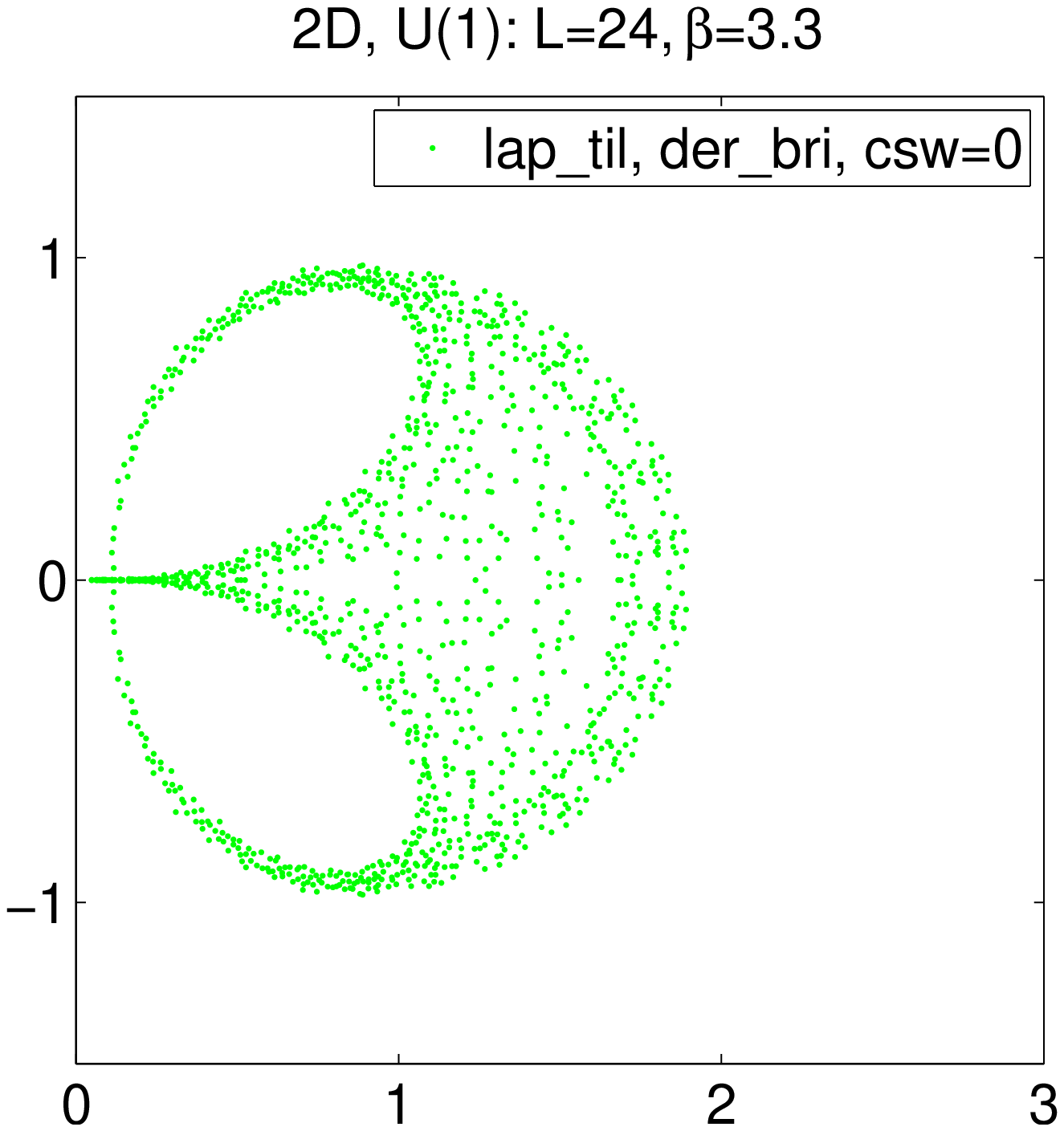,height=5.7cm}
\epsfig{file=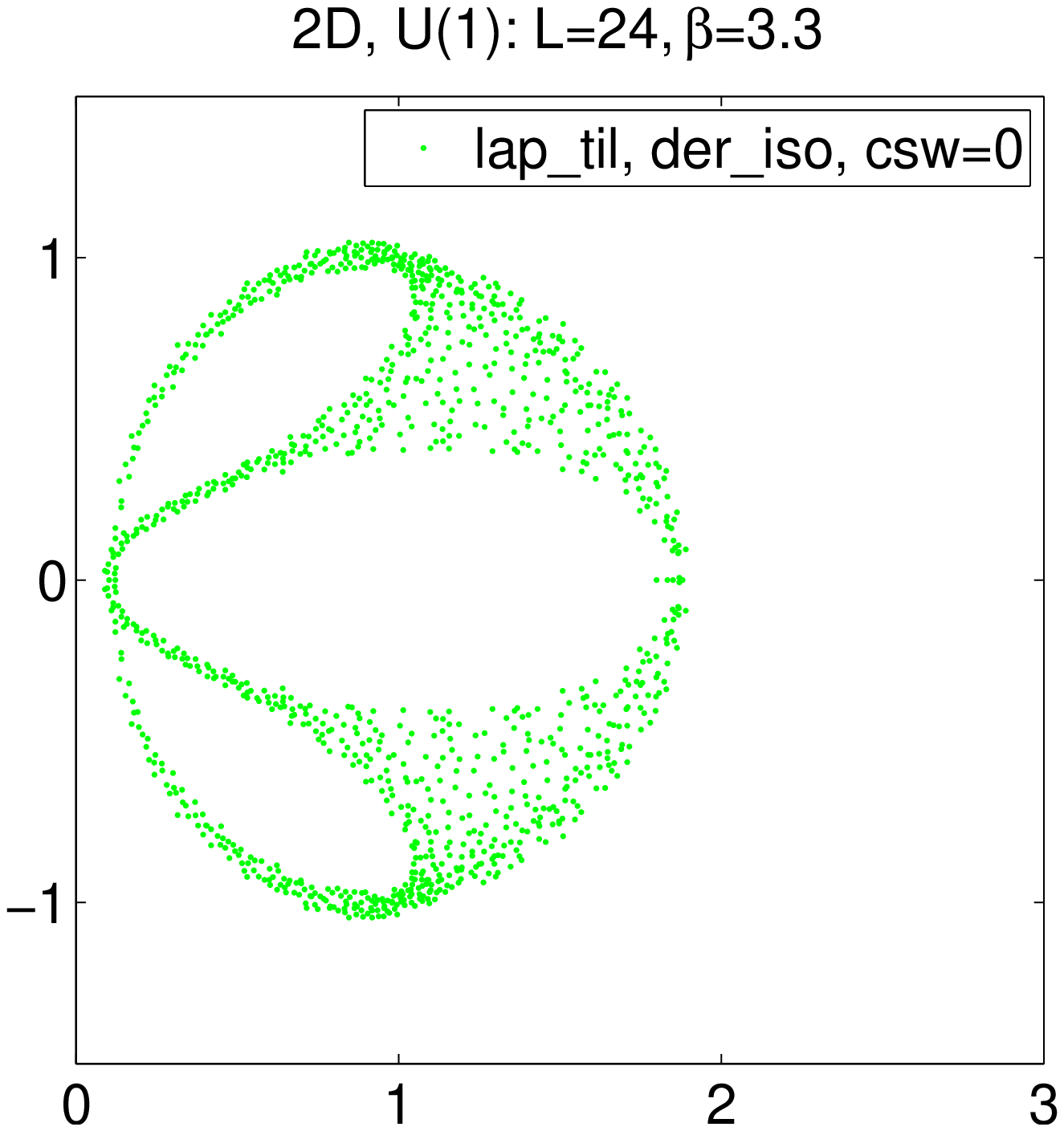,height=5.7cm}\\
\epsfig{file=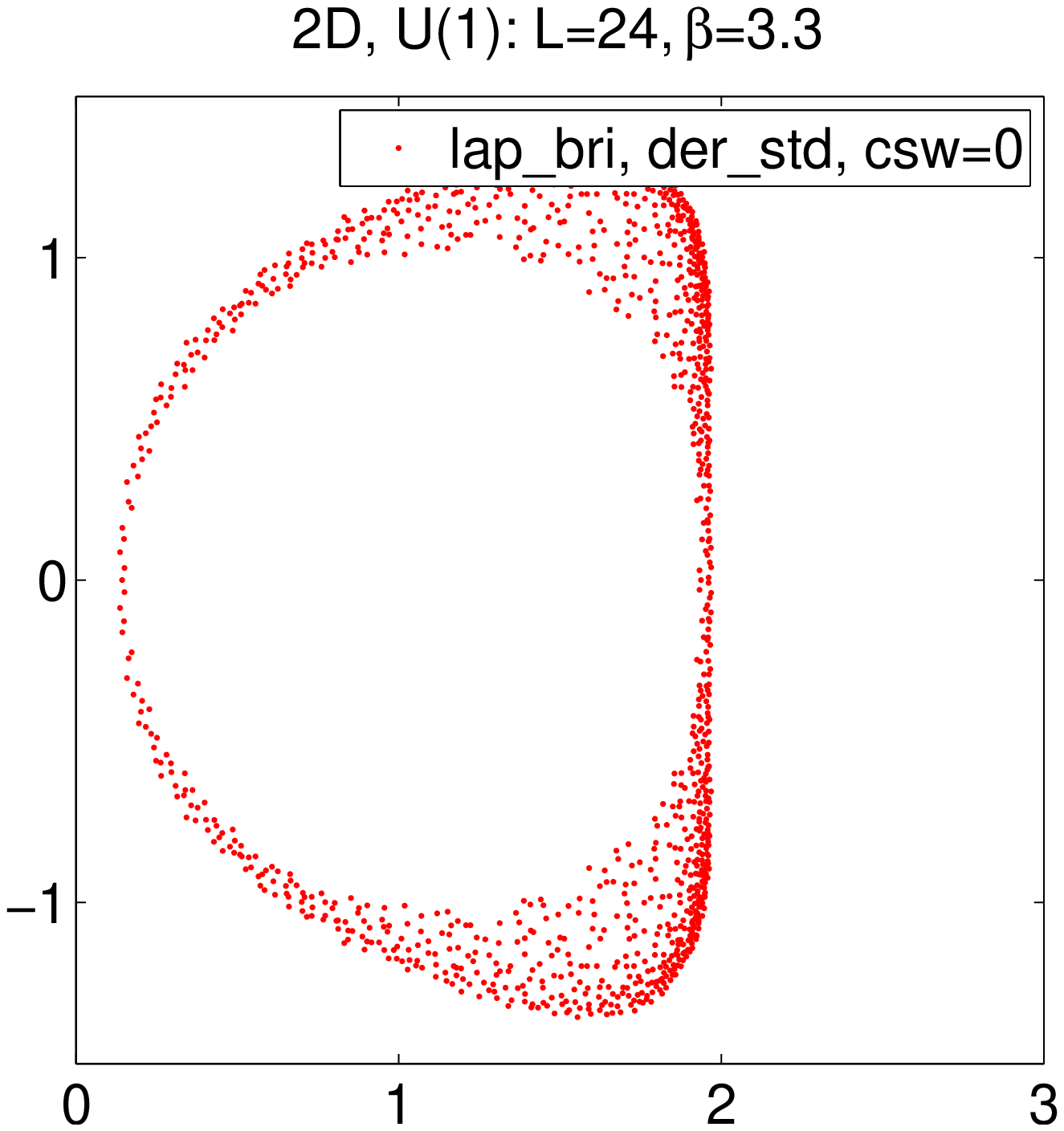,height=5.7cm}
\epsfig{file=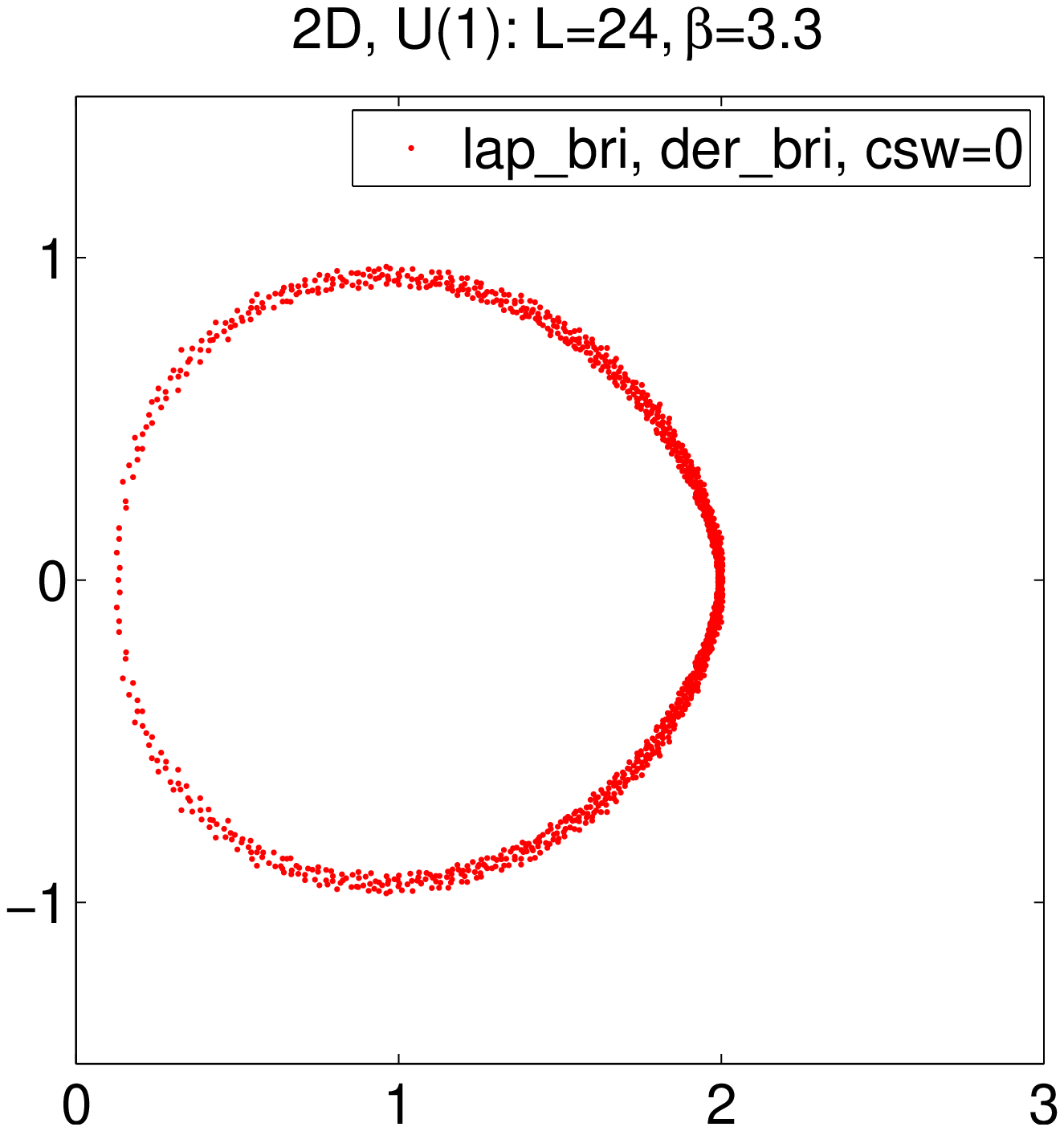,height=5.7cm}
\epsfig{file=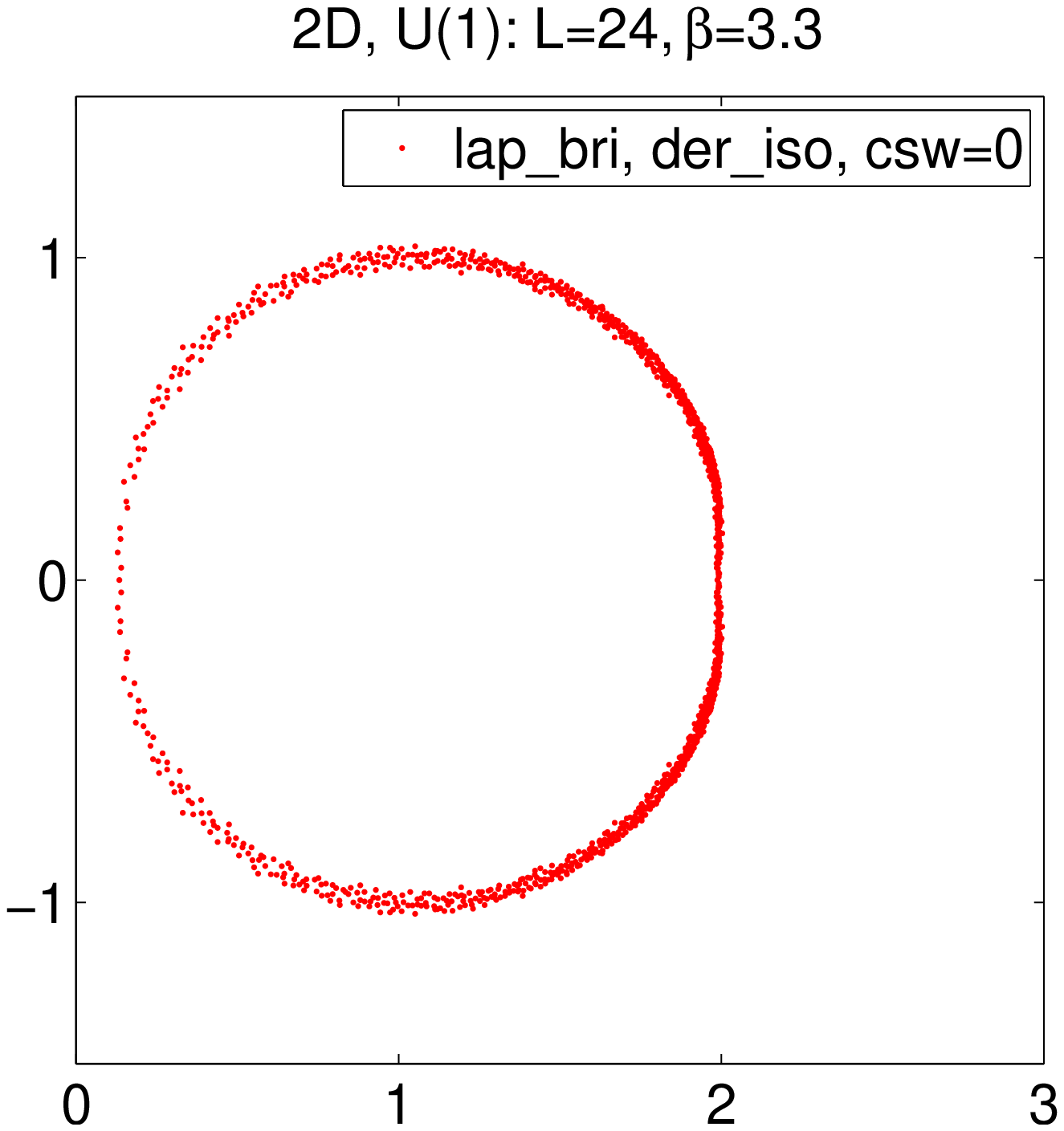,height=5.7cm}\\
\epsfig{file=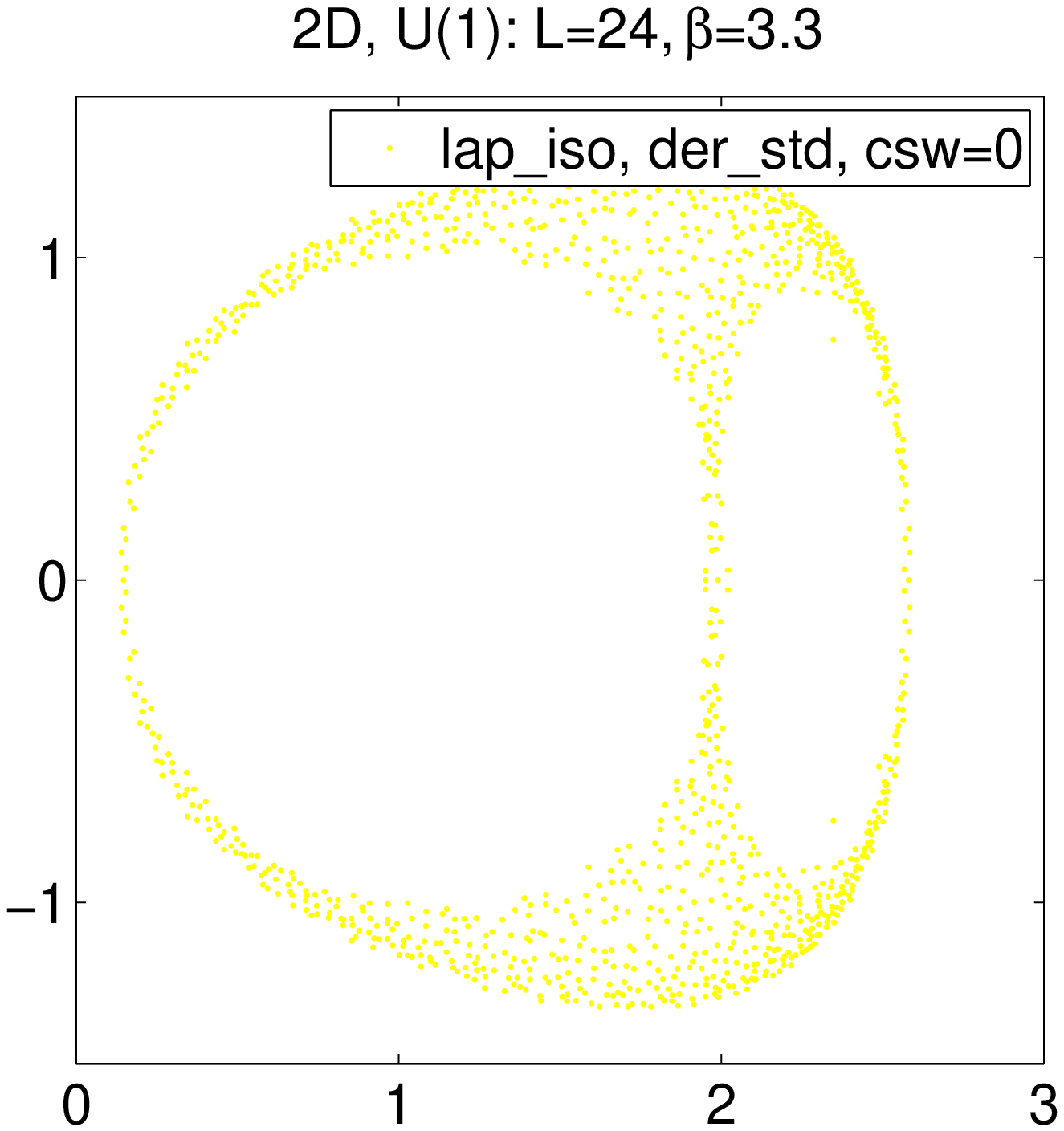,height=5.7cm}
\epsfig{file=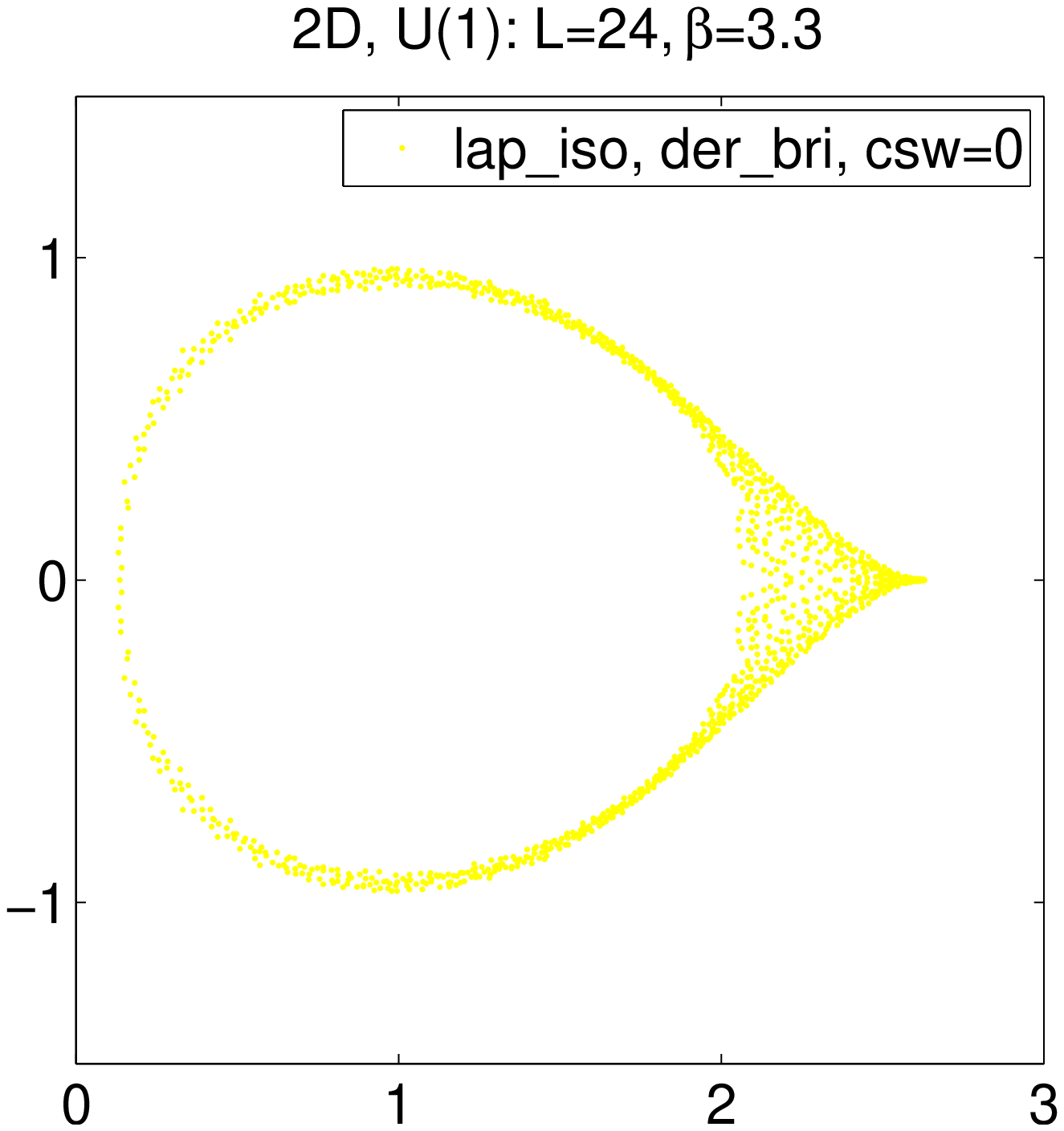,height=5.7cm}
\epsfig{file=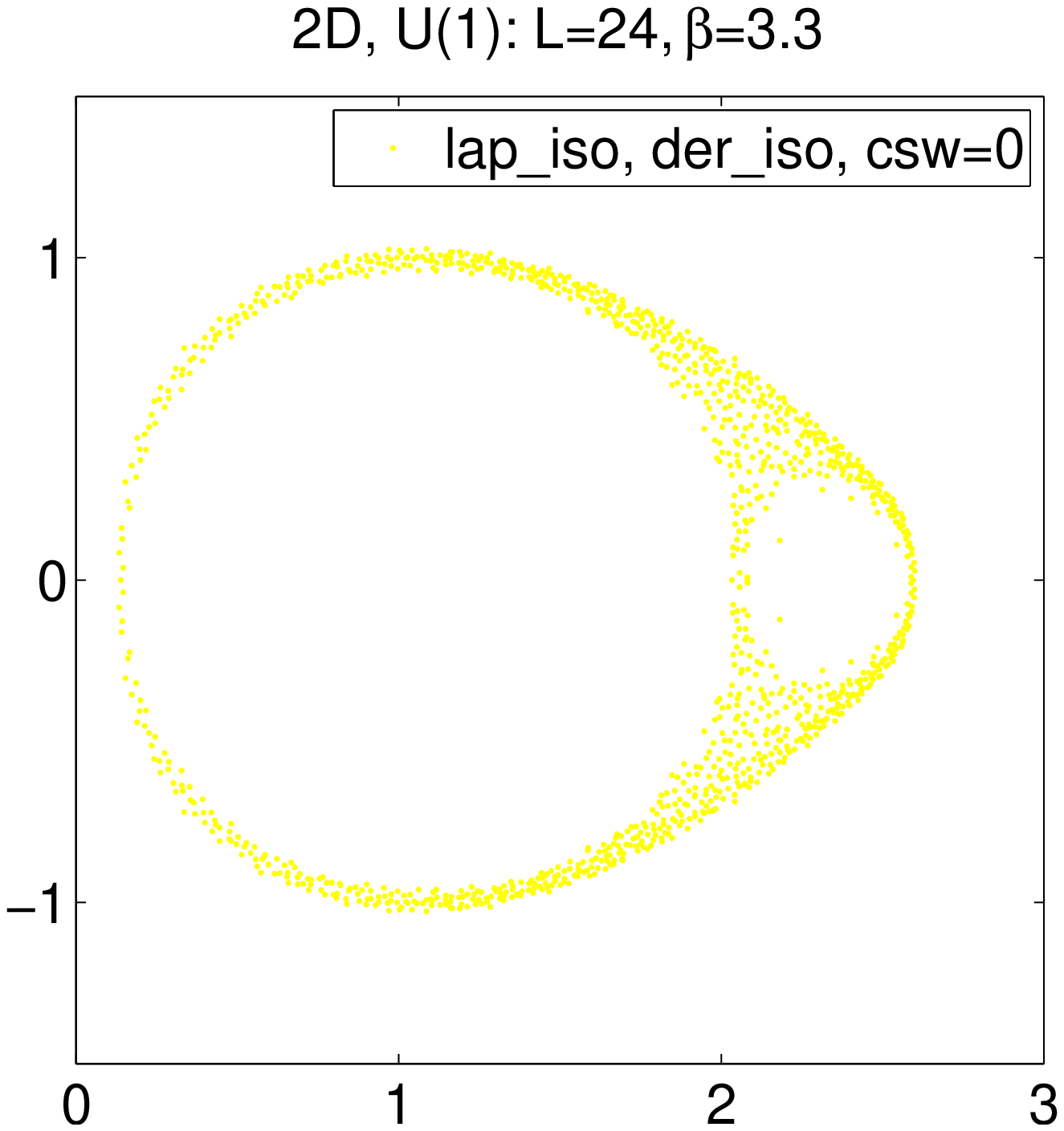,height=5.7cm}
\caption{\label{fig:spec_2D_csw0}\sl
Eigenvalue spectra of all operators considered in 2D with $c_\mr{SW}\!=\!0$.}
\end{figure}

\begin{figure}[!p]
\centering
\epsfig{file=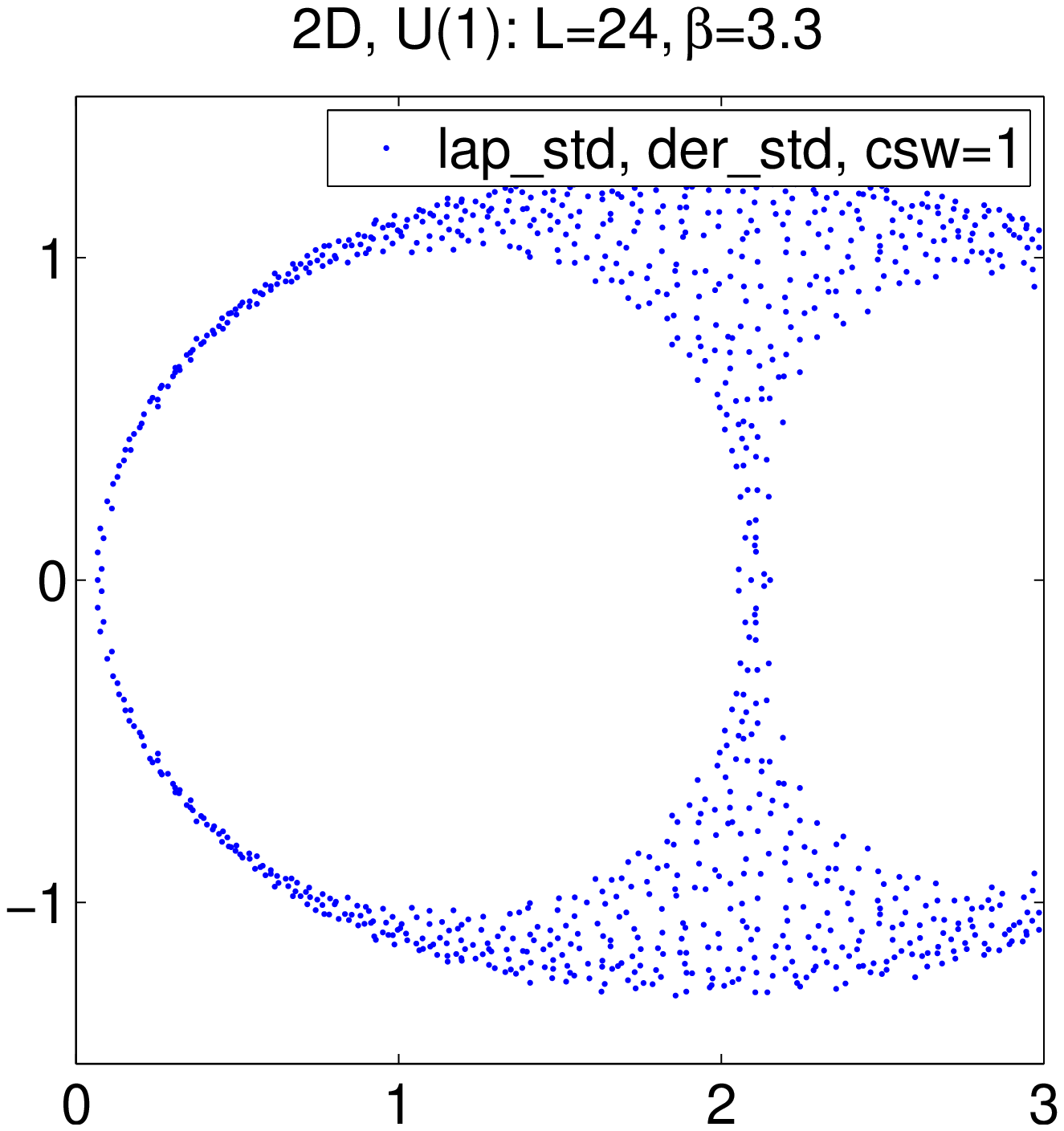,height=5.7cm}
\epsfig{file=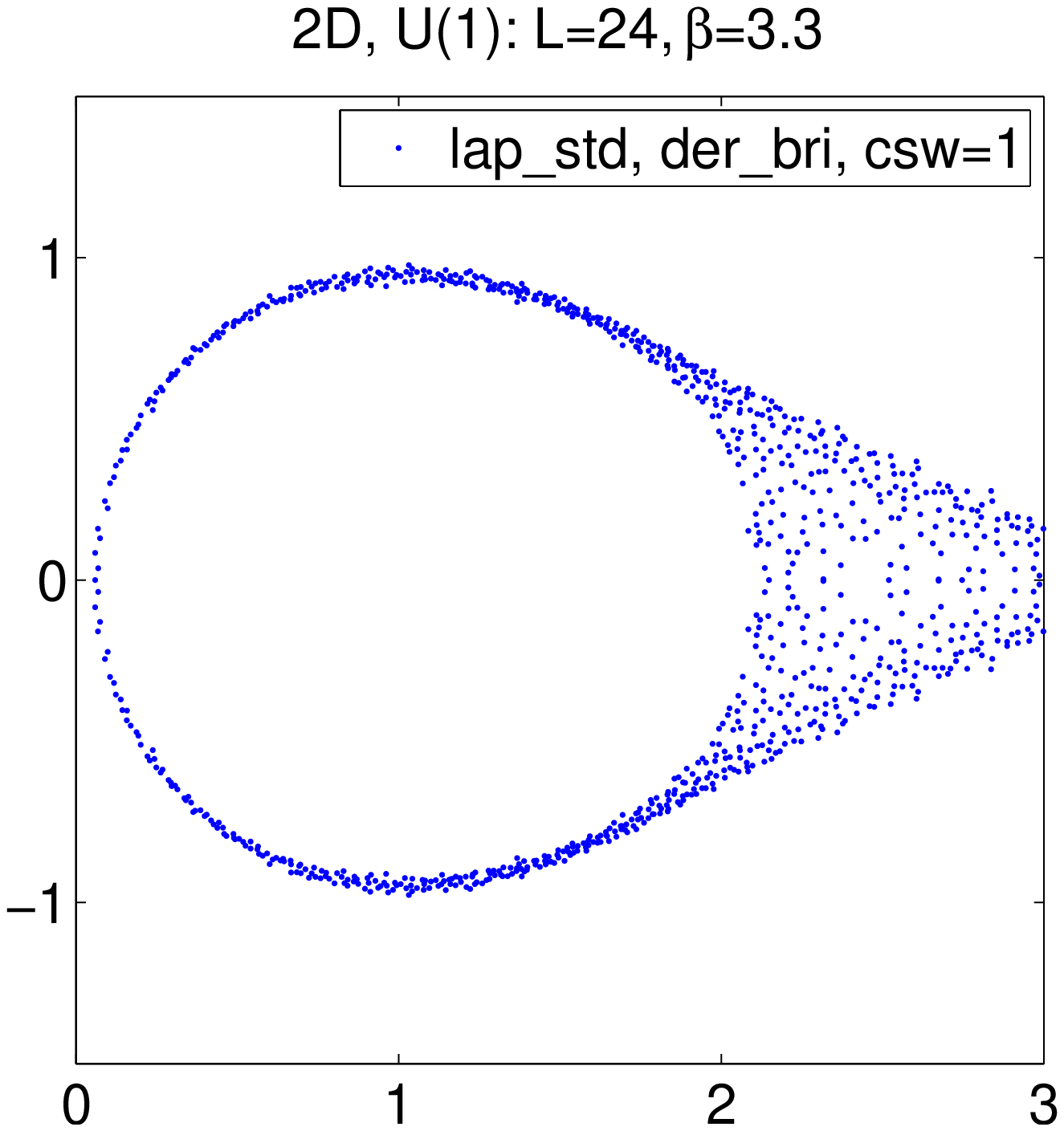,height=5.7cm}
\epsfig{file=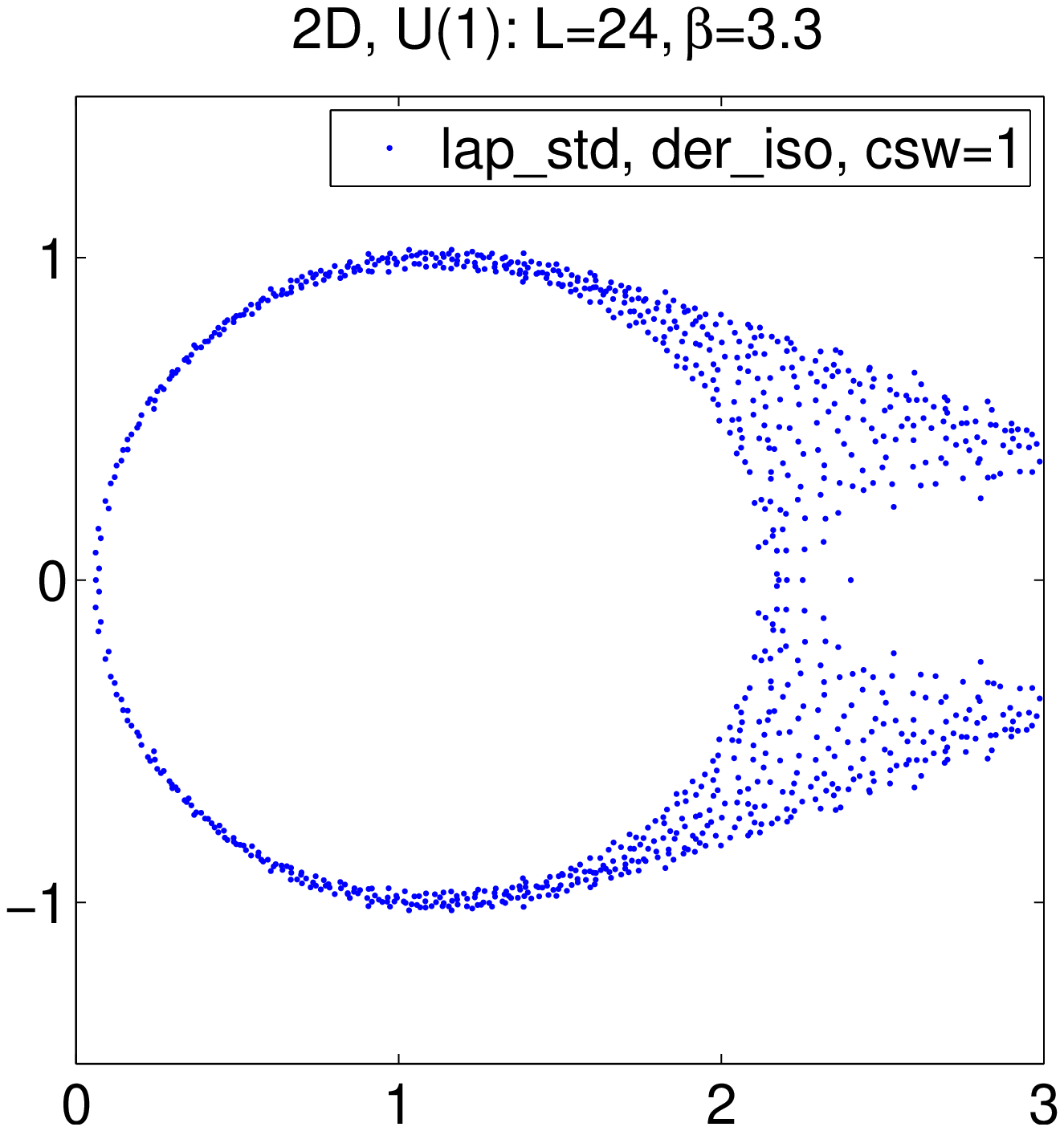,height=5.7cm}\\
\epsfig{file=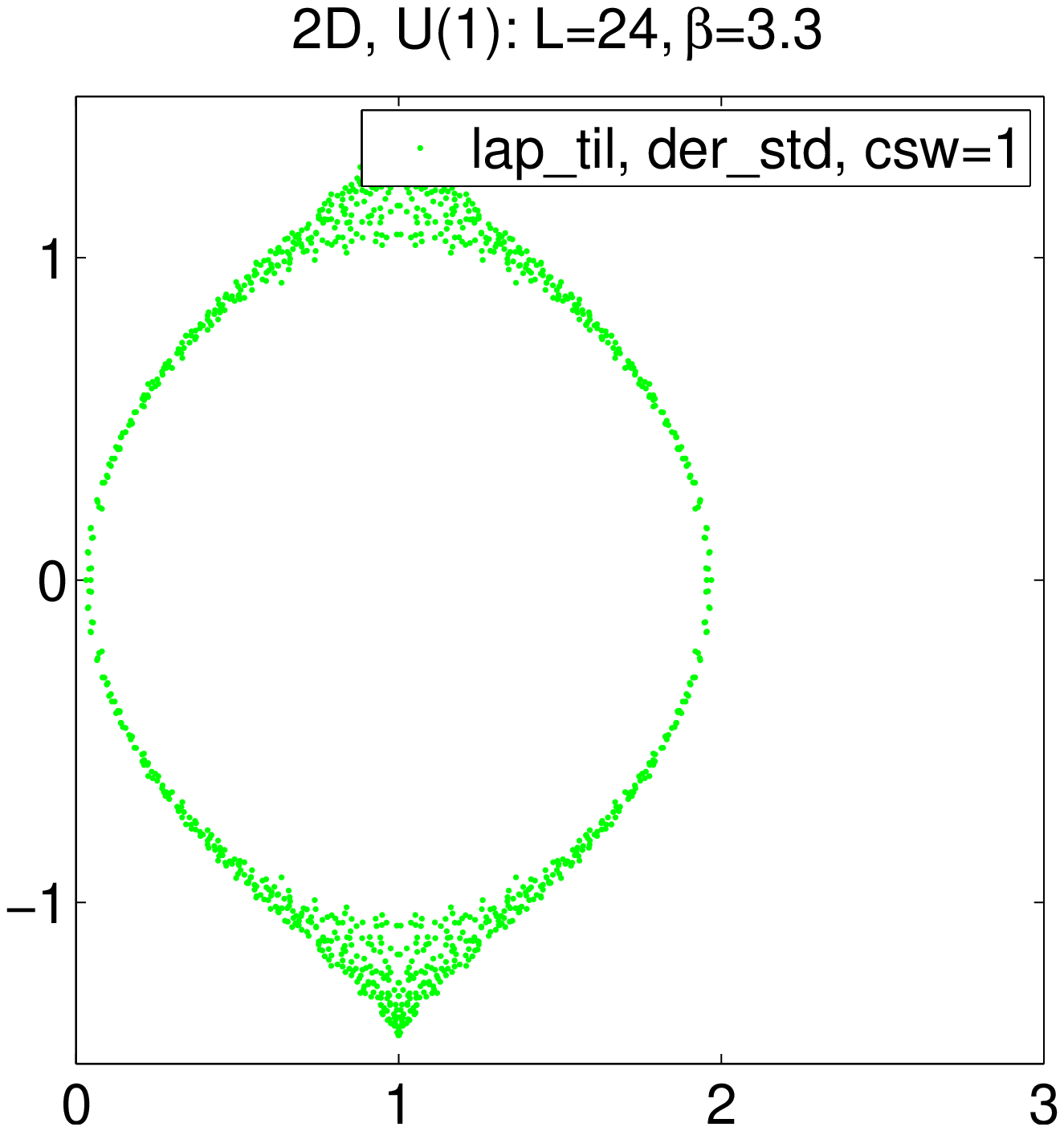,height=5.7cm}
\epsfig{file=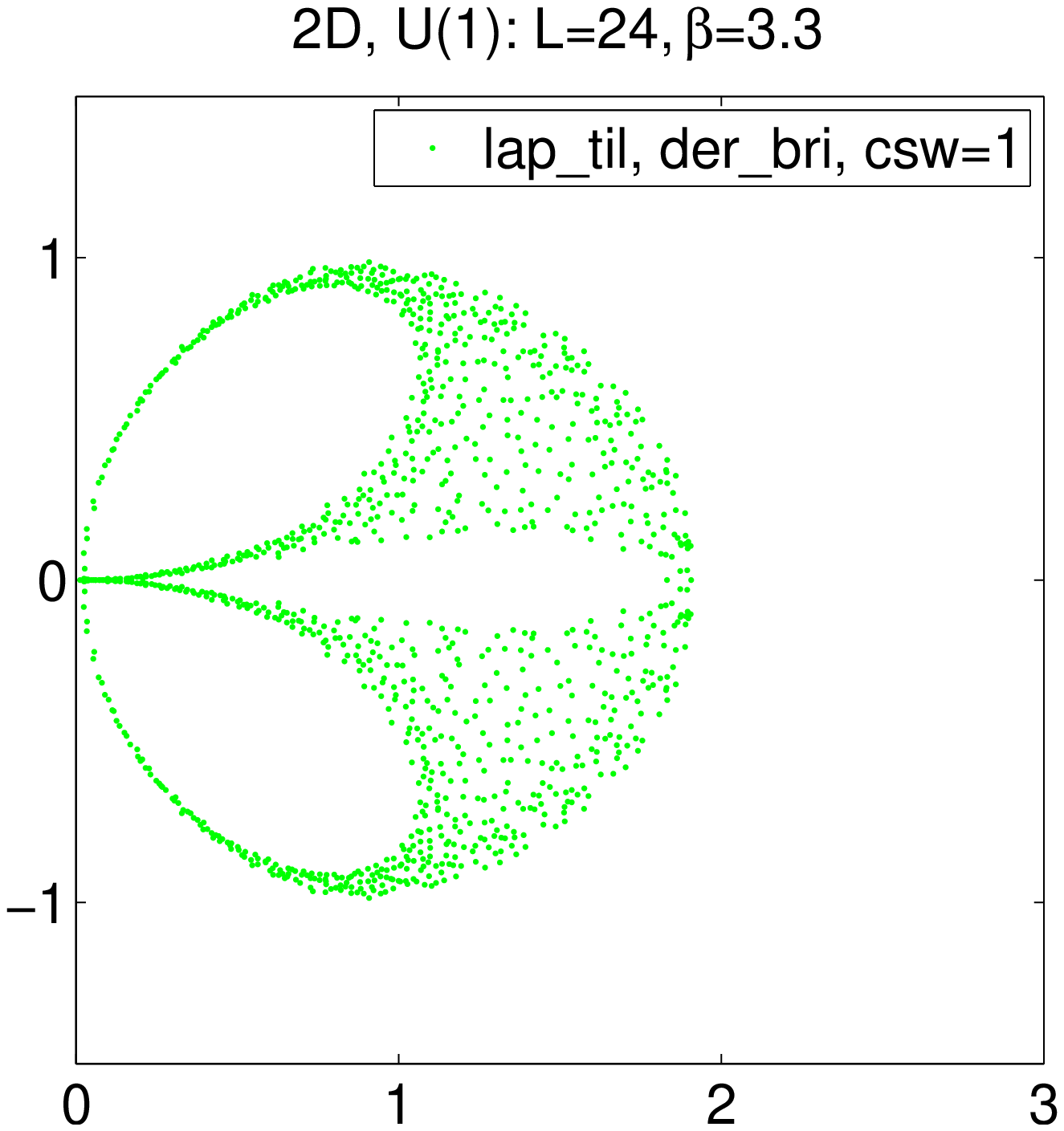,height=5.7cm}
\epsfig{file=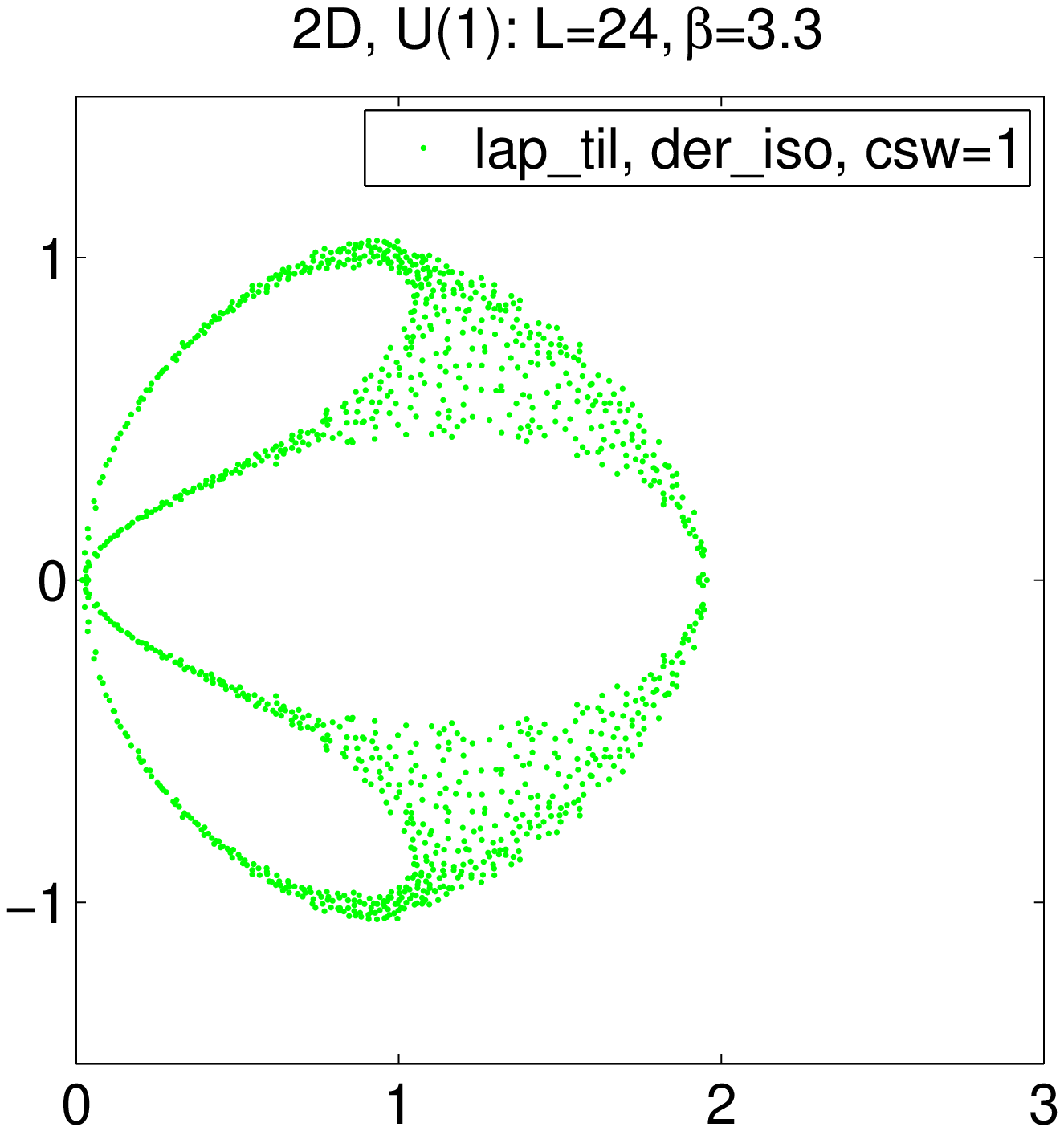,height=5.7cm}\\
\epsfig{file=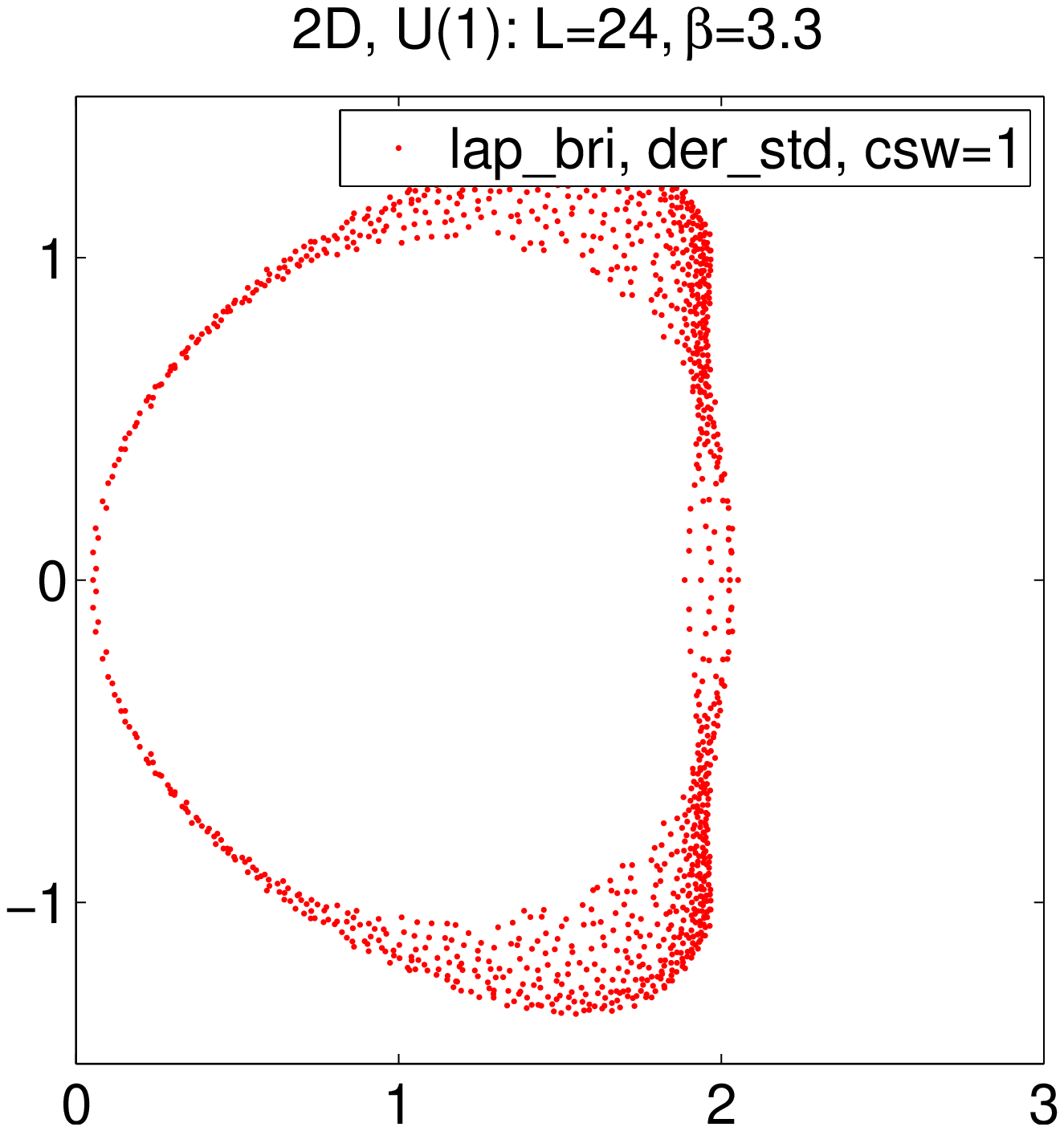,height=5.7cm}
\epsfig{file=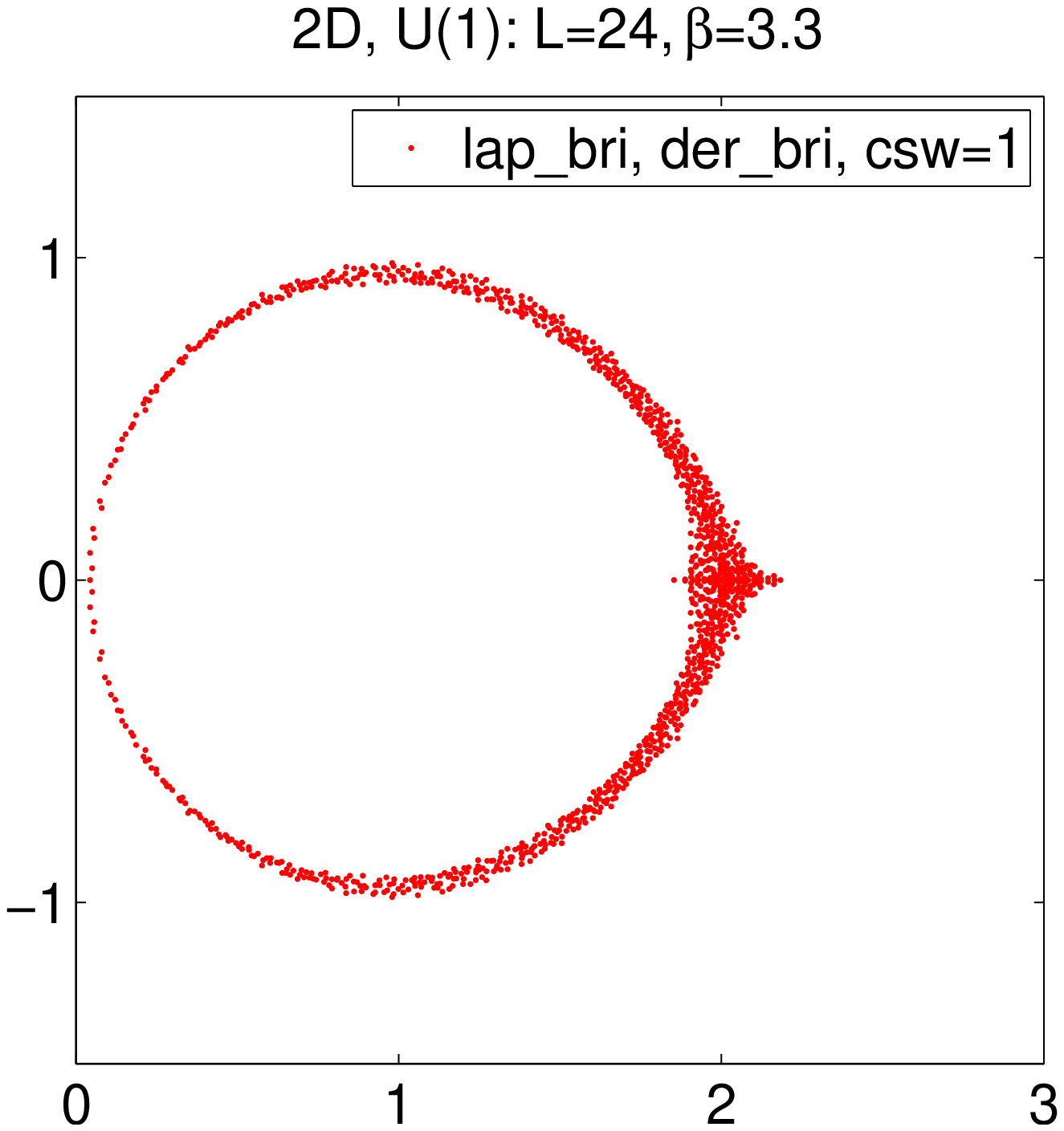,height=5.7cm}
\epsfig{file=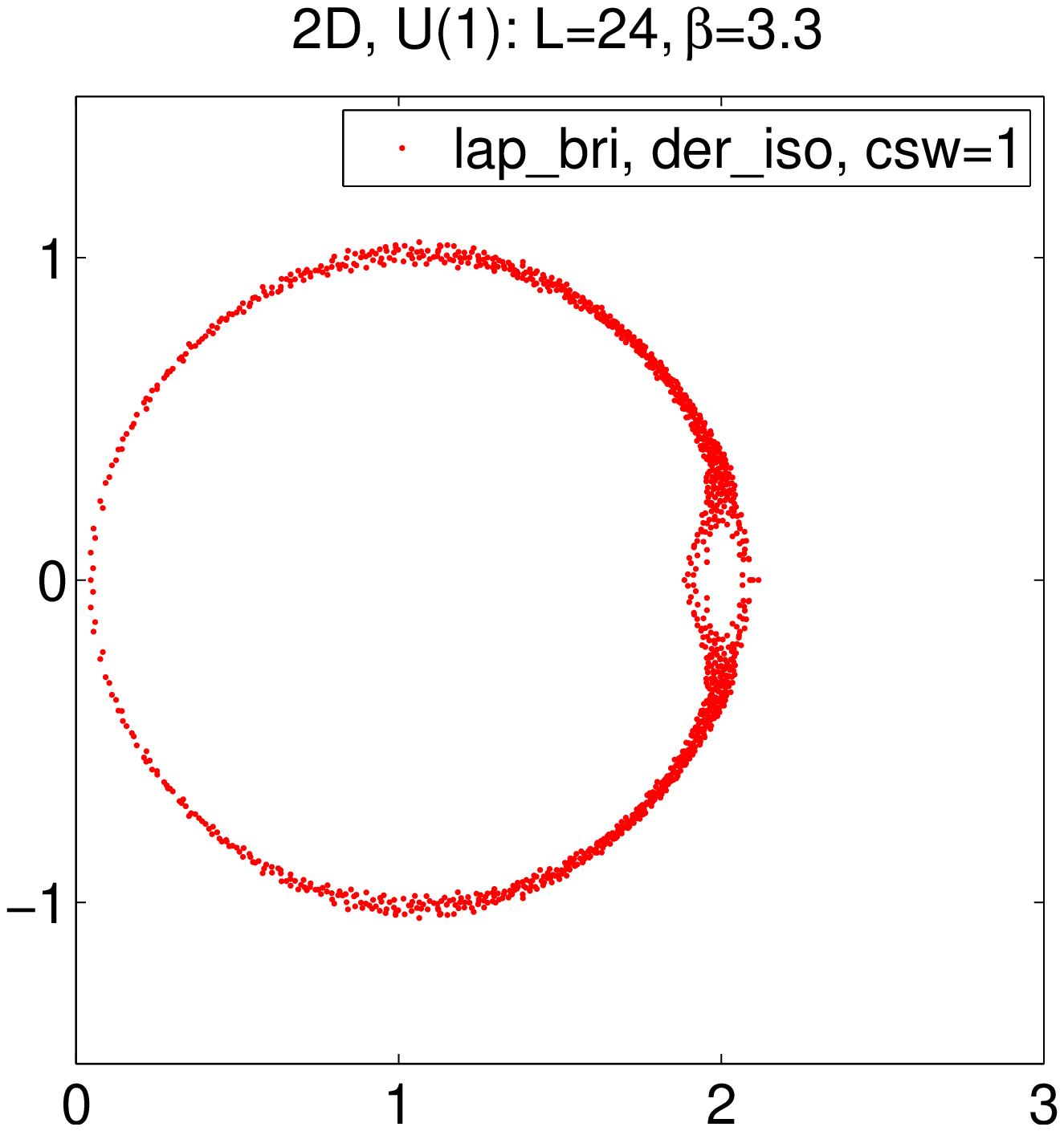,height=5.7cm}\\
\epsfig{file=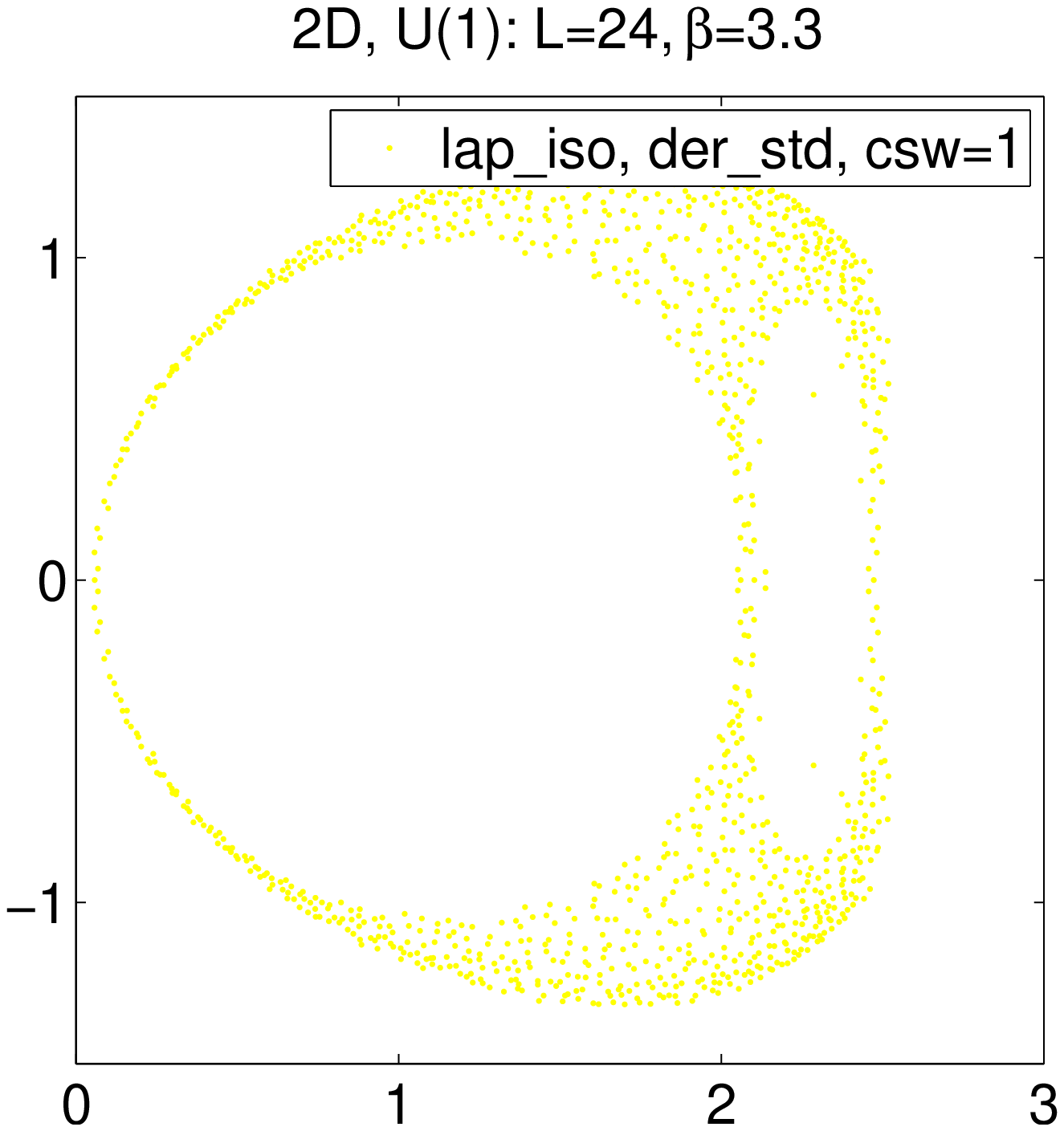,height=5.7cm}
\epsfig{file=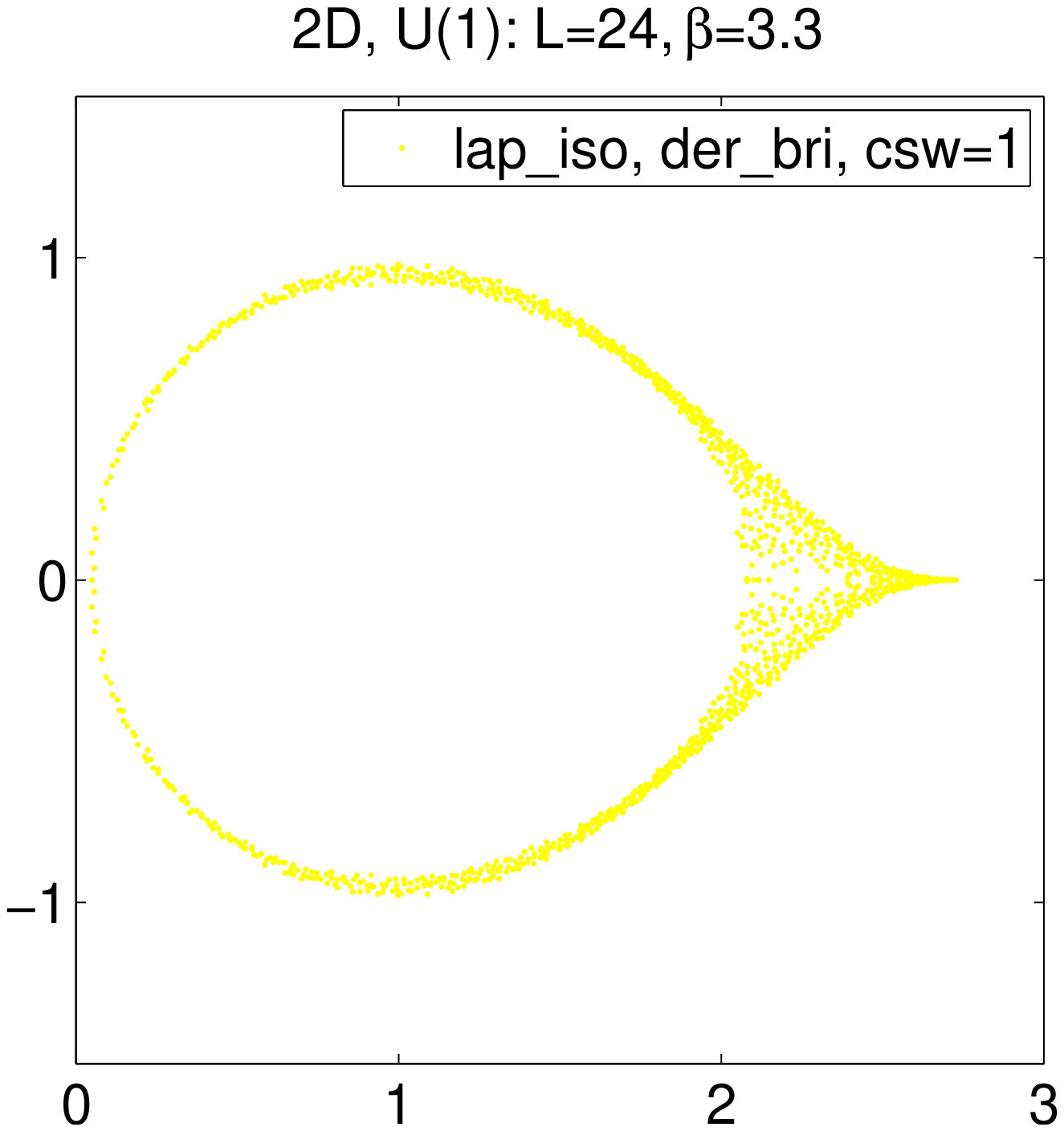,height=5.7cm}
\epsfig{file=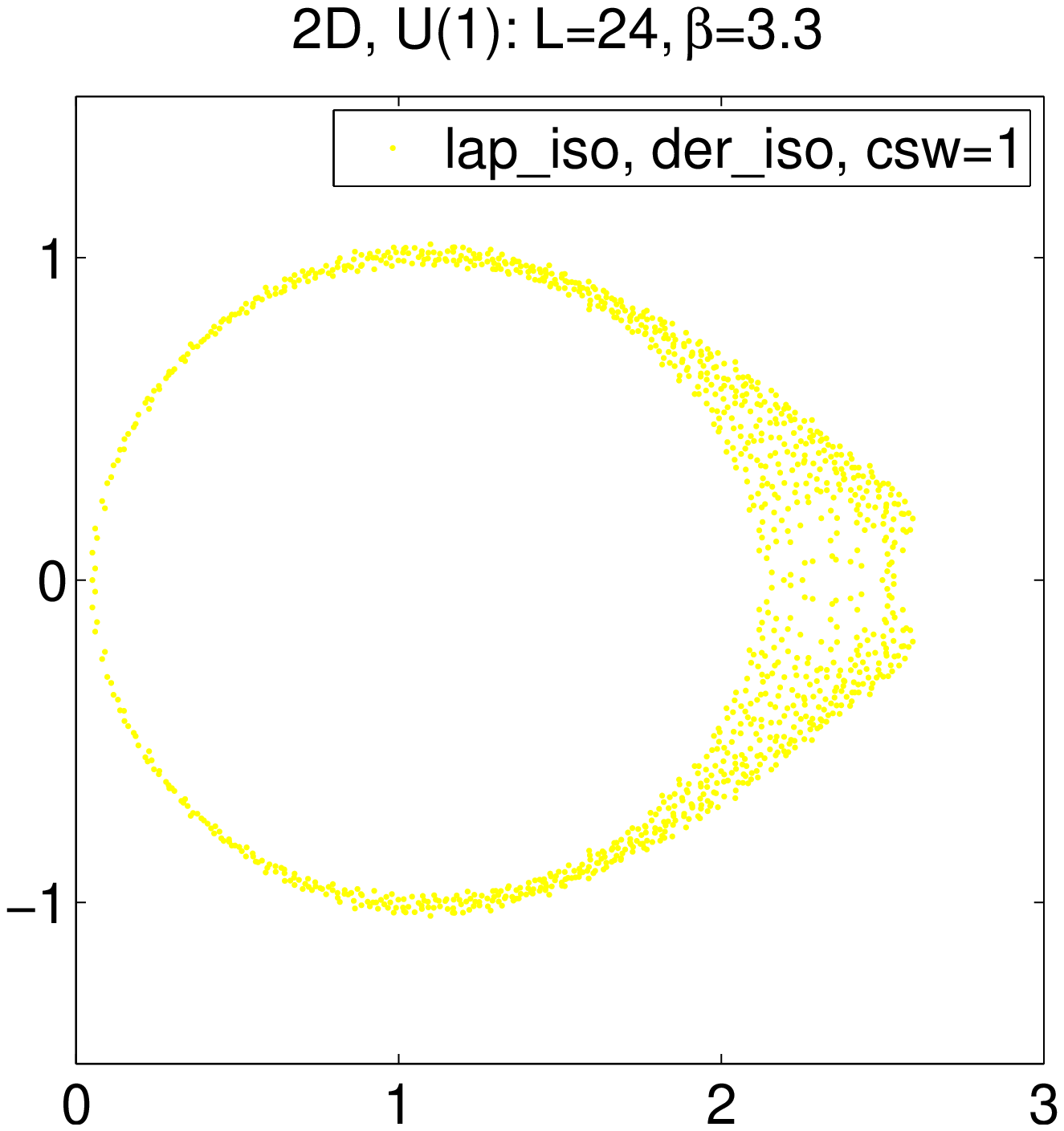,height=5.7cm}
\caption{\label{fig:spec_2D_csw1}\sl
Eigenvalue spectra of all operators considered in 2D with $c_\mr{SW}\!=\!1$.}
\end{figure}

In Fig.\,\ref{fig:spec_2D_csw0} the eigenvalue spectra of the 12 operators
without improvement ($c_\mr{SW}\!=\!0$) are shown.
The Laplacian features as the row index of the panel, and the derivative as the
column index.
Out of these 12 constructions, 9 are undoubled fermion operators, while 3 yield
two species in the continuum limit.
Let us discuss the undoubled operators first.
The three operators with $\lap^\mr{std}$ have three branches, the left-most
physical branch with the correct sensitivity to the topological charge of
the gauge background, a doubly populated branch of wrong-chirality doublers
near $\mr{Re}(z)\!=\!2$ and another species with the correct chirality near
$\mr{Re}(z)\!=\!4$.
Here the choice of derivative affects the spreading of the unphysical branches
in the imaginary direction, but it leaves the topological properties of the
spectrum unaffected.
The three operators with $\lap^\mr{bri}$ have only two branches, the left
(physical) one is undoubled with the correct chirality, the right one includes
three species, two with the wrong chirality and one with the correct chirality.
The three operators with $\lap^\mr{iso}$ have spectra which resemble those
in the first row, except that the lifting of the last branch is reduced -- in
perfect agreement with what one expects on the basis of
Fig.\,\ref{fig:fourier_lap}.
The field-theoretically most interesting spectra are those of the operators
with $\lap^\mr{til}$.
Naively, one would expect that they yield a legal 2-flavor operator (in 2D), as
the second row of Fig.\,\ref{fig:fourier_lap} shows that this Laplacian has two
zeros.
With the naive derivative operator employed, this expectation happens to be
correct; the resulting operator has two \emph{equal chirality} species%
\footnote{Note the difference to staggered fermions in 2D, where the two
species have \emph{opposite chiralities}.}
which survive in the continuum limit and two doublers (again with equal, but
this time wrong chirality) which decouple in the continuum limit.
With any of the two remaining derivatives employed, things are a bit more
involved, as the ``thorn'' or the ``bump'' in the middle or right panel of the
second row illustrate.
The point is that there is an interference%
\footnote{We avoid the word ``mixing'', because this is a phenomenon which
persists in the weak coupling limit.}
between the dimension 5 Laplacian and the dimension 4 derivative; the ``cross
talk'' phenomena in the 2nd and 3rd column of the second row exemplify that
this may affect the structural properties of the fermion operator.
In either case one of the would-be physical species fails to cling nicely on
the imaginary axis for small momenta.
While the operator in the 2nd column is clearly not a legal discretization
(the ``thorn'' violates the property $D\!\sim\!\ri p_\mu\ga_\mu$), the version
in the 3rd column may represent a legal 2-flavor discretization of the Dirac
operator (though, likely, with terrible cut-off effects).
In summary, from this first overview it appears that the four operators towards
the lower right corner of this figure seem most promising.

In Fig.\,\ref{fig:spec_2D_csw1} the same survey is repeated with tree-level
improvement ($c_\mr{SW}\!=\!1$, cf.\ Sec.\,5.3).
Relative to the previous figure, changes seem to be mild.
However, an interesting point is that the clover term shifts correct-type
chirality branches (slightly) to the left and wrong-kind chirality branches
(slightly) to the right.
As a result, for the 9 undoubled operators the additive mass renormalization
(the offset of the physical branch at zero imaginary part) is always reduced.
Also our statements in the previous paragraph regarding the chiralities of the
unphysical branches can now be checked, because they map into a prediction of
the effect of the clover term.
At this point we can probably say that ($\lap^\mr{bri}$, $\nab^\mr{iso}$) fares
best in the sense that its eigenvalue spectrum is closest to that of an
operator satisfying the Ginsparg-Wilson relation%
\footnote{The eigenvalue spectrum of such an operator is in the unit circle
centered at the point $1$ on the real axis \cite{Ginsparg:1981bj}.}.


\subsection{Free field dispersion relations in 2D}

As mentioned in the introduction, the free-field dispersion relation of the
fermion operator is of utmost importance, as this is a property for which there
is no systematic improvement scheme (apart from taking the continuum limit).
With standard $\ga$-matrix identities it follows that the inverse of
$D=\sum\ga_\mu\nab_\mu-\frac{r}{2}\lap+m$ is given by
$D^{-1}=(-\sum\ga_\mu\nab_\mu-\frac{r}{2}\lap+m)/
([\frac{r}{2}\lap-m]^2-\sum\nab_\mu^2)$, where $r$ is the Wilson parameter.
Accordingly, to work out the dispersion relation we have to search for zeros of
$[\frac{r}{2}\lap-m]^2-\sum\nab_\mu^2$, where $\lap$ and $\nab$ denote any one
of the Laplacians or derivatives introduced above.

\begin{figure}[!p]
\centering
\epsfig{file=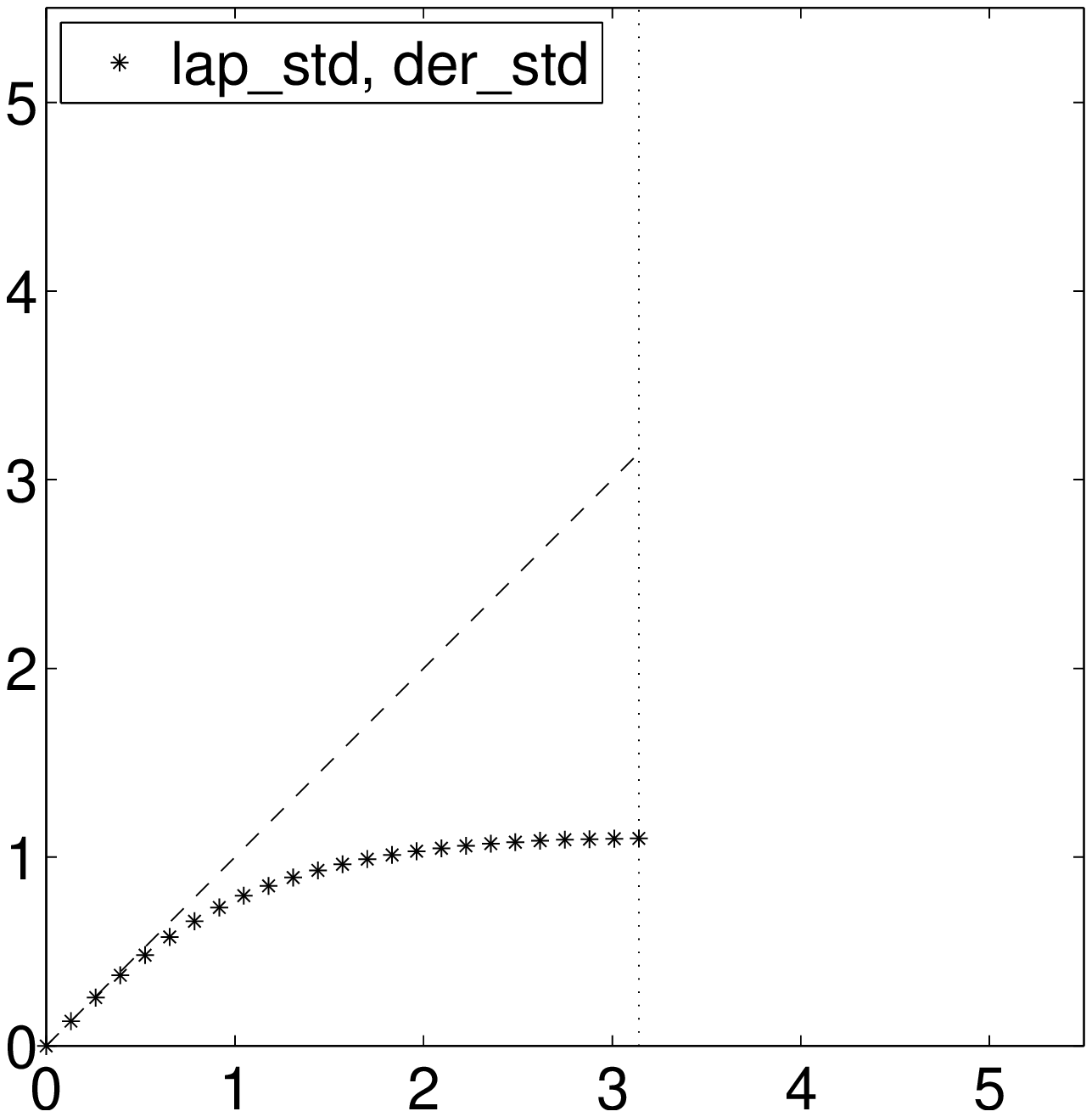,height=5.7cm}
\epsfig{file=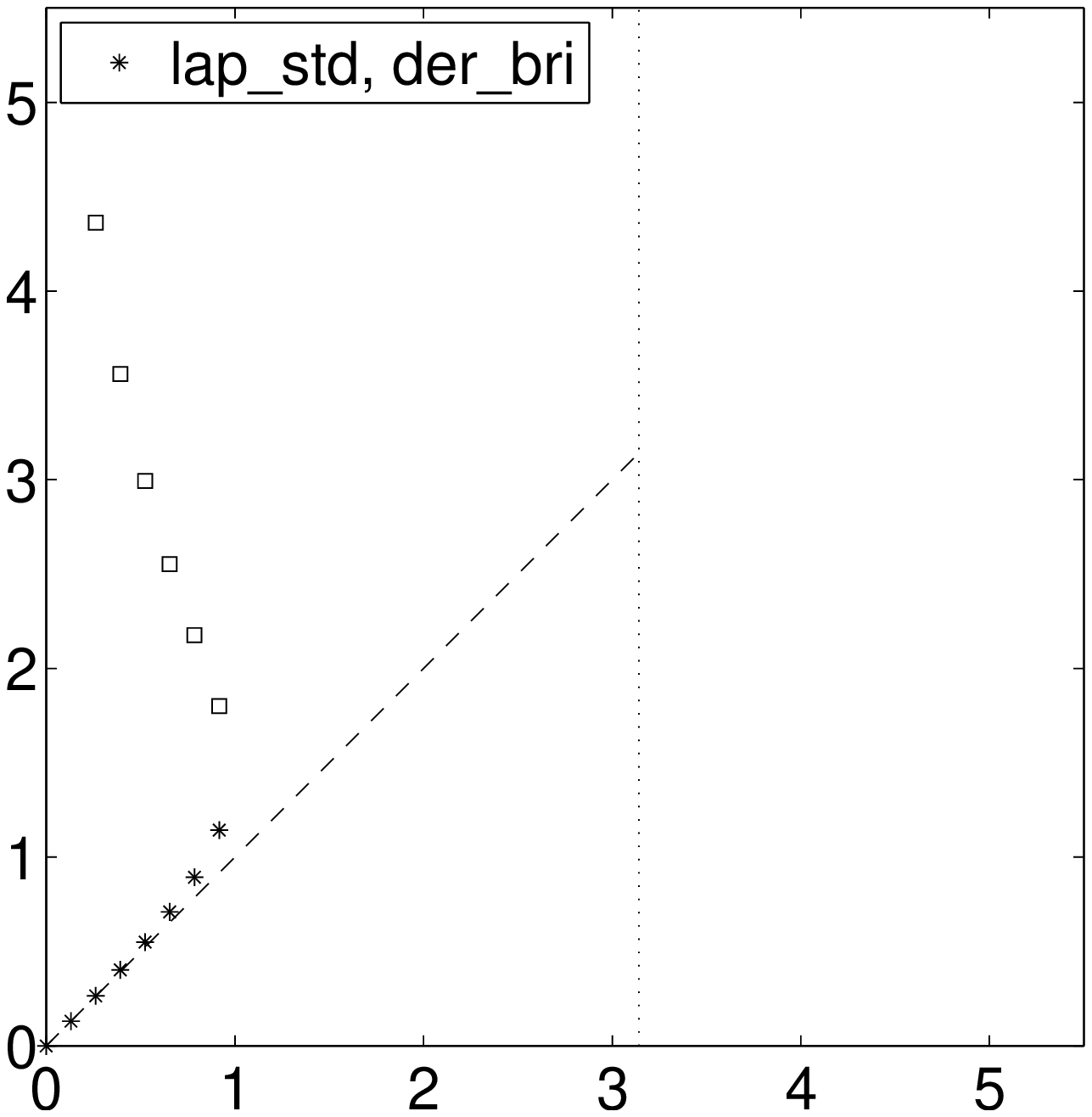,height=5.7cm}
\epsfig{file=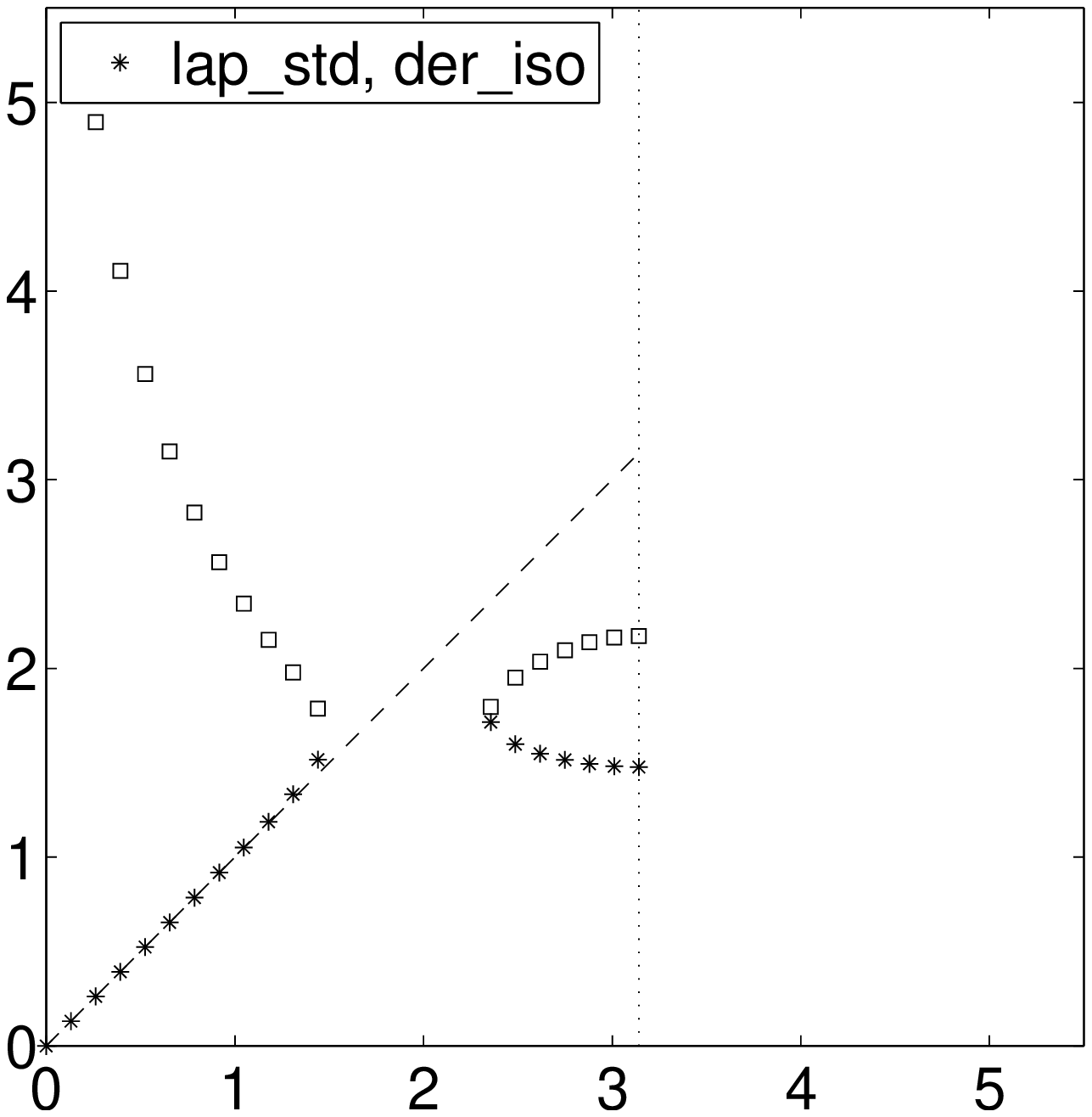,height=5.7cm}\\
\epsfig{file=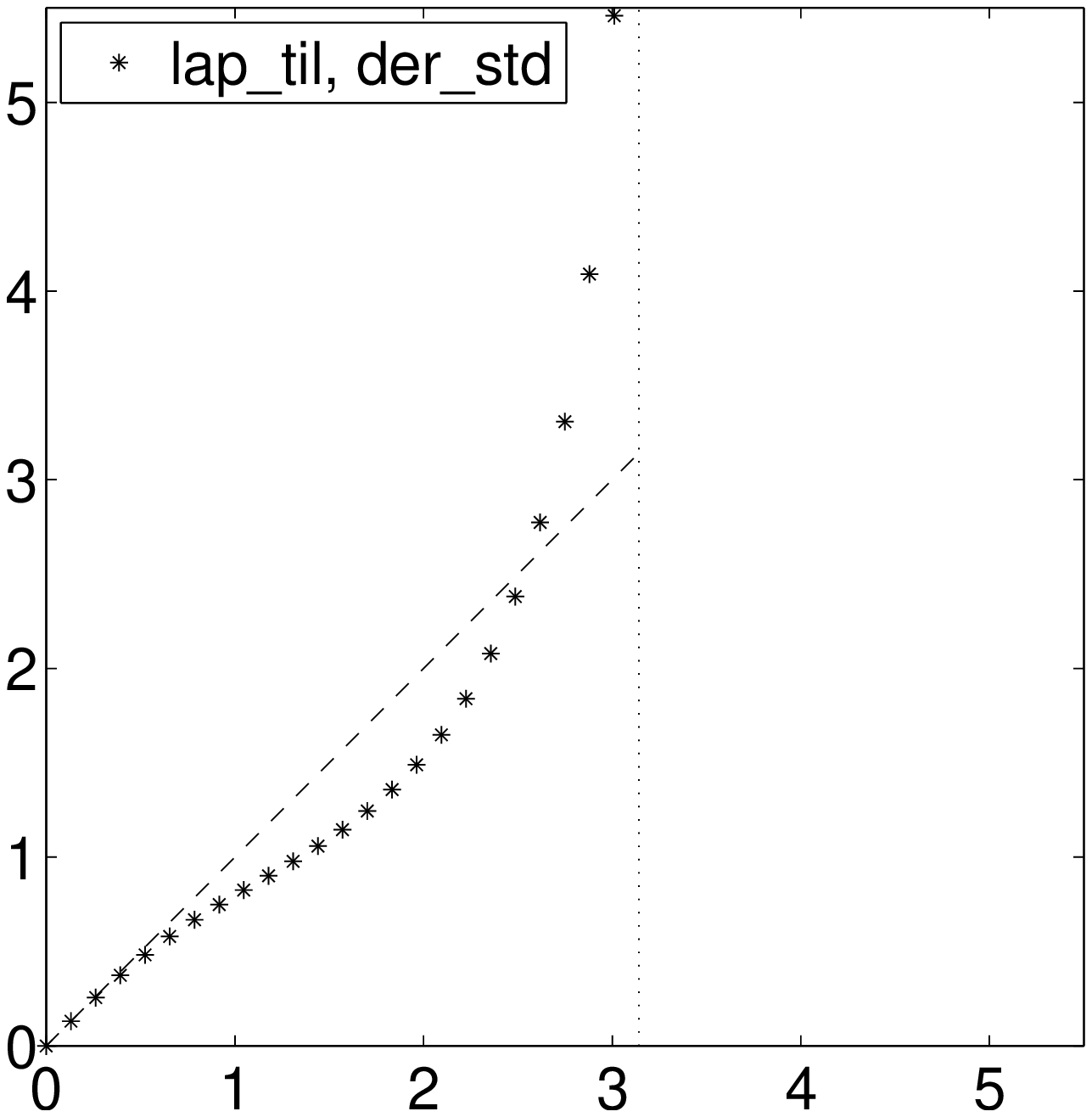,height=5.7cm}
\epsfig{file=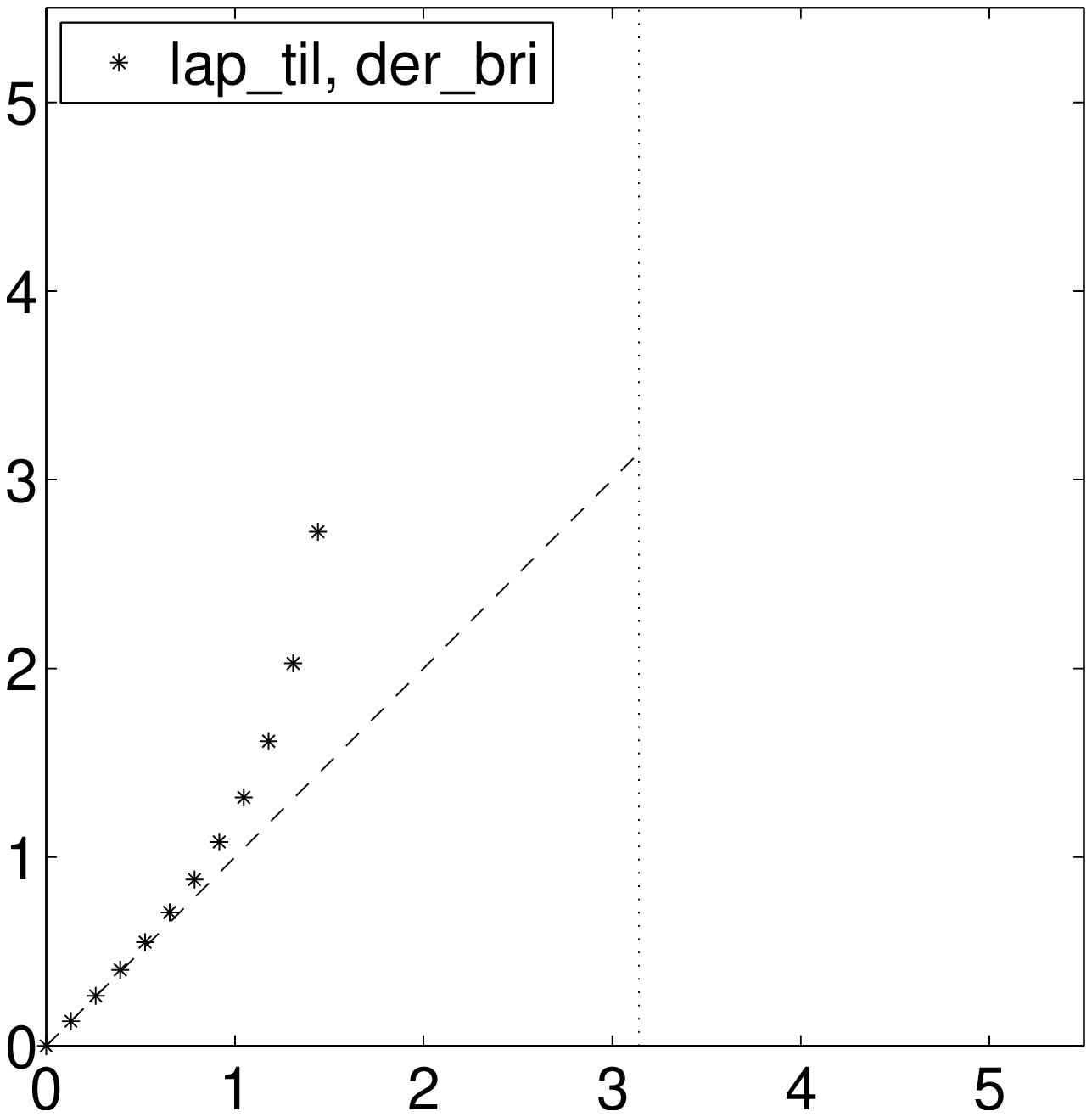,height=5.7cm}
\epsfig{file=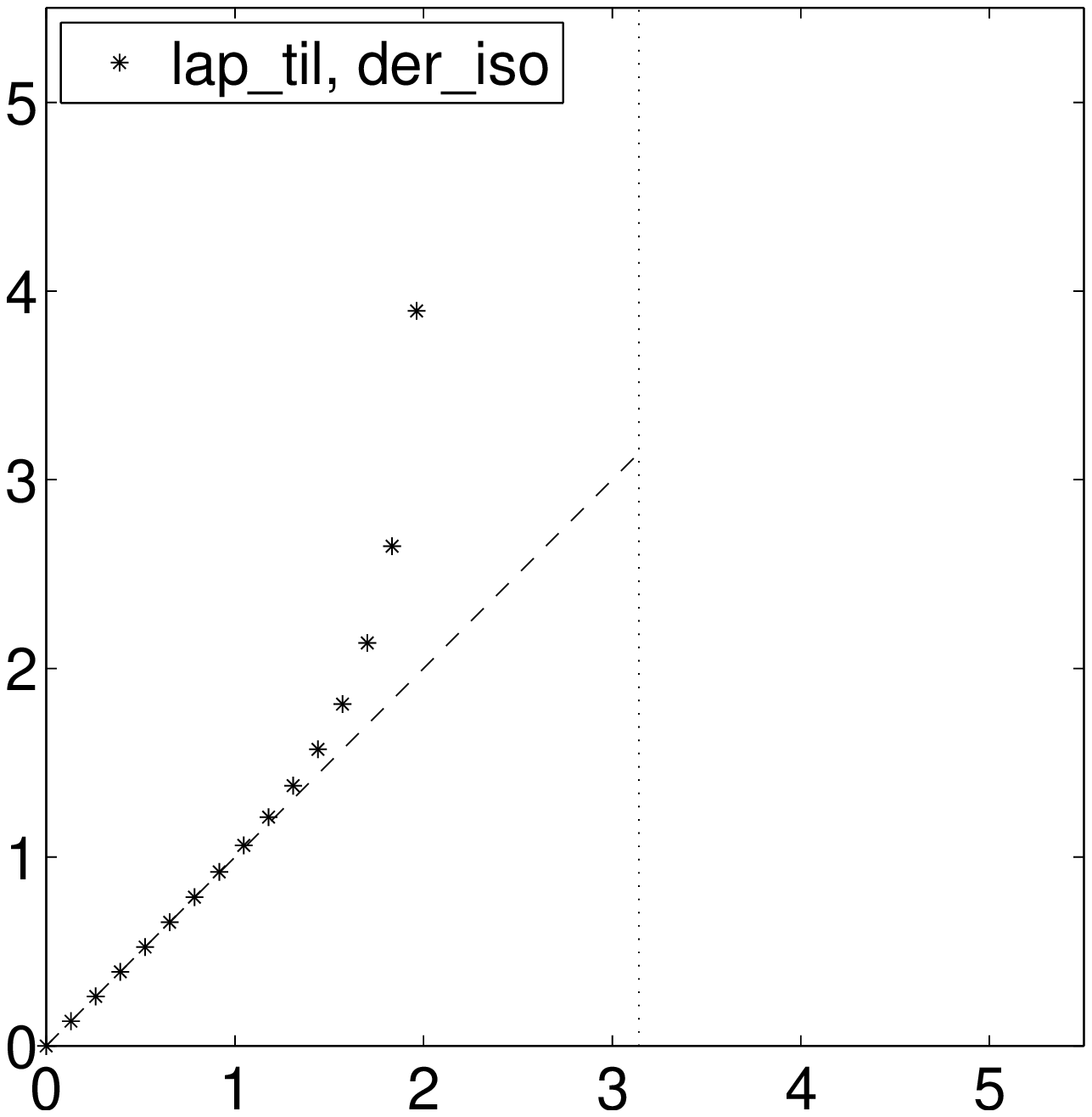,height=5.7cm}\\
\epsfig{file=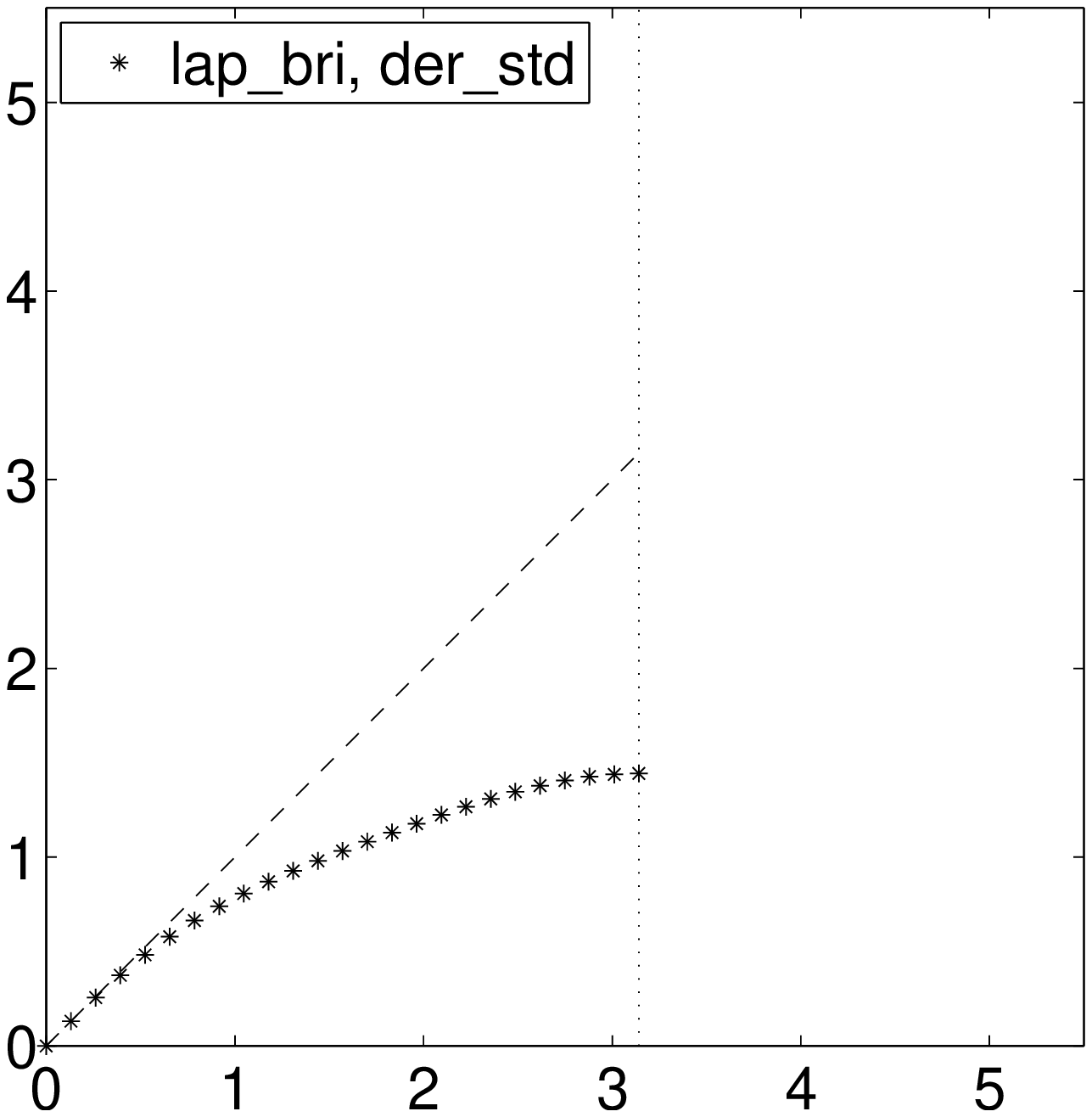,height=5.7cm}
\epsfig{file=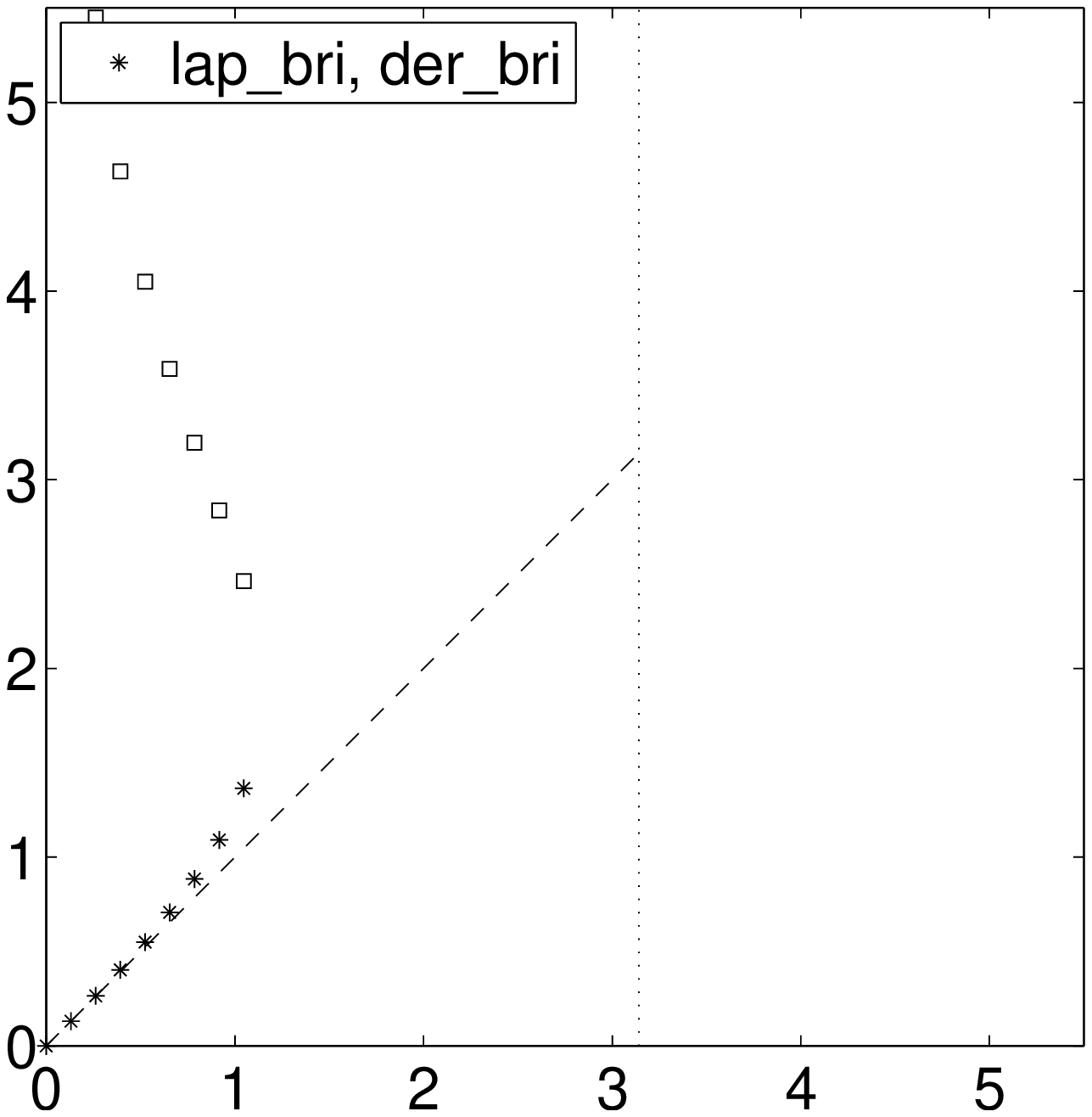,height=5.7cm}
\epsfig{file=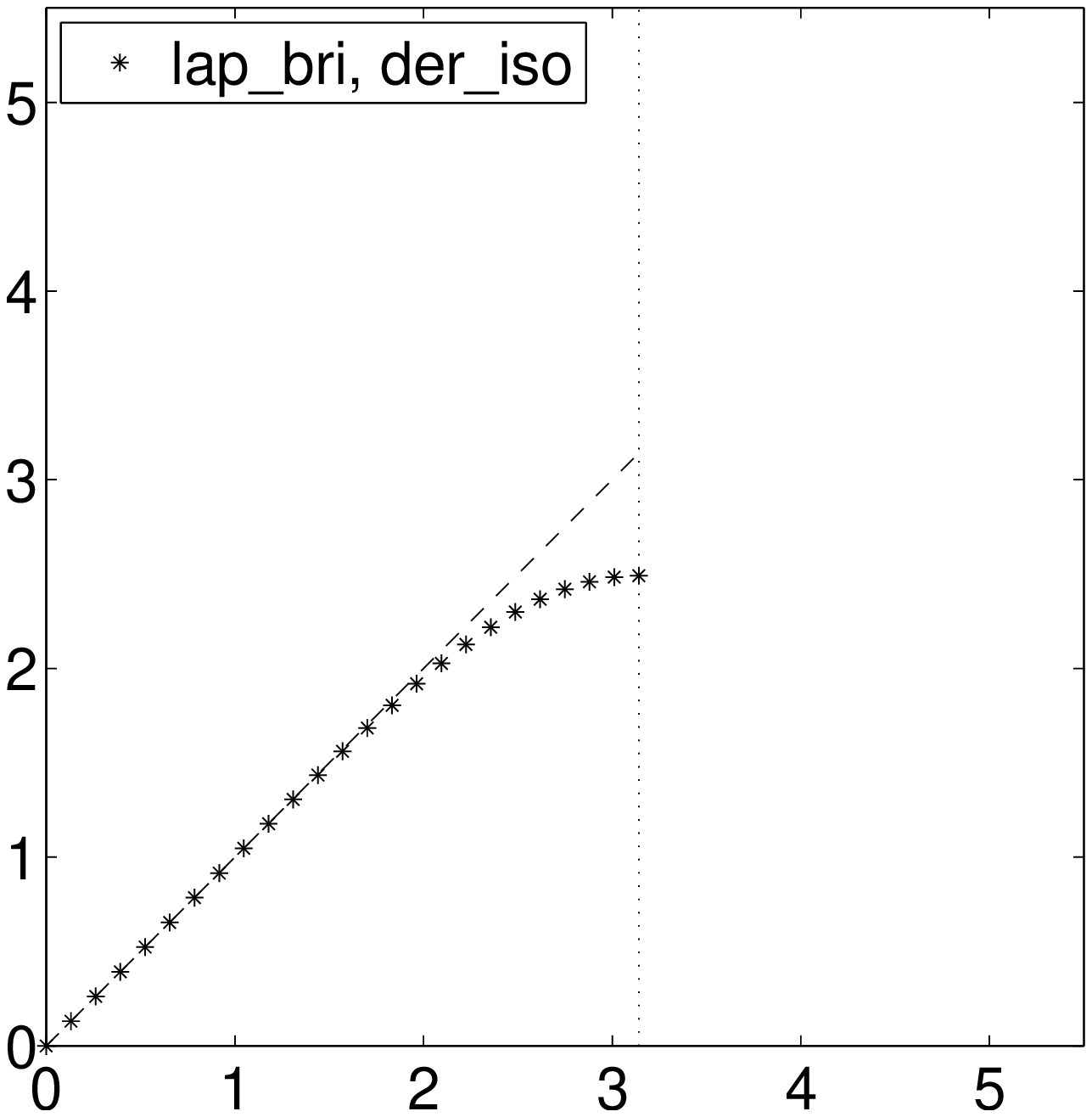,height=5.7cm}\\
\epsfig{file=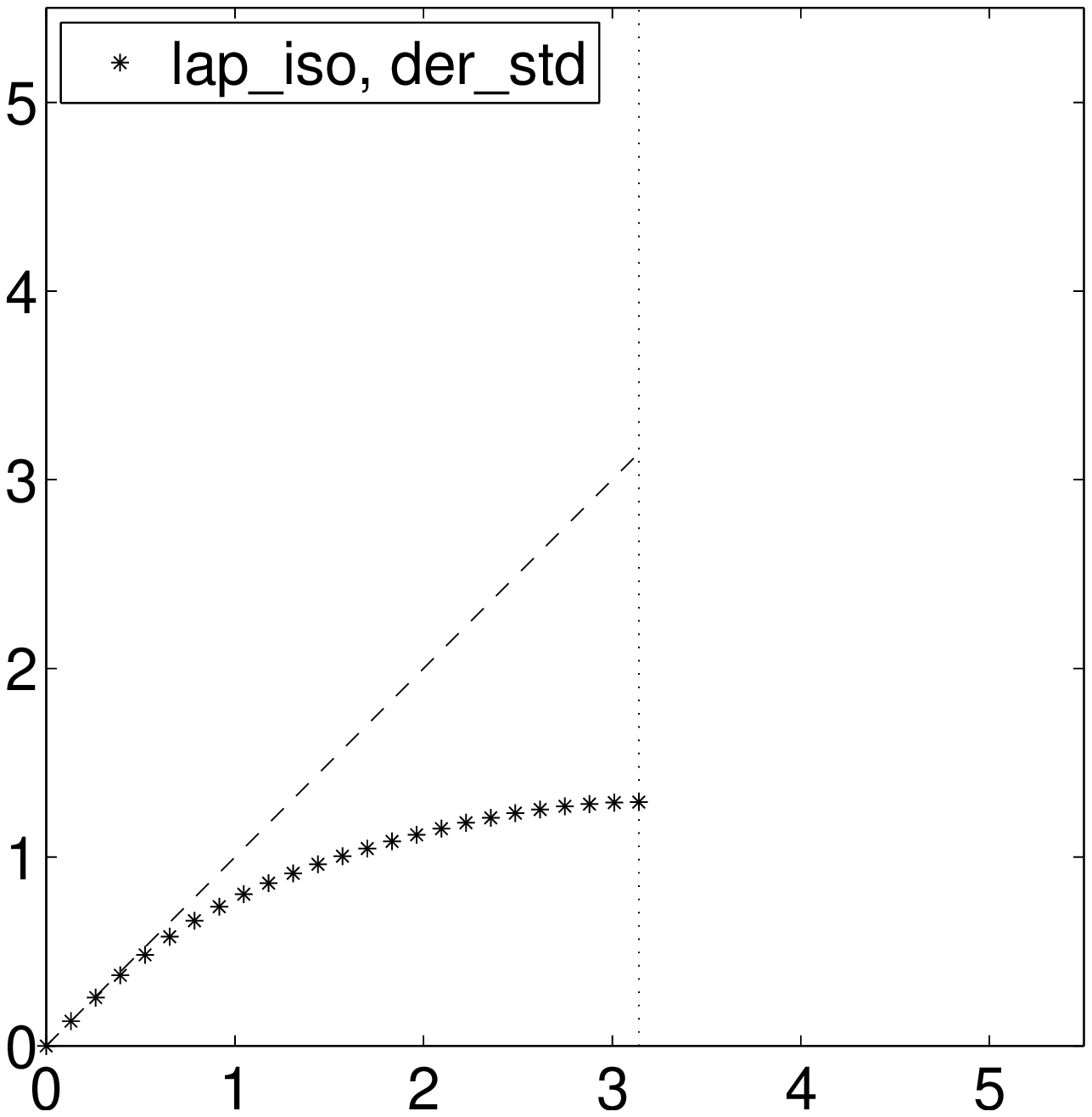,height=5.7cm}
\epsfig{file=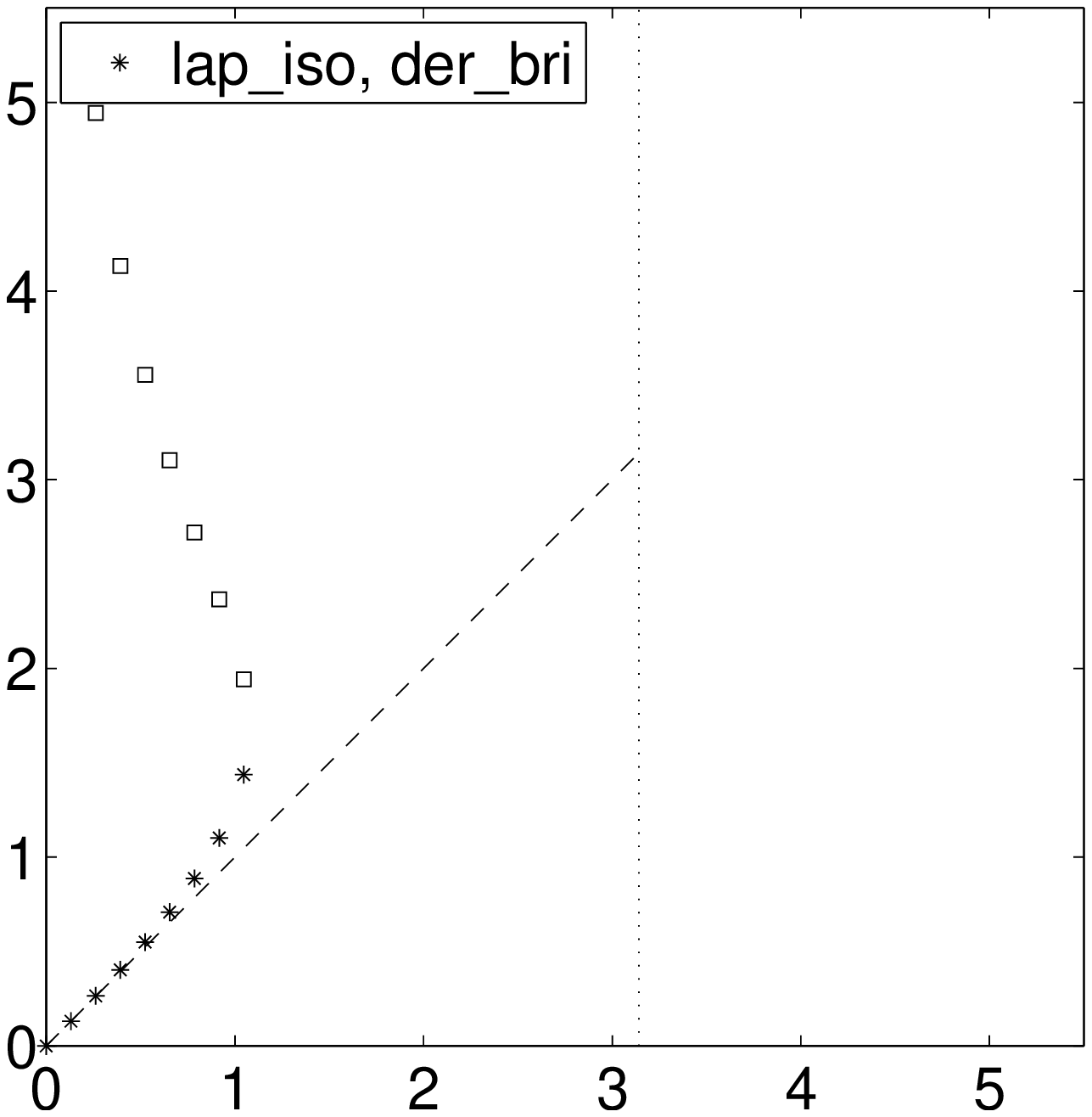,height=5.7cm}
\epsfig{file=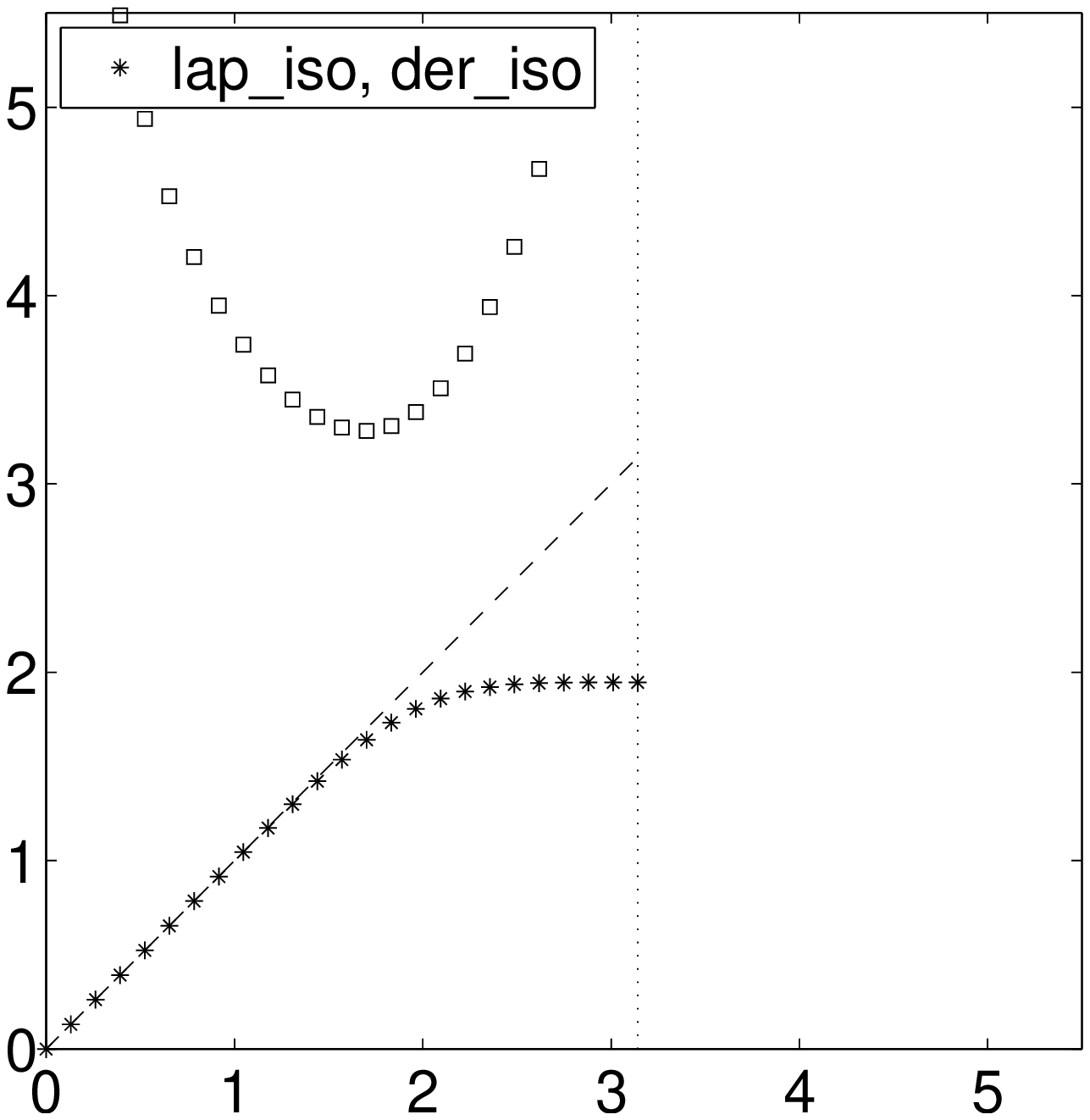,height=5.7cm}
\caption{\label{fig:disp_2D}\sl
Free-field dispersion relations of all operators considered in 2D, where
$|p|_\mr{max}\!=\!\pi/a$.}
\end{figure}

In Fig.\,\ref{fig:disp_2D} we show, for each operator, the real solutions for
$r\!=\!1$ and $m\!=\!0$ over half the Brillouin zone on a 2D lattice with
$L/a\!=\!48$.
The dispersion relation of the standard Wilson operator ($\lap^\mr{std}$,
$\nab^\mr{std}$) deviates soon from the dashed line, which corresponds to the
continuum dispersion relation; in particular towards the boundary of the
Brillouin zone the distortion is significant.
Black boxes indicate a second  real solution.
If sufficiently high, this is harmless, as this branch decouples in the
continuum limit.
Note that (in certain parts of the Brillouin zone) some operators have only
complex roots.
While this proves, again, irrelevant in the continuum limit, it is certainly
not a desirable feature.
Overall, it is clear that the combination ($\lap^\mr{bri}$, $\nab^\mr{iso}$)
fares best in the sense that its dispersion relation is closest to the one in
the continuum.


\section{Construction and main features in 3D}


\subsection{Summary of 3D Laplace stencils}

The ``standard'' stencil of the Laplacian in 3D and the ``tilted'' variety
(as defined in App.\,B) have the Fourier space representation
\bea
a^2\hat\lap^\mr{std}(k_1,k_2,k_3)&=&2\cos(k_1)+2\cos(k_2)+2\cos(k_3)-6
\nonumber
\\
&=&-4\sin^2(k_1/2)-4\sin^2(k_2/2)-4\sin^2(k_3/2)
\label{def_3Dstd}
\\
a^2\hat\lap^\mr{til}(k_1,k_2,k_3)&=&2\cos(k_1)\cos(k_2)\cos(k_3)-2
\nonumber
\\
&=&
16\cos^2(k_1/2)\cos^2(k_2/2)\cos^2(k_3/2)-8\cos^2(k_1/2)\cos^2(k_2/2)-...
\nonumber
\\
&&
+4\cos^2(k_1/2)+...-4
\label{def_3Dtil}
\eea
respectively, with the ellipses denoting cyclic permutations.
From the stencil notation in App.\,B it is easy to see that the former has
only 1-hop contributions, while the latter has only 3-hop contributions
(apart from the central element).
For asymptotically small momenta they both reduce to the continuum relation
$\hat\lap=p_1^2\!+\!p_2^2\!+\!p_3^2$, but the ``tilted'' stencil has three
additional zeros at the boundary of the Brillouin zone [$\hat\lap^\mr{til}$
vanishes at $(k_1,k_2,k_3)-(\pi,\pi,\pi)/2=(\pm\pi,\pm\pi,\pm\pi)/2$ with an
odd number of minus signs].

In 3D the discretizations of the continuum Laplacian which are analogous to
(\ref{def_2Dbri}) and (\ref{def_2Diso}) in 2D are no longer simple linear
combinations of (\ref{def_3Dstd}) and (\ref{def_3Dtil}), because one could
come up with a Laplacian which has only 2-hop contributions (apart from the
center element).
They read
\bea
a^2\hat\lap^\mr{bri}(k_1,k_2,k_3)\!\!&\!=\!&\!\![\cos(k_1)\cos(k_2)\cos(k_3)
+\cos(k_1)\cos(k_2)+...+\cos(k_1)+...-7]/2
\nonumber
\\
\!\!&\!=\!&\!\!
4\cos^2(k_1/2)\cos^2(k_2/2)\cos^2(k_3/2)-4
\label{def_3Dbri}
\\
a^2\hat\lap^\mr{iso}(k_1,k_2,k_3)\!\!&\!=\!&\!\!
[\cos(k_1)\cos(k_2)\cos(k_3)
+3\cos(k_1)\cos(k_2)
+...
+5\cos(k_1)+...-25]/6
\nonumber
\\
\!\!&\!=\!&\!\!
[4\cos^2(k_1/2)\cos^2(k_2/2)\cos^2(k_3/2)
+4\cos^2(k_1/2)\cos^2(k_2/2)+...
-16]/3\quad\phantom{.}
\label{def_3Diso}
\eea
respectively, and their distinctive features are as follows.
The ``Brillouin'' Laplacian (\ref{def_3Dbri}) takes a constant value on the
entire boundary of the Brillouin zone, since
$a^2\hat\lap^\mr{bri}(k_1,k_2,k_3)=-4$ whenever one of the momenta is
$\pm\pi/a$.
On the other hand, the Laplacian (\ref{def_3Diso}) is called ``isotropic''
since $a^2\hat\lap^\mr{iso}(k_1,k_2,k_3)=
-a^2[k_1^2\!+\!k_2^2\!+\!k_3^2]+a^4[k_1^2\!+\!k_2^2\!+\!k_3^2]^2/12+O(a^6)$
has $O(a^4)$ terms which depend only on the combination
$k_1^2\!+\!k_2^2\!+\!k_3^2$.
In other words, $\hat\lap^\mr{iso}(k_1,k_2,k_3)$ respects rotational symmetry
even in the leading term through which it deviates from the continuum.

In 3D there are 3 linearly independent Laplacians with (at most) a 27-point
stencil, and any 3 out of the 4 elements (\ref{def_3Dstd}-\ref{def_3Diso})
form a basis.
A systematic treatment is given in \cite{Patra:2006}.

\begincomment
obj1:=2*cos(k1)+2*cos(k2)+2*cos(k3)-6;
sort(expand(subs(k1=2*k1h,k2=2*k2h,k3=2*k3h,obj1),trig));
series(subs({k1=a*p1,k2=a*p2,k3=a*p3},obj1),a,4);

obj2:=2*cos(k1)*cos(k2)*cos(k3)-2;
sort(expand(subs(k1=2*k1h,k2=2*k2h,k3=2*k3h,obj2),trig));
series(subs({k1=a*p1,k2=a*p2,k3=a*p3},obj2),a,4);

obj3:=4*cos(k1/2)^2*cos(k2/2)^2*cos(k3/2)^2-4;
sort(expand(combine(obj3,trig),trig));
series(subs({k1=a*p1,k2=a*p2,k3=a*p3},obj3),a,4);

obj4:=(cos(k1)-1)*(cos(k2)+5)*(cos(k3)+5)/18:
obj4:=sort(expand(obj4+ \
 subs({k1=k2,k2=k1},obj4)+subs({k1=k3,k3=k1},obj4) \
));
sort(expand(subs(k1=2*k1h,k2=2*k2h,k3=2*k3h,obj4),trig));
factor(series(subs({k1=a*p1,k2=a*p2,k3=a*p3},obj4),a,6));
\endcomment


\subsection{Eigenvalue spectra in 3D}

Like in the preceding section, with four options for $\lap$ and three for
$\nab$, we can construct 12 Dirac operators and study their eigenvalue spectra.
As the gauge group is irrelevant in this step, we prepare a thermalized
background in the $U(1)$ gauge theory with $L/a\!=\!12$ at $\be\!=\!2.2$.
A point worth mentioning is that in odd dimensions there is an ambiguity
regarding the representation of the $\ga$-matrices \cite{Burden:1986by}; we
opt for the 4-dimensional representation (the same one that we will use in 4D).

\begin{figure}[!p]
\centering
\epsfig{file=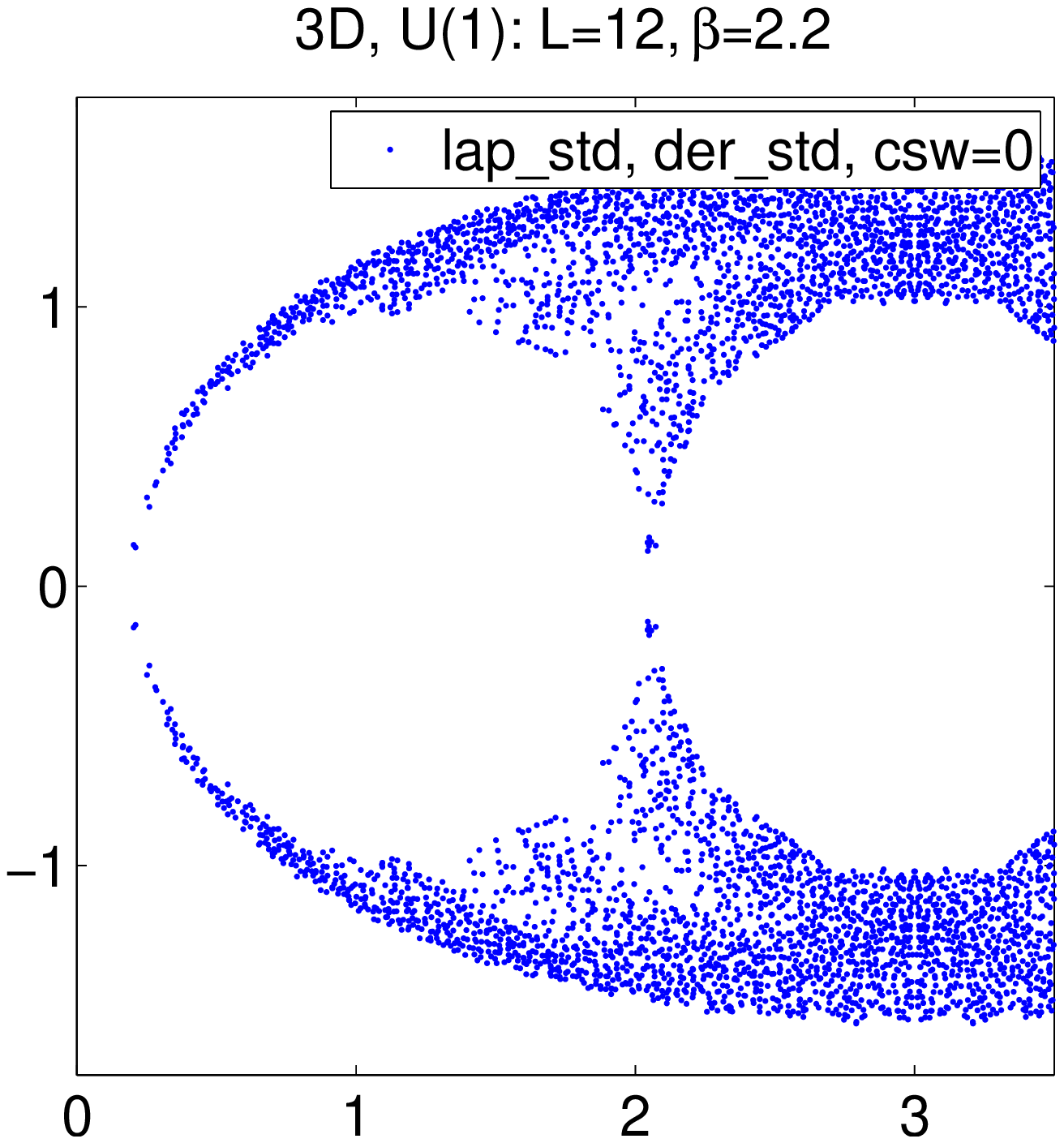,height=5.7cm}
\epsfig{file=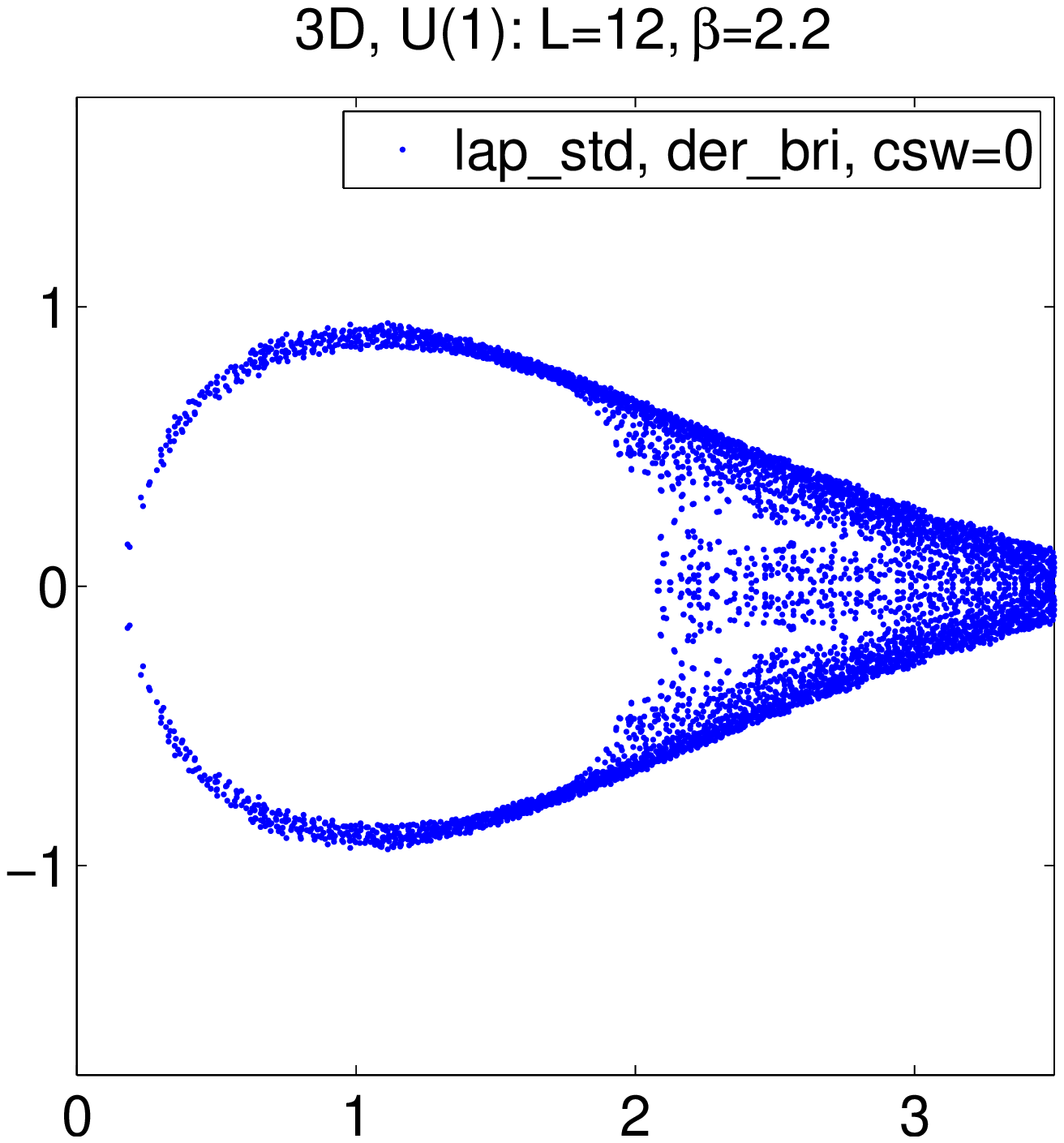,height=5.7cm}
\epsfig{file=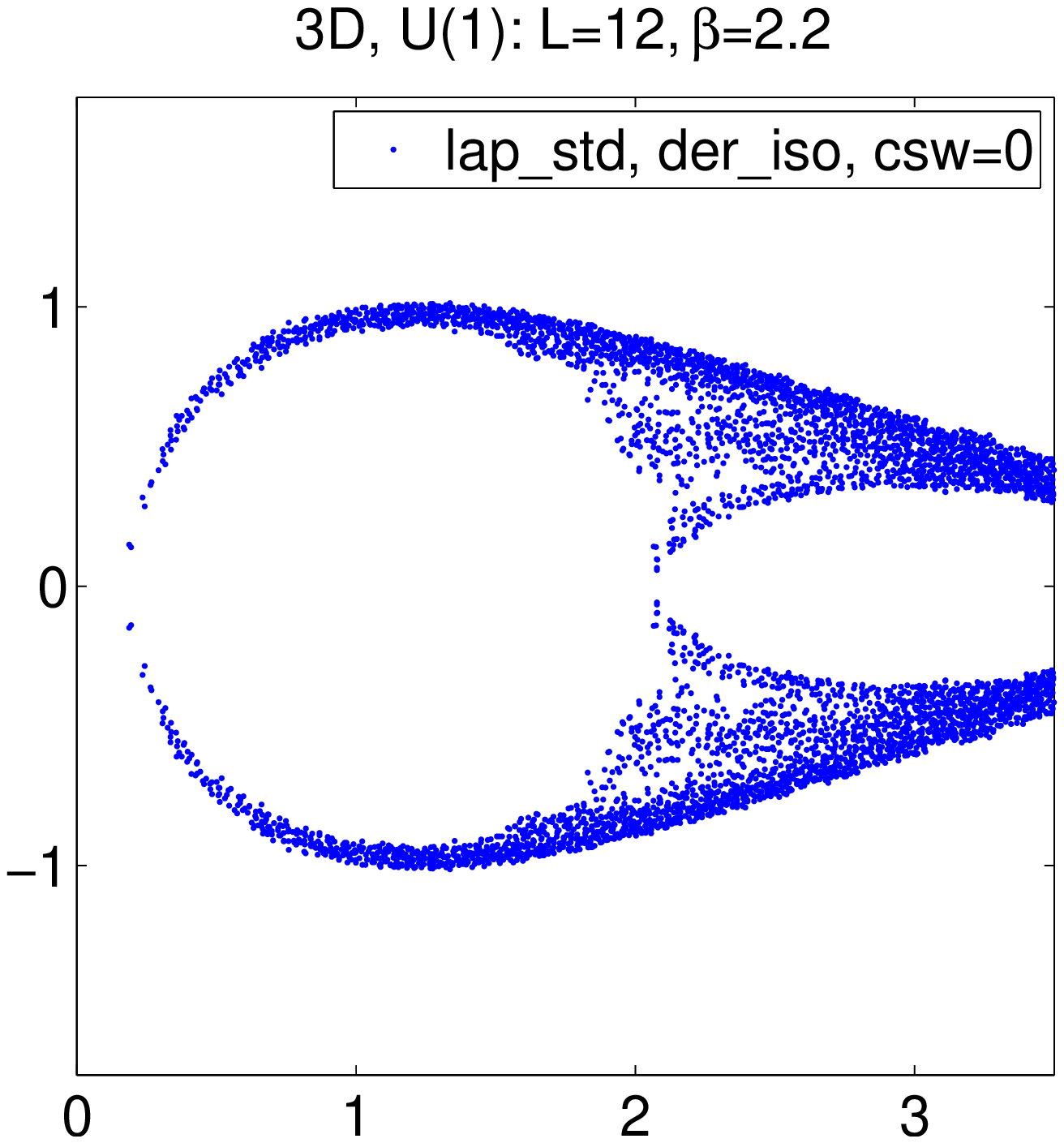,height=5.7cm}\\
\epsfig{file=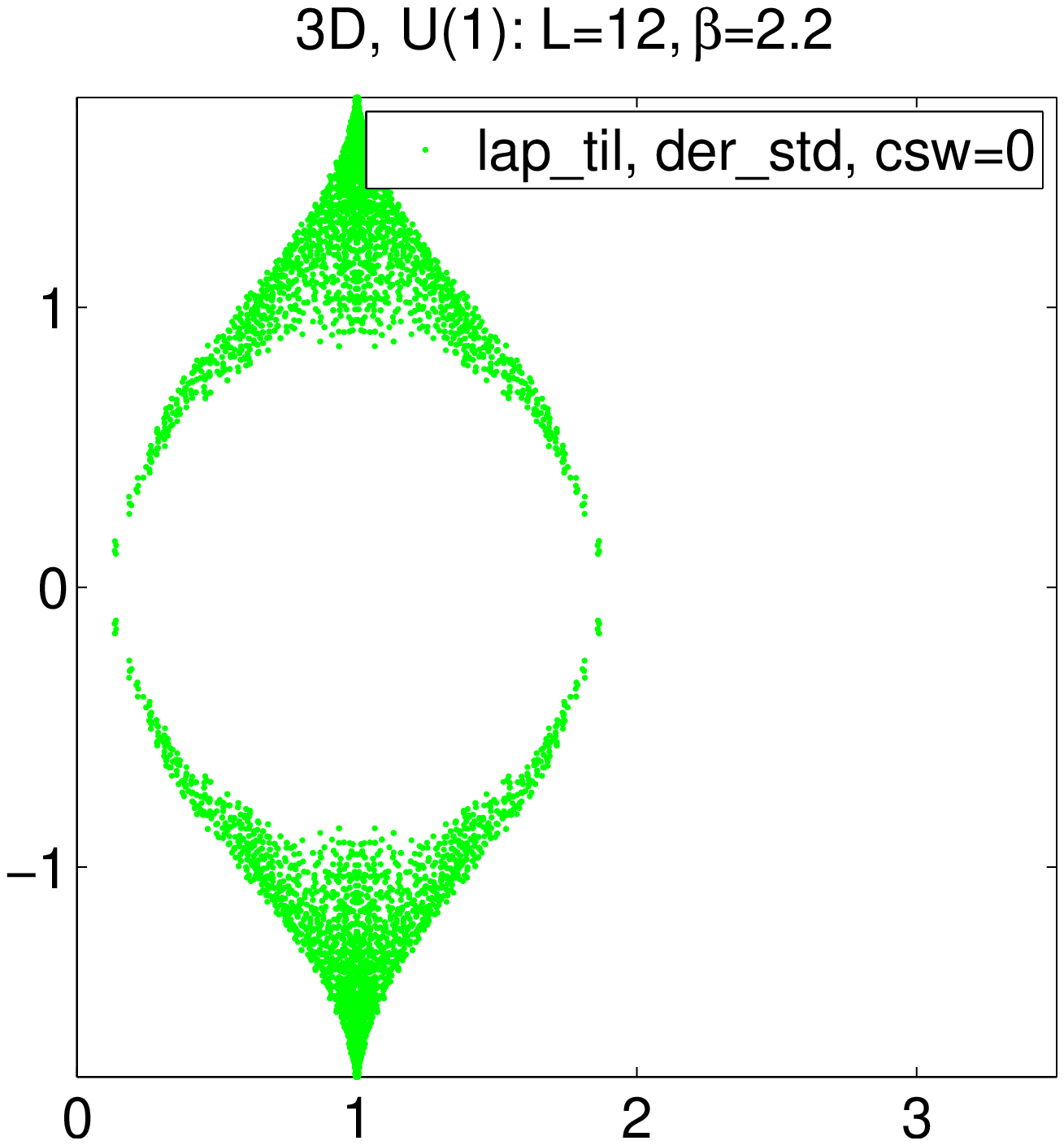,height=5.7cm}
\epsfig{file=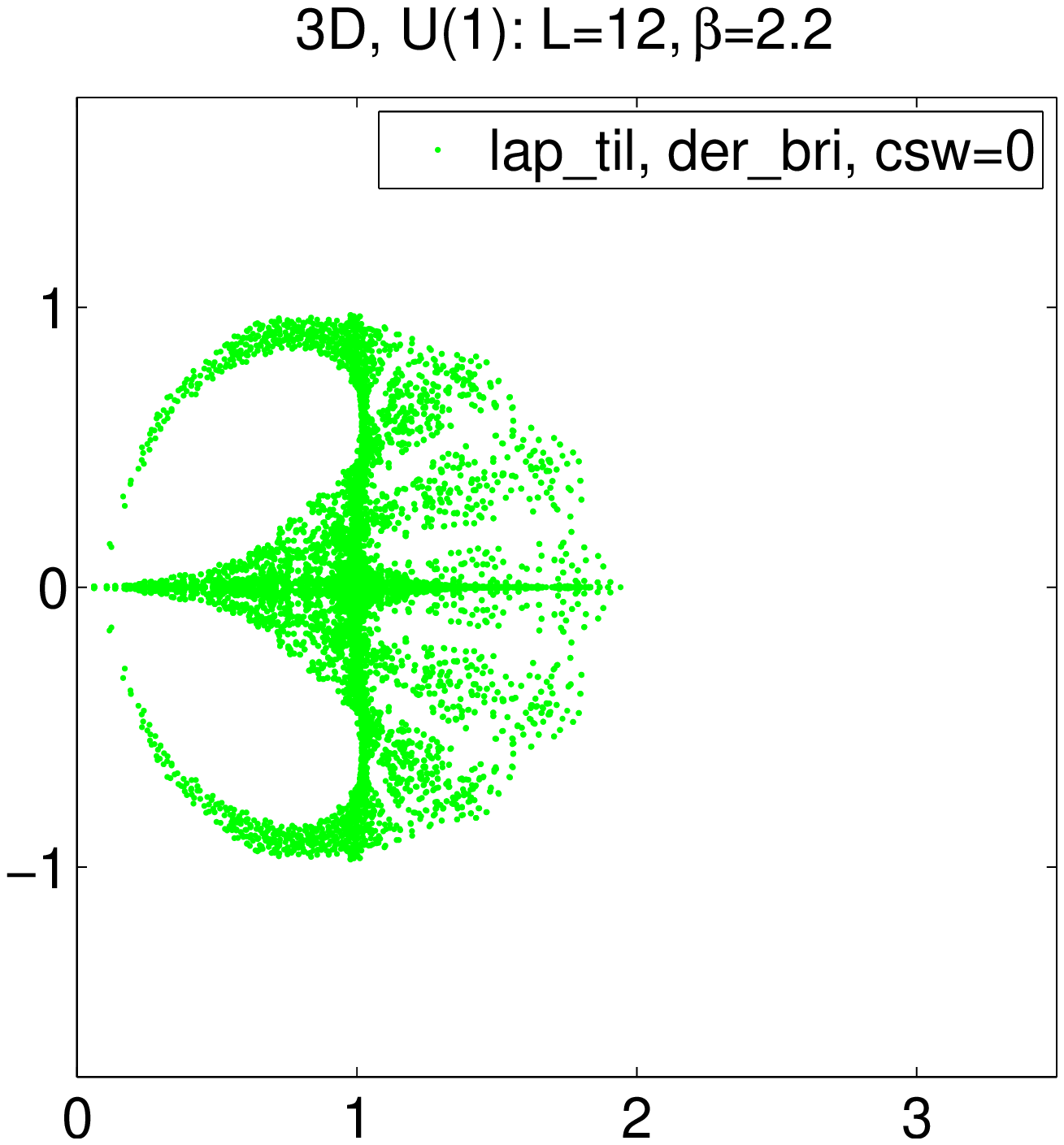,height=5.7cm}
\epsfig{file=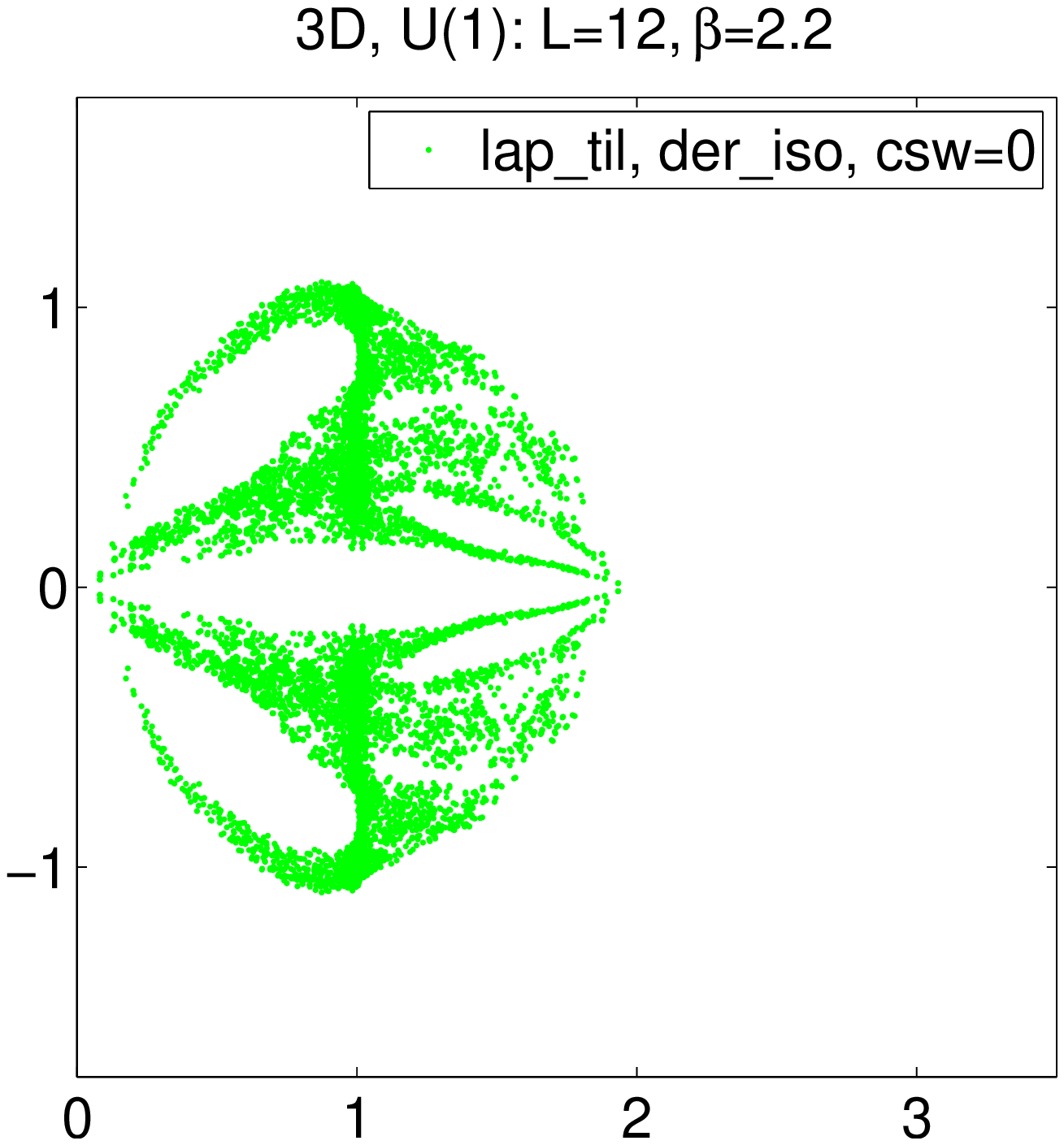,height=5.7cm}\\
\epsfig{file=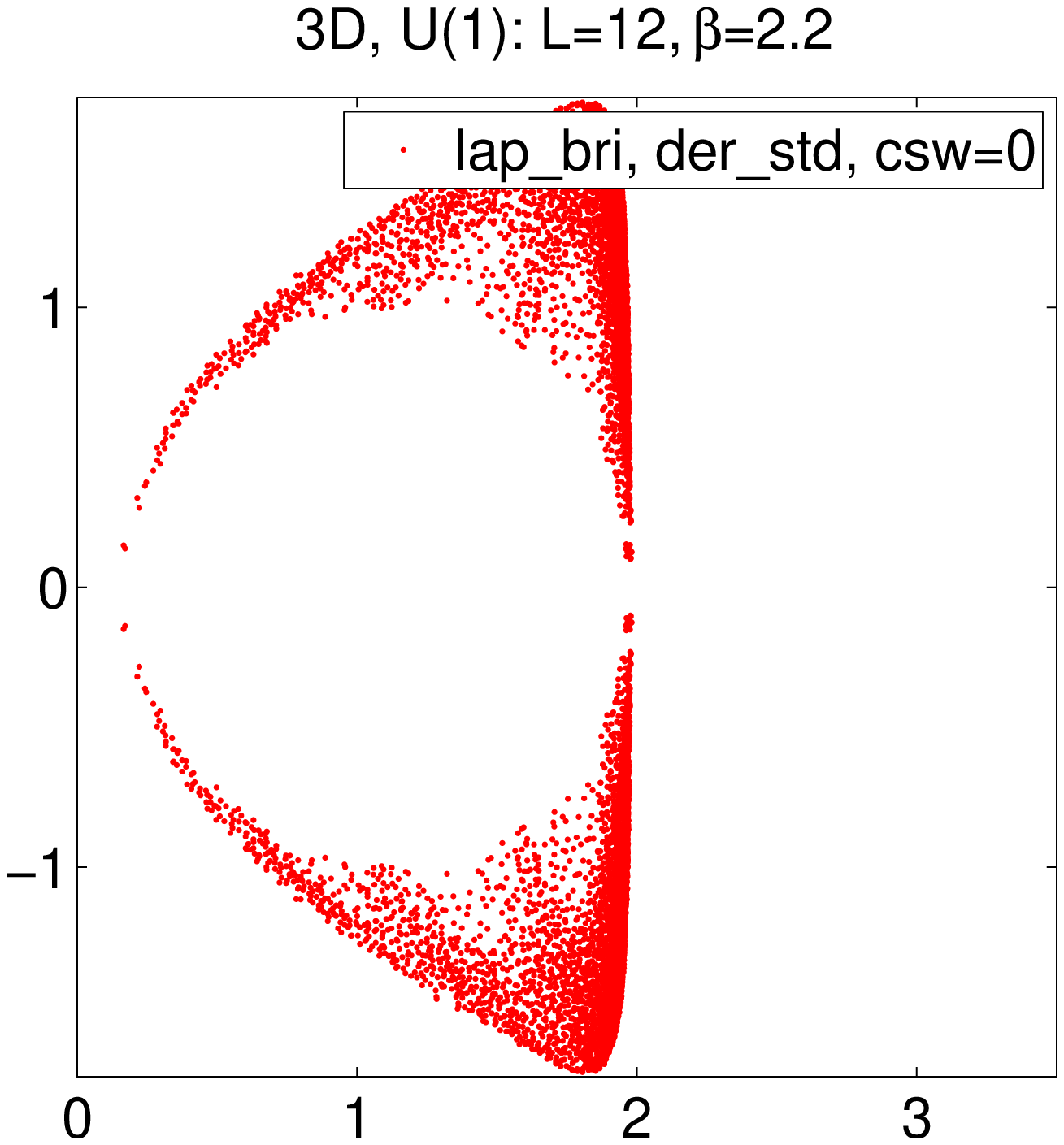,height=5.7cm}
\epsfig{file=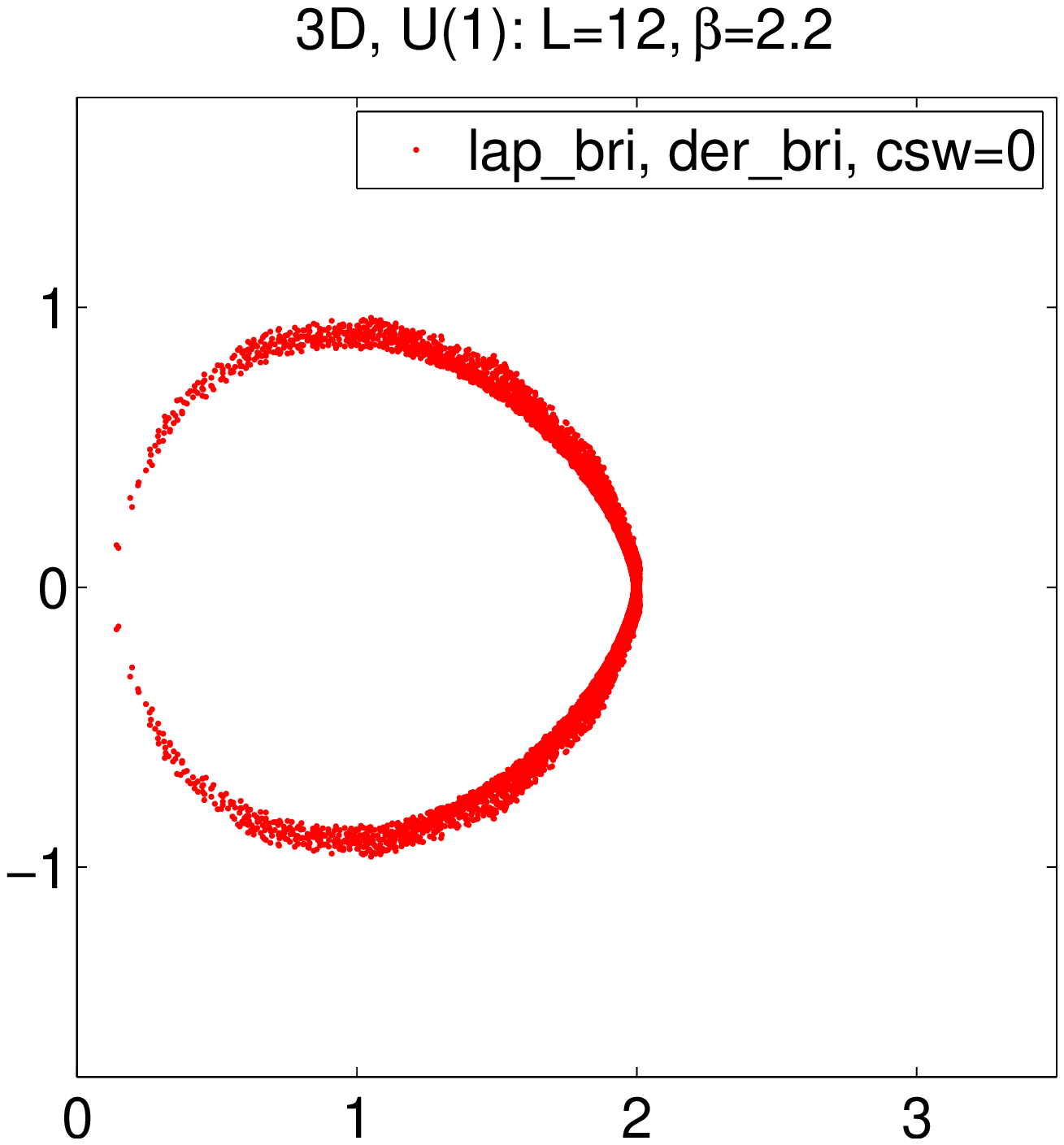,height=5.7cm}
\epsfig{file=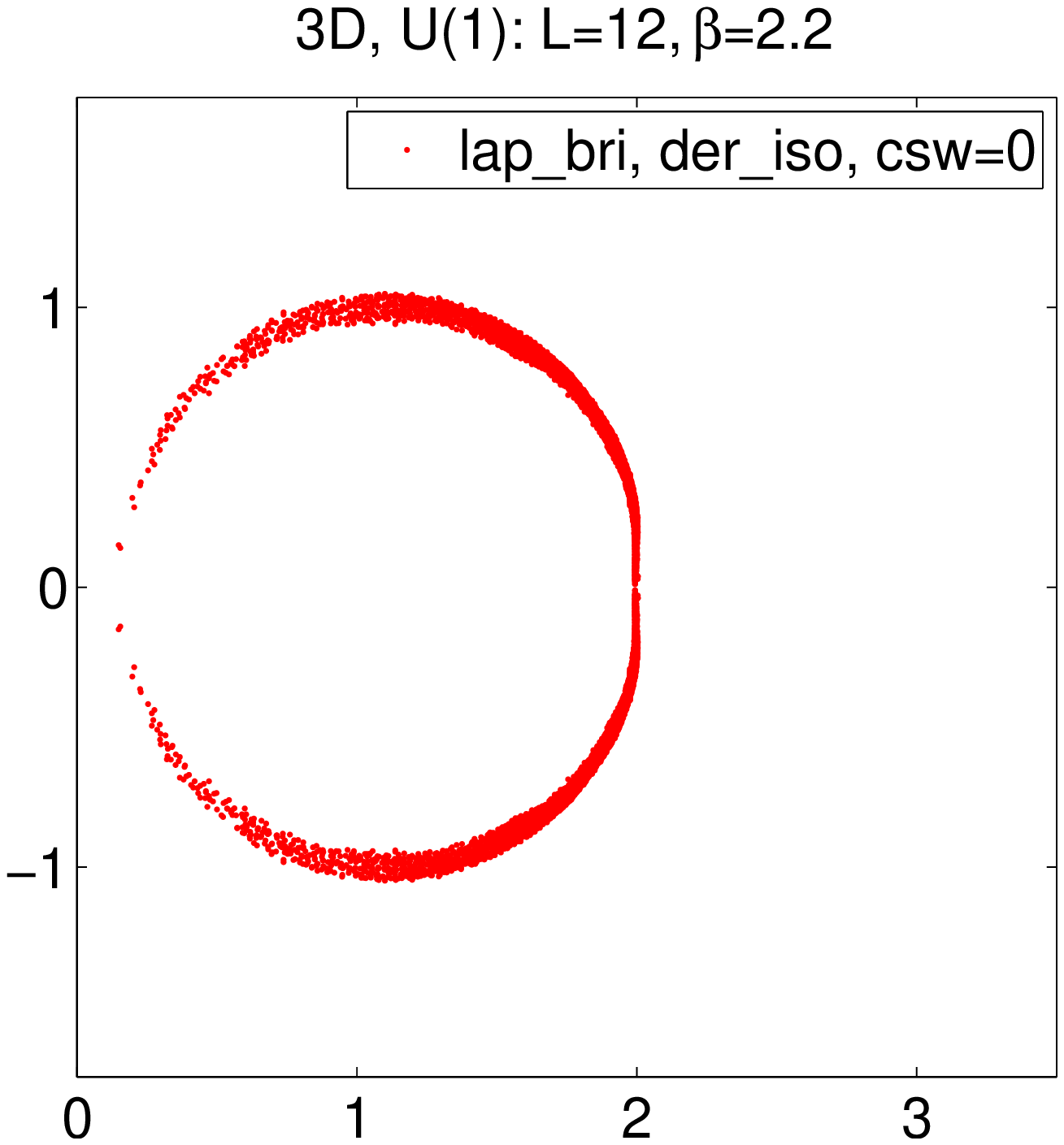,height=5.7cm}\\
\epsfig{file=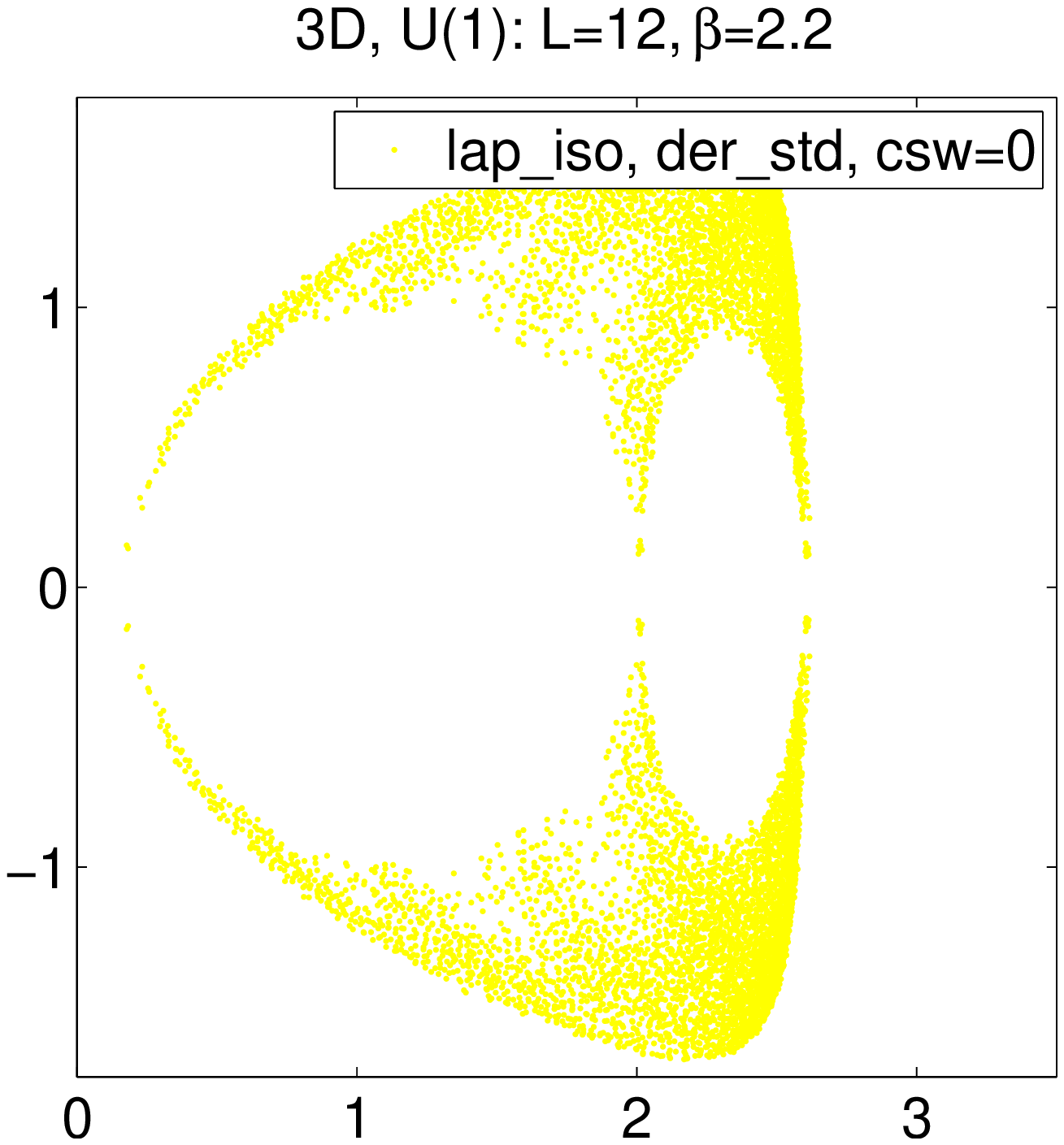,height=5.7cm}
\epsfig{file=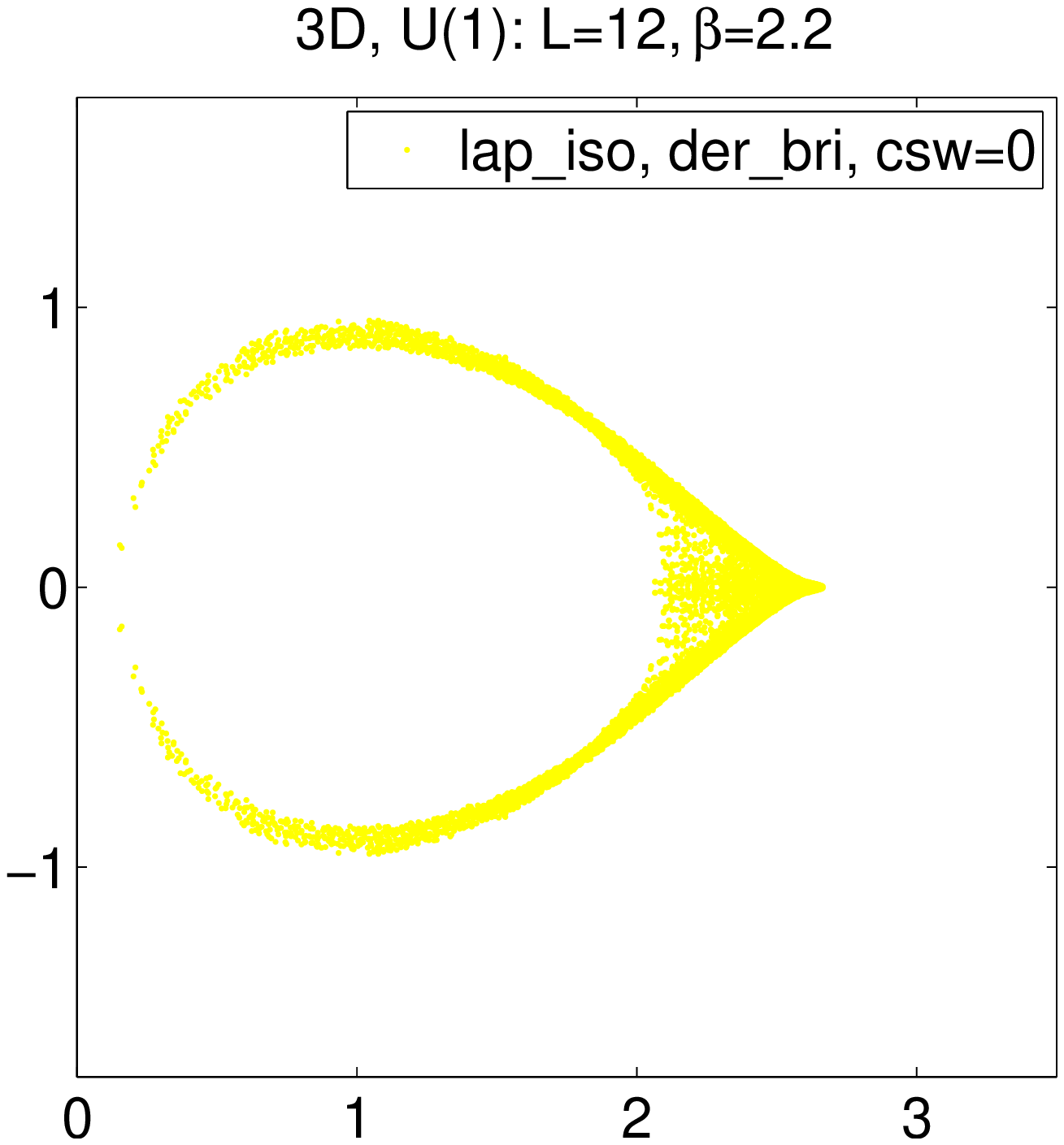,height=5.7cm}
\epsfig{file=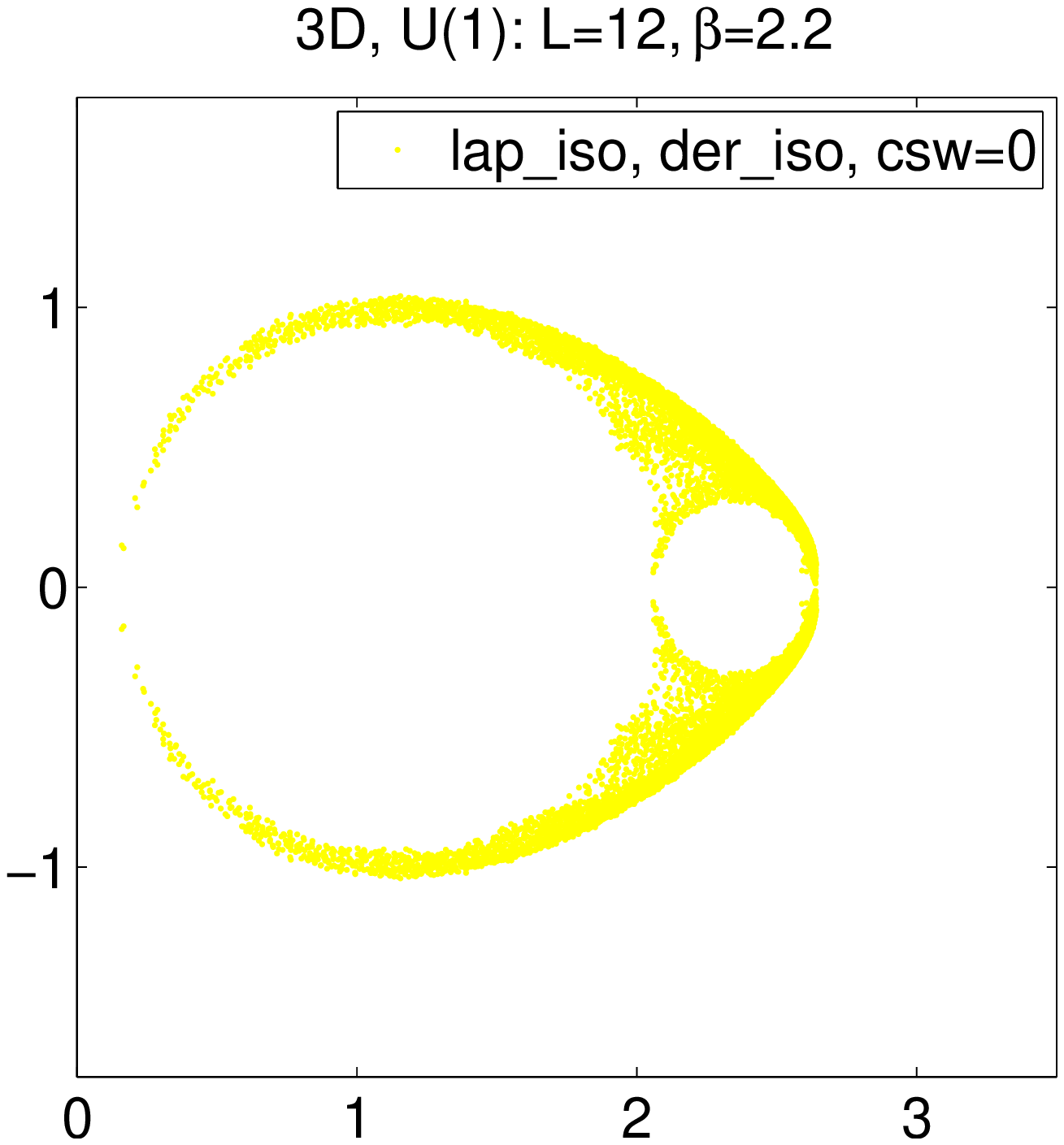,height=5.7cm}
\caption{\label{fig:spec_3D_csw0}\sl
Eigenvalue spectra of all operators considered in 3D with $c_\mr{SW}\!=\!0$.}
\end{figure}

In Fig.\,\ref{fig:spec_3D_csw0} the eigenvalues of the 12 operators without
improvement ($c_\mr{SW}\!=\!0$) are shown.
This time we refrain from showing the counterpart with improvement, as the
difference is (again) minor.
Just as in the previous section, the Laplacian features as the row index of the
panel, and the derivative as the column index.
Out of these 12 constructions, 9 are undoubled fermion operators, while 3 yield
four species in the continuum limit.
The gross features of most operators are rather similar to those in the 2D
case, which was discussed in great detail above.
Perhaps the most significant difference is that the operators with the
standard Laplacian (first row) have branches at $\mr{Re}(z)\!\simeq\!0,2,4,6$,
with multiplicities $1,3,3,1$, respectively.
If the standard Laplacian is replaced by the Brillouin Laplacian (third row) or
the isotropic Laplacian (fourth row), the doublers are lifted more equally;
in particular the former alternative arranges them all near
$\mr{Re}(z)\!\simeq\!2$.
With the tilted Laplacian and the standard derivative, the 8 species arrange
themselves in groups of 4 at $\mr{Re}(z)\!\simeq\!0$ and
$\mr{Re}(z)\!\simeq\!2$, respectively.
As soon as the standard derivative is replaced by the Brillouin or isotropic
variety, some of the 4 would-be-physical modes cross over to the unphysical
side so quickly, that the resulting operator is barely usable.
Looking at the whole figure, one would say that the combination
($\lap^\mr{bri}$, $\nab^\mr{iso}$) fares best in the sense that its eigenvalue
spectrum is reasonably circular.


\subsection{Free field dispersion relations in 3D}

\begin{figure}[!p]
\centering
\epsfig{file=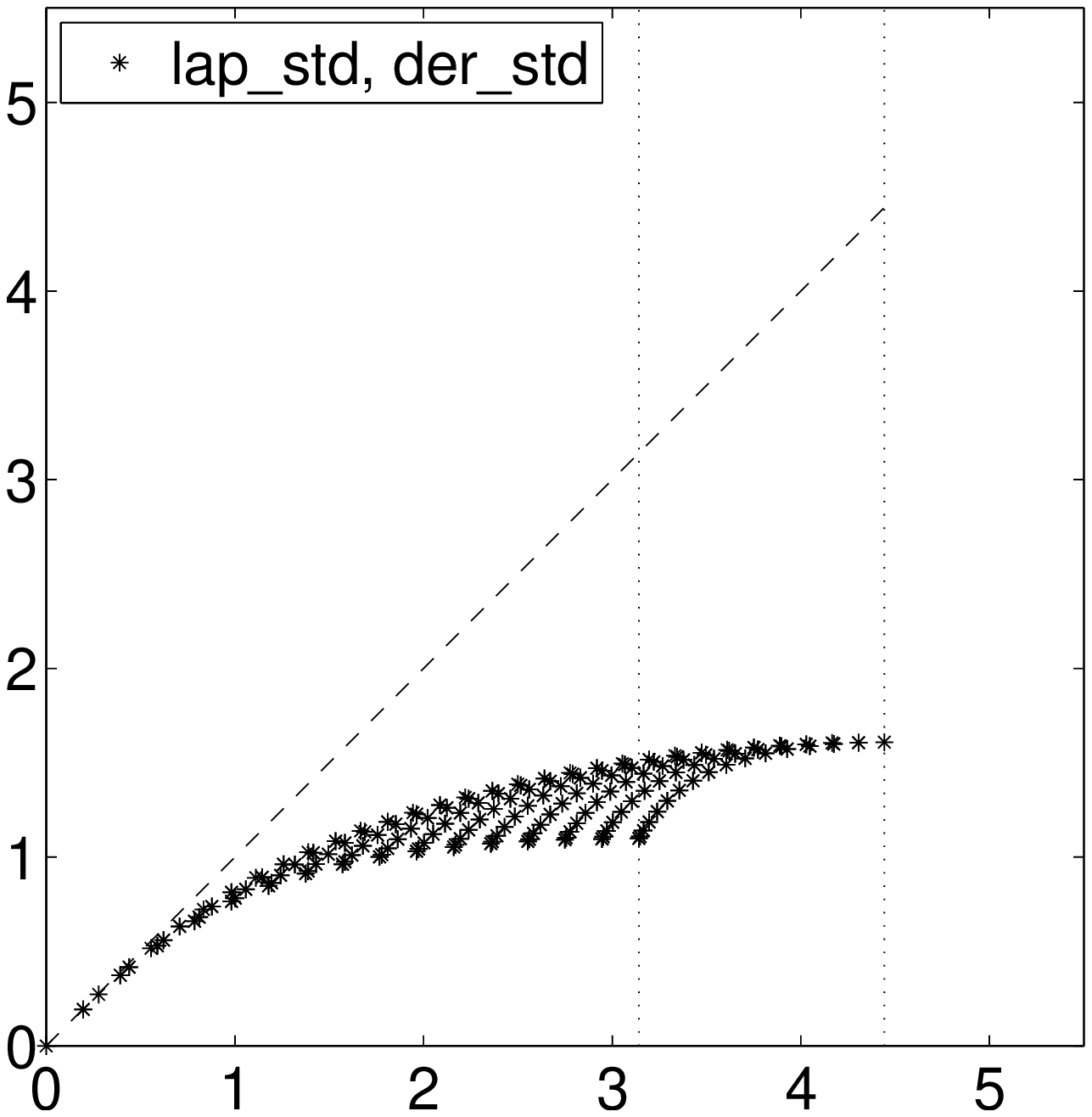,height=5.7cm}
\epsfig{file=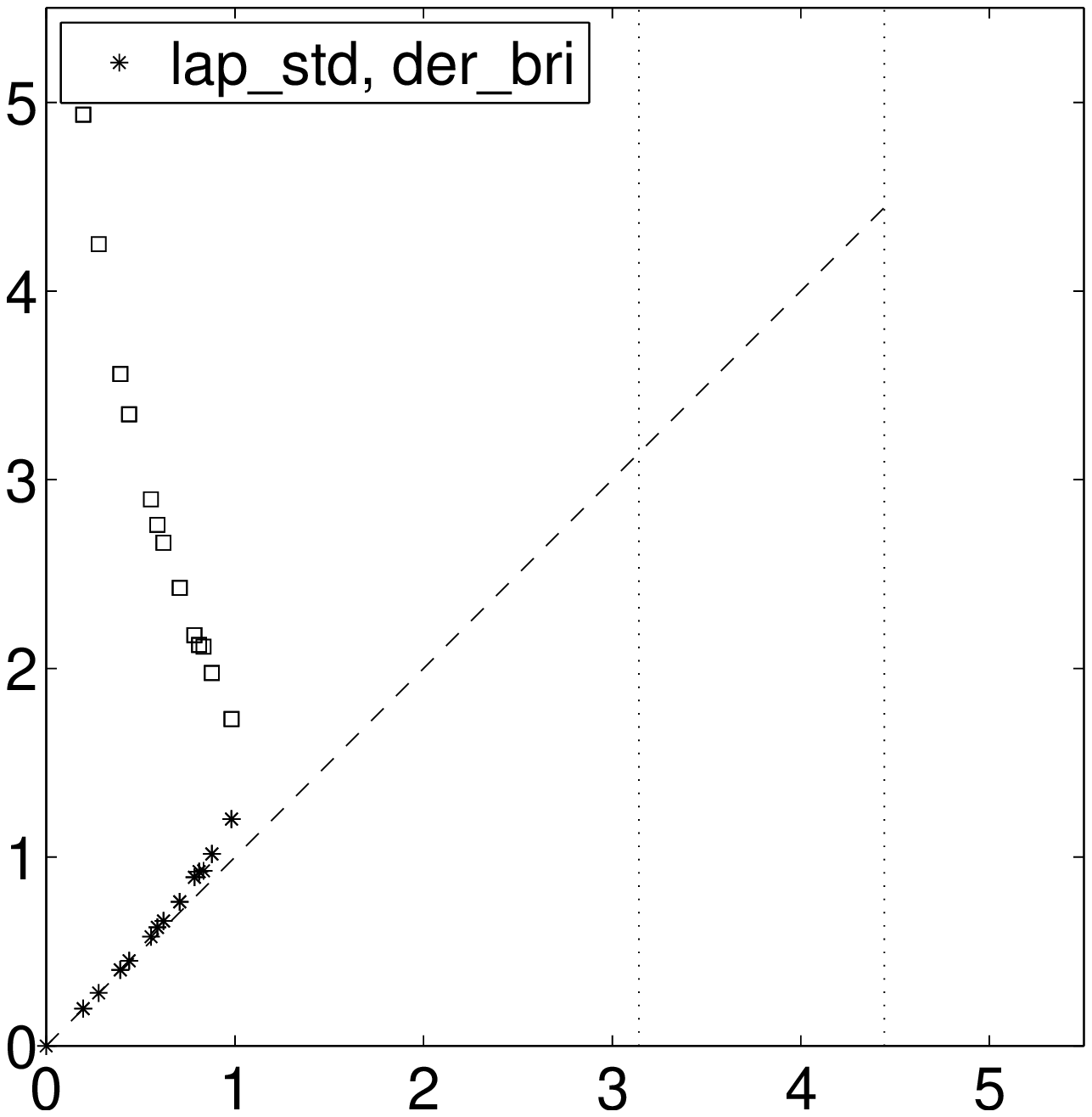,height=5.7cm}
\epsfig{file=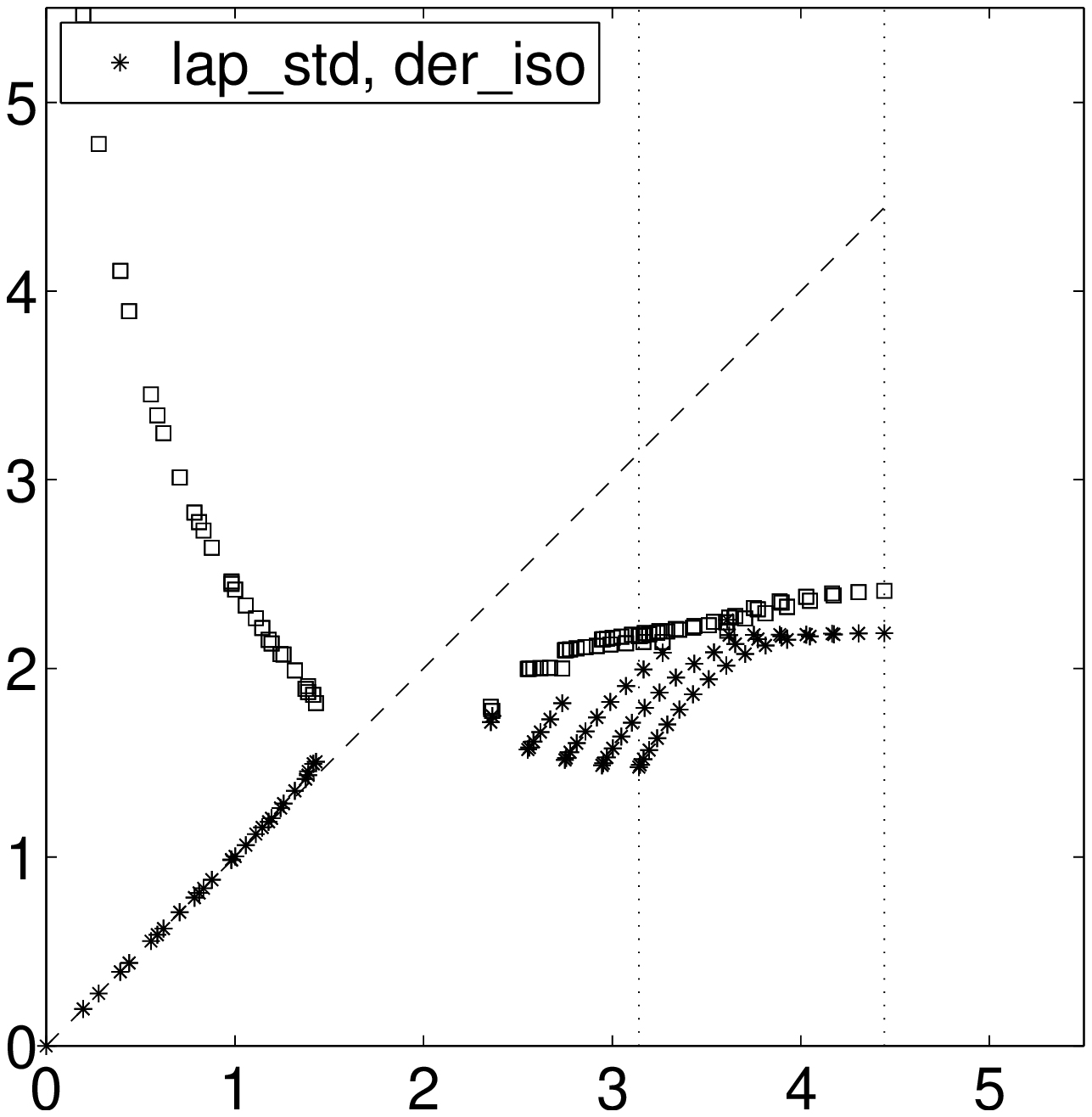,height=5.7cm}\\
\epsfig{file=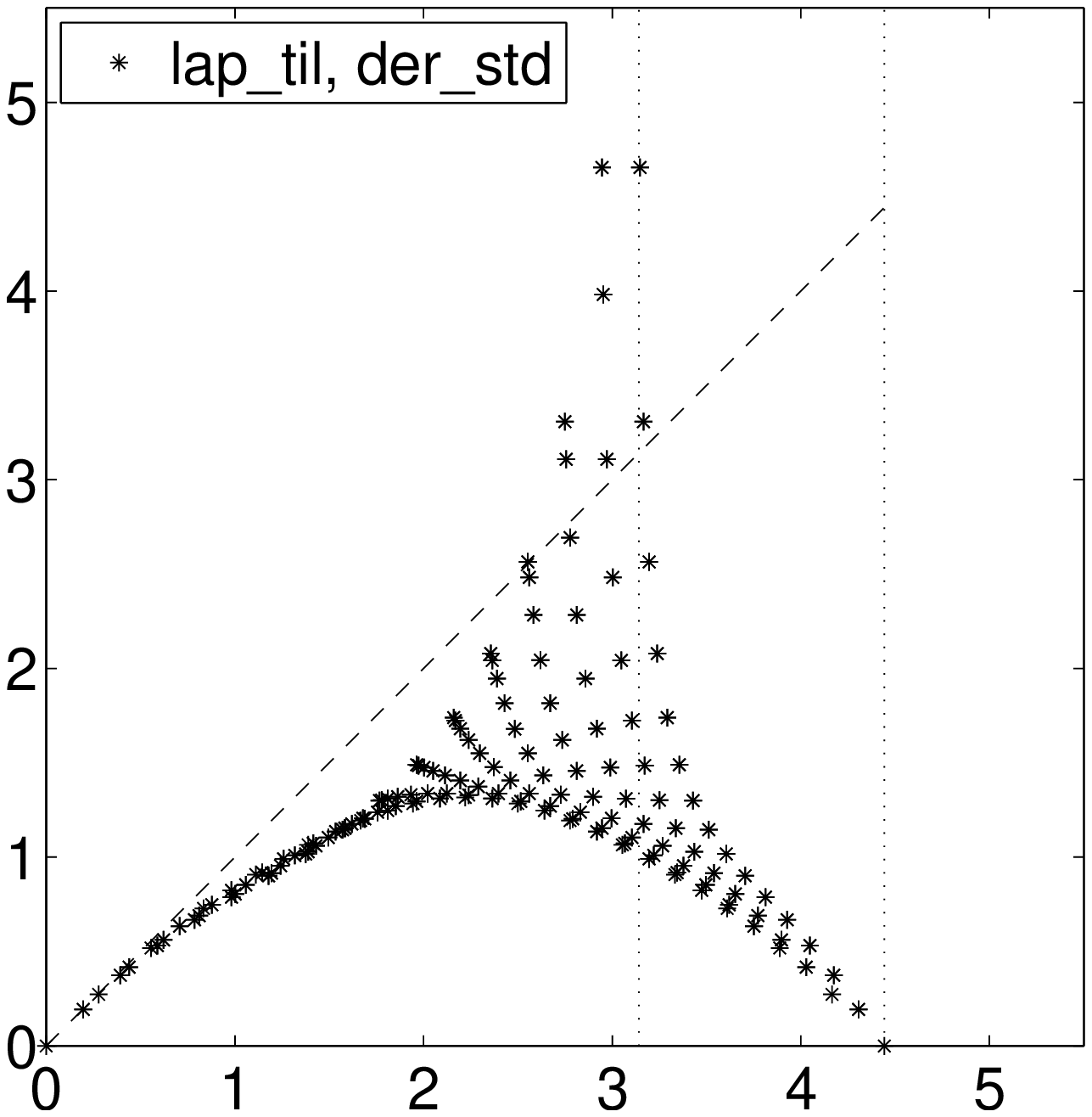,height=5.7cm}
\epsfig{file=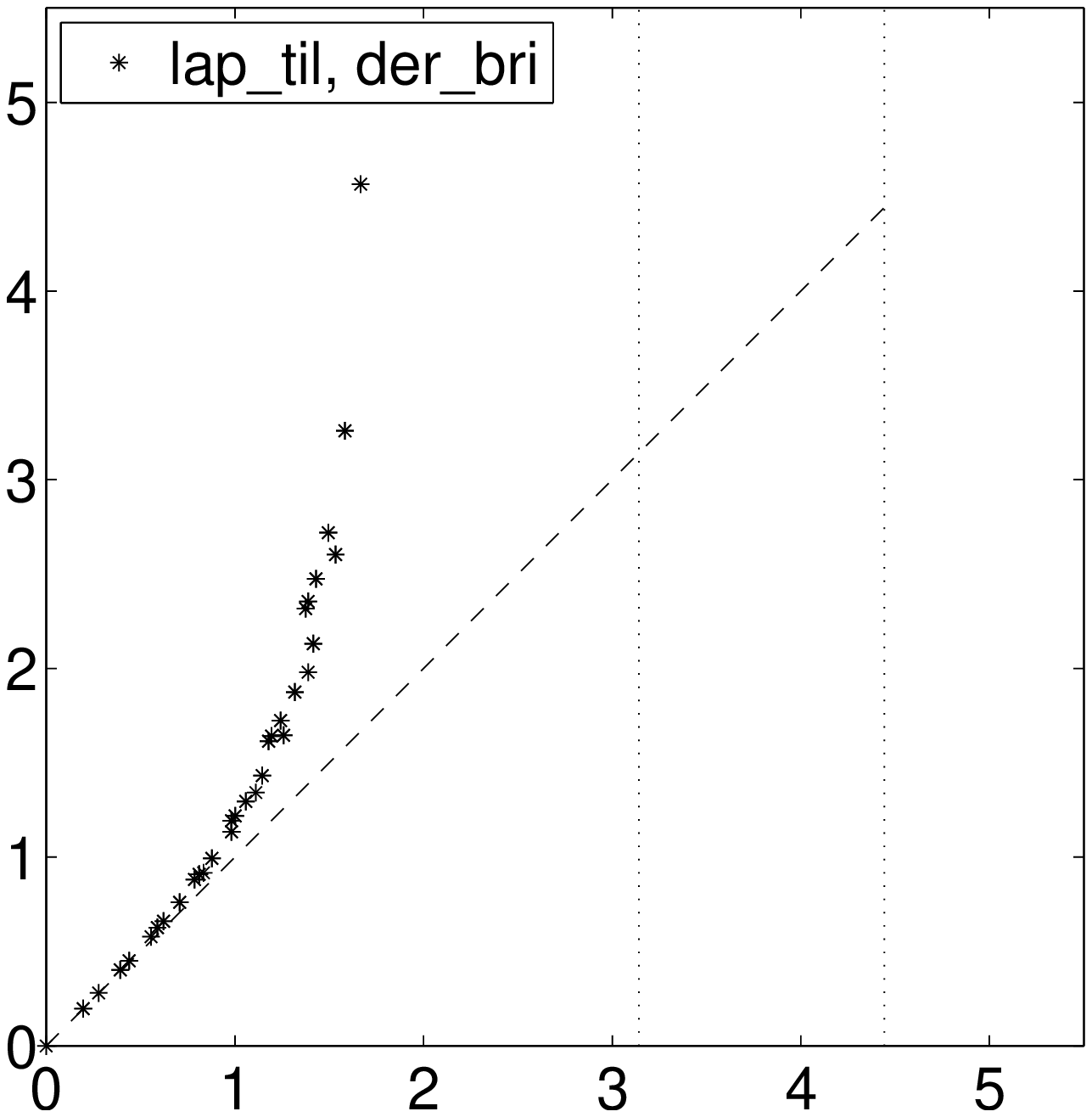,height=5.7cm}
\epsfig{file=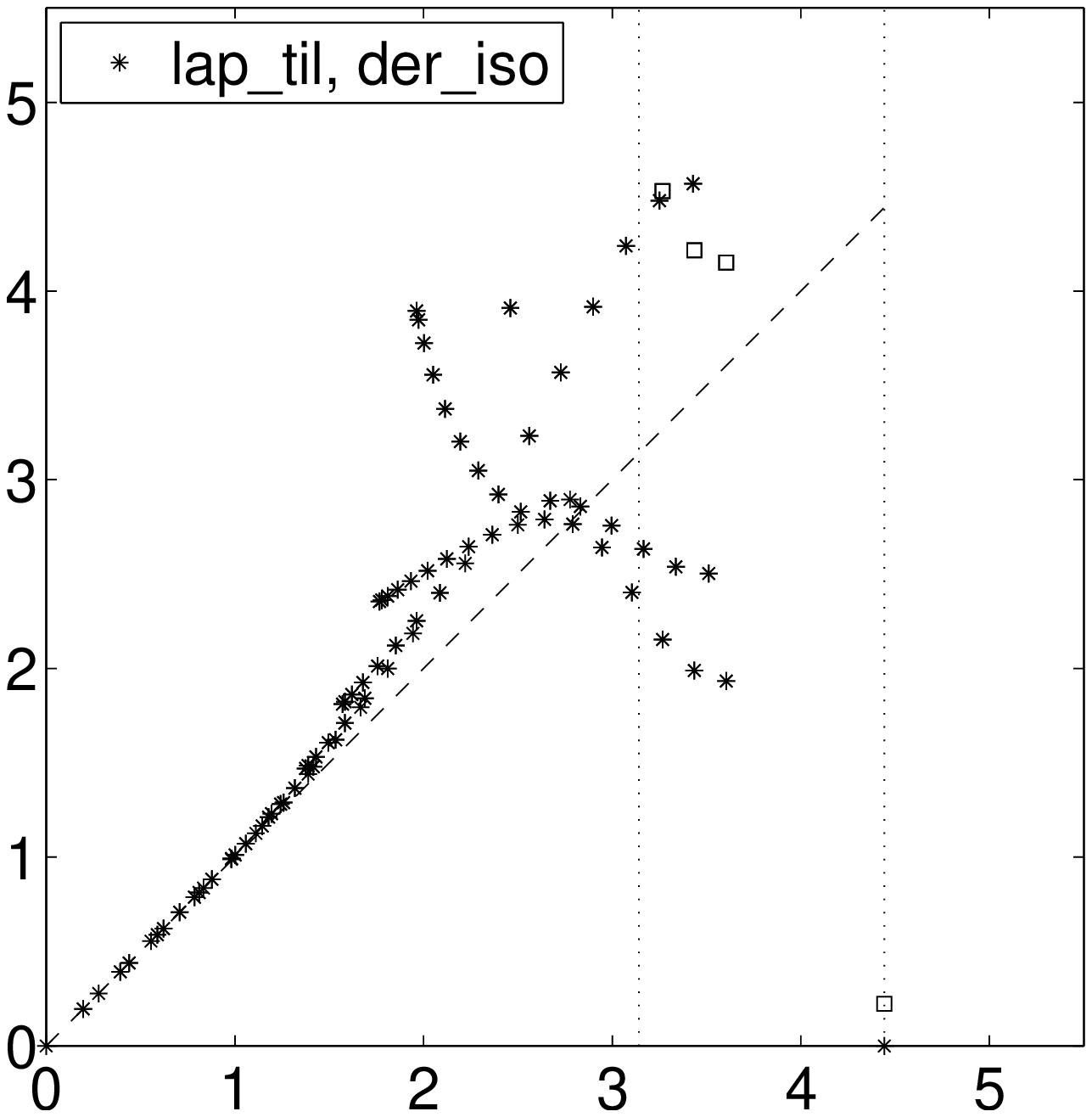,height=5.7cm}\\
\epsfig{file=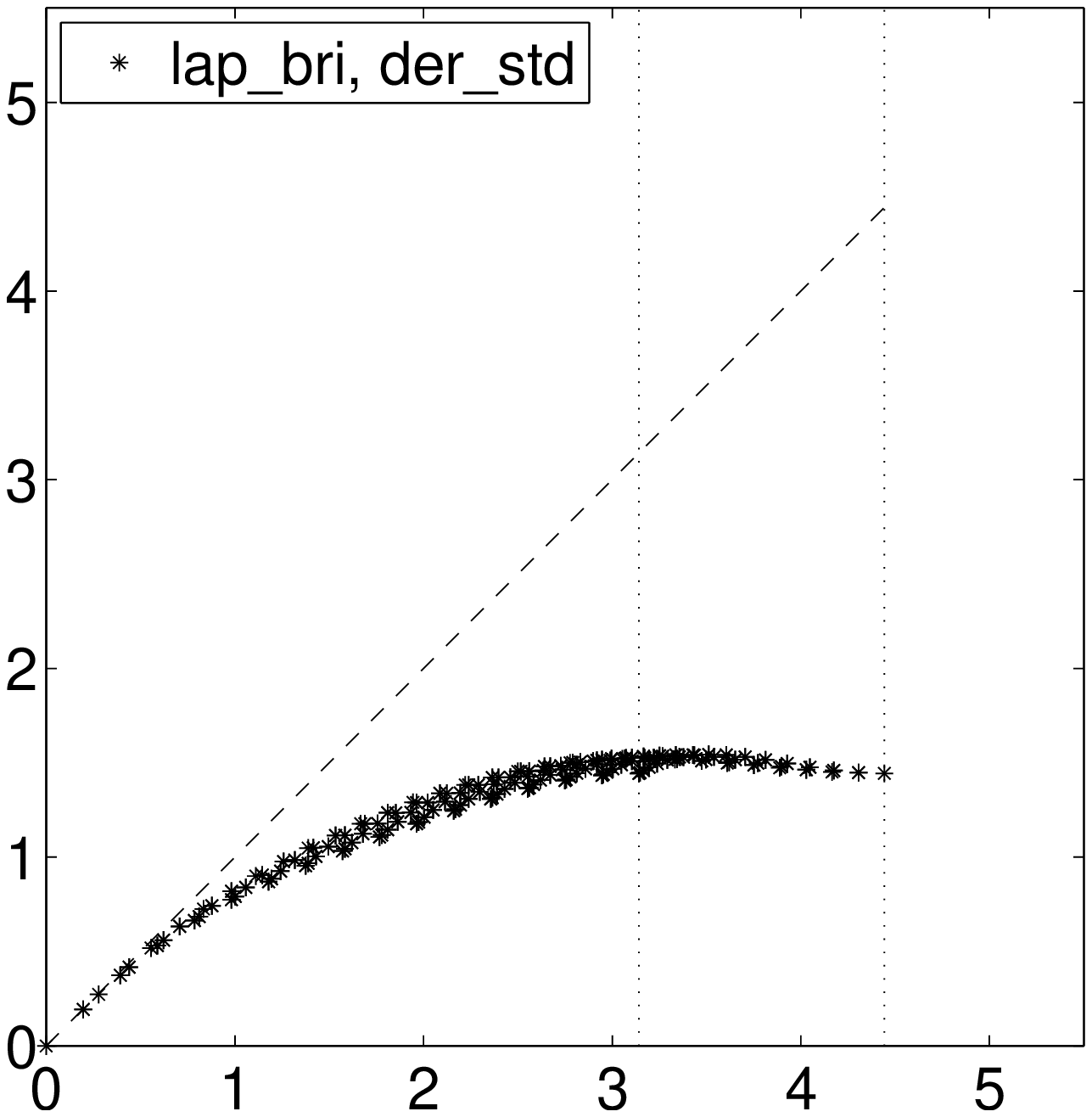,height=5.7cm}
\epsfig{file=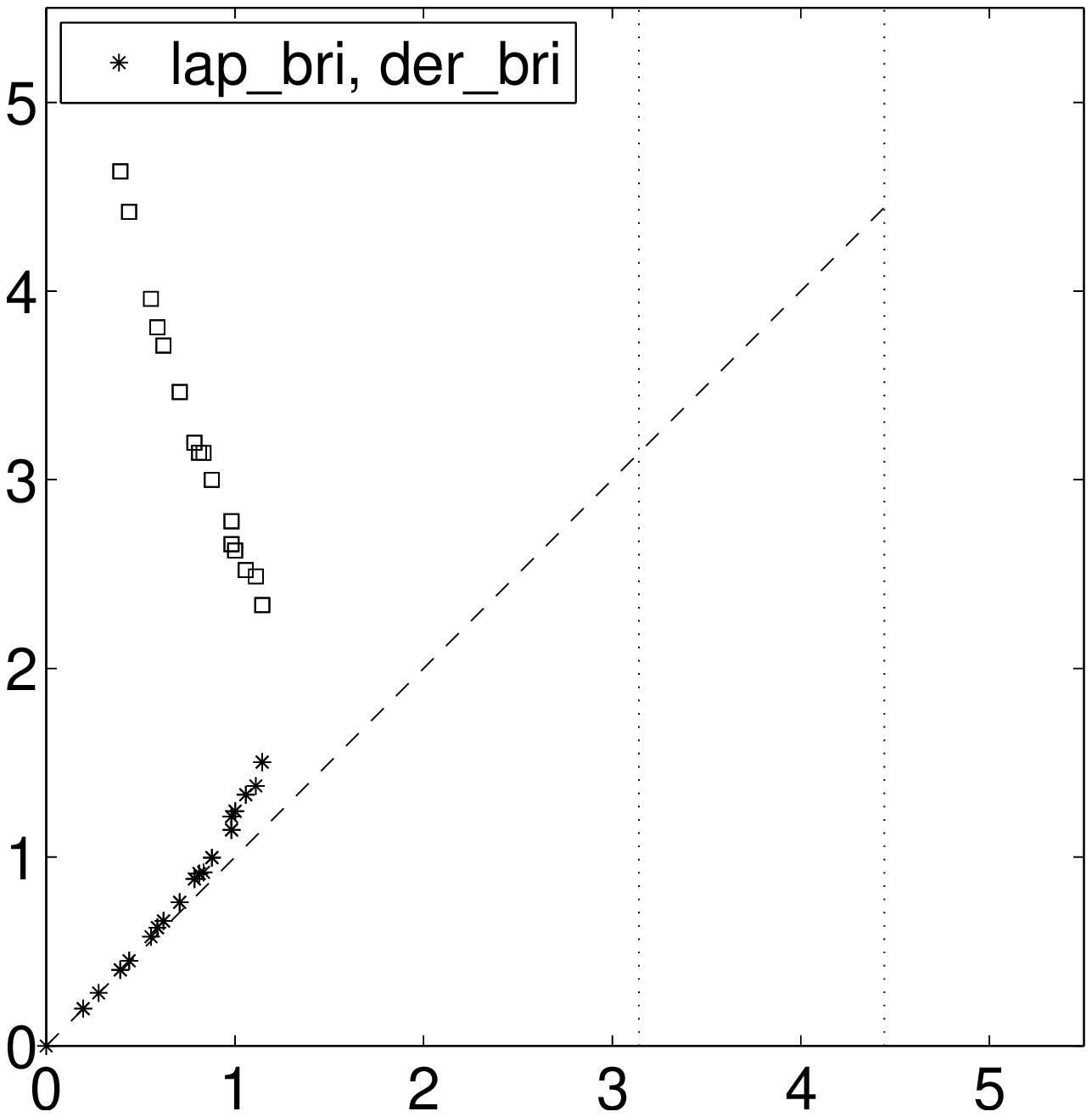,height=5.7cm}
\epsfig{file=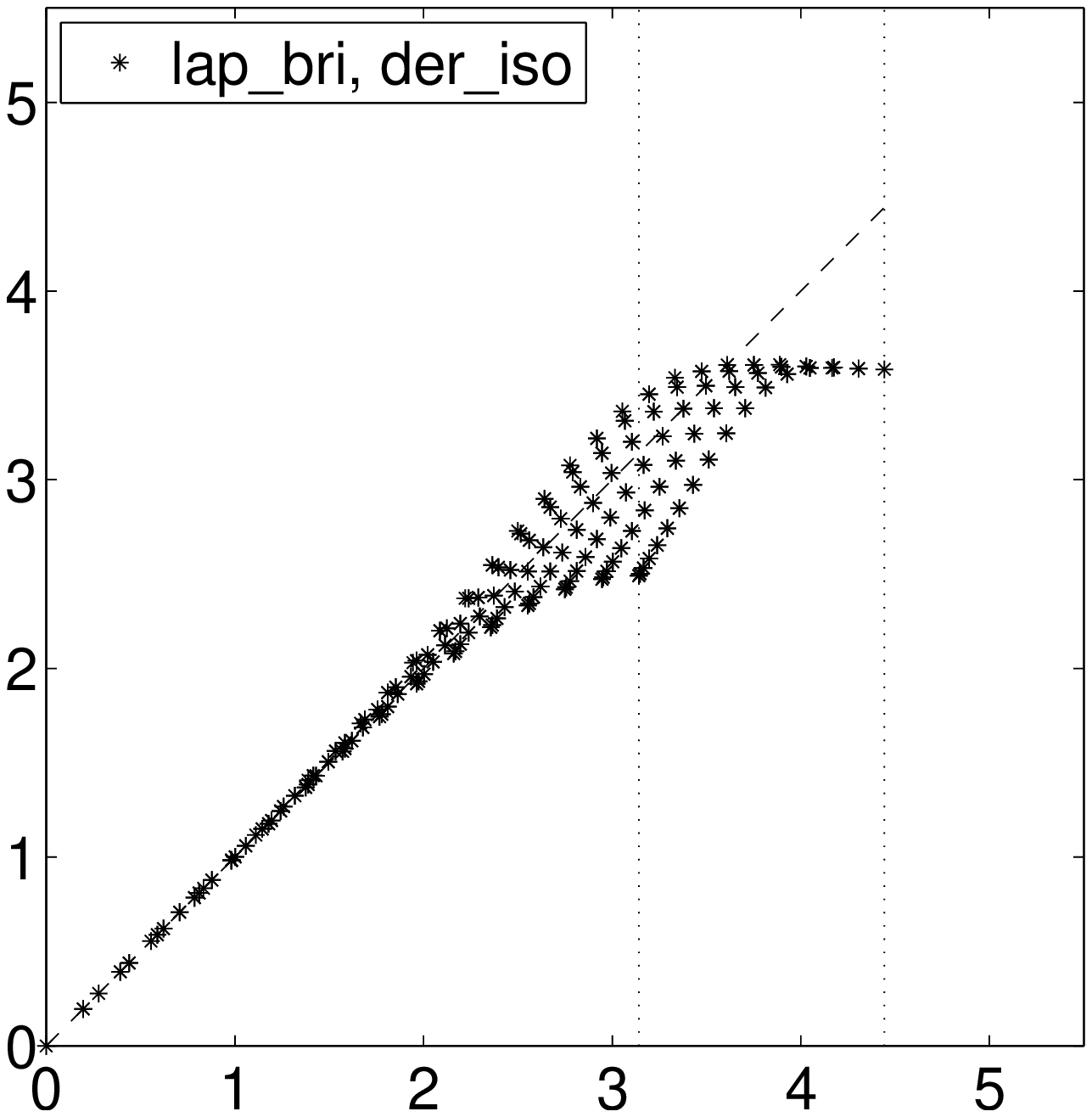,height=5.7cm}\\
\epsfig{file=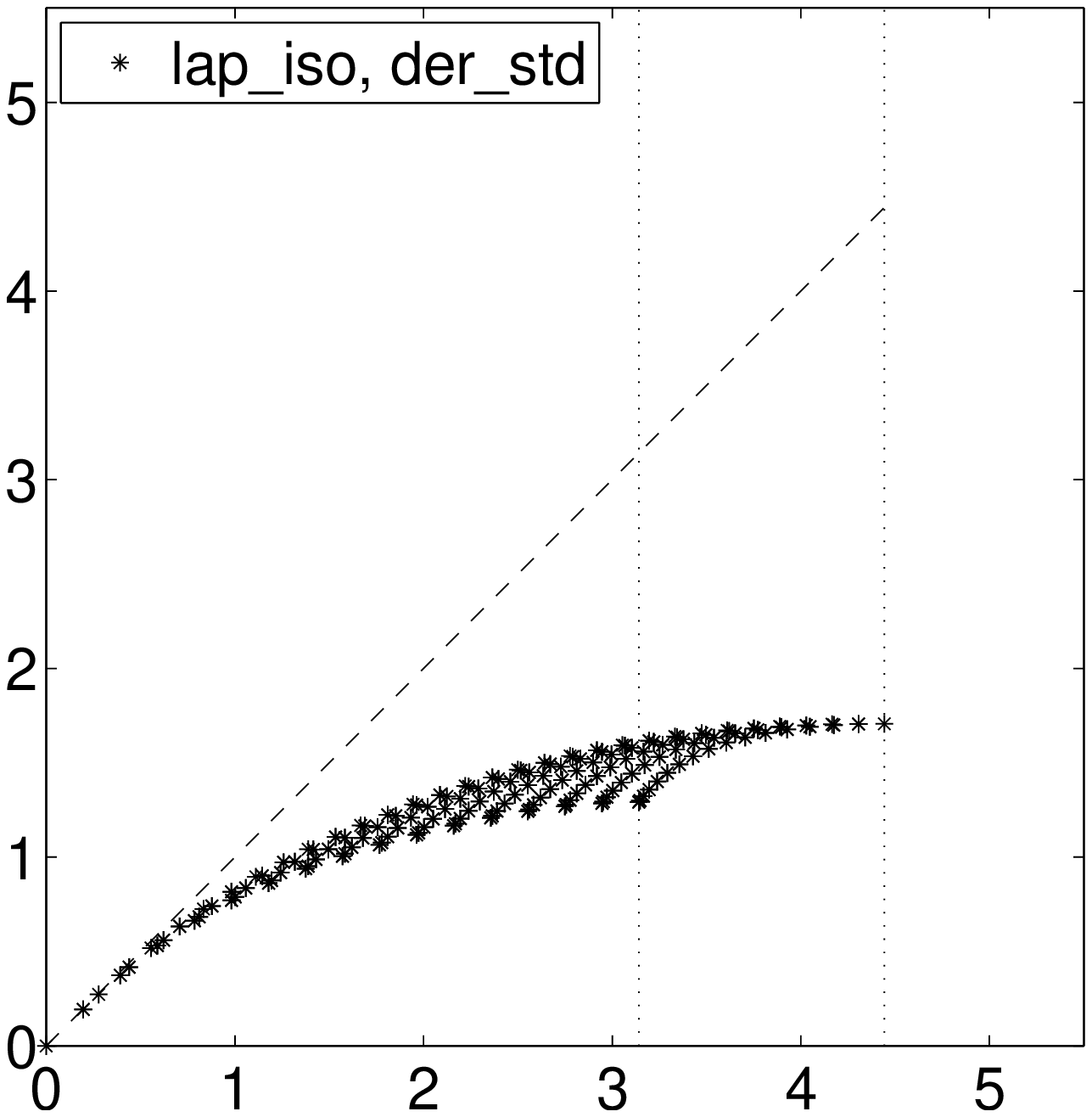,height=5.7cm}
\epsfig{file=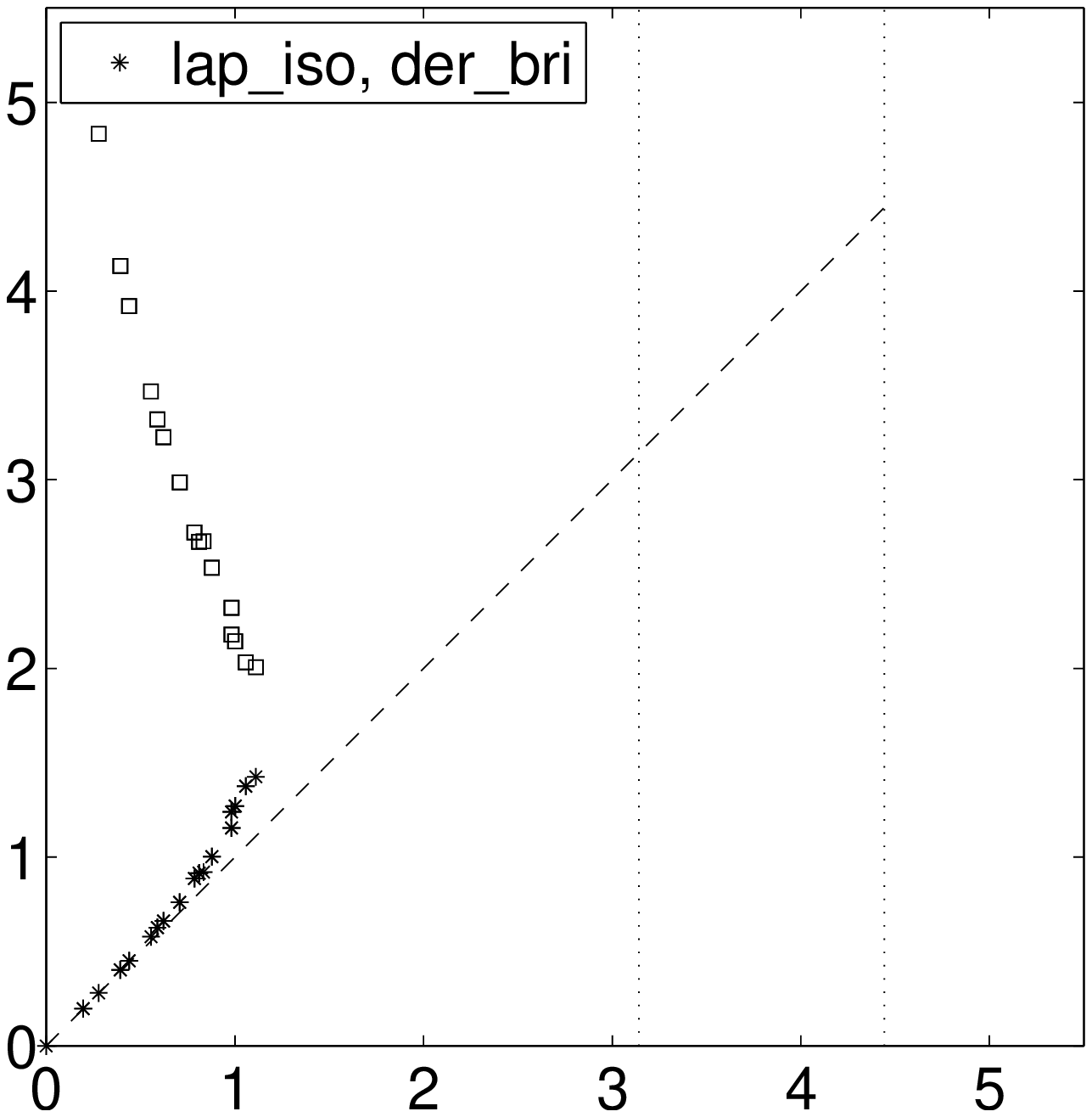,height=5.7cm}
\epsfig{file=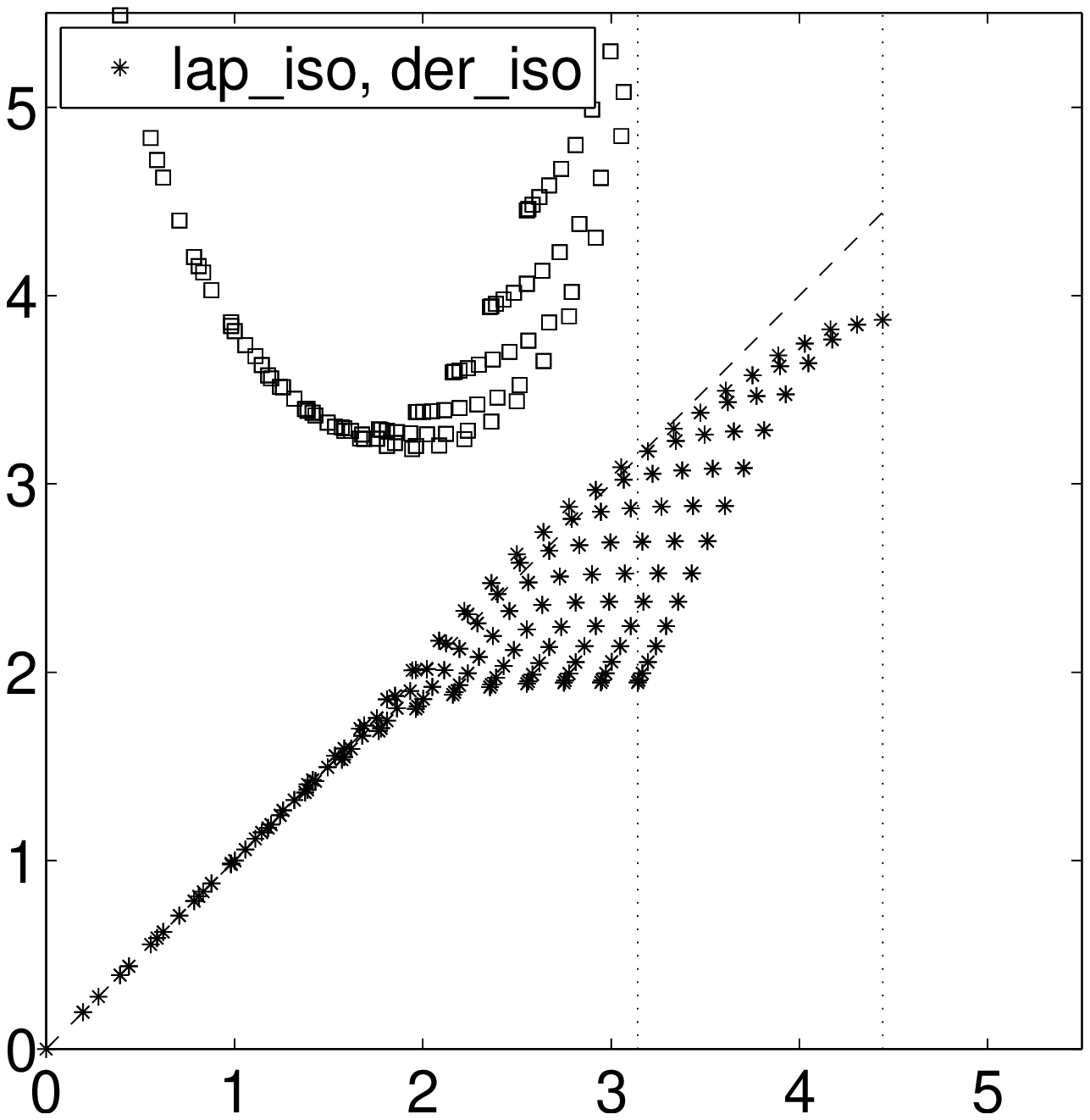,height=5.7cm}
\caption{\label{fig:disp_3D}\sl
Free-field dispersion relations of all operators considered in 3D, where
$|p|_\mr{max}\!=\!\sqrt{2}\pi/a$.}
\end{figure}

In Fig.\,\ref{fig:disp_3D} we show, for each operator, the real solutions for
$r\!=\!1$ and $m\!=\!0$ over half the Brillouin zone on a 3D lattice with
$L/a\!=\!32$.
The dispersion relation of the standard Wilson operator ($\lap^\mr{std}$,
$\nab^\mr{std}$) deviates soon from the dashed line, which corresponds to the
continuum dispersion relation; in particular towards the boundary of the
Brillouin zone the distortion is significant.
In 3D the dispersion relation is no longer a simple curve, it depends on the
orientation of the spatial momentum.
If $\mb{p}$ is chosen on axis, the 2D dispersion relation is reproduced.
The latter ends at $\sqrt{2}\pi/a$ and features as an embedding curve to the 3D
dispersion relation which now reaches out to $\sqrt{3}\pi/a$.
Again, some operators admit a second real solution (open boxes) which decouples
in the continuum, and some operators have, for certain combinations of
$(p_1,p_2)$, only complex solutions.
Overall, it is clear that the combination ($\lap^\mr{bri}$, $\nab^\mr{iso}$)
fares best in the sense that its dispersion relation is closest to the one in
the continuum.


\section{Construction and main features in 4D}


\subsection{Summary of 4D Laplace stencils}

The ``standard'' stencil of the Laplacian in 4D and the ``tilted'' variety
(as defined in App.\,C) have the Fourier space representation
\bea
a^2\!\hat\lap^\mr{\!std\!}(k_1,k_2,k_3,k_4)\!&\!=\!&\!
2\cos(k_1)+2\cos(k_2)+2\cos(k_3)+2\cos(k_4)-8
\nonumber
\\
\!&\!=\!&\!-4\sin^2(k_1/2)-4\sin^2(k_2/2)-4\sin^2(k_3/2)-4\sin^2(k_4/2)
\label{def_4Dstd}
\\
a^2\!\hat\lap^\mr{til}(k_1,k_2,k_3,k_4)\!&\!=\!&\!
2\cos(k_1)\cos(k_2)\cos(k_3)\cos(k_4)-2
\nonumber
\\
\!&\!=\!&\!
32\cos^2(k_1/2)\cos^2(k_2/2)\cos^2(k_3/2)\cos^2(k_4/2)
-16\cos^2(k_1/2)\cos^2(k_2/2)
\nonumber
\\
&&
\cos^2(k_3/2)-...
+8\cos^2(k_1/2)\cos^2(k_2/2)+...
-4\cos^2(k_1)-...
\label{def_4Dtil}
\eea
respectively, with the ellipses denoting cyclic permutations.
From the stencil notation in App.\,C it is easy to see that the former has
only 1-hop contributions, while the latter has only 4-hop contributions
(apart from the central element).
For asymptotically small momenta they both reproduce the continuum relation
$\hat\lap=p_1^2\!+\!p_2^2\!+\!p_3^2\!+\!p_4^2$, but the ``tilted'' stencil has
seven additional zeros at the boundary of the Brillouin zone
[$\hat\lap^\mr{til}$ vanishes at
$(k_1,k_2,k_3,k_4)-(\pi,\pi,\pi,\pi)/2=(\pm\pi,\pm\pi,\pm\pi,\pm\pi)/2$ with
an even number of minus signs].

In 4D the discretizations of the continuum Laplacian which are analogous to
(\ref{def_2Dbri}, \ref{def_2Diso}) read
\bea
a^2\hat\lap^\mr{bri}(k_1,k_2,k_3,k_4)&=&[\cos(k_1)\cos(k_2)\cos(k_3)\cos(k_4)
+\cos(k_1)\cos(k_2)\cos(k_3)+...
\nonumber
\\
&&
+\cos(k_1)\cos(k_2)+...
+\cos(k_1)+...-15]/4
\nonumber
\\
&=&
4\cos^2(k_1/2)\cos^2(k_2/2)\cos^2(k_3/2)\cos^2(k_4/2)-4
\label{def_4Dbri}
\\
a^2\hat\lap^\mr{iso}(k_1,k_2,k_3,k_4)&=&
[2\cos(k_1)\cos(k_2)\cos(k_3)\cos(k_4)
+7\cos(k_1)\cos(k_2)\cos(k_3)+...
\nonumber
\\
&&
+20\cos(k_1)\cos(k_2)+...
+25\cos(k_1)+...-250]/54
\nonumber
\\
&=&[
16\cos^2(k_1/2)\cos^2(k_2/2)\cos^2(k_3/2)\cos^2(k_4/2)
\nonumber
\\
&&
+20\cos^2(k_1/2)\cos^2(k_2/2)\cos^2(k_3/2)+...
\nonumber
\\
&&
+16\cos^2(k_1/2)\cos^2(k_2/2)+...
-16\cos^2(k_1/2)-...
-128]/27
\label{def_4Diso}
\eea
respectively, and their distinctive features are as follows.
The ``Brillouin'' Laplacian (\ref{def_4Dbri}) takes a constant value on the
entire boundary of the Brillouin zone, since
$a^2\hat\lap^\mr{bri}(k_1,k_2,k_3,k_4)=-4$ whenever one of the momenta is
$\pm\pi/a$.
On the other hand, the Laplacian (\ref{def_4Diso}) is called ``isotropic''
since $a^2\hat\lap^\mr{iso}(k_1,k_2,k_3,k_4)=
-a^2[k_1^2\!+\!k_2^2\!+\!k_3^2\!+\!k_4^2]
+a^4[k_1^2\!+\!k_2^2\!+\!k_3^2\!+\!k_4^2]^2/12+O(a^6)$
has $O(a^4)$ terms which depend only on the combination
$k_1^2\!+\!k_2^2\!+\!k_3^2\!+\!k_4^2$.
In other words, $\hat\lap^\mr{iso}(k_1,k_2,k_3,k_4)$ respects rotational
symmetry even in the leading term through which it deviates from the continuum.

In 4D there are 4 linearly independent Laplacians with (at most) an 81-point
stencil, and the 4 elements (\ref{def_4Dstd}-\ref{def_4Diso}) form a basis.
We are unaware of any systematic treatment in the literature.

\begincomment
obj1:=2*cos(k1)+2*cos(k2)+2*cos(k3)+2*cos(k4)-8;
sort(expand(subs(k1=2*k1h,k2=2*k2h,k3=2*k3h,k4=2*k4h,obj1),trig));
series(subs({k1=a*p1,k2=a*p2,k3=a*p3,k4=a*p4},obj1),a,4);

obj2:=2*cos(k1)*cos(k2)*cos(k3)*cos(k4)-2;
sort(expand(subs(k1=2*k1h,k2=2*k2h,k3=2*k3h,k4=2*k4h,obj2),trig));
series(subs({k1=a*p1,k2=a*p2,k3=a*p3,k4=a*p4},obj2),a,4);

obj3:=4*cos(k1/2)^2*cos(k2/2)^2*cos(k3/2)^2*cos(k4/2)^2-4;
sort(expand(combine(obj3,trig),trig));
series(subs({k1=a*p1,k2=a*p2,k3=a*p3,k4=a*p4},obj3),a,4);

obj4:=(cos(k1)-1)*(cos(k2)+5)*(cos(k3)+5)*(cos(k4)+5)/108:
obj4:=sort(expand(obj4+ \
 subs({k1=k2,k2=k1},obj4)+subs({k1=k3,k3=k1},obj4)+subs({k1=k4,k4=k1},obj4) \
));
sort(expand(subs(k1=2*k1h,k2=2*k2h,k3=2*k3h,k4=2*k4h,obj4),trig));
factor(series(subs({k1=a*p1,k2=a*p2,k3=a*p3,k4=a*p4},obj4),a,6));
\endcomment


\subsection{Eigenvalue spectra in 4D}

Like in the previous two sections, with four options for $\lap$ and three for
$\nab$, we can construct 12 Dirac operators and study their eigenvalue spectra.
As the gauge group is irrelevant in this step, we prepare a thermalized
background in the $U(1)$ gauge theory with $L/a\!=\!6$ at $\be\!=\!1.1$.

\begin{figure}[!p]
\centering
\epsfig{file=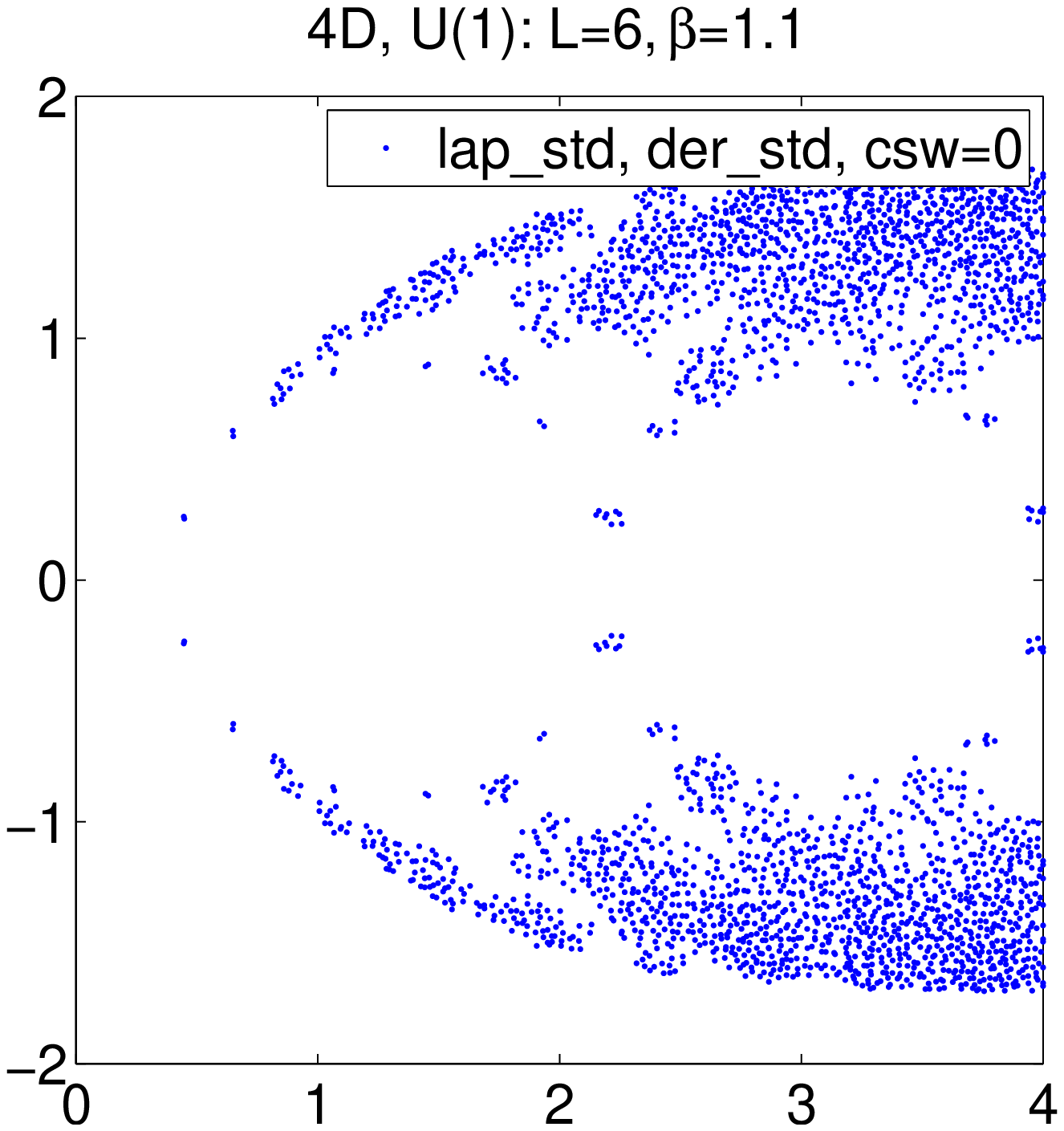,height=5.7cm}
\epsfig{file=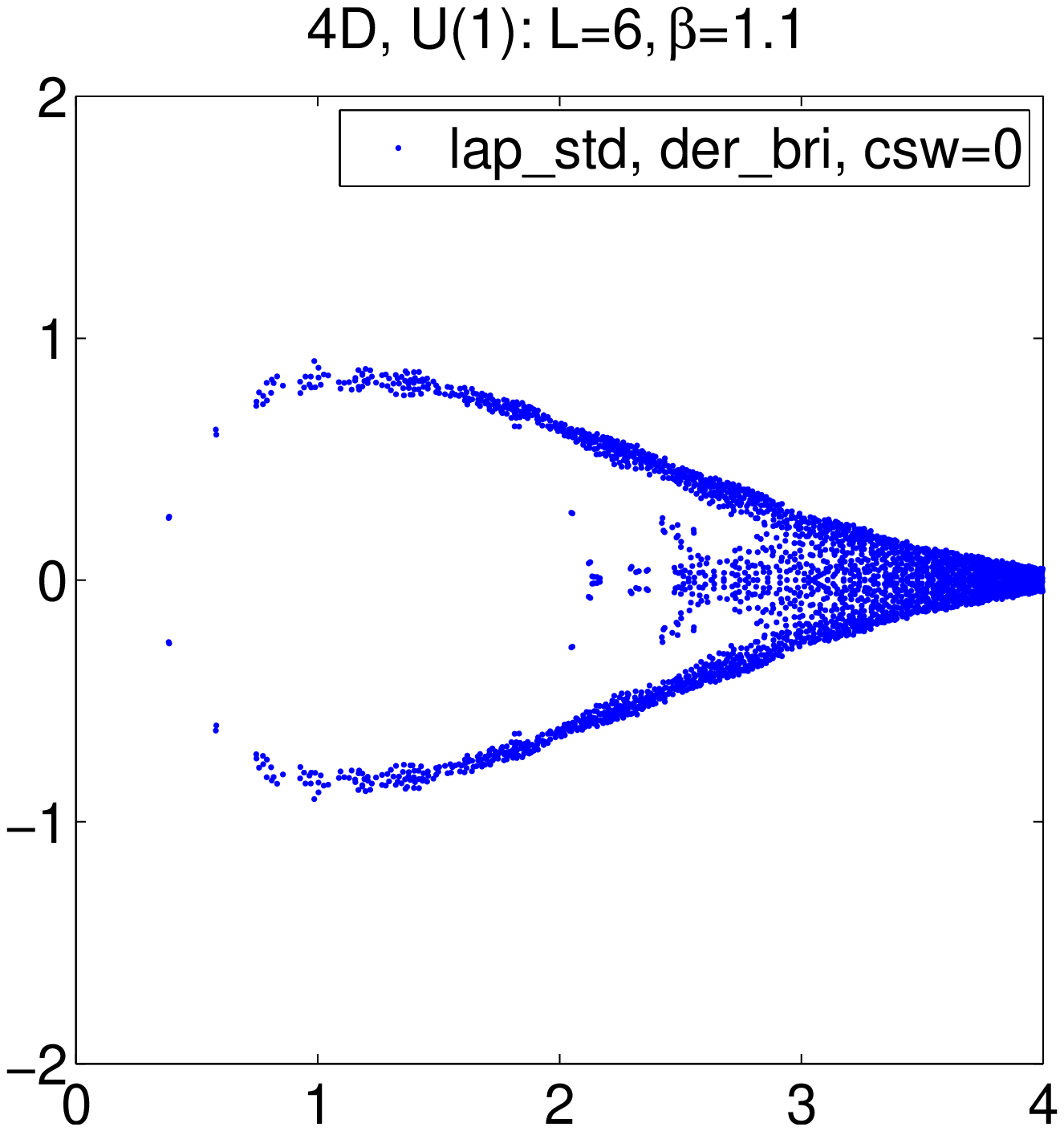,height=5.7cm}
\epsfig{file=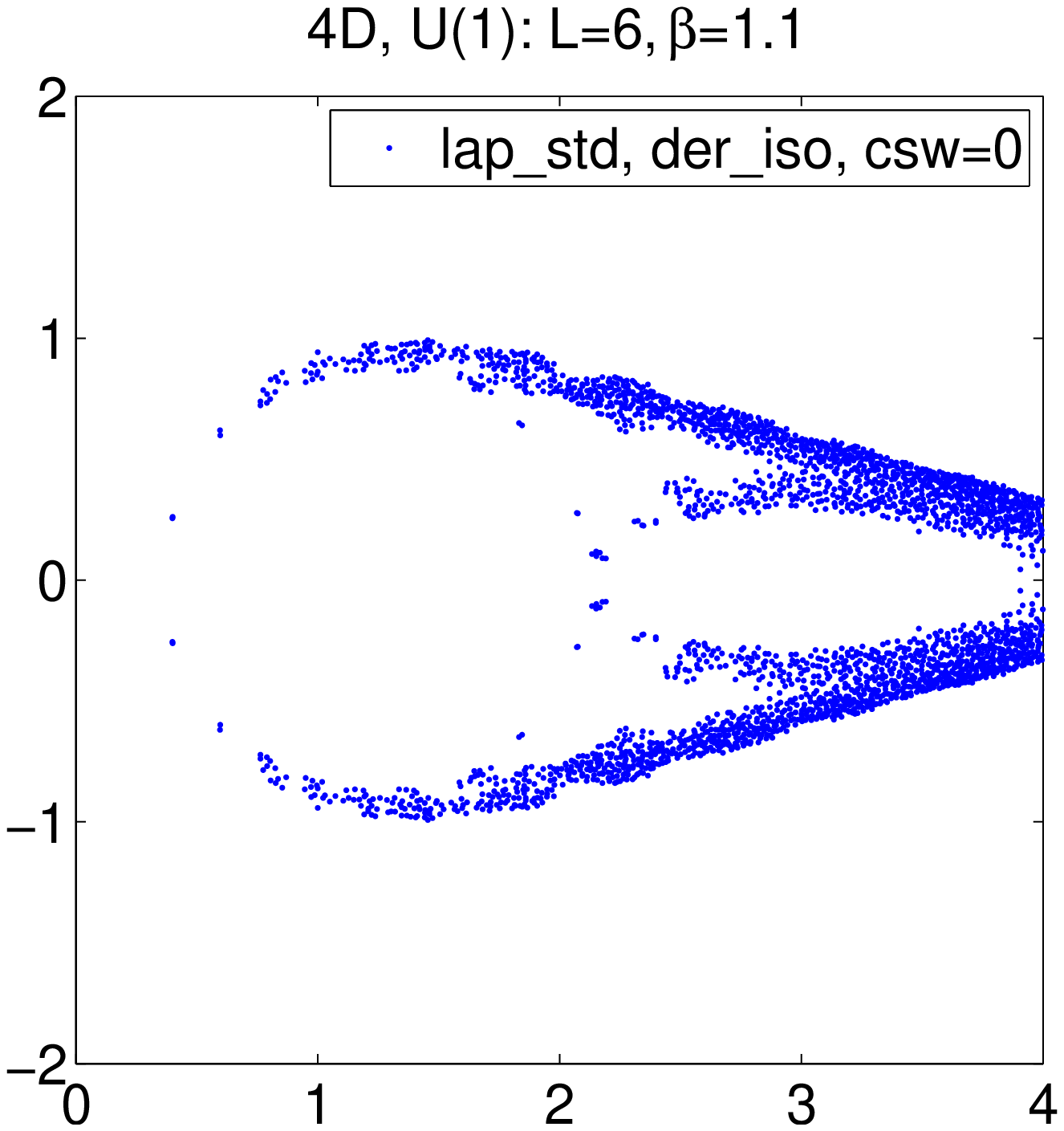,height=5.7cm}\\
\epsfig{file=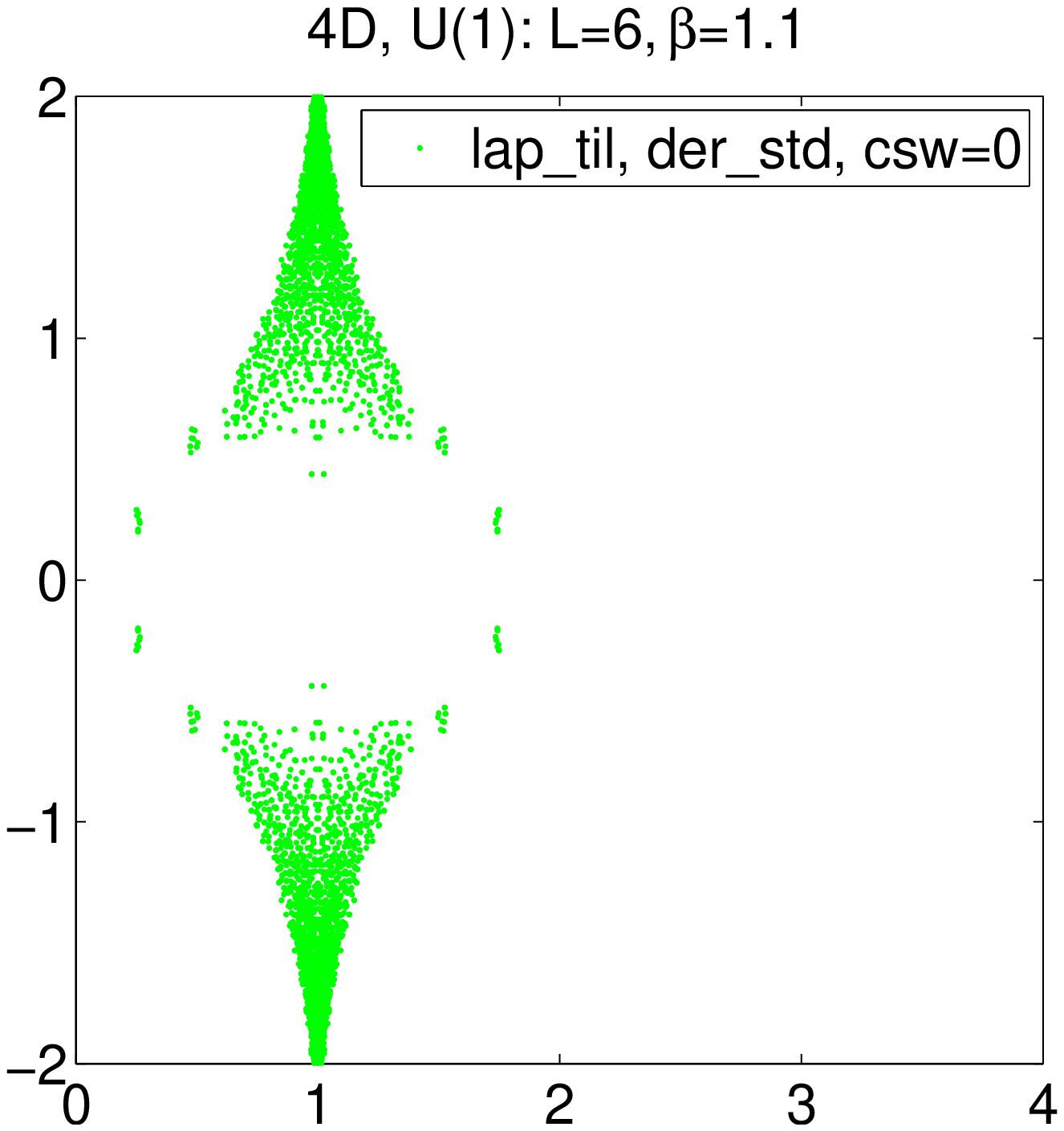,height=5.7cm}
\epsfig{file=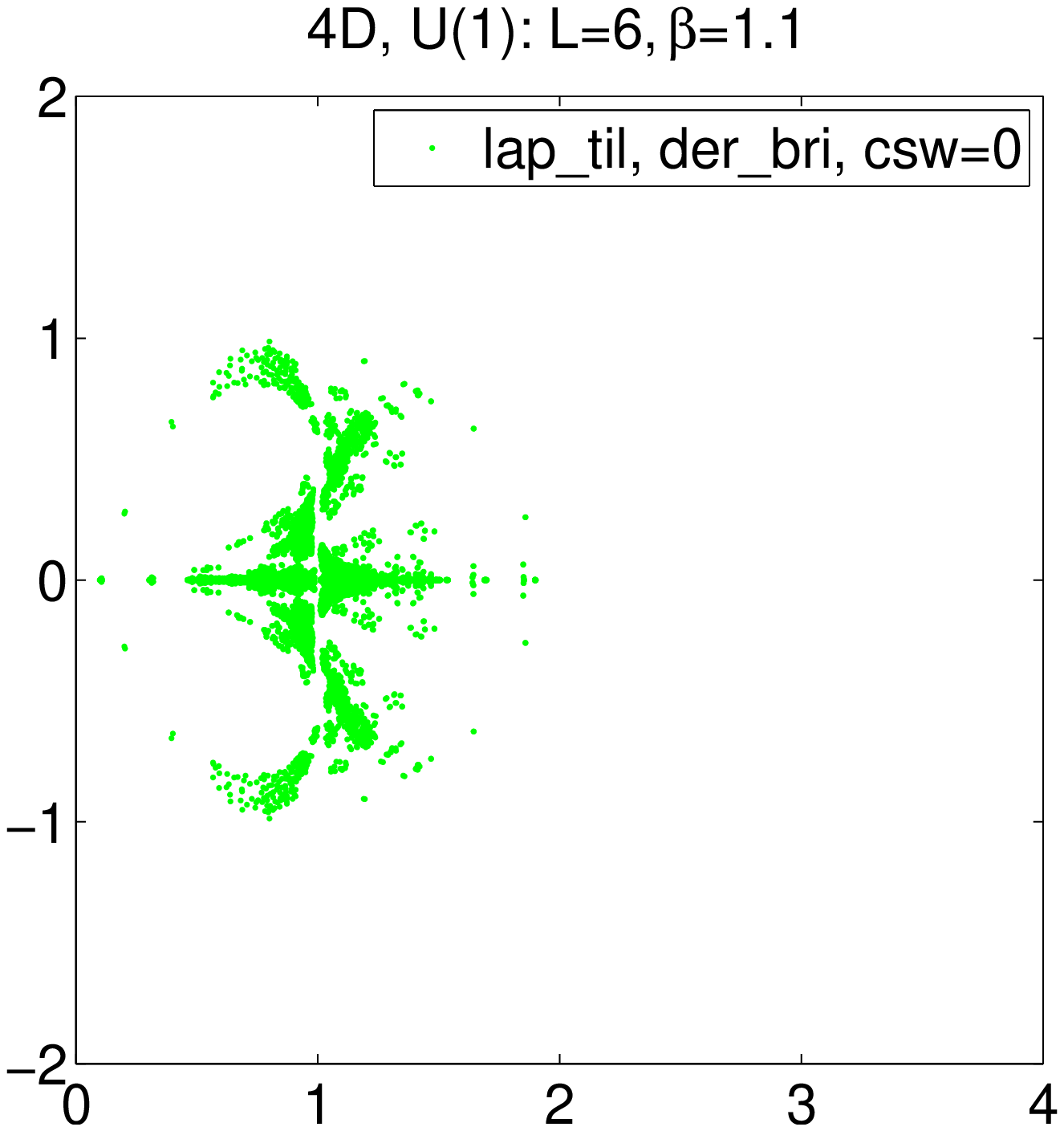,height=5.7cm}
\epsfig{file=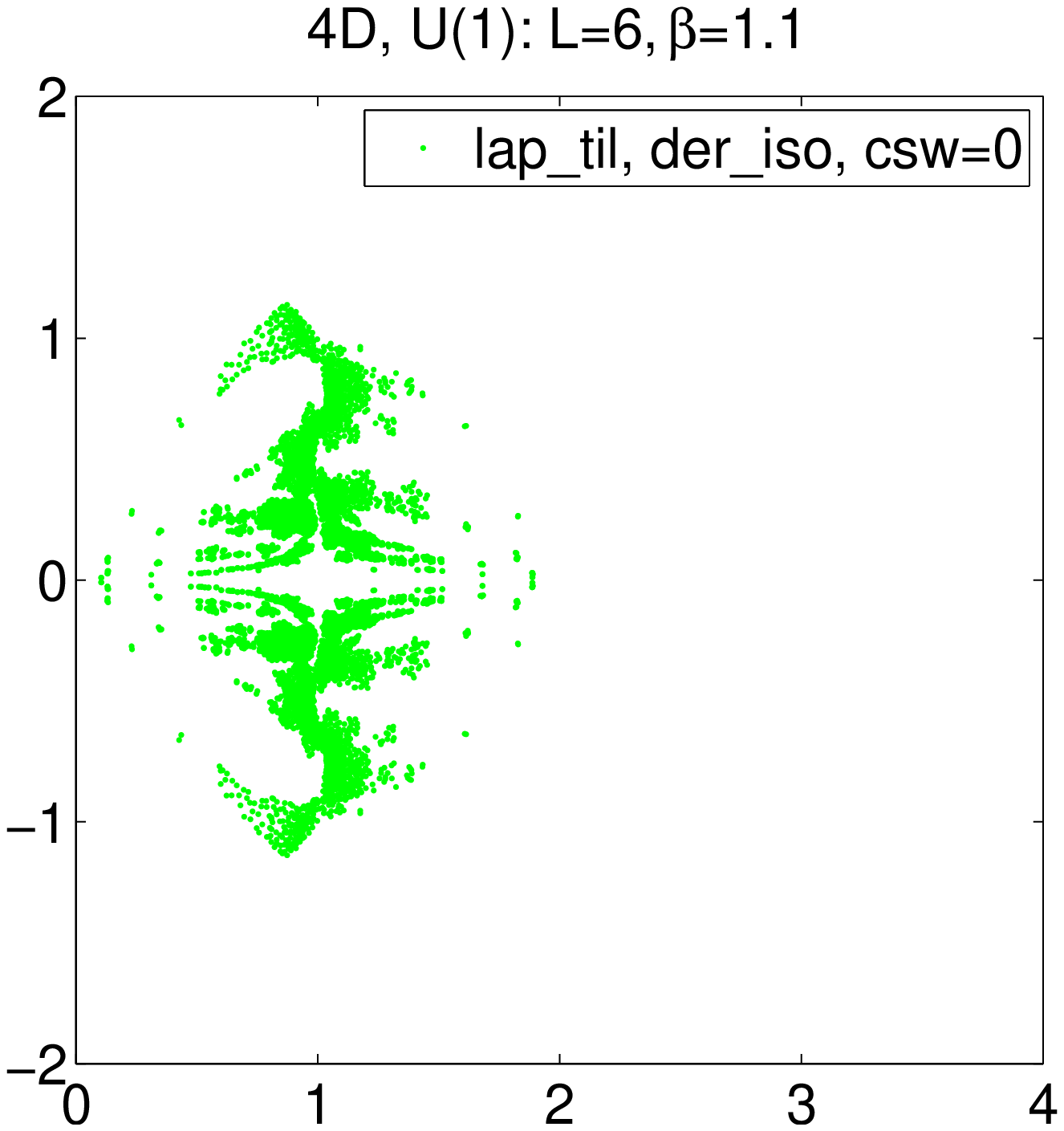,height=5.7cm}\\
\epsfig{file=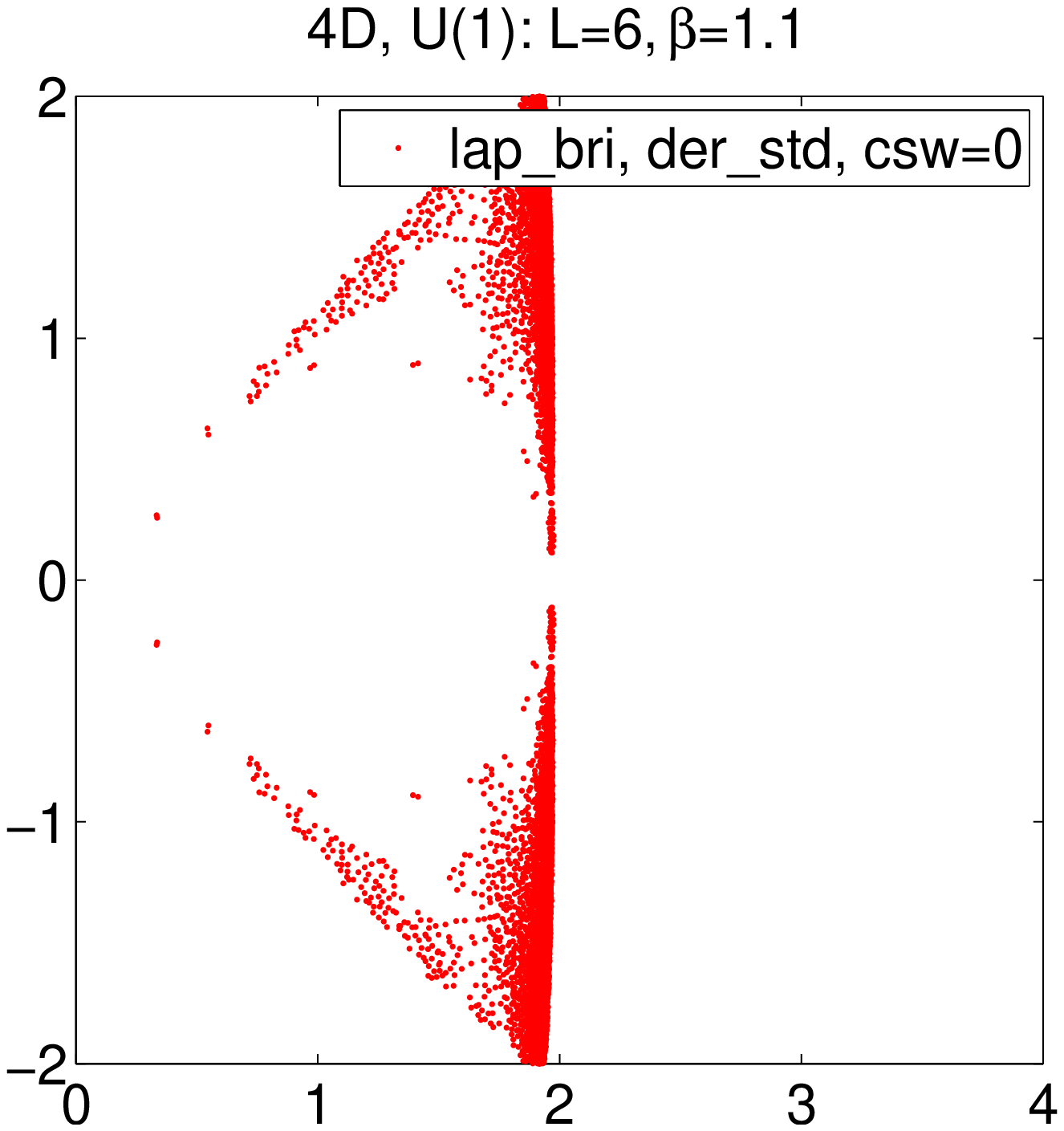,height=5.7cm}
\epsfig{file=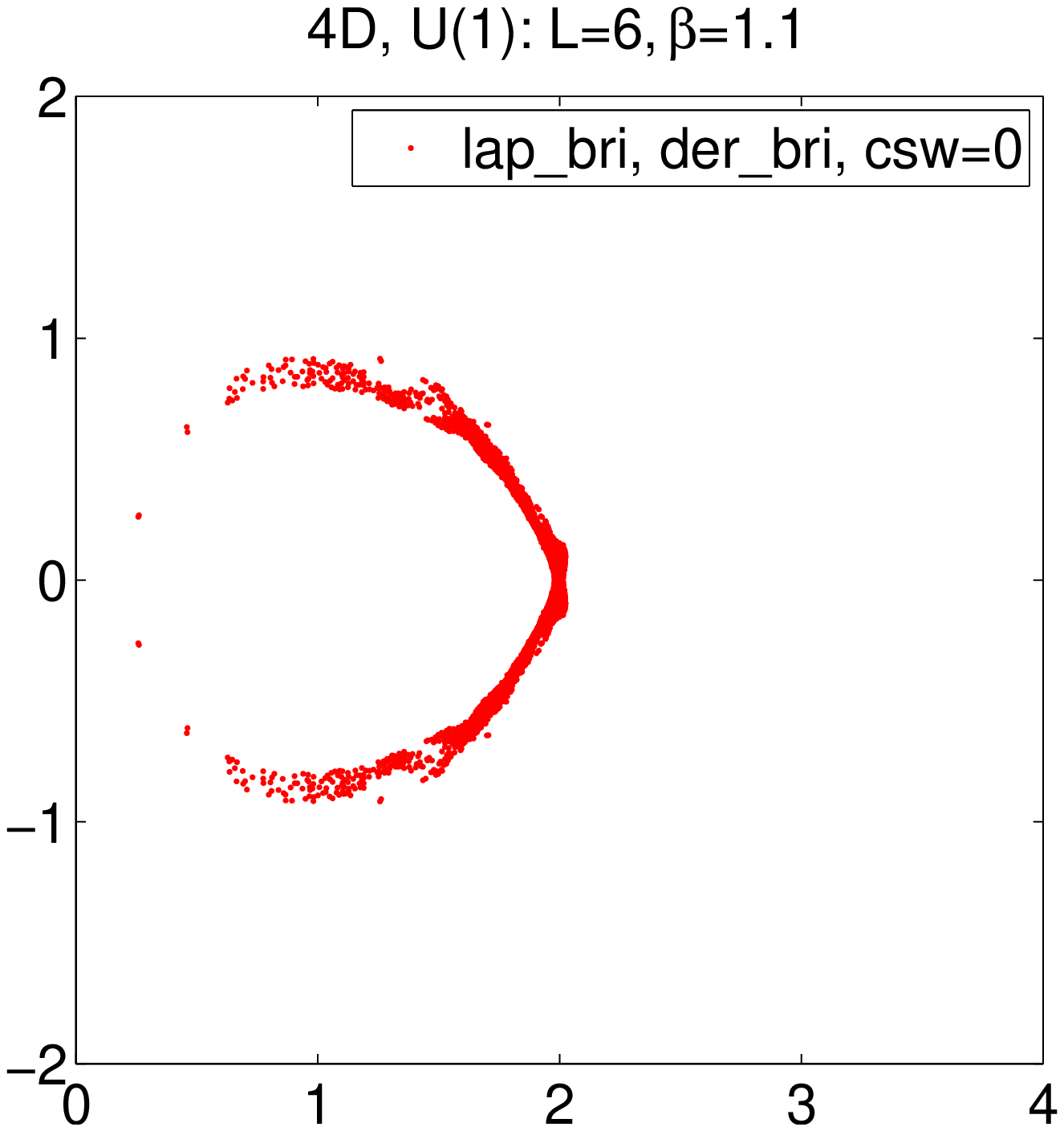,height=5.7cm}
\epsfig{file=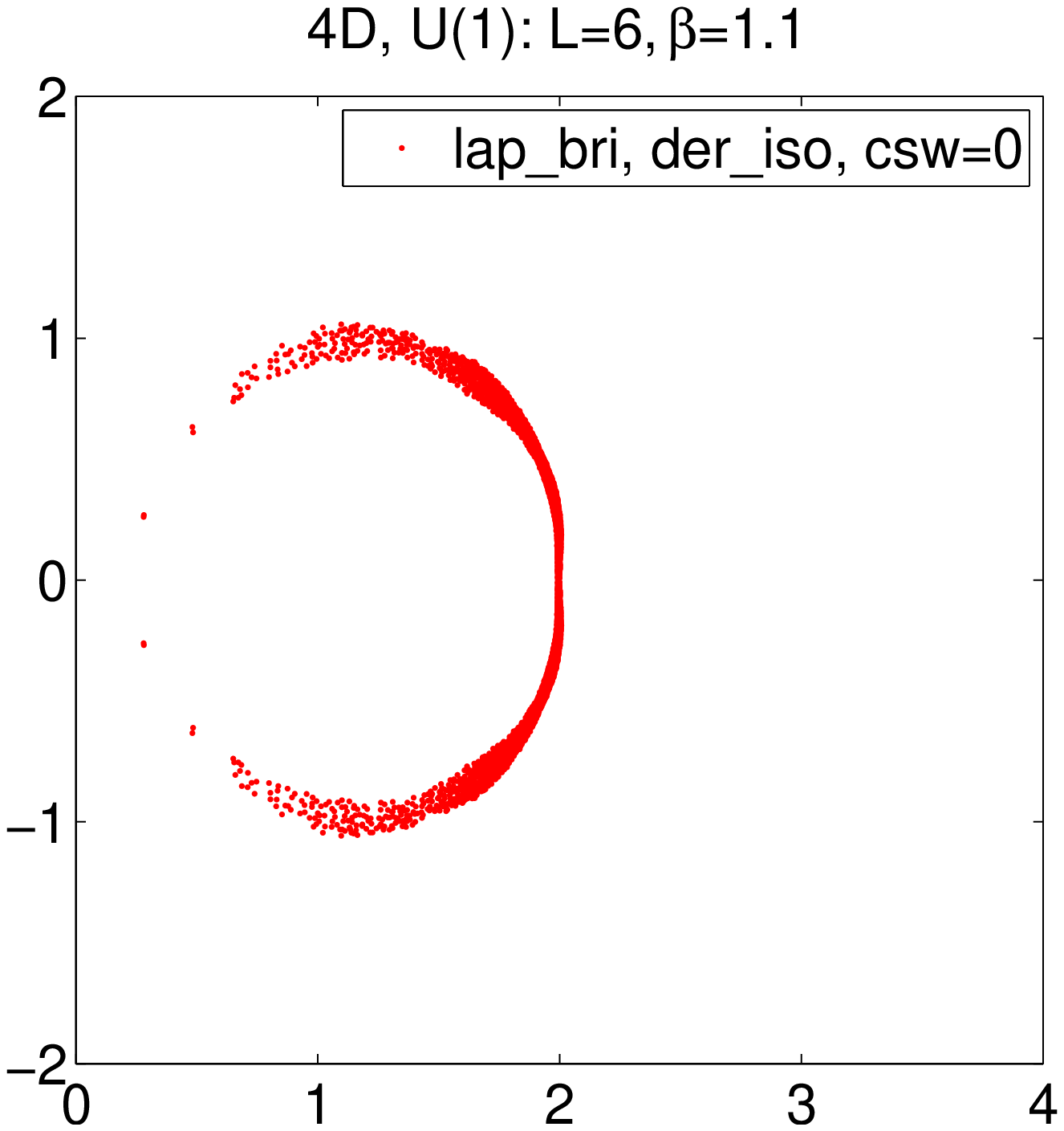,height=5.7cm}\\
\epsfig{file=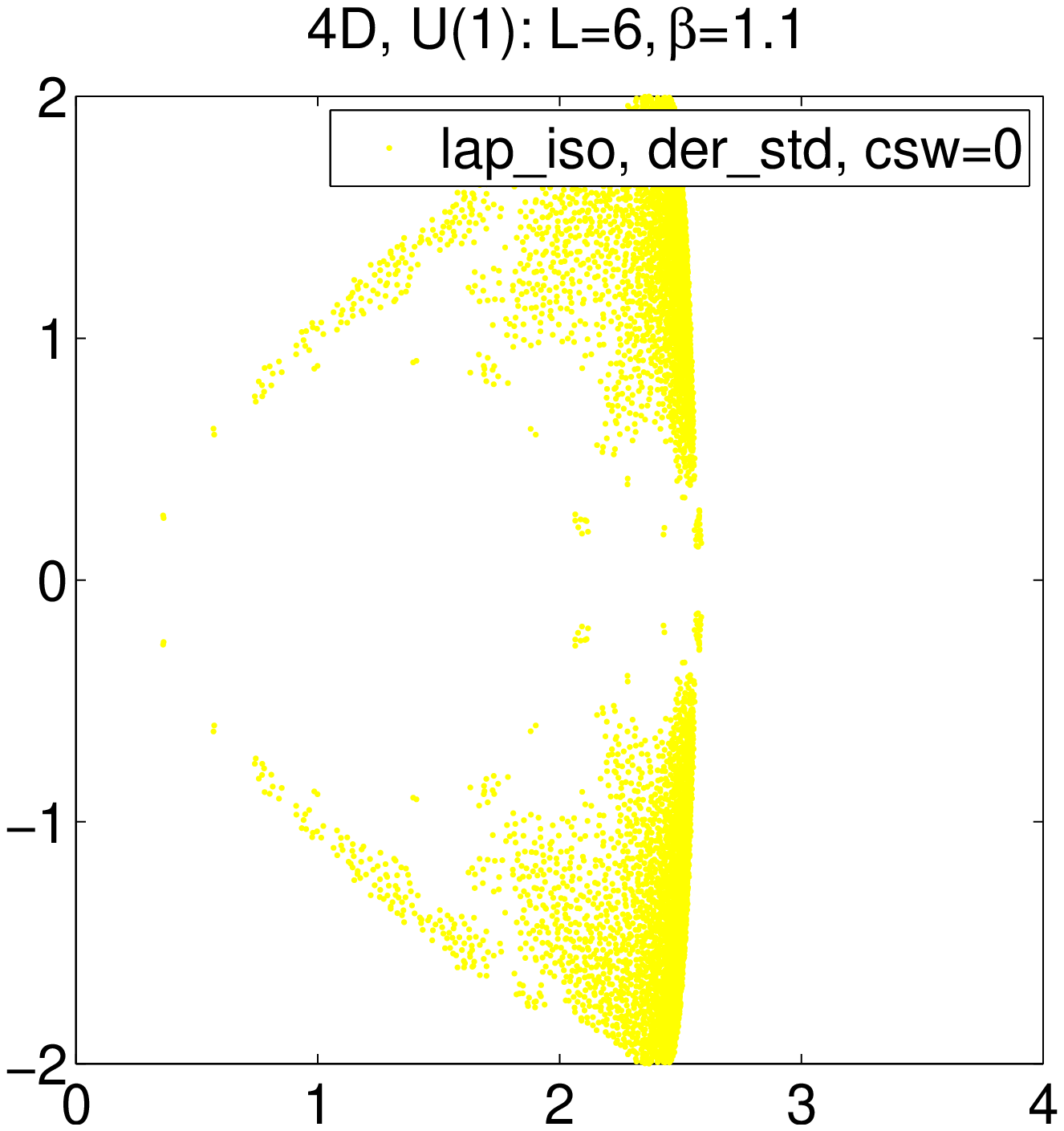,height=5.7cm}
\epsfig{file=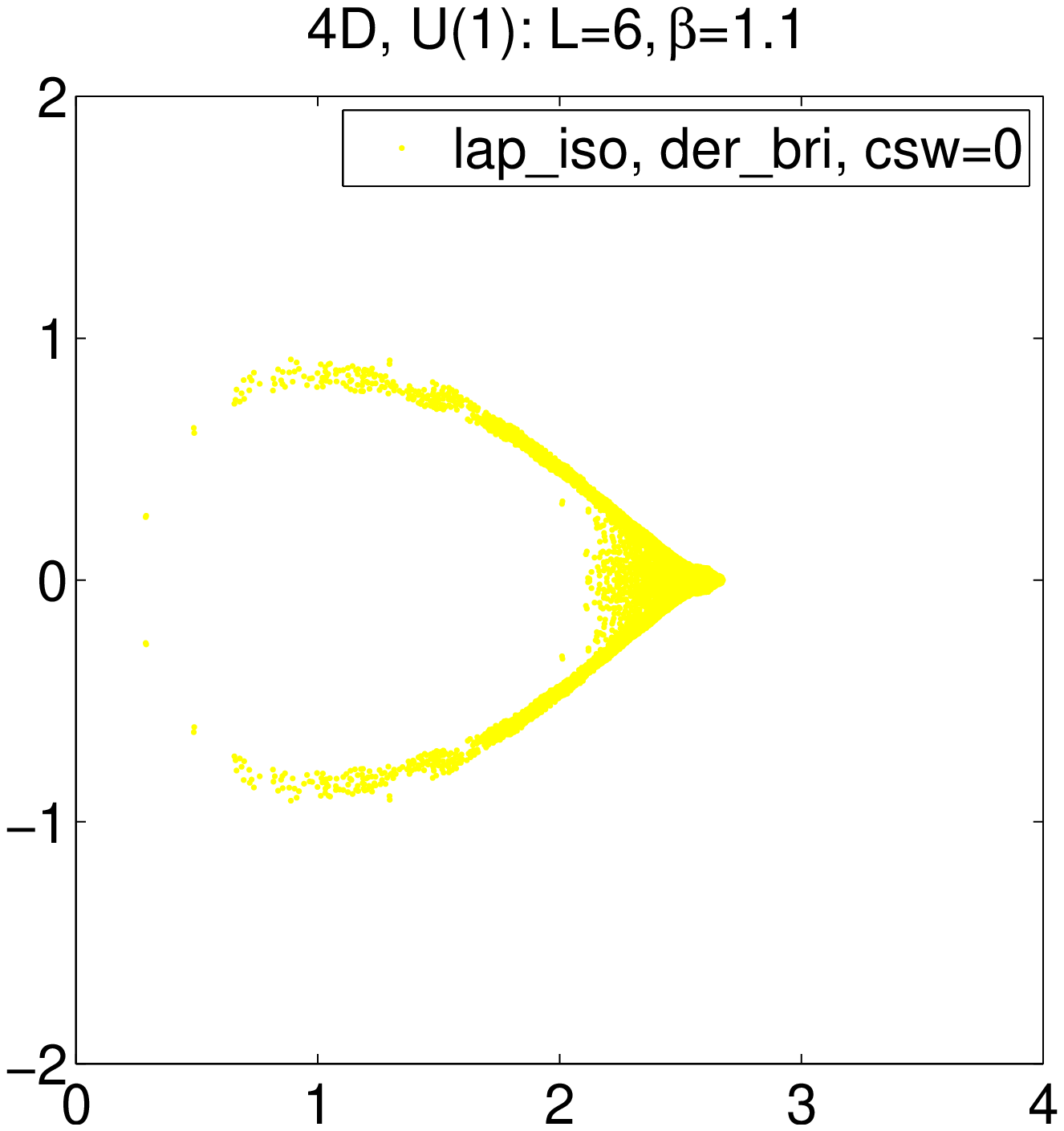,height=5.7cm}
\epsfig{file=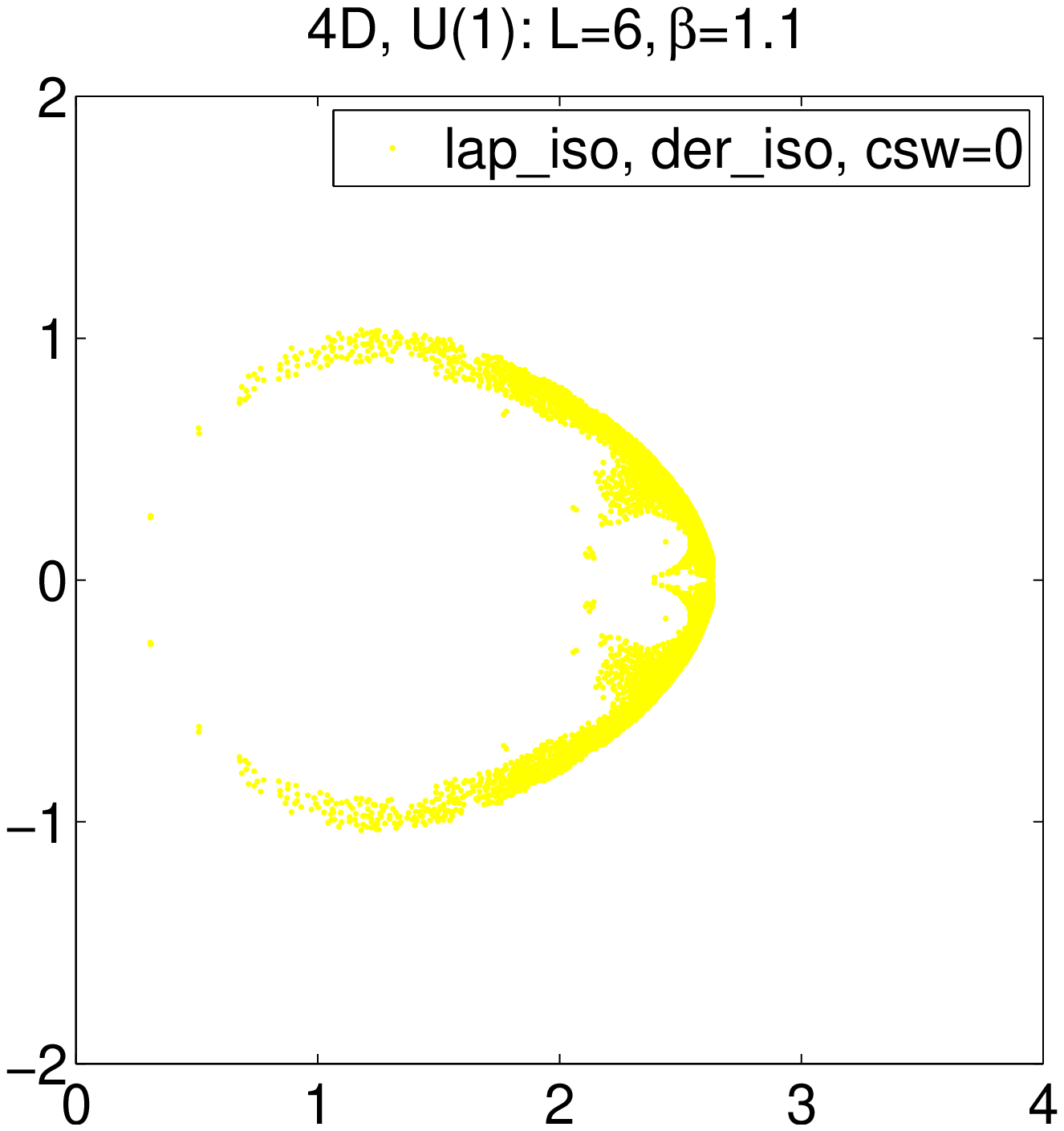,height=5.7cm}
\caption{\label{fig:spec_4D_csw0}\sl
Eigenvalue spectra of all operators considered in 4D with $c_\mr{SW}\!=\!0$.}
\end{figure}

In Fig.\,\ref{fig:spec_4D_csw0} the eigenvalues of the 12 operators without
improvement ($c_\mr{SW}\!=\!0$) are shown.
Again, we refrain from showing the counterpart with improvement, as the
difference is marginal.
Following the tradition of the previous sections, the Laplacian features as the
row index of the panel, and the derivative as the column index.
Out of these 12 constructions, 9 are undoubled fermion operators, while 3 yield
eight species in the continuum limit.
Again, the gross features of these operators are rather similar to their 2D and
3D counterparts.
This time, the operators with the standard Laplacian (first row) have branches
at $\mr{Re}(z)\!\simeq\!0,2,4,6,8$, with multiplicities $1,4,6,4,1$,
respectively, and with alternating chiralities.
Replacing the standard Laplacian by the Brillouin Laplacian (third row) or
the isotropic Laplacian (fourth row), the lifting of the doublers is reduced.
With the tilted Laplacian and the standard derivative, the 16 species arrange
themselves in groups of 8 at $\mr{Re}(z)\!\simeq\!0$ (correct-type chirality)
and $\mr{Re}(z)\!\simeq\!2$ (wrong-kind chirality), respectively%
\footnote{Note the difference to naive or staggered massless fermions in 4D,
where both chiralities sit on top of each other. For the effect of non-standard
staggered mass terms and the resulting eigenvalue spectra see
\cite{Adams:2010gx,Hoelbling:2010jw}.}.
Once the standard derivative is replaced by the Brillouin or isotropic
variety, some ``cross talk'' between the marginal and the irrelevant piece
becomes apparent.
Looking at the whole figure, one would say that the combination
($\lap^\mr{bri}$, $\nab^\mr{iso}$) fares best in the sense that its eigenvalue
spectrum is closest to that of a Ginsparg-Wilson action.


\subsection{Free field dispersion relations in 4D}

\begin{figure}[!p]
\centering
\epsfig{file=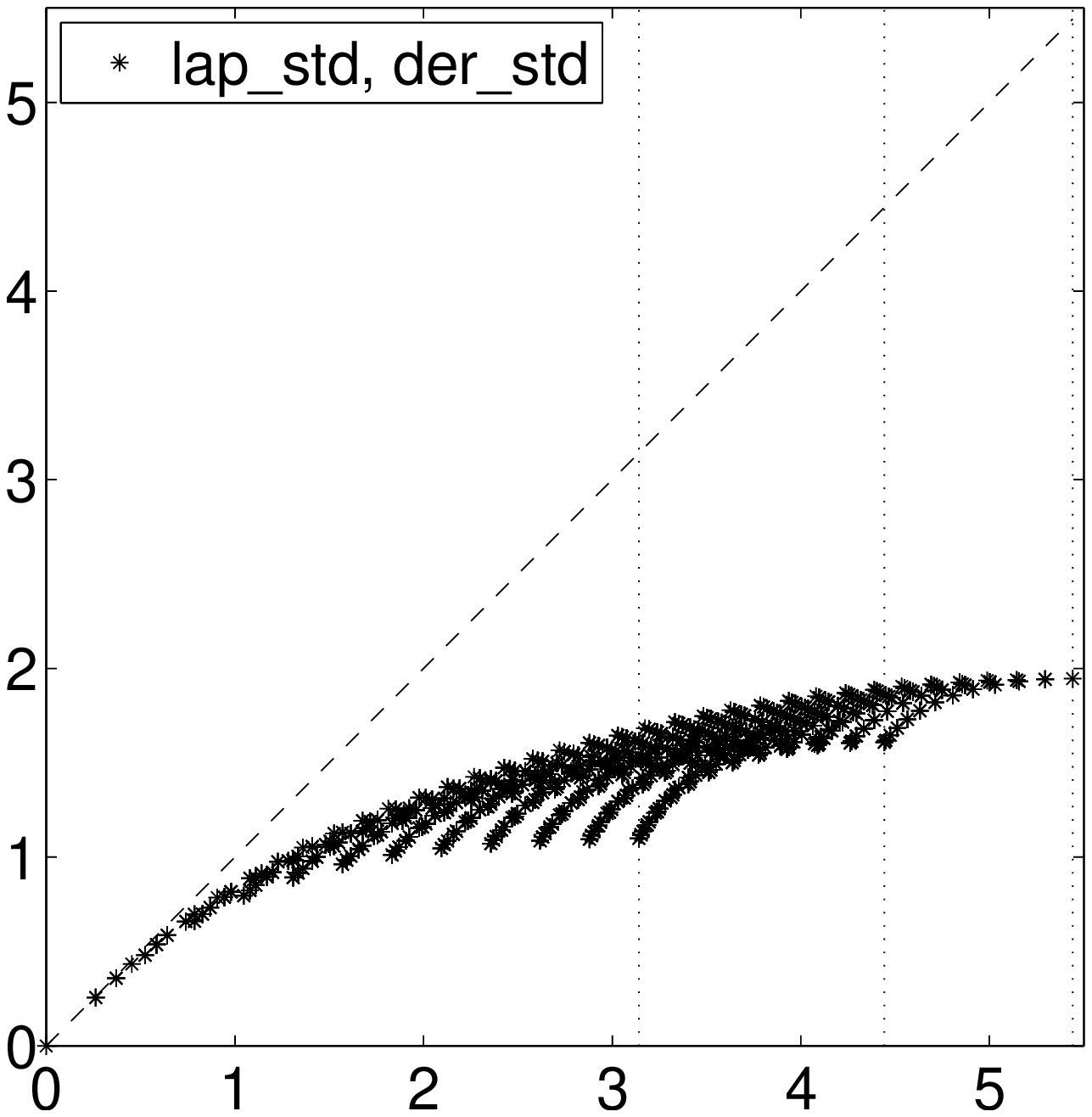,height=5.7cm}
\epsfig{file=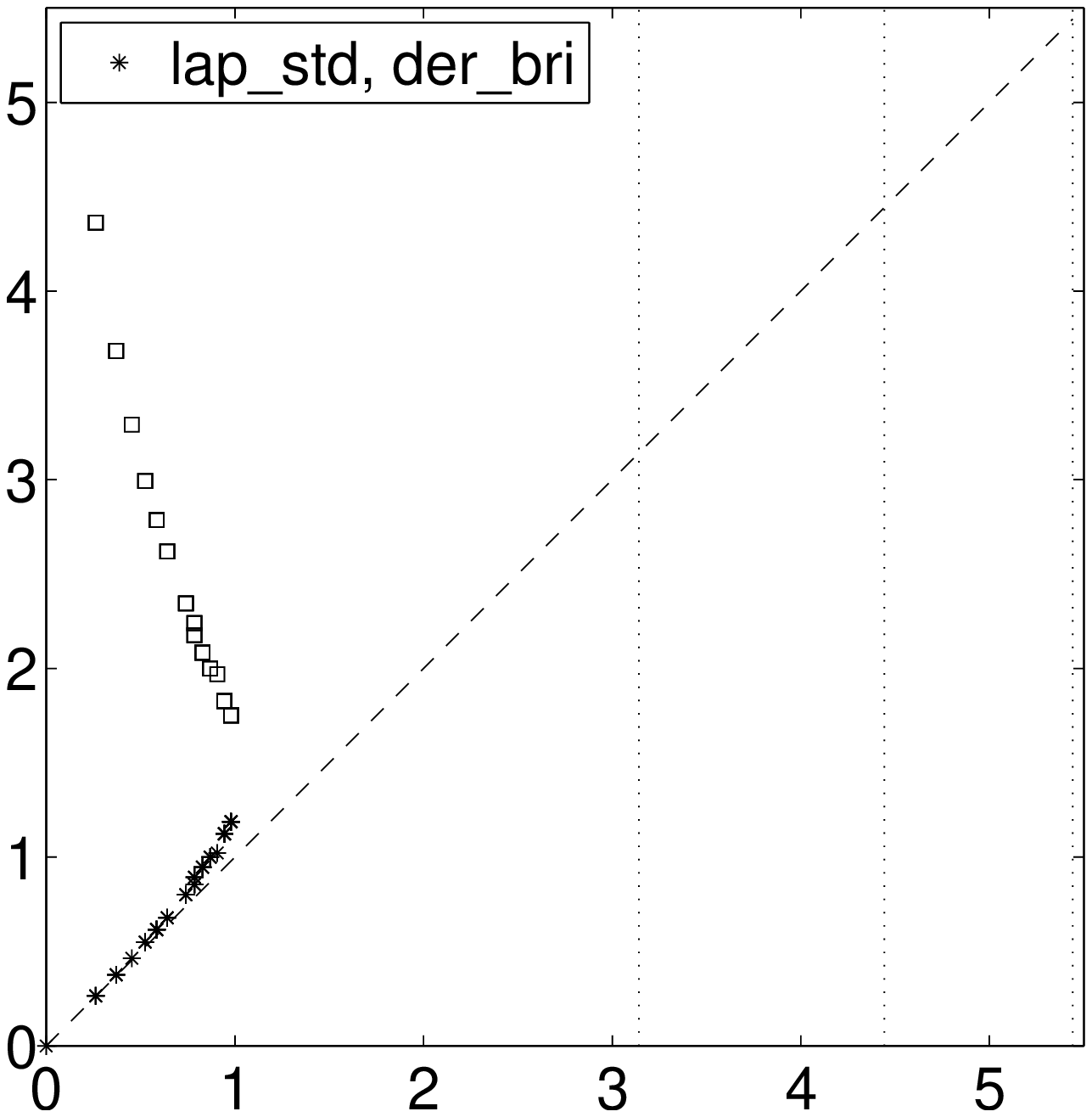,height=5.7cm}
\epsfig{file=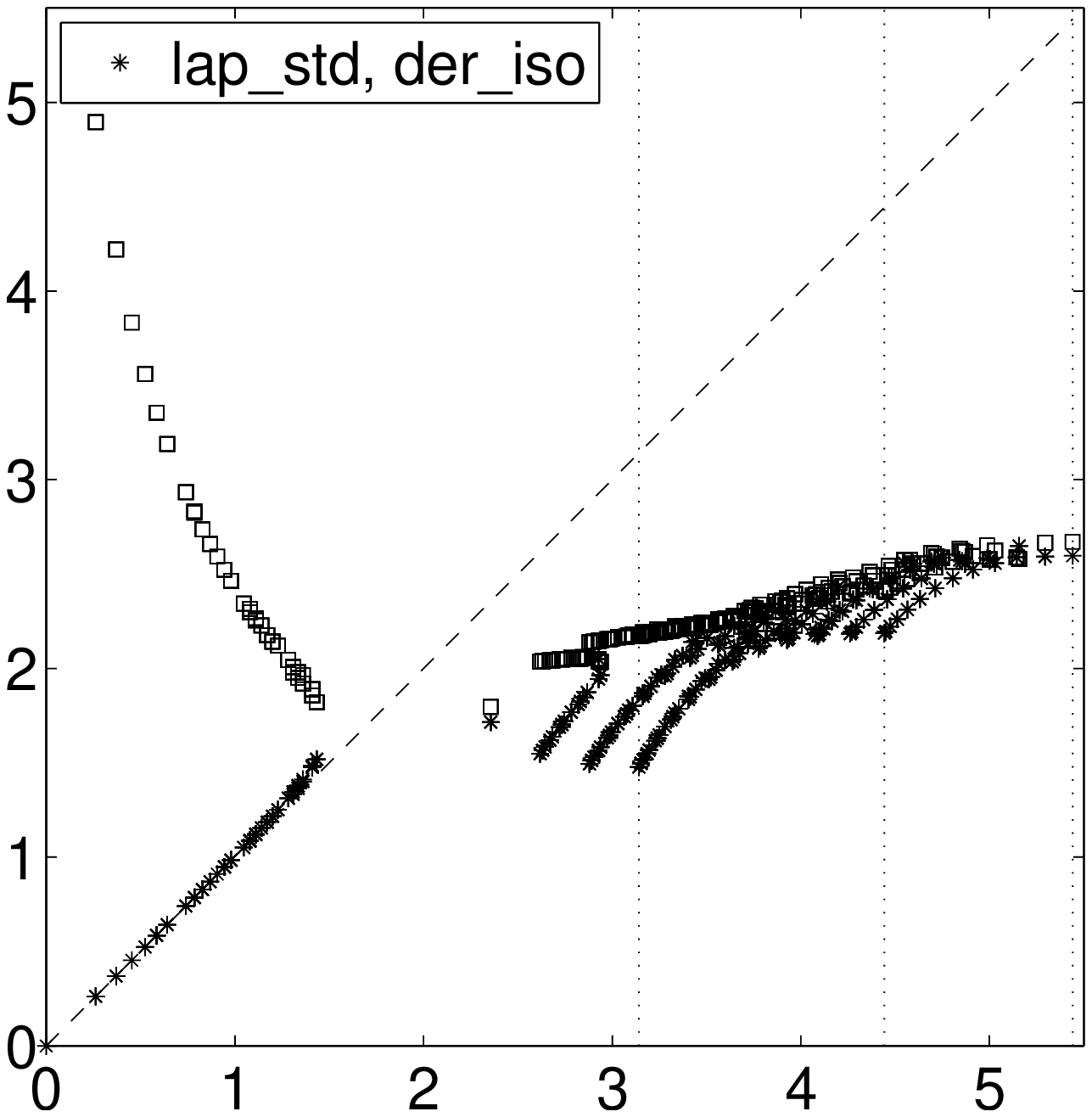,height=5.7cm}\\
\epsfig{file=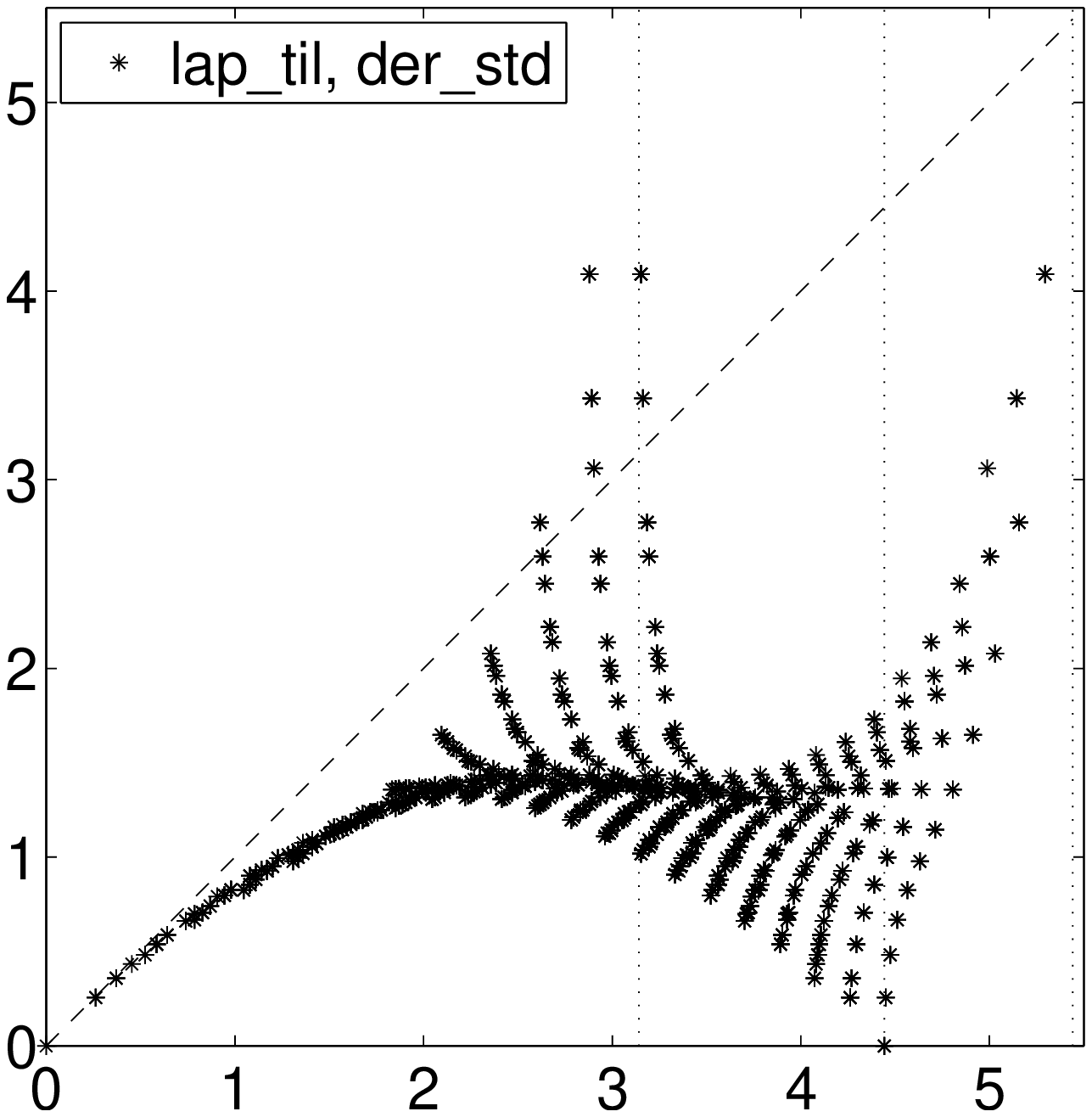,height=5.7cm}
\epsfig{file=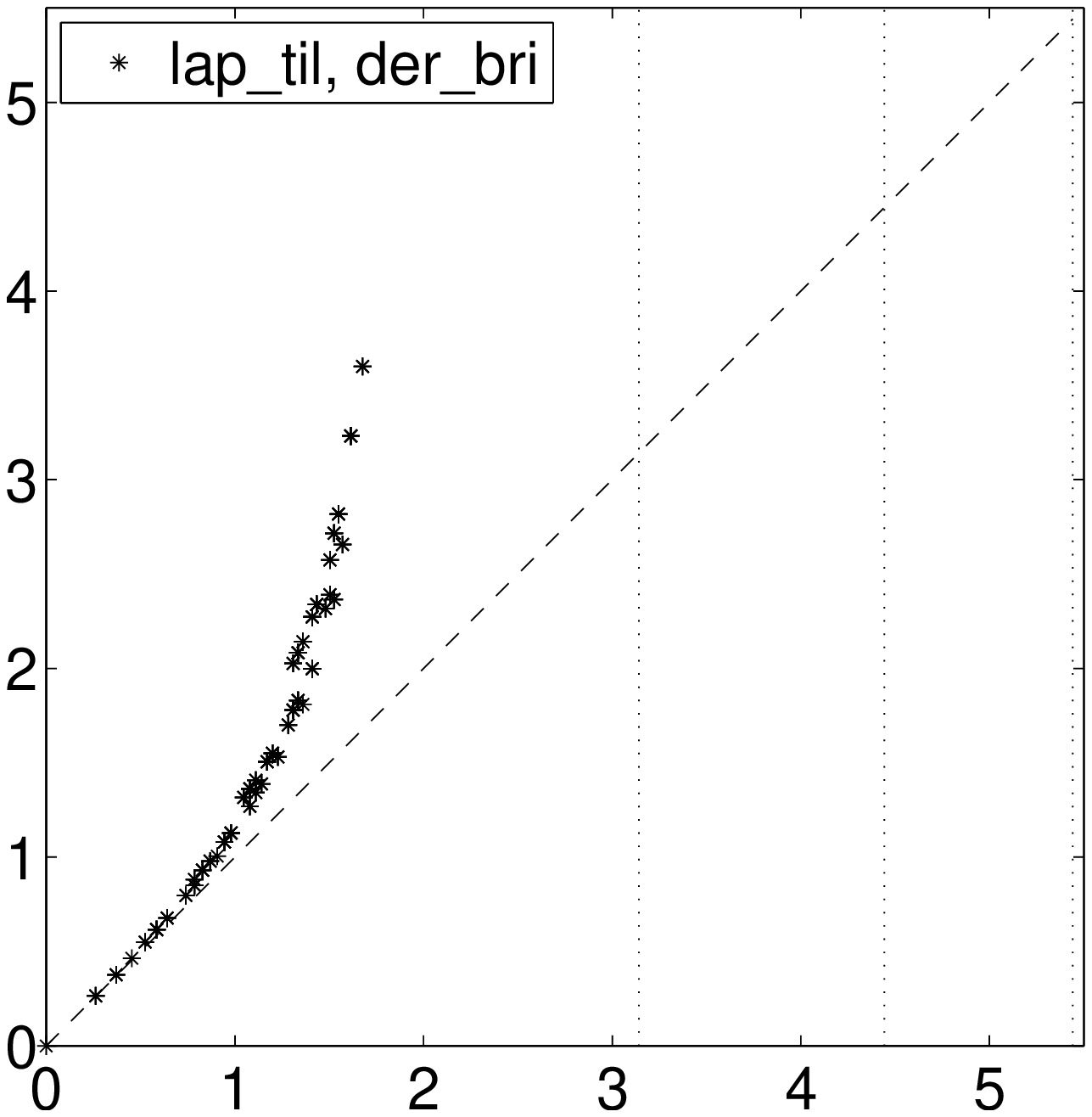,height=5.7cm}
\epsfig{file=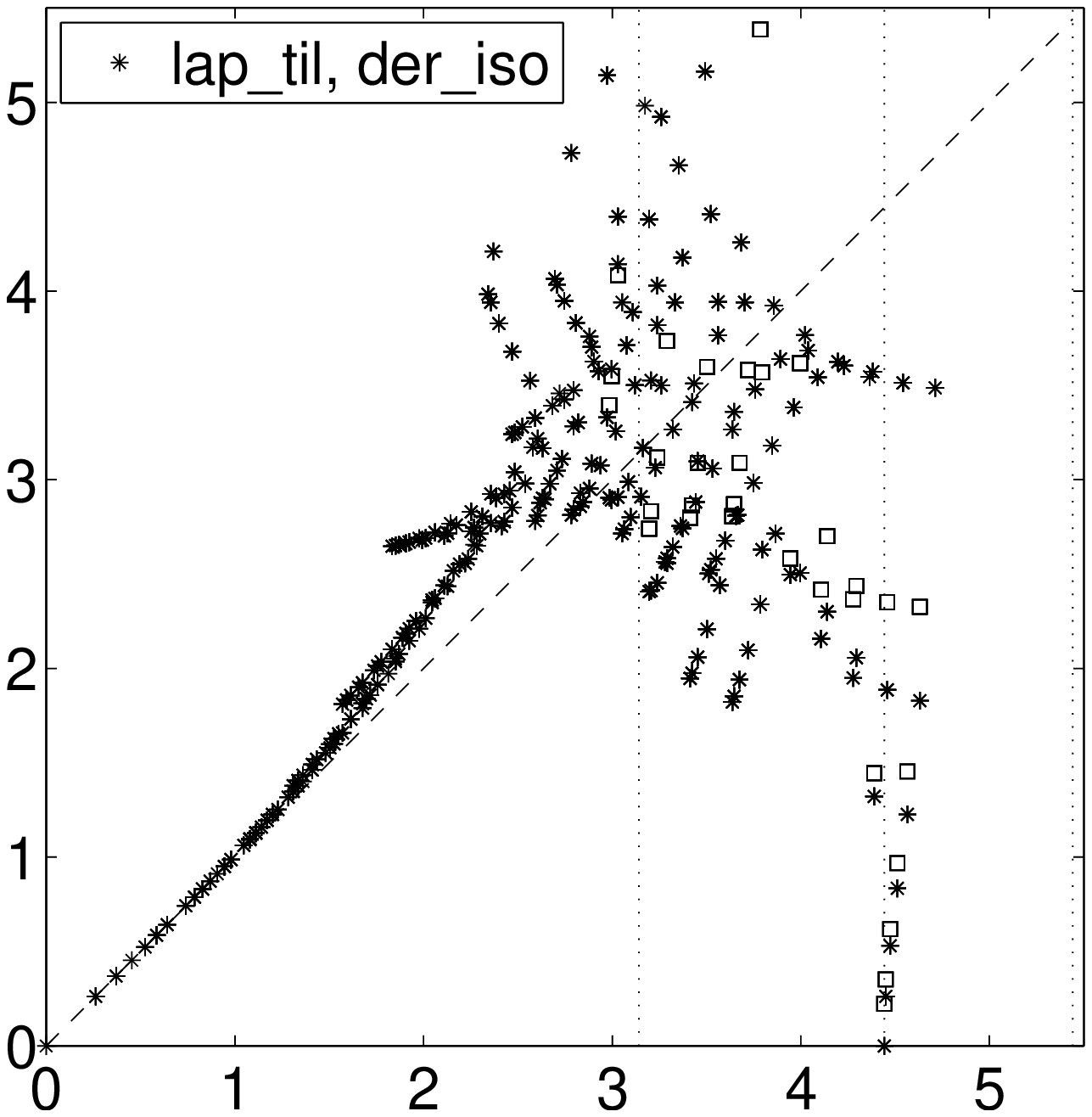,height=5.7cm}\\
\epsfig{file=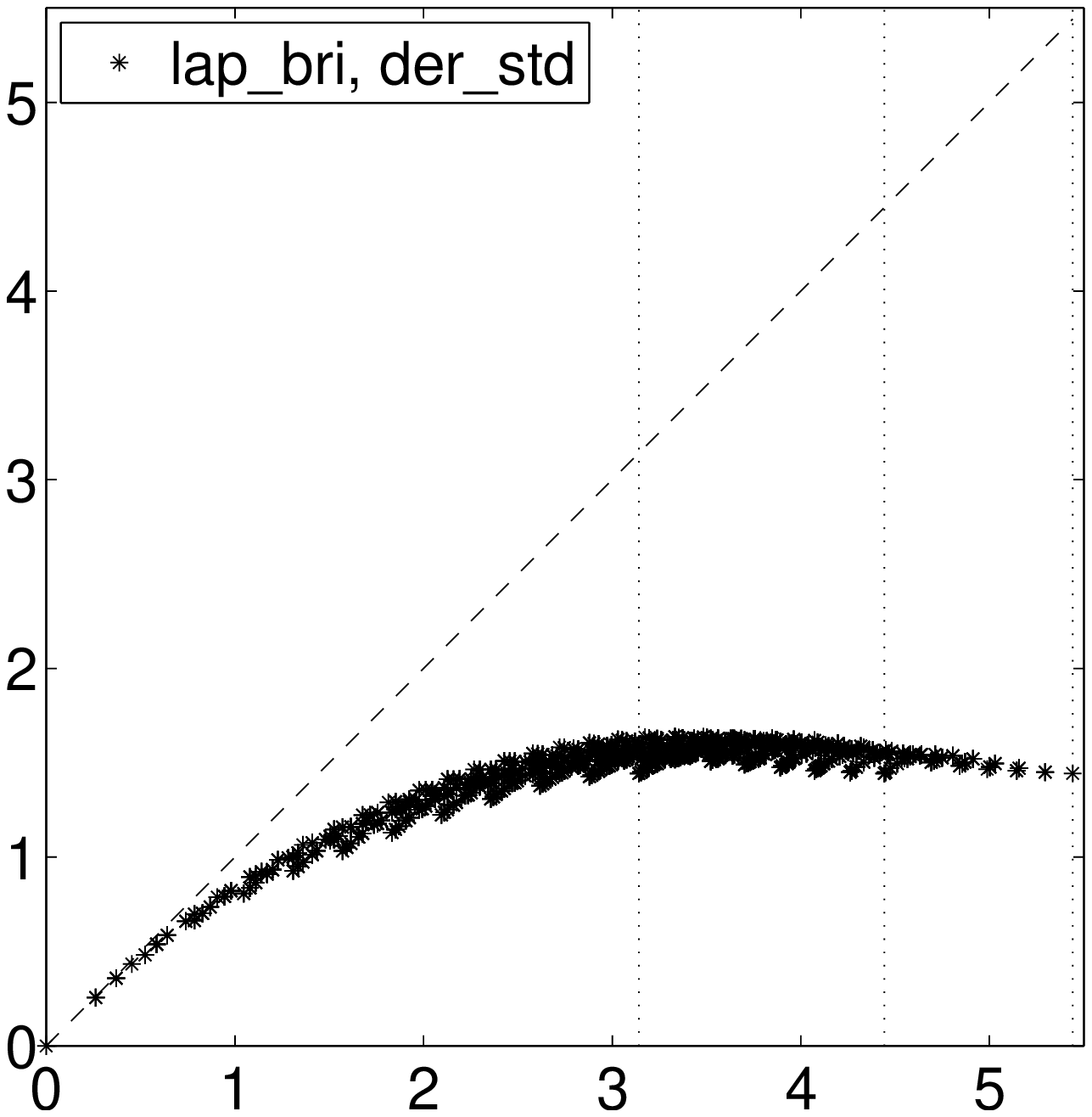,height=5.7cm}
\epsfig{file=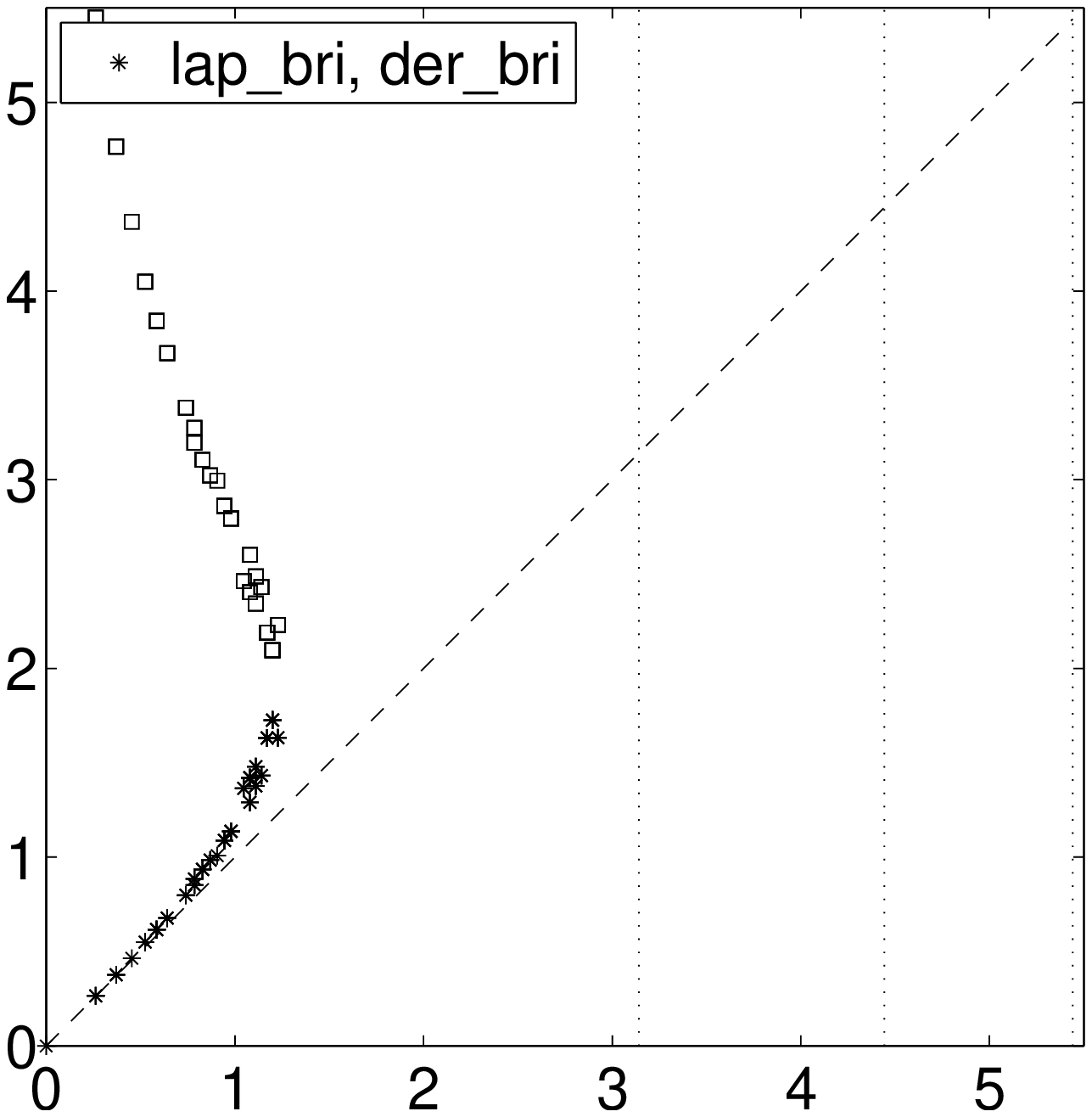,height=5.7cm}
\epsfig{file=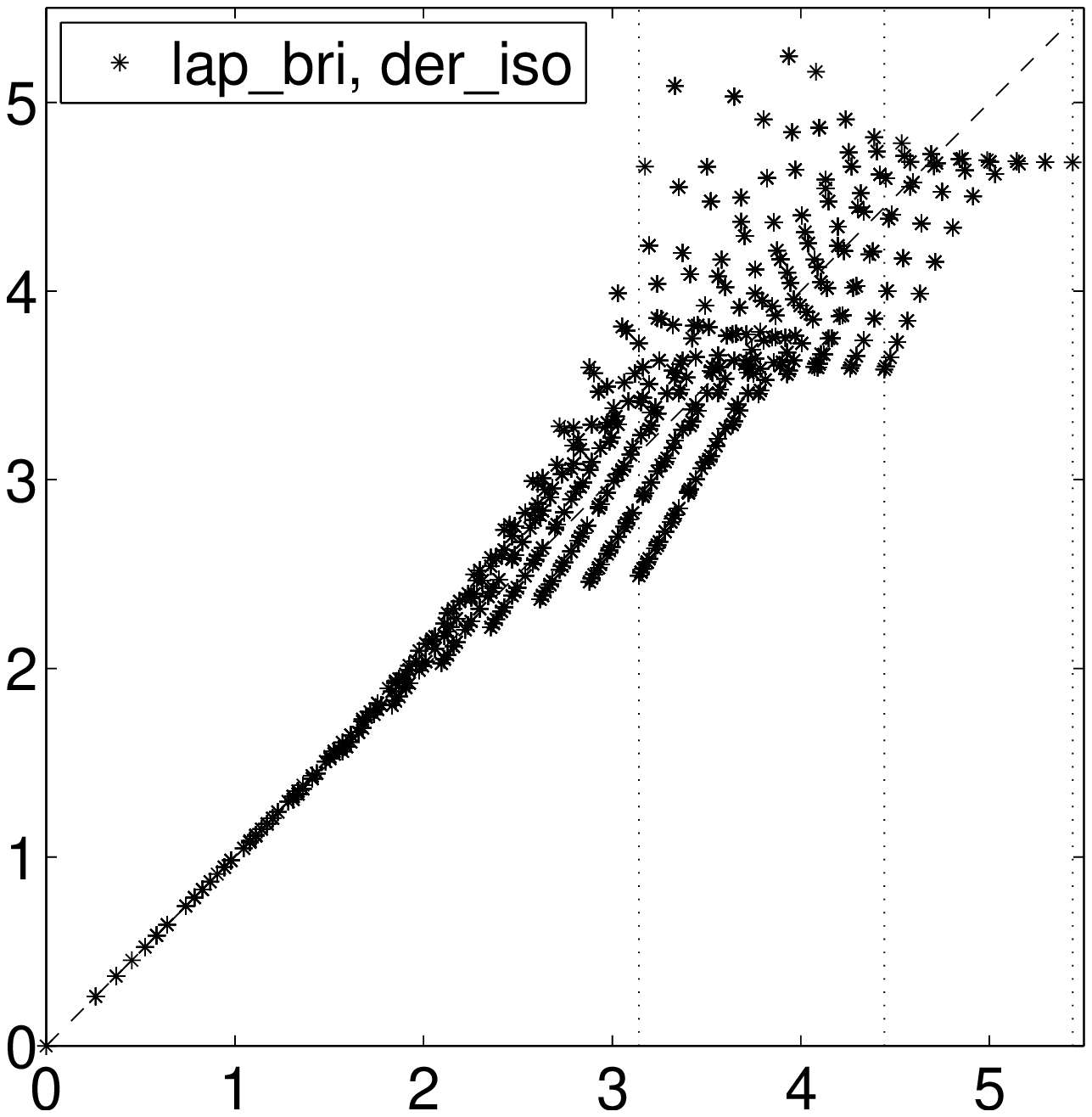,height=5.7cm}\\
\epsfig{file=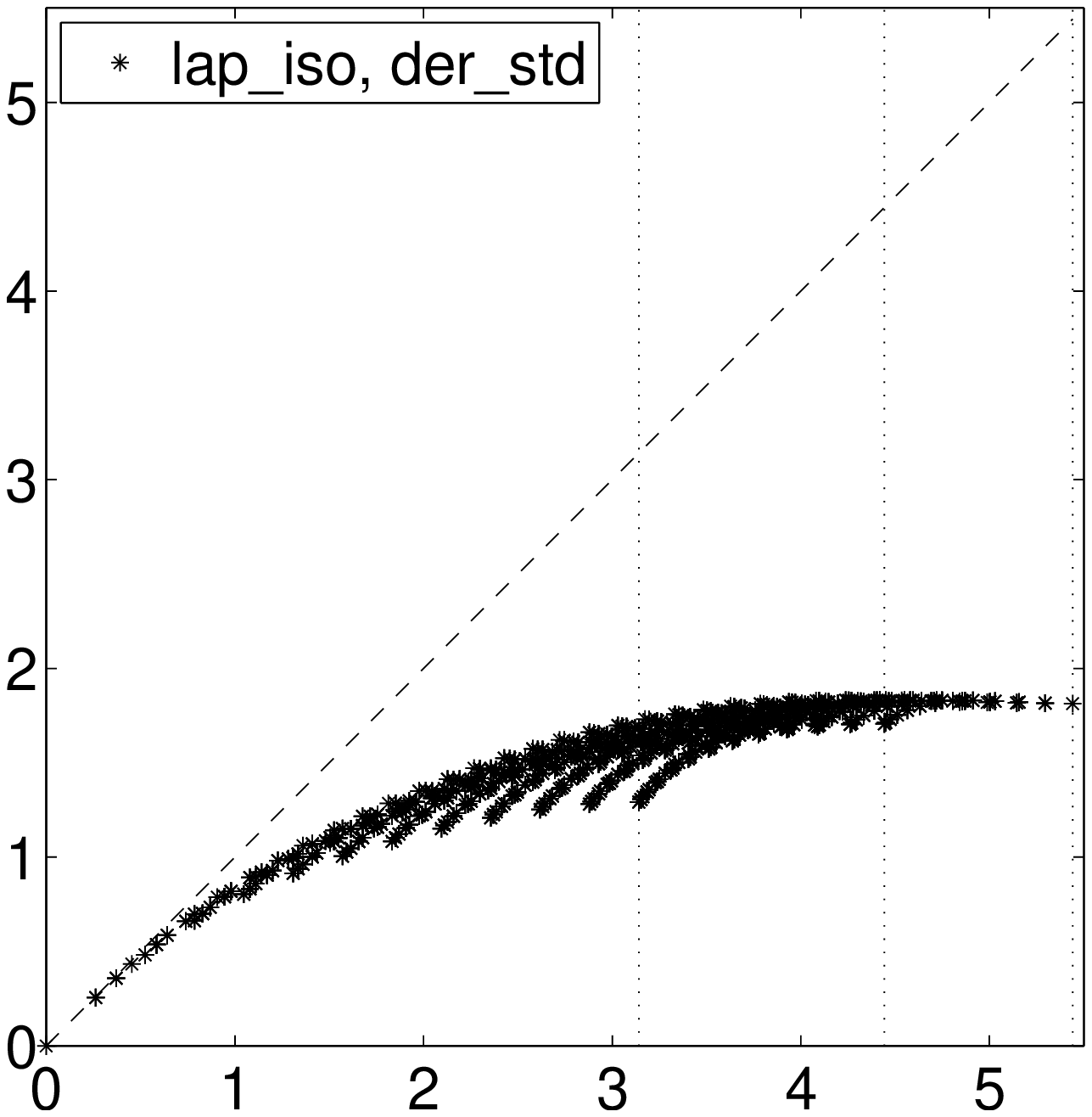,height=5.7cm}
\epsfig{file=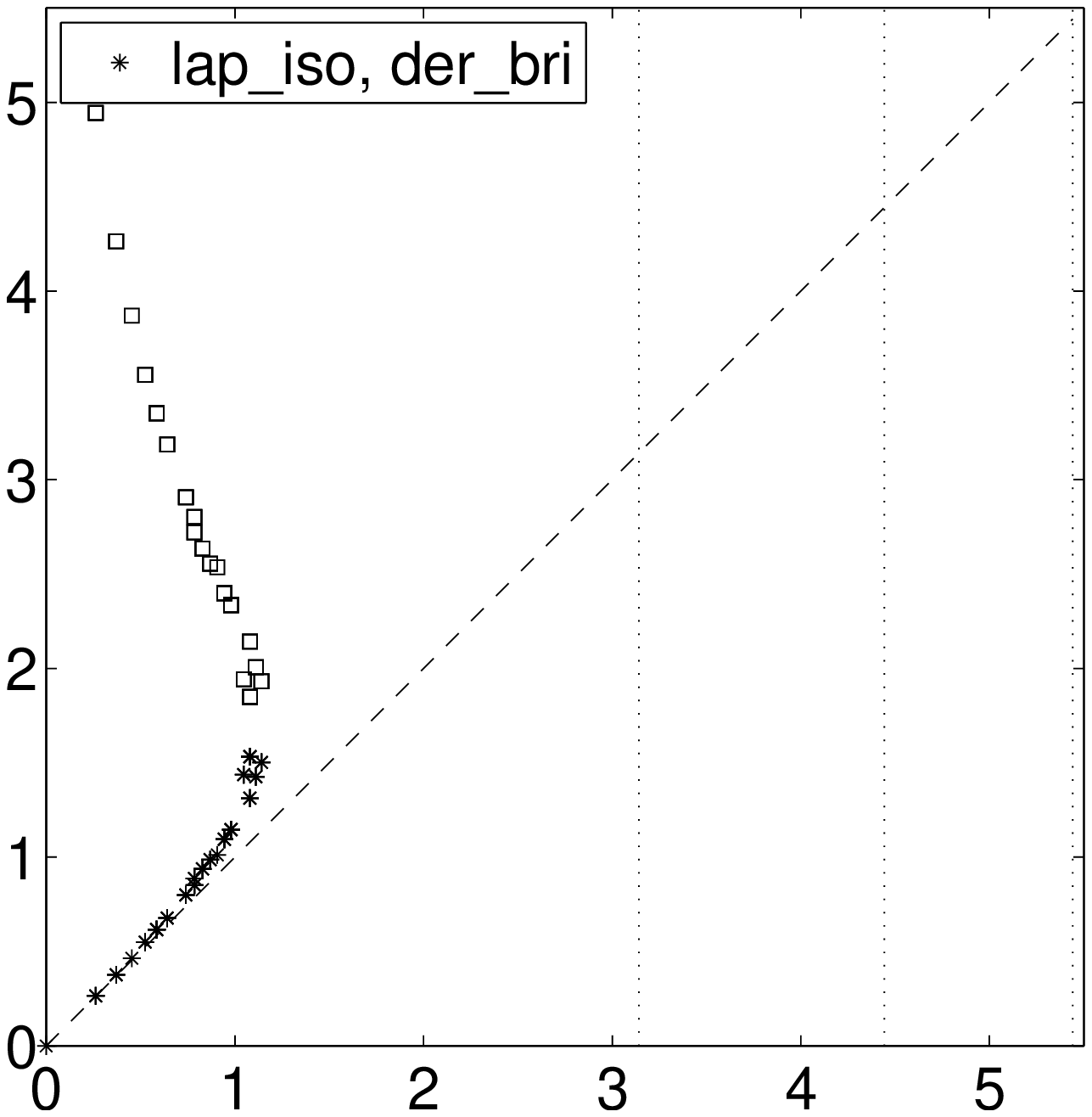,height=5.7cm}
\epsfig{file=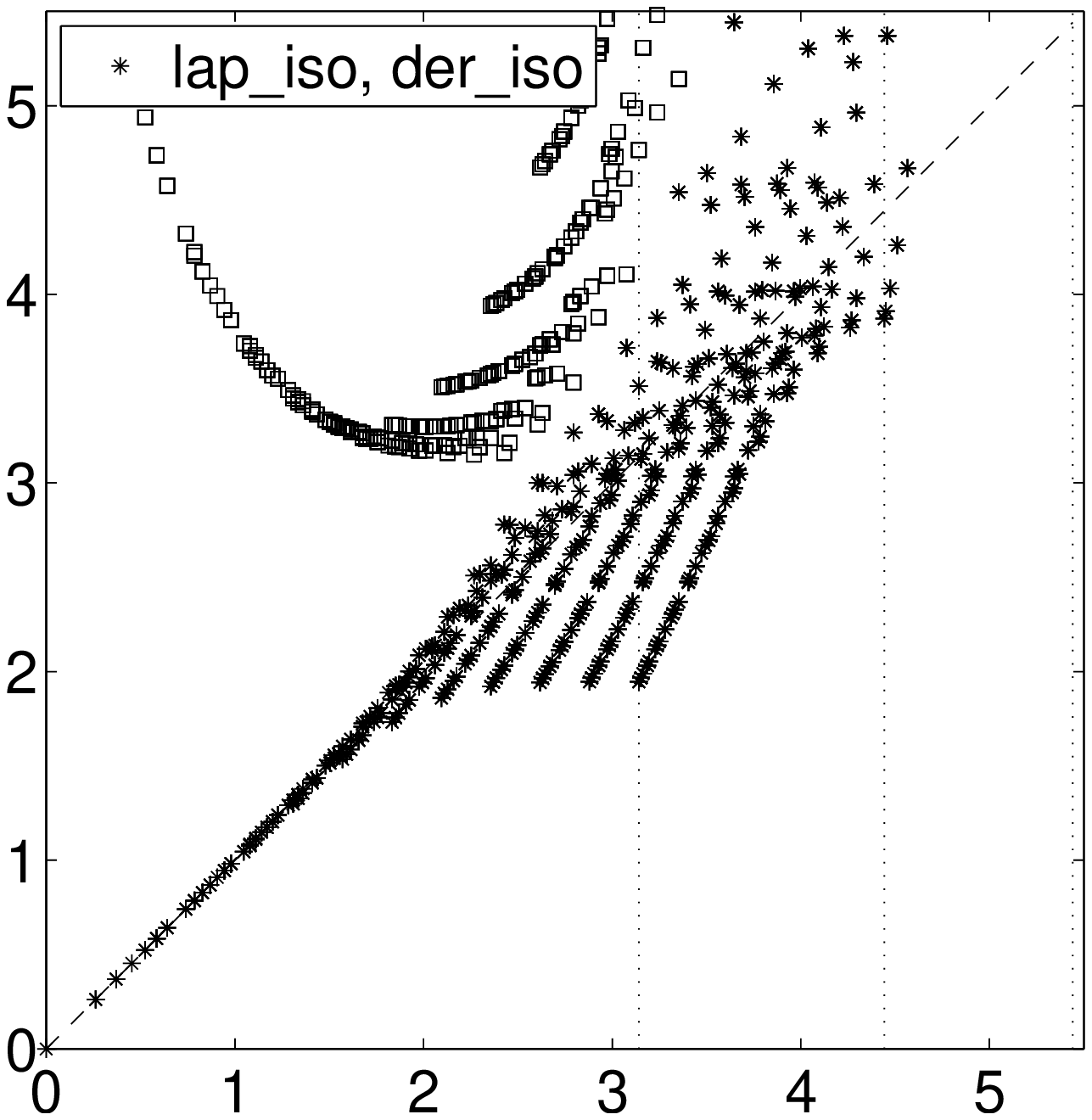,height=5.7cm}
\caption{\label{fig:disp_4D}\sl
Free-field dispersion relations of all operators considered in 4D, where
$|p|_\mr{max}\!\!=\!\!\sqrt{3}\pi/a$.}
\end{figure}

In Fig.\,\ref{fig:disp_4D} we show, for each operator, the real solutions for
$r\!=\!1$ and $m\!=\!0$ over half the Brillouin zone on a 4D lattice with
$L/a\!=\!24$.
The dispersion relation of the standard Wilson operator ($\lap^\mr{std}$,
$\nab^\mr{std}$) deviates soon from the dashed line, which corresponds to the
continuum dispersion relation, and shows large effects of anisotropy.
If $\mb{p}$ is chosen on axis, the 2D dispersion relation is reproduced.
If $\mb{p}$ is chosen as a multiple of $(1,1,0)$, the 3D dispersion relation is
reproduced.
If $\mb{p}$ is chosen as a multiple of $(1,1,1)$ entries at the upper border
are generated; they go out to $\sqrt{3}\pi/a$.
Again, some operators admit a second real solution (open boxes) which decouples
in the continuum, and some operators have, for certain combinations of
$(p_1,p_2,p_3)$, only complex solutions.
Overall, it is clear that the combination ($\lap^\mr{bri}$, $\nab^\mr{iso}$)
fares best in the sense that its dispersion relation is closest to the one in
the continuum.


\section{Specification of operator details in 4D}


\subsection{Overall smearing strategy}

Given the results in the previous three sections, the combination of
``isotropic derivative'' and ``Brillouin Laplacian'' seems most attractive.
In other words, our preferred operator is
\beq
D(x,y)=\sum_\mu \ga_\mu \nab_\mu^\mr{iso}(x,y)
-{a\ovr2}\lap^\mr{bri}(x,y)+m_0\de_{x,y}
-{c_\mr{SW}\ovr2}\sum_{\mu<\nu}\si_{\mu\nu}F_{\mu\nu}\de_{x,y}
\label{def_pref}
\eeq
and below we shall refer to it as the ``Brillouin operator''.

An ingredient which has proven particularly useful in the design of fermion
actions with small cut-off effects is link-smearing, also known under the
label of ``fat links'' \cite{DeGrand:1998jq,Bernard:1999kc,Stephenson:1999ns,
Zanotti:2001yb,DeGrand:2002vu,Capitani:2006ni}.
In the quenched QCD tests reported below a single step of APE smearing
\cite{Albanese:1987ds}
\beq
V_\mu(x)\equiv U_\mu^\mr{APE}(x)=P_{SU(3)}
\Big\{
(1\!-\!\al)I+{\al\ovr6}\sum_{\pm\nu\neq\mu}
U_\nu(x)U_\mu(x\!+\!\hat\nu)U_\nu\dag(x\!+\!\hat\mu)U_\mu\dag(x)
\label{def_ape}
\Big\}
U_\mu(x)
\eeq
with $\al\!=\!0.72$ is applied (for the details of the backprojection to
$SU(3)$ see subsection 5.2).
The result is used as an input in the covariant derivative and the covariant
Laplacian.
The latter operators are made gauge covariant in the simplest possible way,
by summing over all shortest paths, with subsequent backprojection to $SU(3)$.
For instance the hyper-diagonal connections (4 hops) receive contributions
from 24 paths, while the cubic-diagonal connections (3 hops) receive 6
contributions, and the square-diagonals (2 hops) just 2.

We use the same kind of smeared gauge links $V_\mu(x)$ in the construction of
the derivative, the Laplacian, and the field strength tensor.
Since this change is ultralocal and modifies only operators of mass dimension
5 and higher, the universality class of the action is unaffected.
Other smearing strategies are possible, e.g.\ only relevant pieces or only
irrelevant pieces of the action may be smeared.
However, as we are unaware of any advantage of such more complicated schemes,
we prefer to stay with the overall smearing strategy where all ``thin'' links
$U_\mu(x)$ in (\ref{def_pref}), even if within $F_{\mu\nu}$, are replaced by
the same kind of ``fat'' links $V_\mu(x)$.

The goal of our quenched scaling study in Sec.\,6 is to confront
(\ref{def_pref}) with the standard Wilson action.
To compare like with like, we will use the same smearing strategy and the
same kind of clover improvement with $c_\mr{SW}\!=\!1$ (see subsection 5.3
below) in either case.

For completeness let us mention that, in order to simulate full QCD with a
fat-link Brillouin or Wilson action and an HMC algorithm, one would equip
either action with a smearing that is tailored to this purpose (e.g.\
``stout/EXP'' \cite{Morningstar:2003gk}, ``n-APE'' \cite{Hasenfratz:2007rf},
``LOG'' \cite{Durr:2007cy}, ``over-improved stout'' \cite{Moran:2008ra}).


\subsection{Details of the projection to $SU(N)$}

The projection of an arbitrary $N\!\times\!N$ matrix $A$ to $SU(N)$ is usually
defined through a projection to $U(N)$, followed by a projection to unit
determinant.
The first projection is realized as
\beq
P_{U(N)}\{A\}=A(A\dag A)^{-1/2}=(AA\dag)^{-1/2}A
\label{def_projection}
\eeq
where the equivalence of the two representations follows from the singular
value decomposition $A=USV\dag$, with unitary $U,V$ and $S\!>\!0$, resulting in
$P_{U(N)}(A)=UV\dag$.
Since $A\dag A$ and $AA\dag$ are both hermitean, either version of
(\ref{def_projection}) requires only one eigensystem.

The projection to unit determinant is somewhat more involved, even if we
restrict the discussion to unitary arguments, as suggested by the above 2-step
procedure.
The most naive recipe is to divide a given $U\!\in\!U(N)$ by the $N$-th root of
its determinant.
Unfortunately, this is not a valid procedure, since there is a finite
(non-zero) likelihood%
\footnote{The issue arises if $A$ is the sum of only two $SU(N)$ matrices. The
determinant of $A\!=\!U\!+\!V$ is real, since
\bdm
\det(U\!+\!V)=
\det(U(V\dag\!+\!U\dag)V)=\det(V\dag\!+\!U\dag)=
[\det(U\!+\!V)]^*
\edm
with no generalization if $A$ is a sum of three and more unitary matrices.
Hence, with arbitrary $U,V\!\in\!SU(N)$ it may happen that $\det(U\!+\!V)$ lies
on the negative real axis. Taking a look at (\ref{def_projection}) one realizes
that $\det(A)\!<\!0$ implies $\det(B)\!=\!-1$, where $B\!=\!P_{U(N)}\{A\}$ and
$A$ the original sum of two $SU(N)$ matrices.}
that the argument has $\det(U)=-1$, which lies on the branch cut.
It is thus necessary, in general, to distribute the phase rotation (to go from
$\det\!=\!e^{\ri\ph}$ to $\det\!=\!1$) unevenly among the $N$ eigenvalues.

In our opinion a particularly compelling option for fixing this ambiguity
is to notice that the $U(N)$ projection defined in (\ref{def_projection}) can
be understood
as the result of the recipe
\beq
P_{U(N)}\{A\}=\min_{X\in U(N)}\mr{tr}\{(A\!-\!X)\dag (A\!-\!X)\}
\label{proj_UN}
\eeq
and to define the complete projection by using this recipe for the $SU(N)$
group, i.e.\ via
\beq
P_{SU(N)}\{A\}=\min_{Y\in SU(N)}\mr{tr}\{(A\!-\!Y)\dag (A\!-\!Y)\}
\label{proj_SU}
\eeq
in a \emph{single step}, where an algorithmic solution has been proposed in
\cite{deForcrand:2005xr}:
\begin{enumerate}
\itemsep-2pt
\item
Perform a singular value decomposition $A\!=\!USV\dag$ with $U,V\!\in\!U(N)$
and $S\!>\!0$ a diagonal matrix with positive entries.
Indeed, $X\!=\!UV\dag$ is the projection to $U(N)$, but $\det(X)\!\neq\!1$.
\item
Compute $\det(A)\!=\!\rh\exp(\ri\ph)$. Incidentally, $\det(S)\!=\!\rh$ and
$\det(UV\dag)\!=\!\exp(\ri\ph)$.
The matrix $U\exp(-\ri\ph/N)V\dag$ is in $SU(N)$, but it is, in general, not
the one which is closest to $A$.
\item
Find the solution $\{\th_i\}$ for the phases of the matrix
$D\!=\!\mr{diag}(\exp(\ri\th_1),...,\exp(\ri\th_N))$, subject to the constraint
$\sum\th_i\!+\!\ph\!=\!0$ (mod $2\pi$), which maximizes
$\mr{ReTr}(A\dag UDV\dag)$.
By means of the original singular value decomposition, the latter expression
equals $\mr{ReTr}(SD)$, and the expression to be maximized%
\footnote{Note that the global maximum is required. A possible strategy is to
perform a scan, in regularly spaced intervals, over all unconstrained $\th_i$,
followed by a local maximization starting from the largest value obtained in
the survey. Thus, staying content with intervals $\Delta\th_i\!=\!\pi/3$, the
global scan requires $6^{N-1}$ function calls.}
is $s_1\cos(\th_1)+...+s_N\cos(\th_N)$, still subject to the constraint
$\sum\th_i\!+\!\ph\!=\!0$ (mod $2\pi$).
The matrix $Y\!=\!UDV\dag$ is the desired solution.
\end{enumerate}
From (\ref{proj_SU}) it is clear that, if $A$ is subject to a random gauge
transformation $A\to g_1Ag_2\dag$ with $g_{1,2}\!\in\!SU(N)$, the effect must
be that the solution $Y$ transforms as $Y\to g_1Yg_2\dag$.
Up to a set of gauge configurations of measure zero, the singular value
decomposition $A\!=\!USV\dag$ is unique (here we assume a specific ordering
in $S$, e.g.\ $s_1\!>\!...\!>\!s_N\!>\!0$).
As a result, the effect of the random gauge transformation is just $U\to g_1U$,
$V\to g_2V$, while $S$ and the expression to be maximized are unchanged, and
the net effect is thus $UDV\dag\to g_1UDV\dag g_2\dag$, as expected.


\subsection{Tree-level improvement}

The Symanzik effective field theory of cut-off effects of undoubled lattice
Dirac operators is based on an analysis of all local mass dimension 5 operators
consistent with the symmetries of the theory \cite{Sheikholeslami:1985ij,
Heatlie:1990kg,Martinelli:1990ny,Luscher:1996sc,Luscher:1996ug}.
As long as one is content with perturbative or non-perturbative
$O(a)$-improvement of on-shell Green's functions, contact terms can be ignored
and it suffices to add a single improvement term, the so-called clover term
which is included in (\ref{def_wils},\,\ref{def_pref}).
Going through the arguments of \cite{Sheikholeslami:1985ij,Heatlie:1990kg,
Martinelli:1990ny,Luscher:1996sc,Luscher:1996ug}, one realizes that the leading
contribution, in the weak coupling expansion, to the coefficient in front
is independent of the details of the covariant derivative and Laplacian
occurring in the operator.
In other words, $c_\mr{SW}\!=\!1$ holds true, at tree level, also for our
Brillouin operator (\ref{def_pref}), while subleading contributions are, of
course, different.

With $c_\mr{SW}\!=\!1$ the leading cut-off effects are $O(\al a)$, and the overall
smearing strategy does not change this.
However, due to the smearing the coefficient of the $O(\al a)$
term might be so small, that the formally subleading $O(a^2)$ cut-off effects
might prove numerically dominant.


\section{Practical tests in quenched QCD}


\subsection{Scale setting and overall tuning strategy}

We use the Wilson gauge action and a parametrization of $r_0/a$ consistent
with asymptotics \cite{Durr:2006ky}
\beq
\log(r_0/a)={4\pi^2\ovr33}\,\be\,
{1-8.2384/\be+15.310/\be^2\ovr1-2.7395/\be-11.526/\be^2}
\eeq
which is based on data from \cite{Necco:2001xg}.
Upon choosing $L/a=10,12,16,20,24$ and requesting that $L/r_0=3.2653$
(tantamount to $L\!\simeq\!1.6\fm$ if $r_0\!\simeq\!0.49\fm$ is assumed), we
find that we should use the $\be$ values listed in Tab.\,\ref{tab:kappa_crit}.
Note that the quenched scale ambiguity of $\sim\!5\%$ does not limit our
ability to match boxes in terms of $L/r_0$.


We aim for comparisons at a fixed value of the light and the strange quark mass.
This will be achieved by tuning the pion and the $\bar{s}s$ mass (without
disconnected contributions) to $(r_0\Mpi)^2\!\simeq\!1.56\!\simeq\!1.25^2$ and
$(r_0\Mss)^2\!\simeq\!4.56$, respectively.
Given that $\Mss^2\!\simeq\!2\Mka^2\!-\!\Mpi^2$, this will correspond to
$(r_0\Mka)^2\!\simeq\!3.06\!\simeq\!1.75^2$, and hence to a pion mass of about
$500\MeV$ and a kaon mass of about $700\MeV$.
Finite volume effects are expected to be small, since $\Mpi L\!\simeq\!4.08$.


\subsection{Determination of $\ka_\mr{crit}$ and choice of
$\ka_\mr{light},\ka_\mr{strange},\ka_\mr{charm}$}

\begin{figure}[!tb]
\centering
\epsfig{file=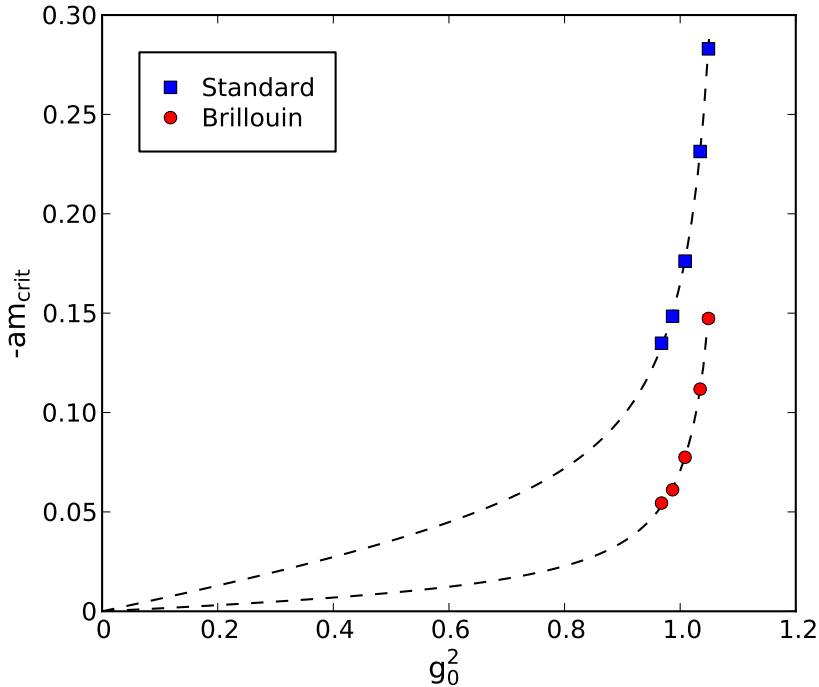,width=0.7\textwidth}\\
\vspace*{-4mm}
\caption{\label{fig:kappa_crit}\sl
Additive mass renormalization versus $g_0^2$ for the Wilson operator and the
Brillouin operator (both with $c_\mr{SW}\!=\!1$ and $\al_\mr{APE}\!=\!0.72$).
In either case a rational fit with the ansatz (\ref{mcrit_oneloop}) is
included. Error bars are significantly smaller than the size of the symbols.}
\end{figure}

As a first step we determine $\ka_\mr{crit}$ for either operator (with 1\,APE
step and $c_\mr{SW}\!=\!1$) over a range of $\be$-values, with results given in
Tab.\,\ref{tab:kappa_crit} and Fig.\,\ref{fig:kappa_crit}.
The perturbative formula reads ($\be=6/g_0^2$)
\beq
am_\mr{crit}=\Sigma_0=-{g_0^2\ovr16\pi^2}C_FS+O(g_0^4)\quad[<0]
\eeq
with $C_F=4/3$ and $am_\mr{crit}=1/(2\ka_\mr{crit})-4$.
At 1-loop order one finds $S=31.98644$ for thin-link Wilson operator with
$c_\mr{SW}\!=\!1$, and $S=4.07175$ for the 1\,APE ($\al\!=\!0.72$) variety with
$c_\mr{SW}\!=\!1$ \cite{Capitani:2006ni}, while for the Brillouin operator no
perturbative information is available.

\begin{table}[!tb]
\centering
\begin{tabular}{|ccc|cc|cc|}
\hline
$\be$ & geom. & $a^{-1}$[GeV] &
\multicolumn{2}{c|}{$\ka_\mr{crit}^\mr{std/std}$ (``Standard'') } &
\multicolumn{2}{c|}{$\ka_\mr{crit}^\mr{iso/bri}$ (``Brillouin'')} \\
\hline
5.72 & $10^3\!\times\!20$ & 1.236 & 0.134516(65) & 0.134533 & 0.129780(64) & 0.129798 \\
5.80 & $12^3\!\times\!24$ & 1.479 & 0.132673(47) & 0.132650 & 0.128594(30) & 0.128582 \\
5.95 & $16^3\!\times\!32$ & 1.978 & 0.130760(59) & 0.130769 & 0.127469(48) & 0.127471 \\
6.08 & $20^3\!\times\!40$ & 2.463 & 0.129818(45) & 0.129864 & 0.126940(30) & 0.126973 \\
6.20 & $24^3\!\times\!48$ & 2.964 & 0.129362(57) & 0.129303 & 0.126725(42) & 0.126676 \\
\hline
\end{tabular}
\caption{\label{tab:kappa_crit}\sl
Summary of $\ka_\mr{crit}$ for either operator with 1\,APE step and
$c_\mr{SW}\!=\!1$, as determined from direct measurements at these couplings
(with error bars) and through the fit (\ref{mcrit_oneloop}).}
\end{table}

To see how far from the perturbative regime we are, we fit our data to the
rational ansatz
\beq
-am_\mr{crit}={c_1g_0^2+c_2g_0^4\ovr1+c_3g_0^2}
\label{mcrit_oneloop}
\eeq
with 2 degrees of freedom, and compare the fitted $c_1$ to the 1-loop
prediction $S/(12\pi^2)$, where available.
For the Wilson operator $c_1\!=\!0.0623(82)$ deviates significantly from the
prediction $0.0344$, while for the Brillouin operator $c_1\!=\!0.0143(54)$,
without a perturbative prediction to compare to.
Given the quality of these fits, the interpolated $\ka_\mr{crit}$ are more
accurate than the direct measurements, and this is why we include these values
in Tab.\,\ref{tab:kappa_crit}.

\begin{figure}[!tb]
\centering
\epsfig{file=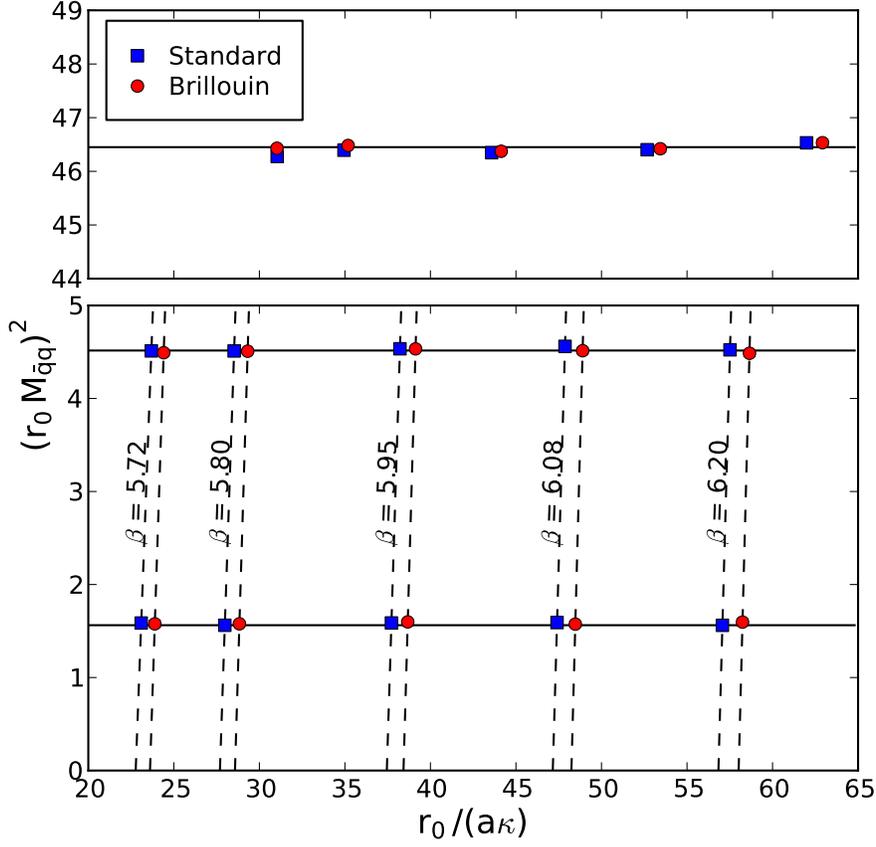,width=0.7\textwidth}
\vspace*{-4mm}
\caption{\label{fig:tuning}\sl
Summary of the final $(r_0\Mpi)^2$, $(r_0\Mss)^2$, $(r_0\Mcc)^2$ to test how
accurately the target values 1.56, 4.56, 46.5 were reached. Error bars are
smaller than the size of the symbols.}
\end{figure}

In order to perform a quenched scaling study we define, for each $\be$, three
reference $\ka$-values which realize $(r_0\Mpi)^2\!=\!1.56$,
$(r_0\Mss)^2\!=\!4.56$ and $(r_0\Mcc)^2\!=\!46.5$.
We call them $\ka_\mr{light}$, $\ka_\mr{strange}$, and $\ka_\mr{charm}$,
respectively (even though the first two are heavier than the respective
physical flavors, and the last one is lighter than the physical charm quark).
These values are determined, for each coupling, by interpolating the results of
a few tuning runs.
The three reference $\ka$-values are then evaluated on the full ensembles, and
the resulting $(r_0M_P)^2$ are compared to the target values in
Fig.\,\ref{fig:tuning}.
It seems the tuning is accurate enough, so that we can proceed with a study
of the scaling of the decay constants.


\subsection{Comparing the scaling of decay constants at fixed $r_0M_P$}

The decay constants $\Fpi,\Fss,\Fcc$ are determined from the improved
and renormalized current
\beq
A^\mr{ren}_\mu=Z_A(1\!+\!b_Aam^\mr{W})(A_\mu\!+\!ac_A\bar\pad_\mu P)
\label{def_aimpren}
\eeq
where $A_\mu$ and $P$ denote the naive axial-vector current and pseudo-scalar
density, respectively, and $m^\mr{W}\!=\!m_0\!-\!m_\mr{crit}$.
In practice, $m^\mr{W}$ in (\ref{def_aimpren}) is often replaced by the PCAC
quark mass
\beq
m^\mr{PCAC}\!=\!\frac
{\sum_{\mb{x}}\<\bar\pad_4[A_4(x)\!+\!ac_A\bar\pad_4 P(x)]O(0)\>}
{2\sum_{\mb{x}}\<P(x)O(0)\>}
\label{def_mpcac}
\eeq
where $\bar\pad$ denotes the symmetric derivative, and usually $O\!\equiv\!P$
is chosen to get maximal signal.
Here it is assumed that the two quark masses are equal; in general the
improvement factor in (\ref{def_aimpren}) is
$(1\!+\!b_Aa(m_j^\mr{W}\!+\!m_k^\mr{W})/2)$ for flavors $j,k$, and the l.h.s.\
of (\ref{def_mpcac}) is $(m_j^\mr{PCAC}\!+\!m_k^\mr{PCAC})/2$.

\begin{figure}[!tb]
\centering
\epsfig{file=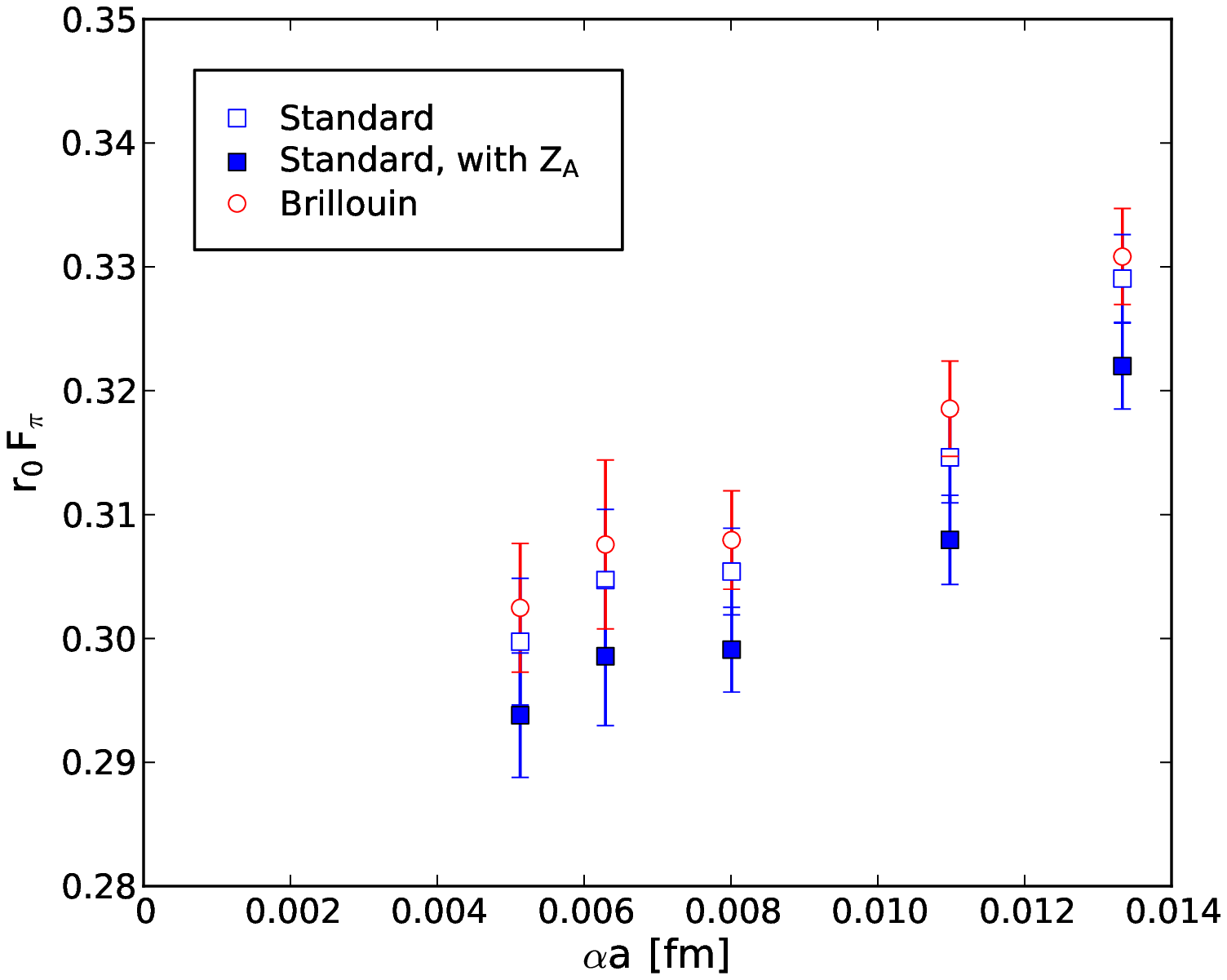,width=0.49\textwidth}
\epsfig{file=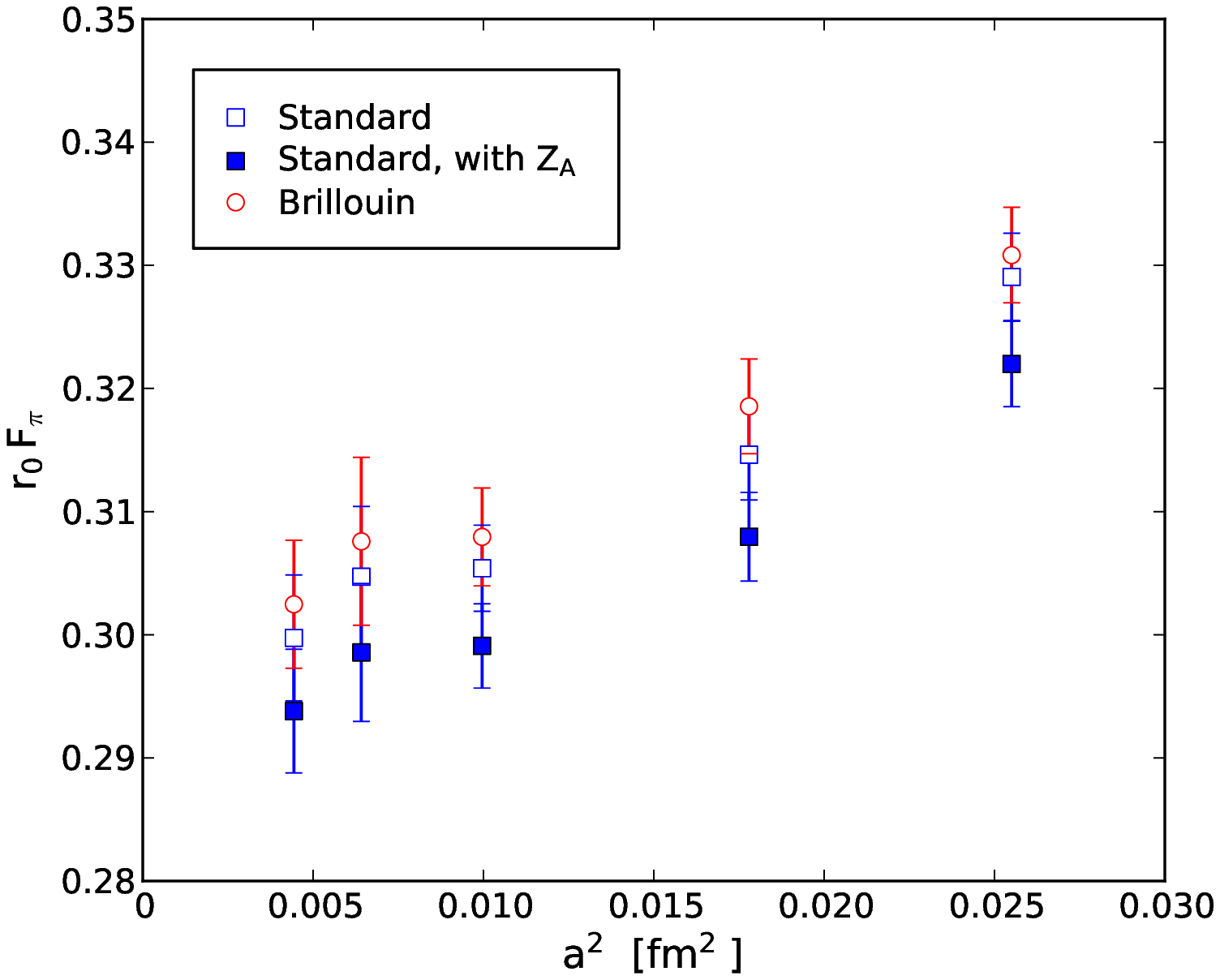,width=0.49\textwidth}\\
\epsfig{file=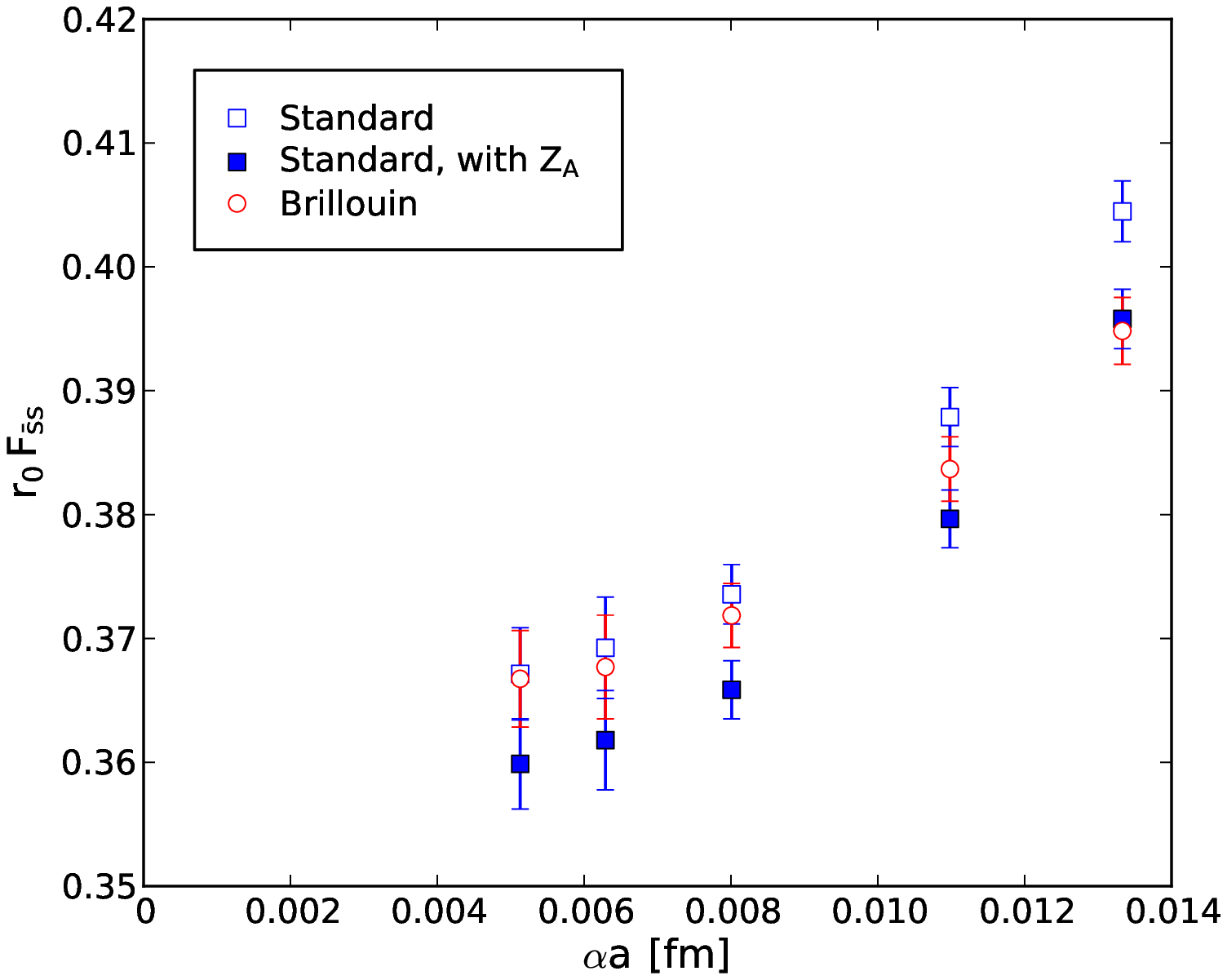,width=0.49\textwidth}
\epsfig{file=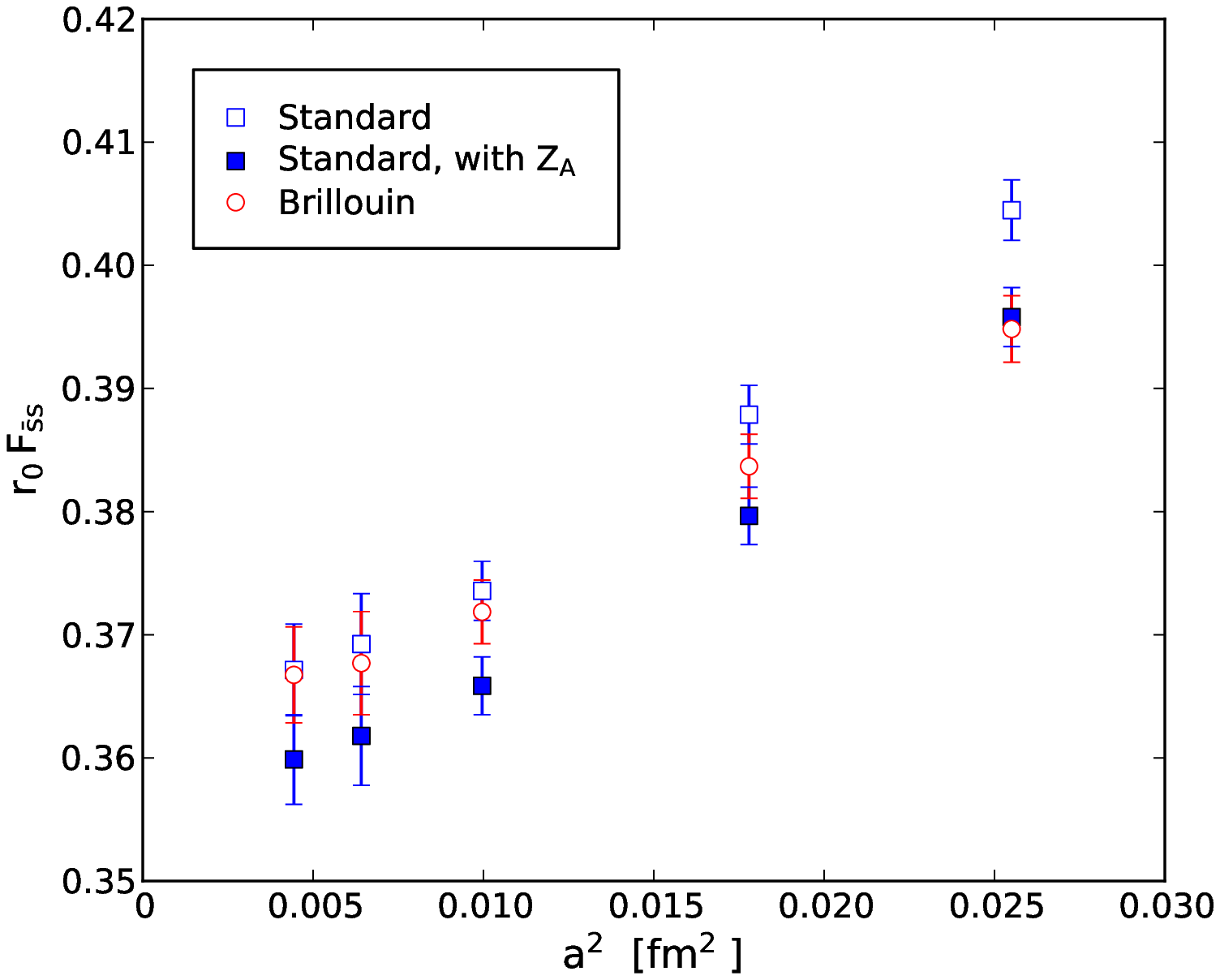,width=0.49\textwidth}\\
\epsfig{file=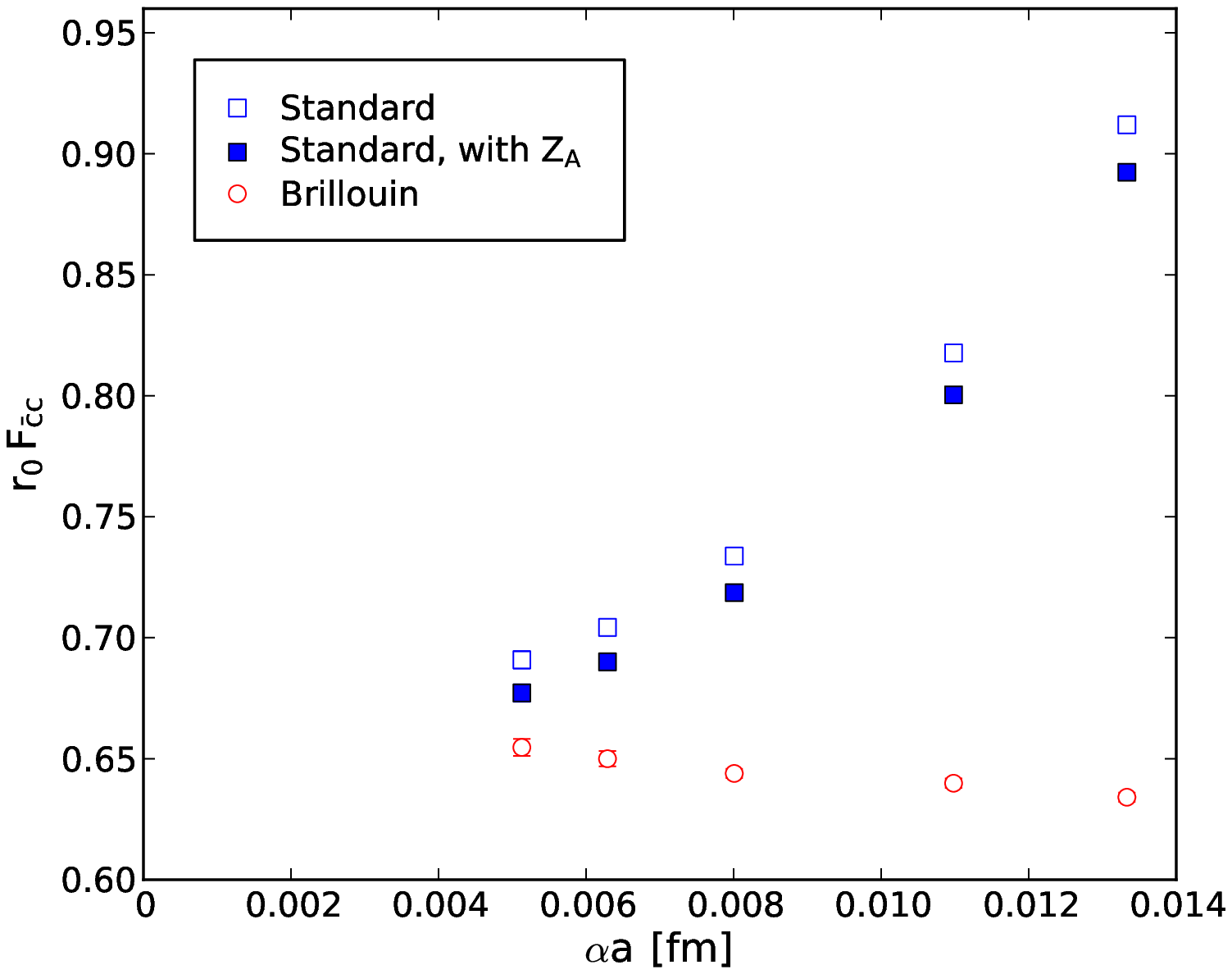,width=0.49\textwidth}
\epsfig{file=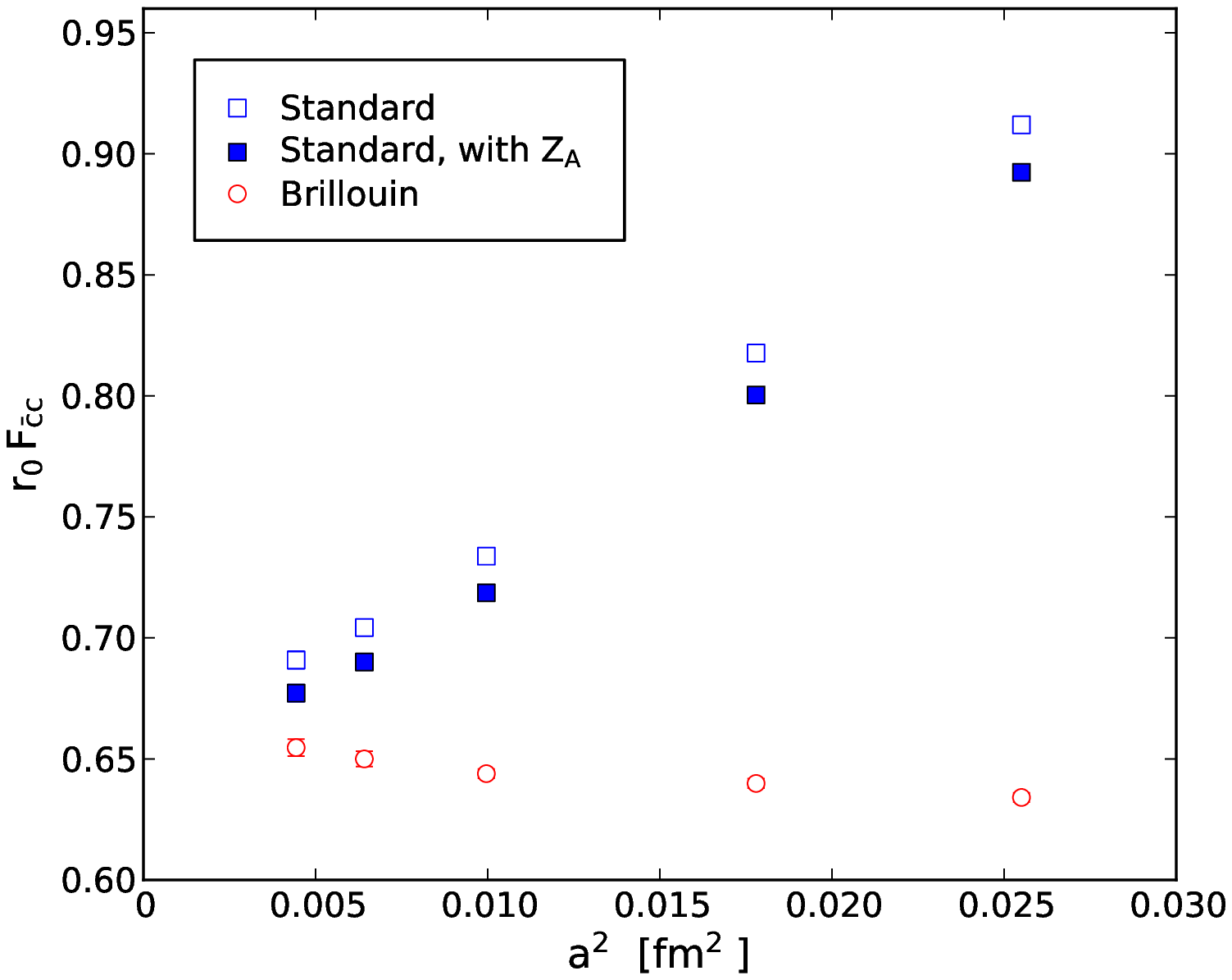,width=0.49\textwidth}
\vspace*{-4mm}
\caption{\label{fig:decay_ind}\sl
Decay constants $\Fpi$ (top), $\Fss$ (middle), $\Fcc$ (bottom) in $r_0$ units
versus $\al a$ (left) and $a^2$ (right). Open symbols indicate the bare values,
filled symbols include the 1-loop $Z_A$.}
\end{figure}

We use the tree-level improvement coefficients $b_A\!=\!1$, $c_A\!=\!0$.
The 1-loop renormalization constant $Z_A\!=\!1\!-\!g_0^2z_A/(12\pi^2)$, which
is needed for consistency, is known for the Wilson operator ($z_A\!=\!2.42423$
with 1 step of $\al_\mr{APE}\!=\!0.72$ smearing and $c_\mr{SW}\!=\!1$ is found
in \cite{Capitani:2006ni}), but not for the Brillouin operator.
In Fig.\,\ref{fig:decay_ind} we plot the decay constants $\Fpi$, $\Fss$, $\Fcc$
versus $\al a$ (left) and $a^2$ (right).
Here everything is made dimensionless through $r_0$.
In case of the Wilson operator the lattice-to-continuum matching factor $Z_A$
is included, but it brings a rather small shift, since it is already close to
1 in the range of couplings where we have data, and it approaches 1 as $a\to0$.
With hindsight we can thus anticipate that also for the Brillouin operator the
data without $Z_A$ are indicative of the approach to the continuum.
Comparing the two operators without $Z_A$, we see little difference in the
light and strange pseudo-scalar data (top and middle), while there is a
pronounced difference in the charm sector (bottom row).
Hence, for the scaling of $r_0\Fcc$ the Brillouin operator seems to bring a
significant improvement.

\begin{figure}[!tb]
\centering
\epsfig{file=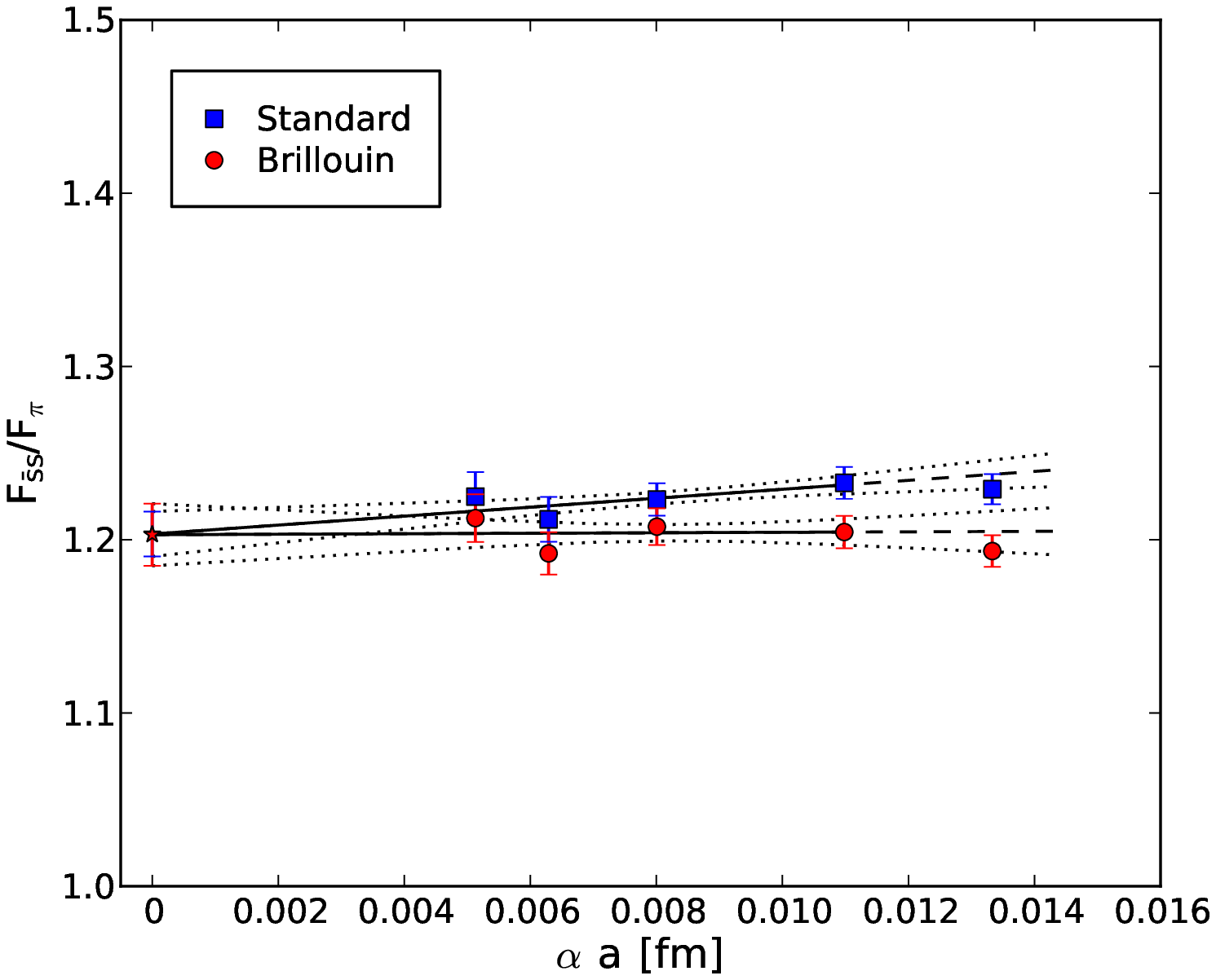,width=0.49\textwidth}
\epsfig{file=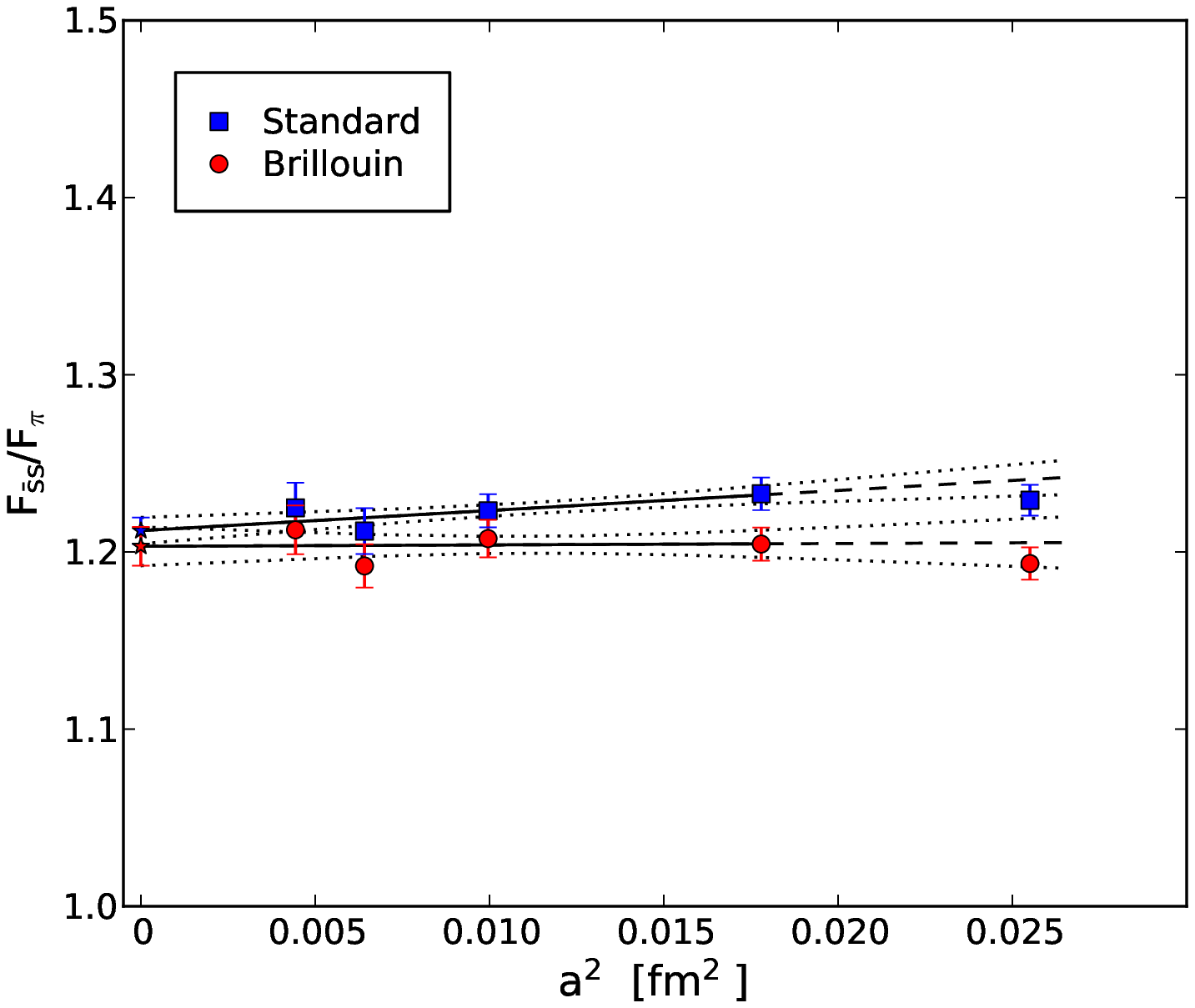,width=0.49\textwidth}\\
\epsfig{file=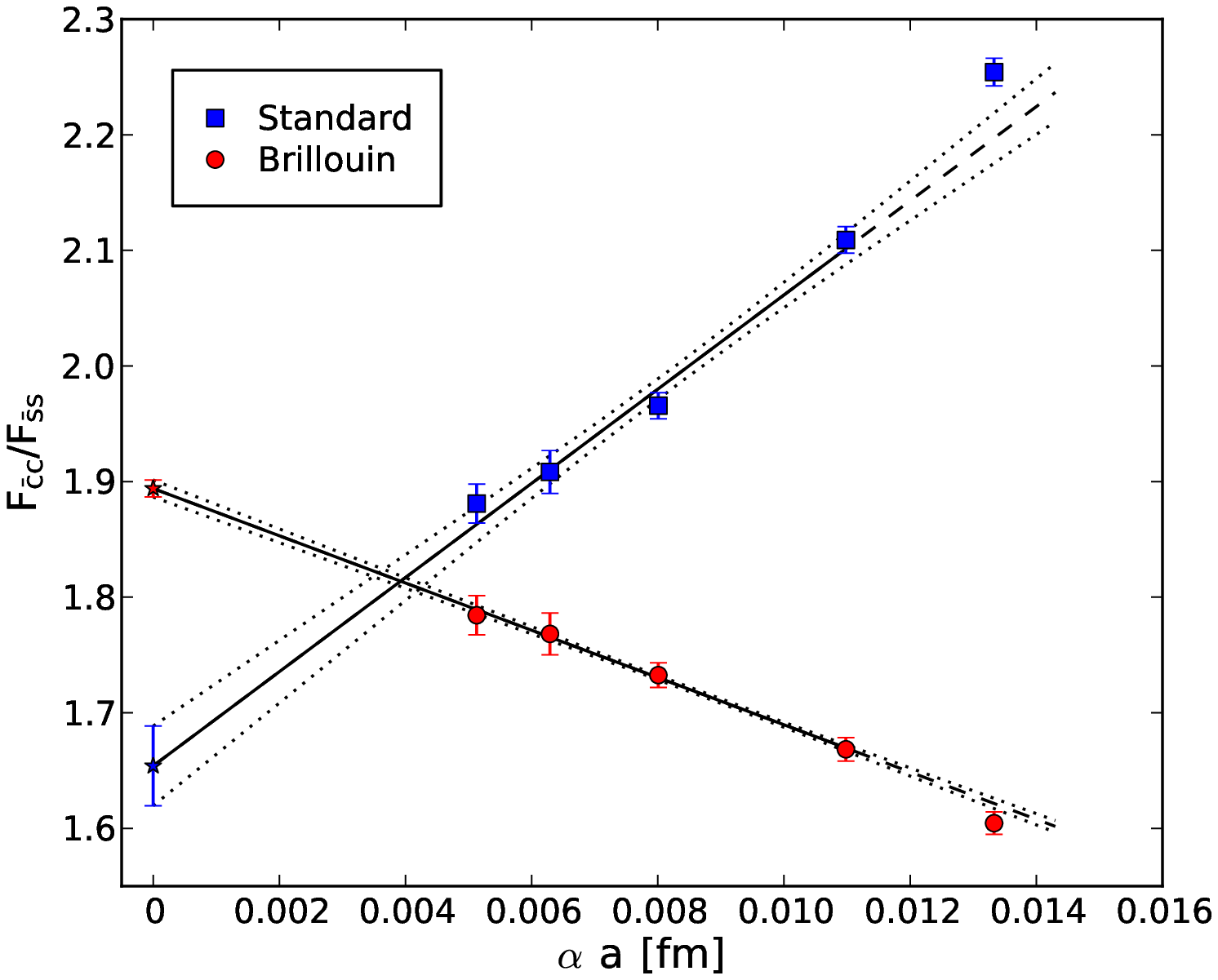,width=0.49\textwidth}
\epsfig{file=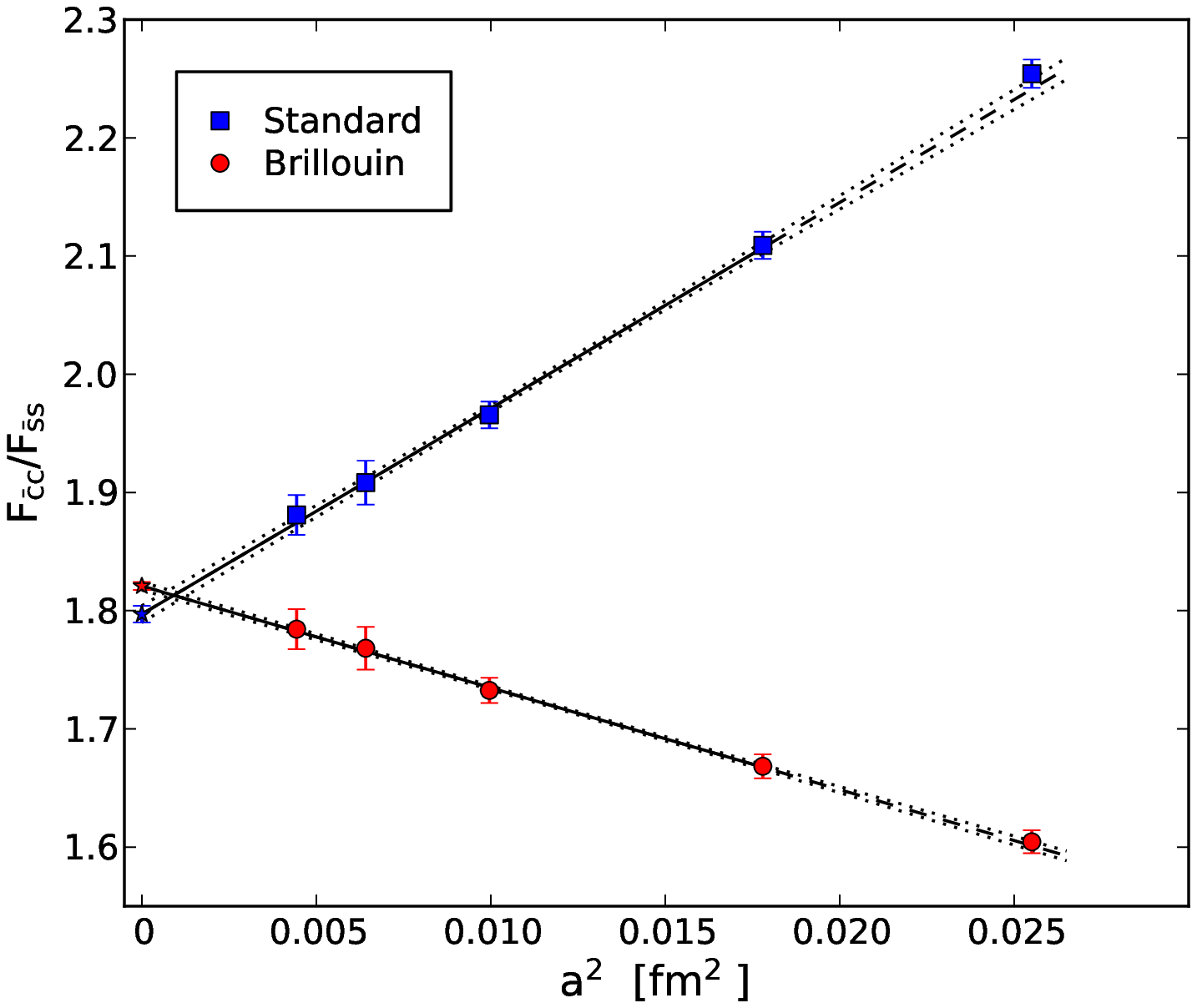,width=0.49\textwidth}
\vspace*{-4mm}
\caption{\label{fig:decay_rat}\sl
The ratios $\Fss/\Fpi$ (top) and $\Fcc/\Fss$ (bottom) versus $\al a$ (left) and
$a^2$ (right). The linear fits include only 4 lattice spacings, and favor the
pure $a^2$ over the pure $\al a$ extrapolation.}
\end{figure}

\begin{figure}[!tb]
\centering
\epsfig{file=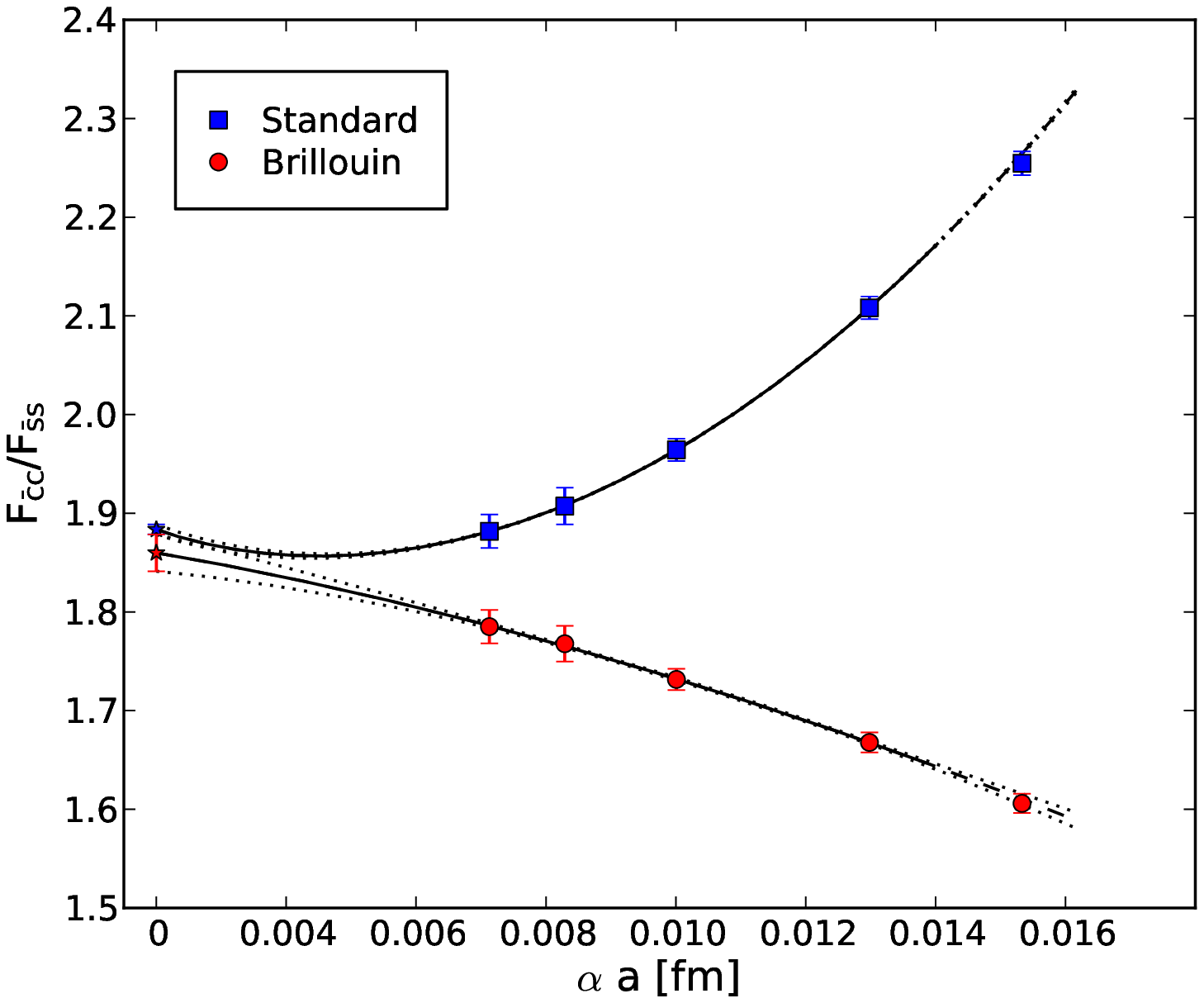,width=0.49\textwidth}
\epsfig{file=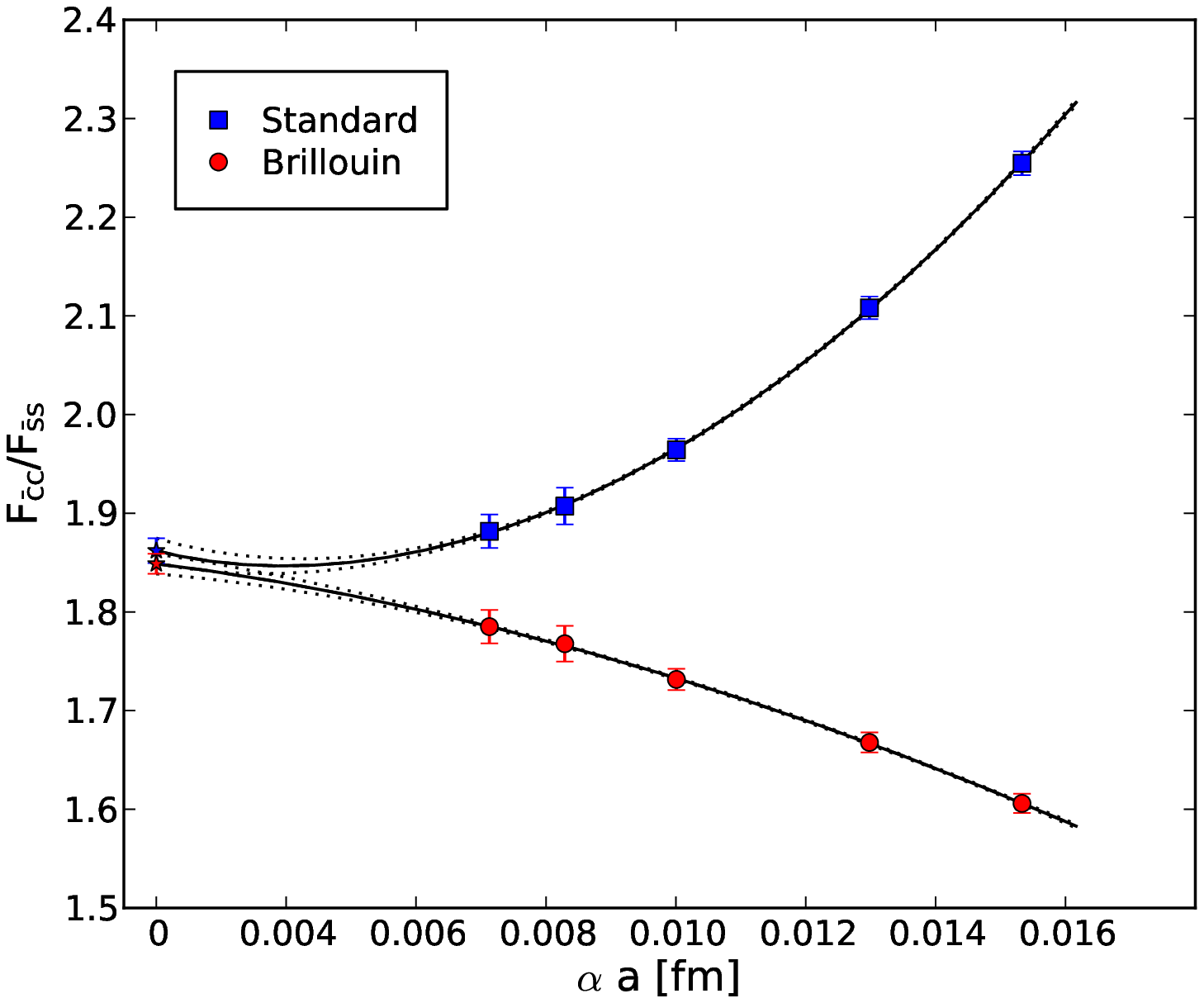,width=0.49\textwidth}
\vspace*{-4mm}
\caption{\label{fig:decay_fit}\sl
Fit of the mixed $\al a$ plus $a^2$ ansatz (\ref{alphatimesa_asq}) to the ratio
$\Fcc/\Fss$ with 4 (left) or 5 (right) lattice spacings included.}
\end{figure}

To get rid of the $Z_A$ factors, we also consider the scaling of the ratios
$\Fss/\Fpi$ and $\Fcc/\Fss$, as shown in Fig.\,\ref{fig:decay_rat}.
Again, we plot the data against $\al a$ (left) and $a^2$ (right).
For the strange-to-light ratio all data happen to be essentially flat, so there
is no advantage of one operator over the other.
For the charm-to-strange ratio, the situation is different.
Fitting the data on the four finer lattices with a pure $\al a$ ansatz yields
two continuum extrapolated results which are not consistent (lower left panel).
Fitting the same data with a pure $a^2$ ansatz leads to two continuum
extrapolated results which are almost consistent (lower right panel).
If we restrict the fits to the three finest lattice spacings, the values
obtained with the pure $\al a$ hypothesis stay inconsistent, while the
continuum results with the pure $a^2$ hypothesis become consistent.
To prevent any misunderstanding, let us emphasize that we think that both
operators have a contribution in $\al a$ and $a^2$ at accessible lattice
spacings.
Still, to the best of our knowledge, this is the first figure which indicates
that, for a tree-level improved operator with some link-smearing, the pure
$a^2$ hypothesis might be closer to the truth than the (formally correct) pure
$\al a$ hypothesis.
Of course, with infinitely precise data one could separate the two
contributions.
To see how far we are from this ideal world, we try a fit of the ratio
$\Fcc/\Fss$ with the ansatz
\beq
\Fcc/\Fss=d_0+d_1\al(a)a+d_2a^2
\label{alphatimesa_asq}
\eeq
giving results shown in Fig.\,\ref{fig:decay_fit}.
The fitted $d_1,d_2$ of the Brillouin operator are significantly smaller than
those of the Wilson operator.
Also by looking at the fits one would say that the Brillouin data alone leave
little doubt that the correct continuum value is somewhere near 1.85, while
with the Wilson data alone this is far from obvious.


\subsection{Comparing the $1/n_\mr{BiCGstab}$ distributions at fixed $r_0\Mpi$}

In quenched QCD with Wilson fermions so-called exceptional configurations (on
which the massive Dirac operator $D_m$ could not be inverted) hindered the
approach to light quark masses.
In full QCD the functional measure suppresses configurations on which $D_m$ has
near-zero modes.
Still, the issue persists in the form of instabilities in the HMC evolution.

\begin{figure}[!tb]
\centering
\epsfig{file=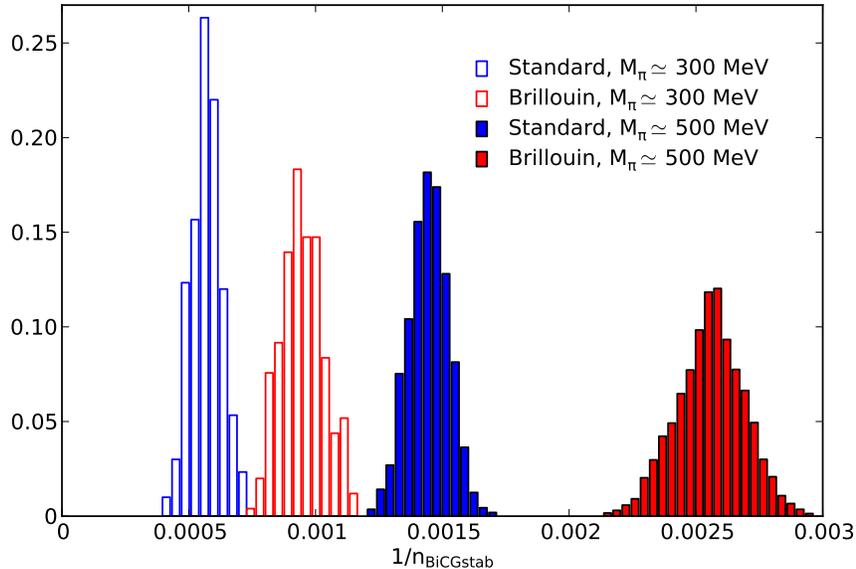,width=0.7\textwidth}
\vspace*{-4mm}
\caption{\label{fig:hist_iter}\sl
Distribution of the inverse iteration count $1/n_\mr{BiCGstab}$ to reach a
norm $\ep\!=\!10^{-7}$ of the residual at $\be\!=\!6.20$ with $\ka$ tuned
to have $(r_0\Mpi)^2\!=\!1.56$ (``500\,MeV'') or $0.56$ (``300\,MeV'').}
\end{figure}

In \cite{DelDebbio:2005qa} it was shown that the stability of these simulations
is linked to the distribution of the lowest eigenvalue of $D_m\dag D_m^{}$.
The latter is roughly Gaussian distributed, and the simulation is deemed safe
as long as the center of the distribution is at least four standard deviations
away from zero.
The BMW collaboration noticed that the smallest eigenvalue of $D_m\dag D_m^{}$
is directly related to the number of iterations in the inversion, and used the
inverse iteration count $1/n_\mr{CG}$ in the monitoring \cite{Durr:2008rw}.
In Fig.\,\ref{fig:hist_iter} we present $1/n_\mr{BiCGstab}$ for either operator
at the values $(r_0\Mpi)^2\!=\!1.56$ and $0.56$ ($\Mpi\!\sim\!500\MeV$ and
$300\MeV$).
In either case an inversion with the Brillouin operator requires about 60\% of
the forward applications%
\footnote{For fixed $\Mpi$ the smallest eigenvalues of the two $A=D_m\dag D_m$
are approximately equal, while the largest eigenvalue is near $2.5^2$ for the
Brillouin operator and near $7.5^2$ for the Wilson operator. Since
$n_\mr{CG}\propto\sqrt{\mr{CN}(A)}$ one would expect the relative iteration
count to be around $1/3$ for CG and around $1/\sqrt{3}\simeq0.6$ for BiCGstab.}
of the Wilson operator.
Finally, according to the safety criterion mentioned above, there seems to be
a slight advantage for the Brillouin operator at low $\Mpi$.
Using another $\be$-value did not bring any major change.


\subsection{Comparing the statistical fluctuations at fixed $r_0\Mpi$}

\begin{figure}[!tb]
\centering
\epsfig{file=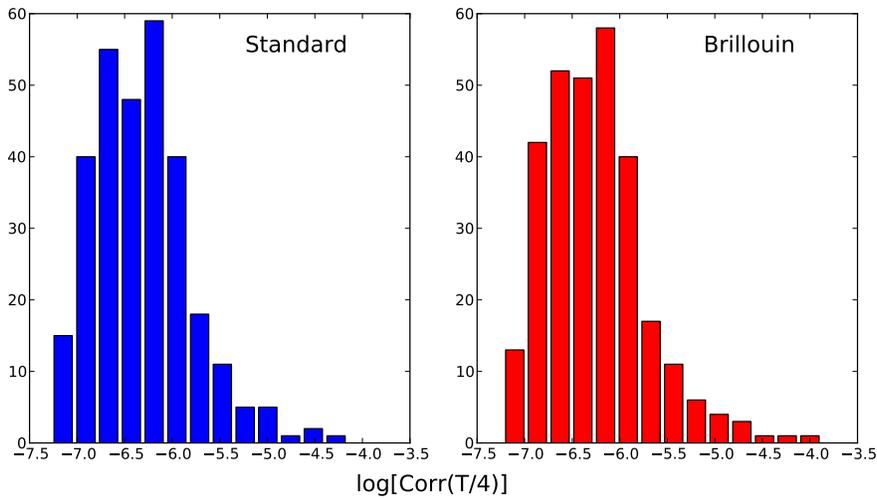,width=0.7\textwidth}
\vspace*{-6mm}
\caption{\label{fig:hist_corr}\sl
Distribution of the logarithm of the correlator $P(T/4)\bar{P}(0)$ at
$\be\!=\!6.20$ with $\ka$ tuned to have $(r_0\Mpi)^2\!=\!1.56$ in either case.}
\end{figure}

Reaching the same statistics for Wilson and Brillouin data may or may not be
a good guide to obtain equally precise physics results.
To compare the fluctuations with either operator we compare the variance in
the correlator $C_{P\bar{P}}(t)$ at $t\!=\!T/4$.
The result is shown in Fig.\,\ref{fig:hist_corr} in the form of a histogram,
with either $\ka$ tuned to realize $(r_0\Mpi)^2\!=\!1.56$.
Essentially, there is no noticeable difference between the two operators.
Looking at other $\be$-values we arrived at the same conclusion


\section{Suitability as overlap kernel}


\subsection{Details of the overlap action}

Given any undoubled (flavor symmetry respecting) ``kernel'' Dirac operator
$D_{\mr{kn},m}$ at a quark mass $m$, the massless overlap operator is defined
through \cite{Neuberger:1997fp}
\beq
D_\mr{ov}=
D_{\mr{ov},0}=
{\rh\ovr a}\,
[1+D_{\mr{kn},-\rh/a}(D_{\mr{kn},-\rh/a}\dag D_{\mr{kn},-\rh/a})^{-1/2}]=
{\rh\ovr a}\,
[1+\gaf\,\mr{sign}(\gaf D_{\mr{kn},-\rh/a})]
\label{def_over}
\eeq
with $0\!<\!\rh\!<\!2$.
Traditionally, the kernel parameter $\rh$ was tuned to a value above 1 to
maximize the locality of $D_\mr{ov}$ on coarse lattices
\cite{Hernandez:1998et}.
However, on fine lattices (and with some link smearing or filtering of the
kernel also on relatively coarse ones) maximal locality is obtained for
$\rh\!<\!1$ \cite{Kovacs:2002nz,Durr:2005an}.
As $\rh$ is part of the action definition, it is desirable to keep it fixed,
and we stay with the canonical choice $\rh\!=\!1$.

The massive overlap operator follows by adding a ``chirally rotated'' scalar
term \cite{Neuberger:1997bg}
\beq
D_{\mr{ov},m}=D_\mr{ov}+m(1\!-{a\ovr2\rh}D_\mr{ov})=
(1\!-{am\ovr2\rho})D_\mr{ov}+m
\label{def_mass}
\eeq
which yields an operator with a circular eigenvalue spectrum of radius
$\rh\!-\!am/2$ around the point $(\rh\!+\!am/2,0)$ in the complex plane.

There is still a choice to be made regarding the filtering of the underlying
kernel operator (we use 1 and 5 APE steps) and whether one wants to equip it
with a clover term.


\subsection{Comparing the near-normality of the kernels}

Wilson-type operators are usually non-normal, i.e.\ $[D,D\dag]\!\neq\!0$
\cite{Hip:2001hc}.
This means that the spectral representation takes the form
$D=\sum\la_n|\ph_n\>\<\ps_n|$, with no simple connection
between $|\ph_n\>$ and $\<\ps_n|$ and, as a result of this, no simple
connection between the eigenvalue spectra of $D$ and $D\dag D$.
Chiral operators are usually normal, i.e.\ $[D,D\dag]\!=\!0$ for a staggered or
overlap Dirac operator.
This means that $D=\sum\la_n|\ps_n\>\<\ps_n|$, with $|\ps_n\>$ and $\<\ps_n|$
being the complex conjugate transpose of each other, and the spectrum of
$D\dag D$ can be inferred from the one of $D$.

\begin{figure}[!tb]
\centering
\epsfig{file=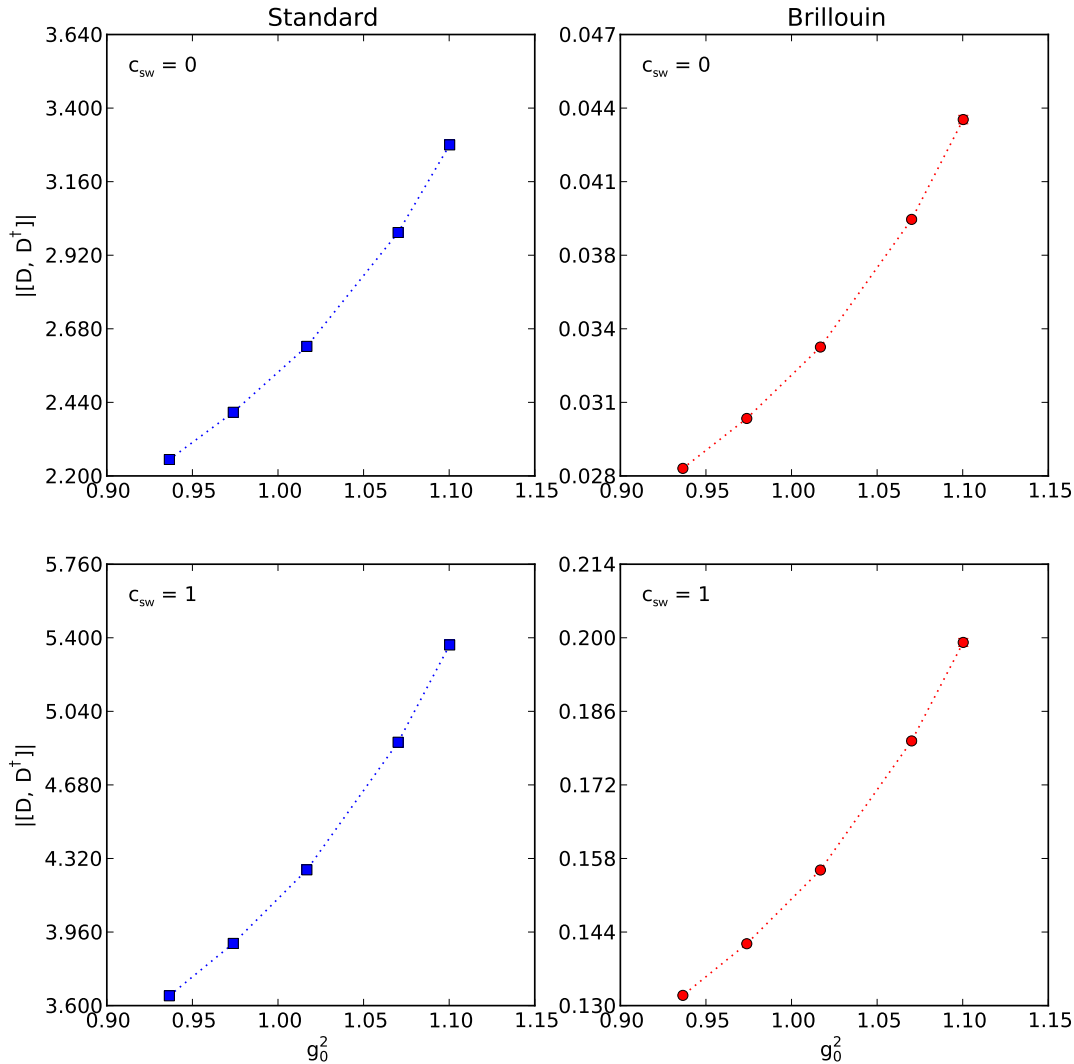,width=0.95\textwidth}
\vspace*{-8mm}
\caption{\label{fig:normality}\sl
Non-normality $||\,[D,D\dag]\,||$ versus $g_0^2$, after 1 APE step, for the
Wilson (left) and Brillouin (right) operator, with $c_\mr{SW}\!=\!0$ (top) and
$1$ (bottom). Note the difference in scale.}
\end{figure}

In this sense one may understand the non-normality of a Wilson-type fermion,
defined as the norm of the commutator, as a measure of ``how far'' it still is
from a formulation with continuum-like features.
Therefore, we measure $||(D\dag\!D-DD\dag)\et||$ for a few dozen normalized
Gaussian random vectors $\et$, with $D$ being the Wilson or the Brillouin
operator.
By doing this on 15 configurations for each ensemble, we obtain the data shown
in Fig.\,\ref{fig:normality}.
Unsurprisingly, with either operator the non-normality decreases towards the
continuum, but the Brillouin operator fares significantly better.
Surprisingly, switching on the clover term increases the non-normality, but it
remains true that with the Brillouin operator the norm of the commutator is
about an order of magnitude smaller than with the Wilson operator.

\subsection{Comparing the Ginsparg-Wilson violation of the kernels}

\begin{figure}[!tb]
\centering
\epsfig{file=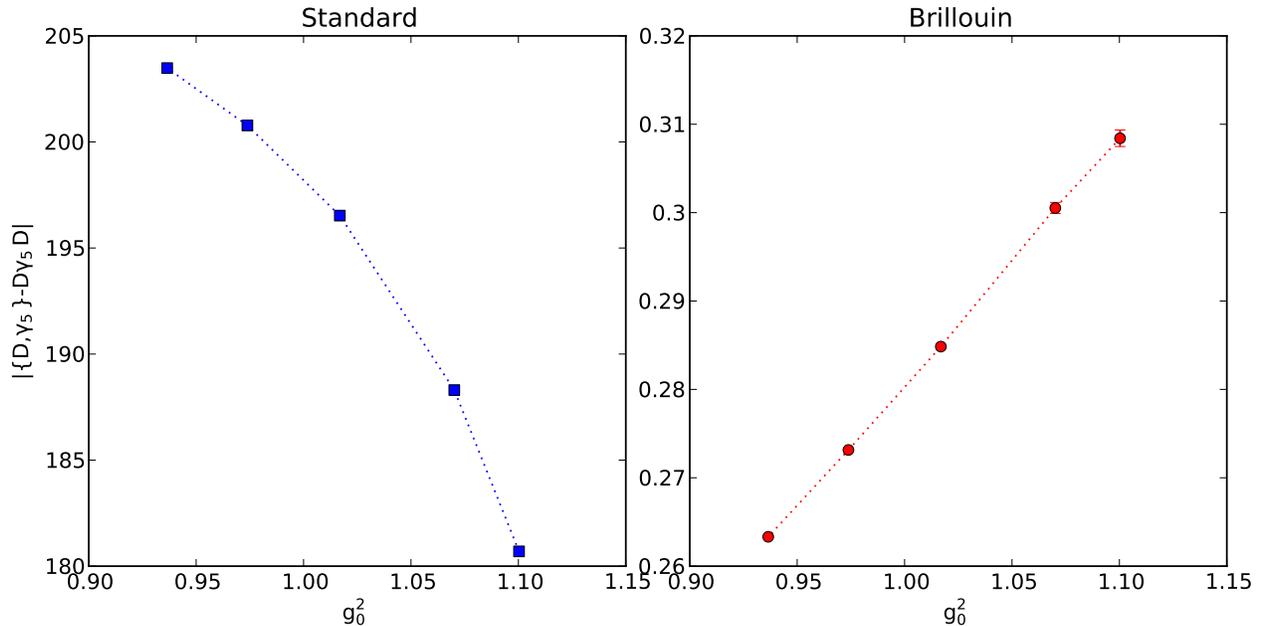,width=\textwidth}
\vspace*{-8mm}
\caption{\label{fig:ginswils}\sl
Violation of the Ginsparg-Wilson relation (\ref{def_ginswils}) with $\rh\!=\!1$
as a function of $6/\be$ for either operator (see text for details). Note the
difference in scale.}
\end{figure}

The Ginsparg-Wilson relation for a massless $D$ reads \cite{Ginsparg:1981bj}
\beq
\gaf D+D\gaf=\frac{a}{\rh}D\gaf D
\label{def_ginswils}
\eeq
and we intend to plug in our operators with $\ka$ set to $\ka_\mr{crit}$.
Since the latter are known only for $c_\mr{SW}\!=\!1$
(cf.\ Tab.\,\ref{tab:kappa_crit})
we do this with improvement.
We measure $||(D\gaf+\gaf D-D\gaf D)\et||$ for a few dozen
normalized Gaussian random vectors $\et$ on 15 configurations of each ensemble.
The result is shown in Fig.\,\ref{fig:ginswils}.
A priori it is not clear whether it makes sense to plug a distinctly
non-chiral operator into the Ginsparg-Wilson relation (\ref{def_ginswils}), but
the result of our experiment seems to suggest that at least for the Brillouin
operator it does.

\subsection{Comparing the condition number of the hermitean kernels}

The cost of the overlap construction is determined by the smallest mode (in
absolute magnitude) of the \emph{shifted} hermitean kernel
$H_{\mr{kn},-\rh/a}\!=\!\gaf D_{\mr{kn},-\rh/a}\!=\!\gaf(D_\mr{kn}\!-\!\rh/a)$.
Equivalently, one can look at the condition number of the squared operator
$A\!=\!D_{\mr{kn},-\rh/a}\dag D_{\mr{kn},-\rh/a}^{}$.
In practice, one considers the so-called Ritz eigenvalues, i.e.\ the
eigenvalues of the symmetric tridiagonal matrix that emerges from the Lanczos
process on $A$.
They approximate the extremal eigenvalues of $A$.

\begin{figure}[!p]
\centering 
\epsfig{file=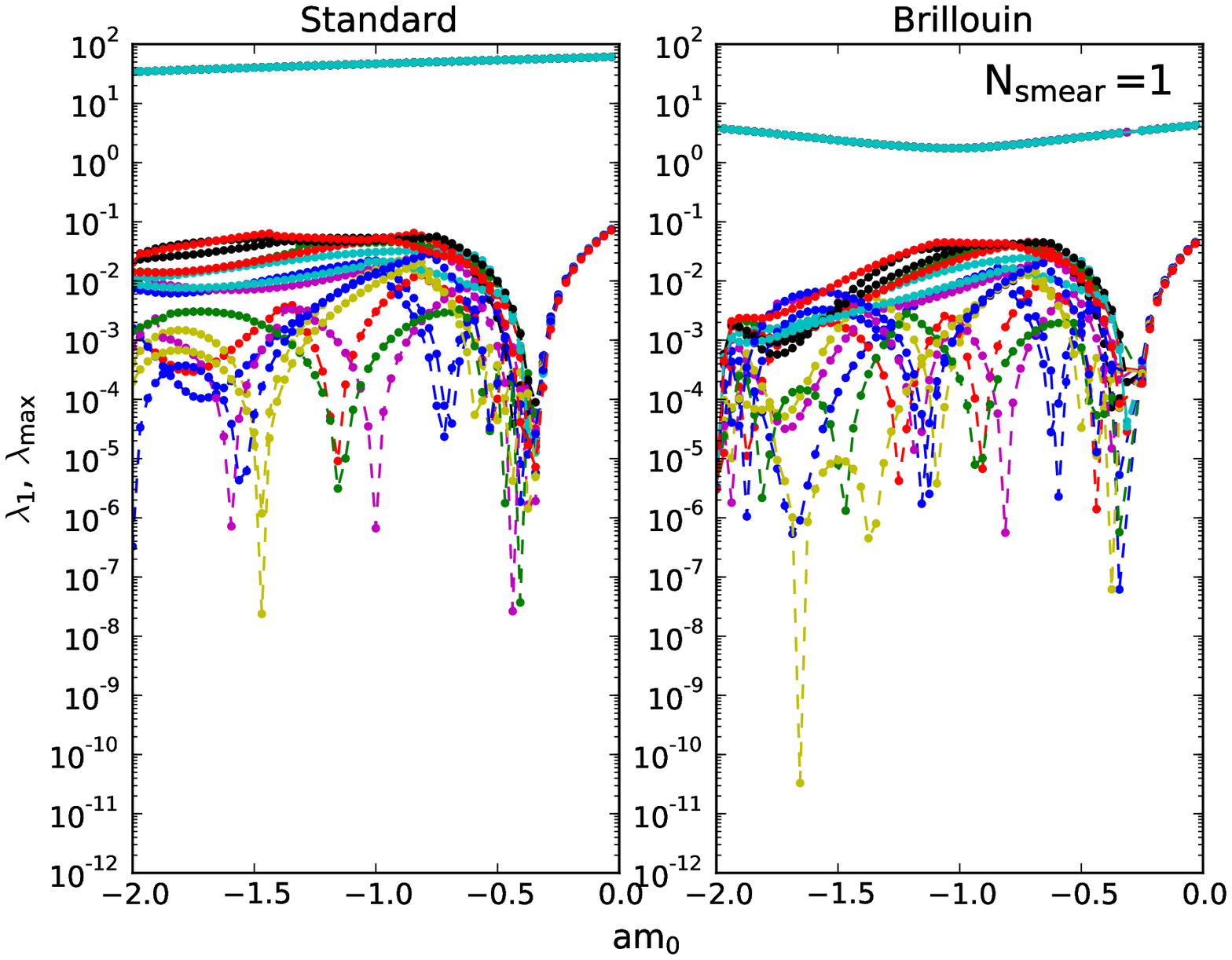,height=0.46\textheight}\\
\epsfig{file=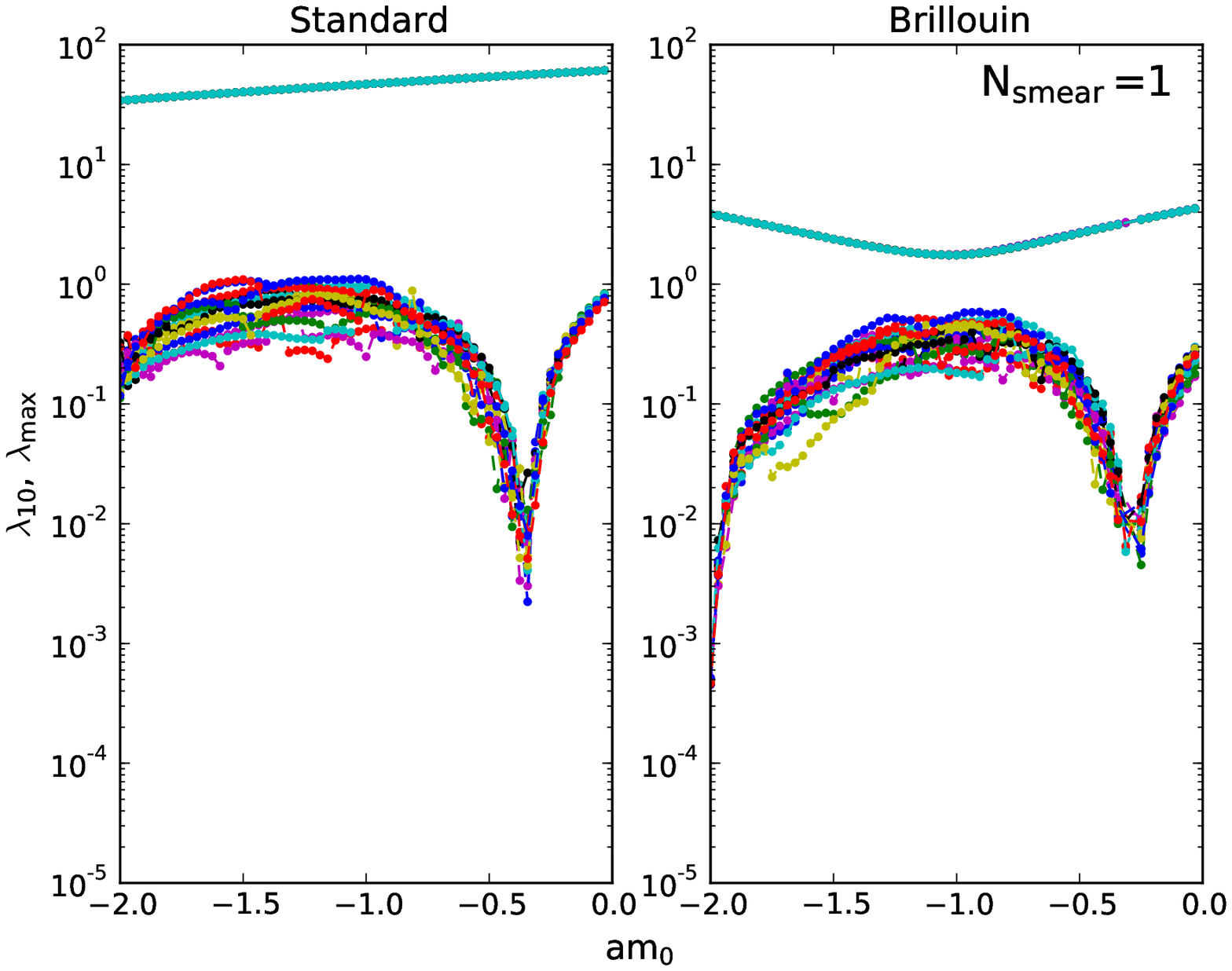,height=0.46\textheight}\\
\vspace*{-4mm}
\caption{\label{fig:ev_smear1}\sl
The 1st (top), 10th (bottom) and largest Ritz eigenvalue of $D_m\dag D_m^{}$
for the Wilson and the Brillouin operator, on 16 configs at $\be\!=\!6.20$,
with $c_\mr{SW}\!=\!0$ and 1 step of APE smearing.}
\end{figure}

\begin{figure}[!p]
\centering 
\epsfig{file=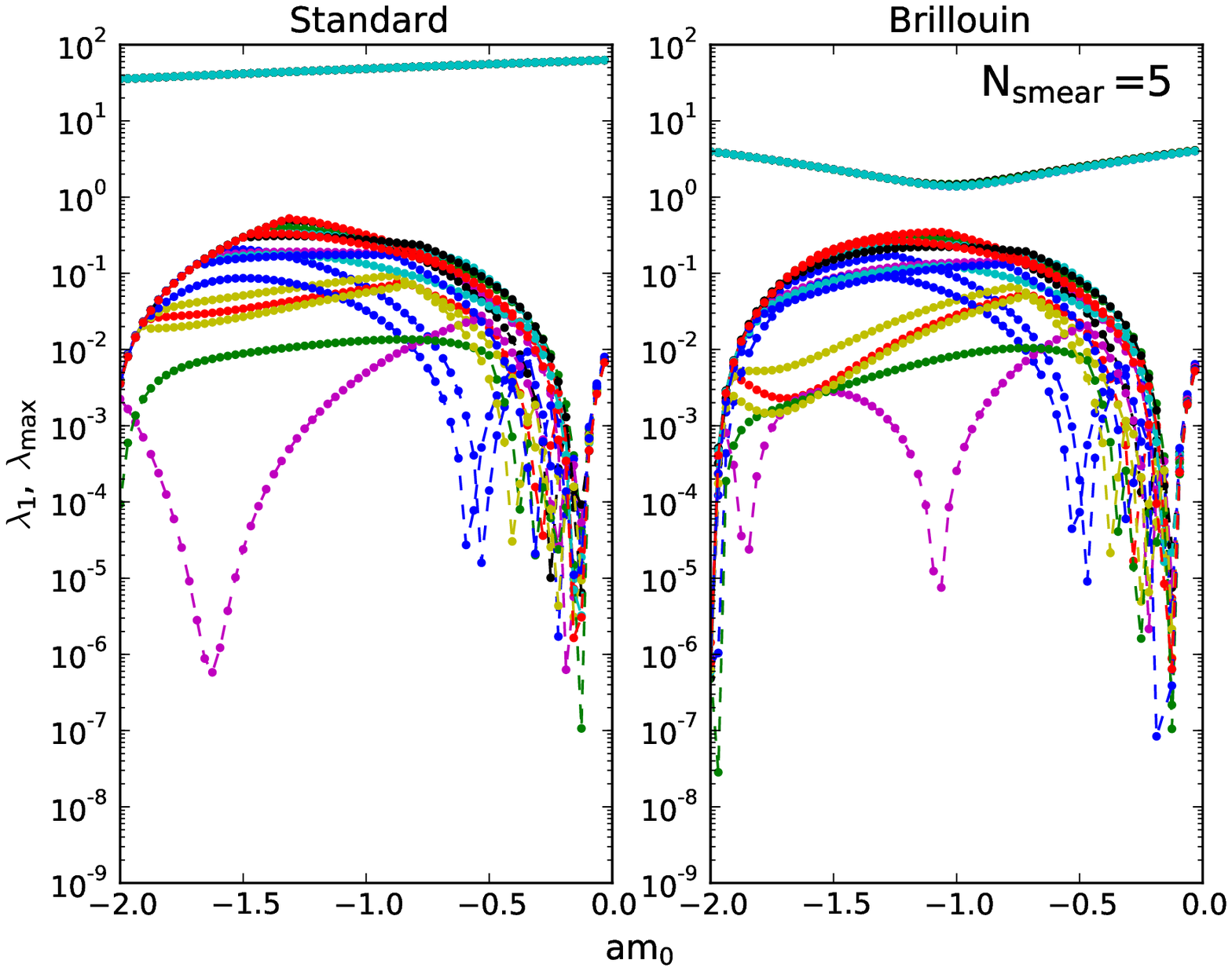,height=0.46\textheight}\\
\epsfig{file=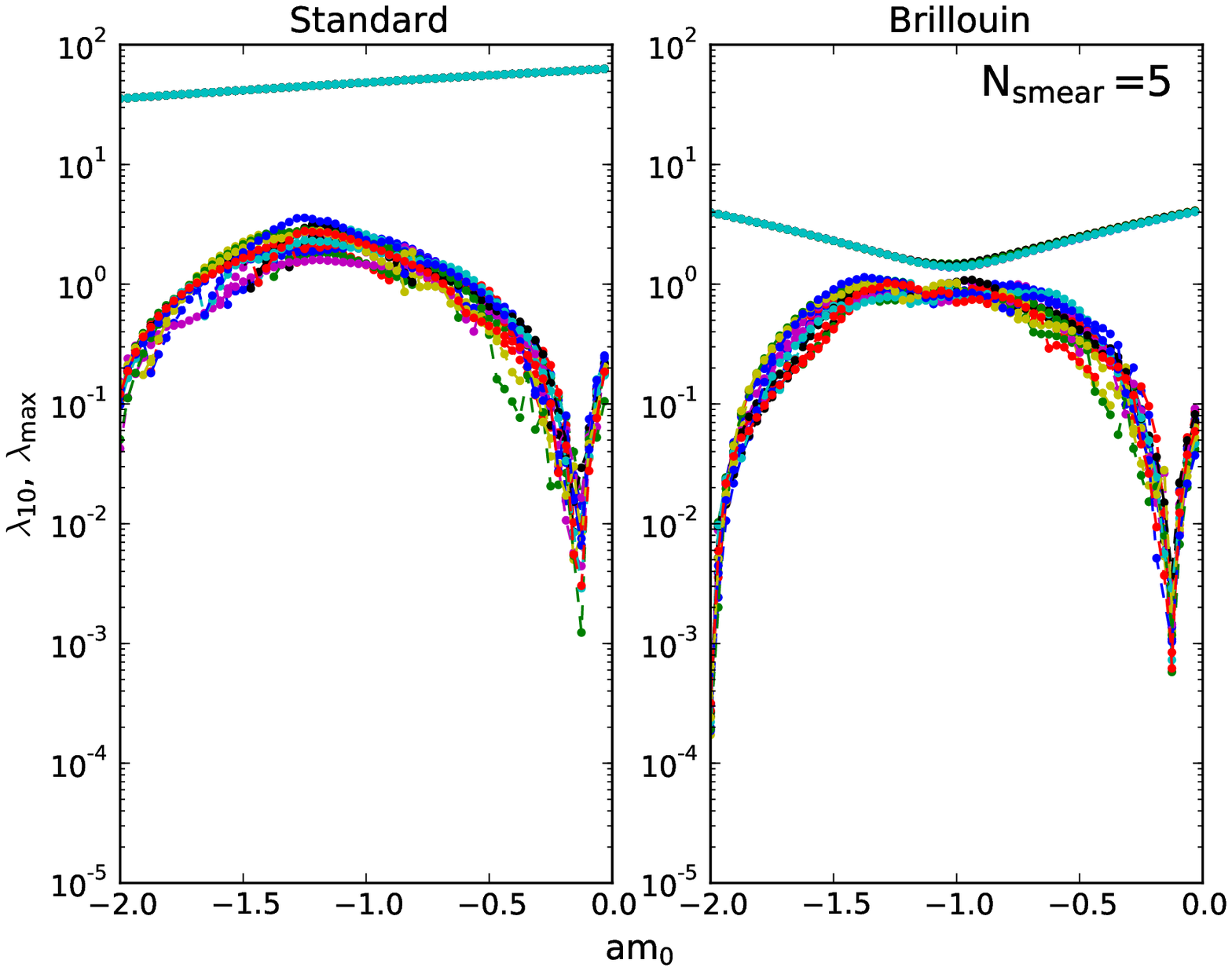,height=0.46\textheight}
\vspace*{-4mm}
\caption{\label{fig:ev_smear5}\sl
The 1st (top), 10th (bottom) and largest Ritz eigenvalue of $D_m\dag D_m^{}$
for the Wilson and the Brillouin operator, on 16 configs at $\be\!=\!6.20$,
with $c_\mr{SW}\!=\!0$ and 5 steps of APE smearing.}
\end{figure}

In the top panel of Fig.\,\ref{fig:ev_smear1} we plot the smallest and the
largest Ritz eigenvalue of $A$ made from the standard Wilson kernel (left) or
the Brillouin kernel (right) as a function of the mass $am_0\!=\!-\rh$.
In the bottom panel the 10th and (again) the largest eigenvalues are shown.
In either panel, 16 configurations of our finest ensemble ($\be\!=\!6.20$) are
used, after 1 step of APE smearing is applied, and the clover coefficient is
set to zero.
Note that the gap between the largest and the smallest eigenvalue is just the
condition number of $A$, and the gap between the largest and the 10th
eigenvalue is the condition number of $A$ restricted to the subspace orthogonal
to the lowest 9 eigenmodes.
Hence, after $O(10)$ eigenmodes are projected, the Brillouin kernel allows for
a reduced order of the polynomial or rational representation of the sign
function, since its spectral range is 1-2 orders of magnitude smaller.

In Fig.\,\ref{fig:ev_smear5} the same exercise is repeated with 5 steps of
APE smearing.
The overall picture is unchanged; again the resulting condition number of the
Brillouin kernel is significantly smaller.
We also find little impact of 19 (instead of 9) eigenmodes projected, and
whether the kernel is Symanzik improved or not.
In short, the reduction of the condition number comes predominantly from the
lowering of the largest eigenvalue (in line with what one would expect from the
eigenvalue spectra shown in Fig.\,\ref{fig:spec_4D_csw0}).
By chance, one of the configurations used in Fig.\,\ref{fig:ev_smear5} happens
to be close (in configuration space) to a barrier between two topological
sectors.
With the Wilson kernel the crossing occurs near $\rh\!=\!1.7$, with the
Brillouin kernel close to $\rh\!=\!1.1$.

In summary, we find that the shifted Brillouin kernel has a significantly
reduced condition number, in particular with a bit of link smearing and after
$O(10)$ eigenmodes are projected.
This allows for a lower degree polynomial or rational representation of the
sign function.


\subsection{Comparing the locality of the resulting overlap actions}




The locality of the overlap action with standard Wilson kernel was first
studied in \cite{Hernandez:1998et}.
In \cite{Bietenholz:2002ks} it was shown that a nearly chiral (but still
ultralocal) kernel can significantly improve the coordinate-space locality of
the resulting overlap action.
In \cite{Kovacs:2002nz,Durr:2005an} it was shown that even a slight
modification through some link-smearing can lead to a considerable improvement.
Therefore, one may hope that trading the Wilson kernel for the Brillouin kernel
leads to a noticeable improvement of the locality of the overlap operator.

\begin{figure}[!tb]
\centering
\epsfig{file=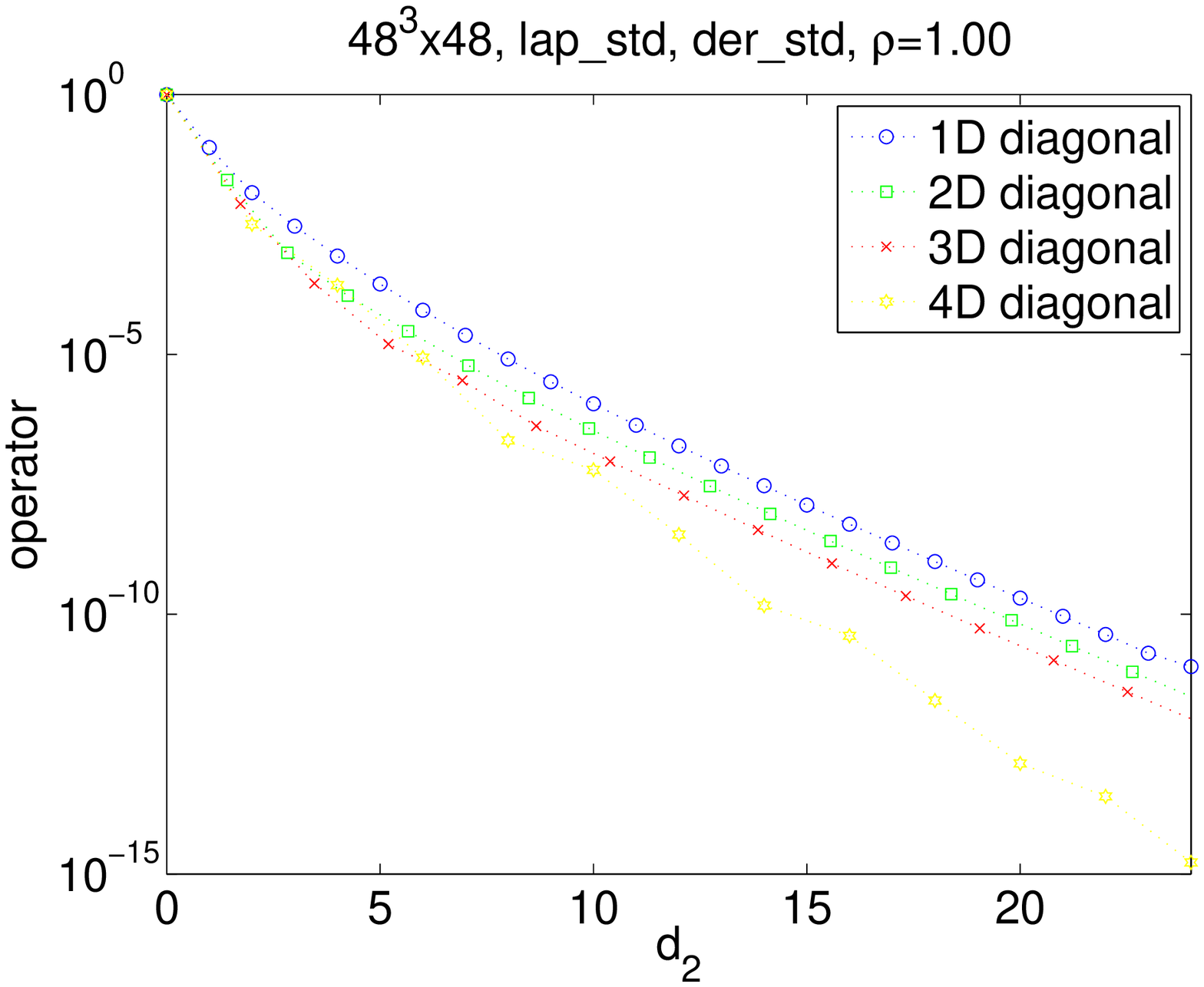,width=0.49\textwidth}
\epsfig{file=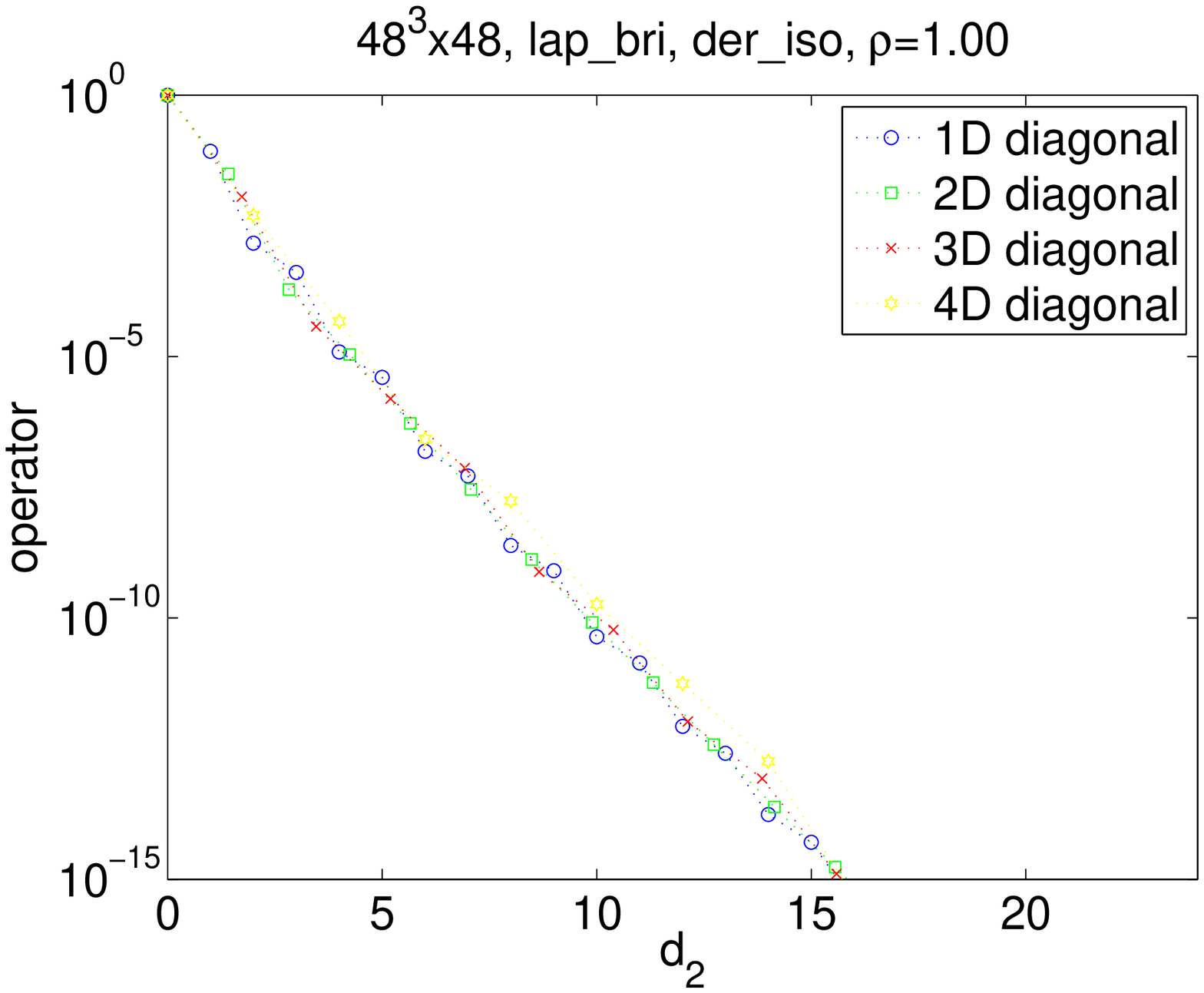,width=0.49\textwidth}
\vspace*{-4mm}
\caption{\label{fig:freeloca}\sl
Localization of the $\rh\!=\!1$ overlap operator with the standard Wilson
kernel (left) or the new Brillouin kernel (right) on a free $48^4$ lattice,
for four directions of the separation.}
\label{freeloca}
\end{figure}

The localization of the overlap made from the Wilson or the Brillouin kernel is
shown for a $48^4$ lattice in the free field case in Fig.\,\ref{fig:freeloca}.
The Frobenius norm of $D(x,y)$ is plotted as a function of the Euclidean
distance $d_2=||x\!-\!y||_2$.
Evidently, the Brillouin kernel diminishes the anisotropy effects and makes the
operator fall off at about twice the rate as before.

\section{Summary}

\begin{figure}[!tb]
\centering
\epsfig{file=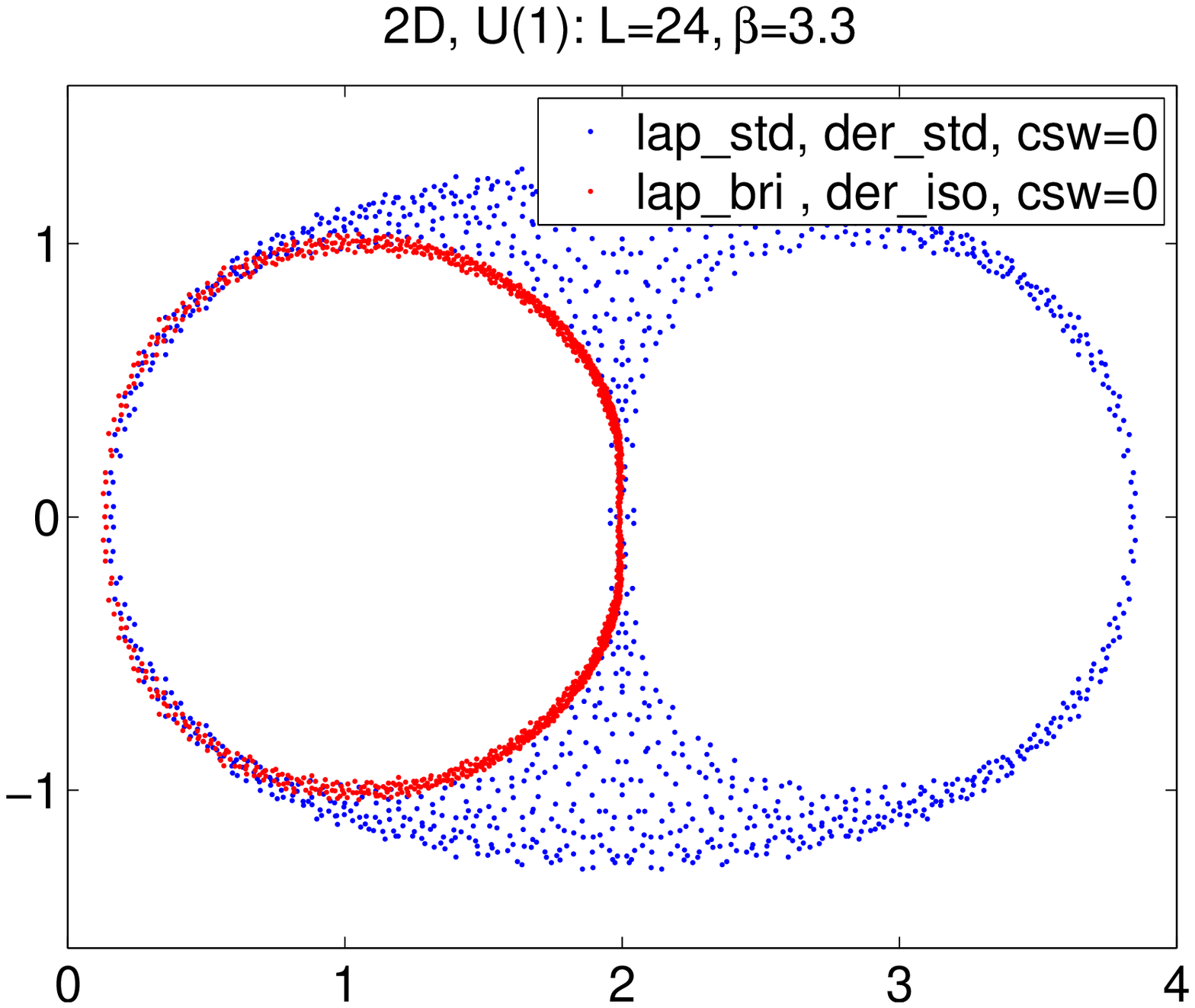,width=8.4cm}
\epsfig{file=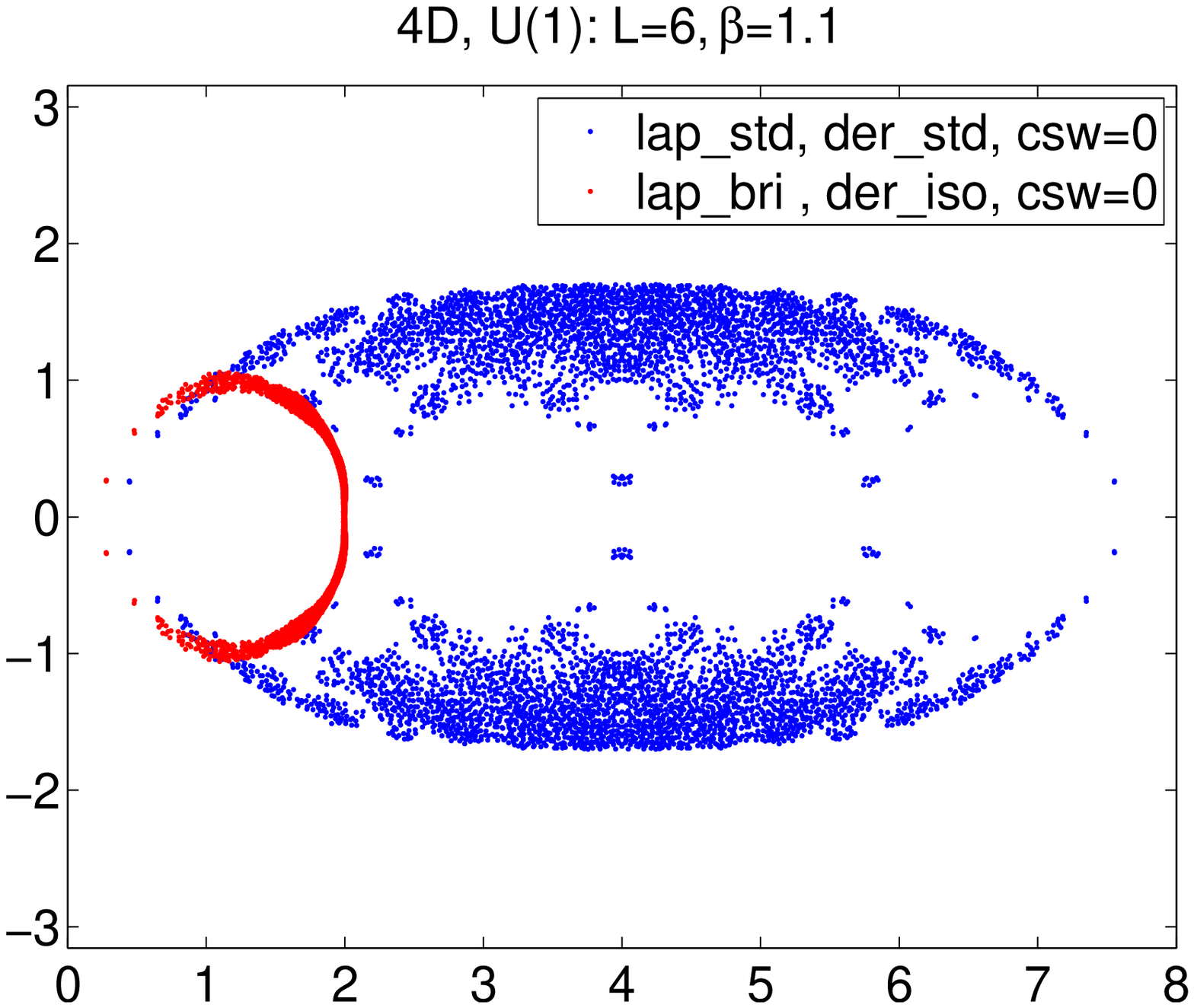,width=8.4cm}
\vspace*{-4mm}
\caption{\label{fig:summary}\sl
Wilson (lap\_std,der\_std) and Brillouin (lap\_bri,der\_iso) eigenvalue spectra
without link smearing and without clover improvement, in 2D (left) and 4D
(right).}
\end{figure}

We have introduced an ultralocal single-flavor lattice Dirac operator, based on
the gauge covariant versions of $\nab^\mr{iso}$ and $\lap^\mr{bri}$ in
(\ref{def_pref}).
Relative to the Wilson operator its eigenvalue spectrum is more Ginsparg-Wilson
like (cf.\ Fig.\,\ref{fig:summary}), and its dispersion relation is more
continuum-like%
\footnote{A similar strategy has been adopted for staggered fermions in
\cite{Heller:1999xz}.}.
As species doubling and global anomalies depend only on topological features of
the dispersion relation \cite{Karsten:1980wd,Wilczek:1987kw}, from the
conceptual viewpoint this is a Wilson-like fermion.

When combined with some link smearing and clover improvement, our action was
found to show good scaling of decay constants even in the physical charm
region, and we expect that the near-agreement between perturbative and
non-perturbative improvement coefficients found with the Wilson operator
\cite{Capitani:2006ni,Hoffmann:2007nm,Shamir:2010cq} carries over to this
action, too.
It appears that lattice perturbation theory is conceptually not any more
difficult than for standard Wilson fermions, but intermediate expressions may
be longer, in particular if several smearing steps are included and the
backgrounds are made from improved glue \cite{Capitani:2002mp}.

Regarding the cost of a simulation with the Brillouin operator it is hard to
make generic statements.
What can be compared is the number of forward applications needed (at a given
value of $\Mpi$, cf.\ Fig.\,\ref{fig:hist_iter}, where our Brillouin operator
is seen to fare better).
However, the cost of an individual forward application depends vey much on the
architecture used.
One extreme case is a serial machine which is CPU-limited; in this case 80
neighbor couplings instead of 8 make each application a factor $O(10)$ slower,
whereupon the advantage is gone.
On the other hand, highly threaded architectures such as GPUs (for an early
application to lattice QCD see \cite{Egri:2006zm}) may be entirely
bandwidth-limited; in such a case clever coding might keep the cost of a
forward application essentially unchanged, relative to the Wilson operator.
In our view, the upshot is that the usefulness of the Brillouin operator should
be tested in phenomenological applications where all aspects of a formulation
play a role, including the onset of the Symanzik scaling regime.
In addition, there is a faint possibility that the Brillouin operator might be
more susceptible to multigrid methods to solve for a given right-hand vector.

Our Brillouin operator is a specific representative of the class of Dirac
operators
\beq
D(x,y)=\sum_\mu\ga_\mu\rh_\mu(x\!-\!y)+\la(x\!-\!y)
\eeq
where the derivative and the Laplacian are expressed as
\bea
\rh_\mu(x\!-\!y)&=&\rh_1[\de_{x+\hat\mu,y}-\de_{x-\hat\mu,y}]
\nonumber
\\
&+&\rh_2\sum_{\nu}
[\de_{x+\hat\mu+\hat\nu,y}-\de_{x-\hat\mu+\hat\nu,y}]
\nonumber
\\[-1mm]
&+&\rh_3\sum_{\nu,\rh}
[\de_{x+\hat\mu+\hat\nu+\hat\rh,y}-\de_{x-\hat\mu+\hat\nu+\hat\rh,y}]
\nonumber
\\[-1mm]
&+&\rh_4\sum_{\nu,\rh,\si}
[\de_{x+\hat\mu+\hat\nu+\hat\rh+\hat\si,y}-\de_{x-\hat\mu+\hat\nu+\hat\rh+\hat\si,y}]
\\
\la(x\!-\!y)&=&\la_0\;\de_{x,y}
\nonumber
\\
&+&\la_1\sum_{\mu}
[\de_{x+\hat\mu,y}-\de_{x-\hat\mu,y}]
\nonumber
\\[-1mm]
&+&\la_2\sum_{\mu,\nu}
[\de_{x+\hat\mu+\hat\nu,y}-\de_{x-\hat\mu+\hat\nu,y}]
\nonumber
\\[-1mm]
&+&\la_3\sum_{\mu,\nu,\rh}
[\de_{x+\hat\mu+\hat\nu+\hat\rh,y}-\de_{x-\hat\mu+\hat\nu+\hat\rh,y}]
\nonumber
\\[-1mm]
&+&\la_4\sum_{\mu,\nu,\rh,\si}
[\de_{x+\hat\mu+\hat\nu+\hat\rh+\hat\si,y}-\de_{x-\hat\mu+\hat\nu+\hat\rh+\hat\si,y}]
\eea
with the understanding that the sums extend over positive and negative
directions mutually orthogonal to each other (and in case of the derivative
terms also to $\hat\mu$).
As discussed in \cite{Hegde:2008nx}, in order to obtain the correct continuum
dispersion relation one requires
\bea
2\rh_1+12\rh_2+24\rh_3+16\rh_4&=&1
\nonumber
\\
\la_0+8\la_1+24\la_2+32\la_3+16\la_4&=&0
\eea
which all operators in Tab.\,\ref{tab:rholambda} obey.
In addition, our Brillouin (der\_iso, lap\_bri) operator satisfies
\bea
\la_0+4\la_1-16\la_3-16\la_4&=&2
\nonumber
\\
\la_0-8\la_2+16\la_4&=&2
\nonumber
\\
\la_0-4\la_1+16\la_3-16\la_4&=&2
\nonumber
\\
\la_0-8\la_1+24\la_2-32\la_3+16\la_4&=&2
\eea
which means that in the weak coupling limit all doublers are lifted by an equal
amount, and
\beq
12\rh_2+48\rh_3+48\rh_4-1=0
\eeq
which ensures that the physical branch of the free field dispersion relation
$E/p$ has no $O(a^2)$ contribution \cite{Hegde:2008nx}, and that the leading
cut-off effects in the deviation of the pressure from the Stefan-Boltzmann
limit are $\propto1/N_t^4$ \cite{Hegde:2008nx}.
In other words, our Brillouin operator is expected to define a discretized
version of QCD with decent bulk thermodynamic properties.

\begin{table}[!tb]
\centering
\begin{tabular}{|c|ccc||c|ccc|}
\hline
        &Wilson&``hypercube''&der\_iso&         & Wilson &``hypercube'' &lap\_bri \\
\hline
   --   &  --  &     --      &   --   & $\la_0$ &    4   &  1.852720547 & 240/128 \\
$\rh_1$ & 1/2  & 0.136846794 & 64/432 & $\la_1$ &  -1/2  & -0.060757866 & -8/128  \\
$\rh_2$ &  0   & 0.032077284 & 16/432 & $\la_2$ &    0   & -0.030036032 & -4/128  \\
$\rh_3$ &  0   & 0.011058131 &  4/432 & $\la_3$ &    0   & -0.015967620 & -2/128  \\
$\rh_4$ &  0   & 0.004748991 &  1/432 & $\la_4$ &    0   & -0.008426812 & -1/128  \\
\hline
\end{tabular}
\caption{\label{tab:rholambda}\sl
The $\rh_i$ and $\la_i$ of the derivative and Laplacian parts of three
Wilson-type fermions (Wilson's version, the hypercube fermion by Bietenholz et
al.\ \cite{Bietenholz:1996pf}, and our Brillouin operator).}
\end{table}


\bigskip

\noindent
{\bf Acknowledgments}:
We thank Zoltan Fodor for useful correspondence.
Computations were performed on the high-performance cluster JUROPA at JSC,
with resources allocated through a NIC grant. 
Both authors were supported by the SFB TR-55.

\clearpage


\section*{App.\,A: Discrete Laplacians and derivatives in 2D}


We give four Laplace stencils in 2D along with their momentum-space
representation:
\begin{itemize}
\itemsep-1pt
\item
Standard Laplacian in 2D:\\[2mm]
\begin{tabular}{|ccc|}
\hline 0&1&0\\1&-4&1\\0&1&0\\ \hline
\end{tabular}\,\,/1\\[1mm]
$\hat\lap=2\cos(k_1)+2\cos(k_2)-4$
\hfill [to be read as $a^2\hat\lap=2\cos(ap_1)+2\cos(ap_2)-4$]
\item
Tilted Laplacian in 2D:\\[2mm]
\begin{tabular}{|ccc|}
\hline 1&0&1\\0&-4&0\\1&0&1\\ \hline
\end{tabular}\,\,/2\\[1mm]
$\hat\lap=2\cos(k_1)\cos(k_2)-2$
\hfill [mind the second zero at $k_1\!=\!k_2\!=\!\pi$]
\item
Brillouin Laplacian in 2D:\\[2mm]
\begin{tabular}{|ccc|}
\hline 1&2&1\\2&$\!$-12$\!$&2\\1&2&1\\ \hline
\end{tabular}\,\,/4\\[1mm]
$\hat\lap=4\cos^2(k_1/2)\cos^2(k_2/2)-4$
\hfill [takes constant value $-4$ at boundary of BZ]
\item
Isotropic Laplacian in 2D:\\[2mm]
\begin{tabular}{|ccc|}
\hline 1&4&1\\4&$\!$-20$\!$&4\\1&4&1\\ \hline
\end{tabular}\,\,/6\\[1mm]
$\hat\lap=[2\cos(k_1)\cos(k_2)+4\cos(k_1)+4\cos(k_2)-10]/3$
\hfill [a.k.a.\ ``Mehrstellen'' Laplacian]
\end{itemize}
We give three $x$-derivative stencils in 2D along with their momentum-space
representation:
\begin{itemize}
\itemsep-1pt
\item
Standard $x$-derivative in 2D:\\[2mm]
\begin{tabular}{|ccc|}
\hline 0&0&0\\-1&0&1\\0&0&0\\ \hline
\end{tabular}\,\,/2\\[1mm]
$\hat\pad_x=\ri\sin(k_1)$
\hfill [to be read as $a\hat\pad_x=\ri\sin(ap_1)$]
\item
Brillouin $x$-derivative in 2D:\\[2mm]
\begin{tabular}{|ccc|}
\hline -1&0&1\\-2&0&2\\-1&0&1\\ \hline
\end{tabular}\,\,/8\\[1mm]
$\hat\pad_x=\ri\sin(k_1)[\cos(k_2)+1]/2$
\item
Isotropic $x$-derivative in 2D:\\[2mm]
\begin{tabular}{|ccc|}
\hline -1&0&1\\-4&0&4\\-1&0&1\\ \hline
\end{tabular}\,\,/12\\[1mm]
$\hat\pad_x=\ri\sin(k_1)[\cos(k_2)+2]/3$
\end{itemize}

\begincomment
lap1:=sort(expand(simplify(( \
 exp(+I*k1)+exp(-I*k1)+exp(+I*k2)+exp(-I*k2) \
 -4)/1)));
lap2:=sort(expand(simplify(( \
 +exp(I*(+k1+k2))+exp(I*(+k1-k2))+exp(I*(-k1+k2))+exp(I*(-k1-k2)) \
 -4)/2)));
lap3:=sort(expand(simplify(( \
 +2*exp(+I*k1)+2*exp(-I*k1)+2*exp(+I*k2)+2*exp(-I*k2)
 +exp(I*(+k1+k2))+exp(I*(+k1-k2))+exp(I*(-k1+k2))+exp(I*(-k1-k2)) \
 -12)/4)));
lap4:=sort(expand(simplify(( \
 +4*exp(I*k1)+4*exp(-I*k1)+4*exp(I*k2)+4*exp(-I*k2)
 +exp(I*(k1+k2))+exp(I*(k1-k2))+exp(I*(-k1+k2))+exp(I*(-k1-k2)) \
 -20)/6)));
der1:=simplify((exp(I*k1)-exp(-I*k1))/2);
der2:=factor(expand(simplify(( \
 +2*exp(+I*k1) \
 +1*exp(I*(+k1+k2))+1*exp(I*(+k1-k2)) \
 -2*exp(-I*k1) \
 -1*exp(I*(-k1+k2))-1*exp(I*(-k1-k2)) \
 )/8)));
der3:=factor(expand(simplify(( \
 +4*exp(+I*k1) \
 +1*exp(I*(+k1+k2))+1*exp(I*(+k1-k2)) \
 -4*exp(-I*k1) \
 -1*exp(I*(-k1+k2))-1*exp(I*(-k1-k2)) \
 )/12)));
\endcomment

\clearpage


\section*{App.\,B: Discrete Laplacians and derivatives in 3D}


We give four Laplace stencils in 3D along with their momentum-space
representation:
\begin{itemize}
\itemsep-1pt
\item
Standard Laplacian in 3D:\\[2mm]
\begin{tabular}{|ccc|}
\hline 0&0&0\\0&1&0\\0&0&0\\ \hline
\end{tabular}
\begin{tabular}{|ccc|}
\hline 0&1&0\\1&$\!$-6$\!$&1\\0&1&0\\ \hline
\end{tabular}
\begin{tabular}{|ccc|}
\hline 0&0&0\\0&1&0\\0&0&0\\ \hline
\end{tabular}\,\,/1\\[1mm]
$\hat\lap=2\cos(k_1)+2\cos(k_2)+2\cos(k_3)-6$
\item
Tilted Laplacian in 3D:\\[2mm]
\begin{tabular}{|ccc|}
\hline 1&0&1\\0&0&0\\1&0&1\\ \hline
\end{tabular}
\begin{tabular}{|ccc|}
\hline 0&0&0\\0&$\!$-8$\!$&0\\0&0&0\\ \hline
\end{tabular}
\begin{tabular}{|ccc|}
\hline 1&0&1\\0&0&0\\1&0&1\\ \hline
\end{tabular}\,\,/4\\[1mm]
$\hat\lap=2\cos(k_1)\cos(k_2)\cos(k_3)-2$
\hfill [mind the three additional zeros]
\item
Brillouin Laplacian in 3D:\\[2mm]
\begin{tabular}{|ccc|}
\hline 1&2&1\\2&4&2\\1&2&1\\ \hline
\end{tabular}
\begin{tabular}{|ccc|}
\hline 2&4&2\\4&$\!\!$-56$\!\!$&4\\2&4&2\\ \hline
\end{tabular}
\begin{tabular}{|ccc|}
\hline 1&2&1\\2&4&2\\1&2&1\\ \hline
\end{tabular}\,\,/16\\[1mm]
$\hat\lap=4\cos^2(k_1/2)\cos^2(k_2/2)\cos^2(k_3/2)-4$
\hfill [takes constant value at boundary of BZ]
\item
Isotropic Laplacian in 3D:\\[2mm]
\begin{tabular}{|ccc|}
\hline 1&6&1\\6&$\!$20$\!$&6\\1&6&1\\ \hline
\end{tabular}
\begin{tabular}{|ccc|}
\hline 6&20&6\\$\!$20$\!$&$\!\!$-200$\!\!$&$\!$20$\!$\\6&20&6\\ \hline
\end{tabular}
\begin{tabular}{|ccc|}
\hline 1&6&1\\6&$\!$20$\!$&6\\1&6&1\\ \hline
\end{tabular}\,\,/48\\[1mm]
$\hat\lap=[\cos(k_1)\cos(k_2)\cos(k_3)
+3\cos(k_1)\cos(k_2)
+...
+5\cos(k_1)+...-25]/6$
\end{itemize}
We give three $x$-derivative stencils in 3D along with their momentum-space
representation:
\begin{itemize}
\itemsep-1pt
\item
Standard $x$-derivative in 3D:\\[2mm]
\begin{tabular}{|ccc|}
\hline 0&0&0\\0&0&0\\0&0&0\\ \hline
\end{tabular}
\begin{tabular}{|ccc|}
\hline 0&0&0\\-1&0&1\\0&0&0\\ \hline
\end{tabular}
\begin{tabular}{|ccc|}
\hline 0&0&0\\0&0&0\\0&0&0\\ \hline
\end{tabular}\,\,/2\\[1mm]
$\hat\pad_x=\ri\sin(k_1)$
\item
Brillouin $x$-derivative in 3D:\\[2mm]
\begin{tabular}{|ccc|}
\hline -1&0&1\\-2&0&2\\-1&0&1\\ \hline
\end{tabular}
\begin{tabular}{|ccc|}
\hline -2&0&2\\-4&0&4\\-2&0&2\\ \hline
\end{tabular}
\begin{tabular}{|ccc|}
\hline -1&0&1\\-2&0&2\\-1&0&1\\ \hline
\end{tabular}\,\,/32\\[1mm]
$\hat\pad_x=\ri\sin(k_1)[\cos(k_2)+1][\cos(k_3)+1]/4$
\item
Isotropic $x$-derivative in 3D:\\[2mm]
\begin{tabular}{|ccc|}
\hline -1&0&1\\-4&0&4\\-1&0&1\\ \hline
\end{tabular}
\begin{tabular}{|ccc|}
\hline -4&0&4\\$\!$-16$\!\!$&0&$\!\!$16$\!$\\-4&0&4\\ \hline
\end{tabular}
\begin{tabular}{|ccc|}
\hline -1&0&1\\-4&0&4\\-1&0&1\\ \hline
\end{tabular}\,\,/72\\[1mm]
$\hat\pad_x=\ri\sin(k_1)[\cos(k_2)+2][\cos(k_3)+2]/9$
\end{itemize}

\begincomment
lap1:=sort(expand(simplify(( \
 exp(+I*k1)+exp(-I*k1)+exp(+I*k2)+exp(-I*k2)+exp(+I*k3)+exp(-I*k3)
 -6)/1)));
lap2:=sort(expand(simplify(( \
 +exp(I*(+k1+k2-k3))+exp(I*(+k1-k2-k3))+exp(I*(-k1+k2-k3))+exp(I*(-k1-k2-k3)) \
 +exp(I*(+k1+k2+k3))+exp(I*(+k1-k2+k3))+exp(I*(-k1+k2+k3))+exp(I*(-k1-k2+k3)) \
 -8)/4)));
lap3:=sort(expand(simplify(( \
 +exp(I*(+k1+k2-k3))+exp(I*(+k1-k2-k3))+exp(I*(-k1+k2-k3))+exp(I*(-k1-k2-k3)) \
 +exp(I*(+k1+k2+k3))+exp(I*(+k1-k2+k3))+exp(I*(-k1+k2+k3))+exp(I*(-k1-k2+k3)) \
 +2*exp(I*(+k1+k2))+2*exp(I*(+k1-k2))+2*exp(I*(-k1+k2))+2*exp(I*(-k1-k2)) \
 +2*exp(I*(+k1+k3))+2*exp(I*(+k1-k3))+2*exp(I*(-k1+k3))+2*exp(I*(-k1-k3)) \
 +2*exp(I*(+k2+k3))+2*exp(I*(+k2-k3))+2*exp(I*(-k2+k3))+2*exp(I*(-k2-k3)) \
 +4*exp(I*(+k1))+4*exp(I*(-k1)) \
 +4*exp(I*(+k2))+4*exp(I*(-k2)) \
 +4*exp(I*(+k3))+4*exp(I*(-k3)) \
 -56)/16)));
lap4:=sort(expand(simplify(( \
 +exp(I*(+k1+k2-k3))+exp(I*(+k1-k2-k3))+exp(I*(-k1+k2-k3))+exp(I*(-k1-k2-k3)) \
 +exp(I*(+k1+k2+k3))+exp(I*(+k1-k2+k3))+exp(I*(-k1+k2+k3))+exp(I*(-k1-k2+k3)) \
 +6*exp(I*(+k1+k2))+6*exp(I*(+k1-k2))+6*exp(I*(-k1+k2))+6*exp(I*(-k1-k2)) \
 +6*exp(I*(+k1+k3))+6*exp(I*(+k1-k3))+6*exp(I*(-k1+k3))+6*exp(I*(-k1-k3)) \
 +6*exp(I*(+k2+k3))+6*exp(I*(+k2-k3))+6*exp(I*(-k2+k3))+6*exp(I*(-k2-k3)) \
 +20*exp(I*(+k1))+20*exp(I*(-k1)) \
 +20*exp(I*(+k2))+20*exp(I*(-k2)) \
 +20*exp(I*(+k3))+20*exp(I*(-k3)) \
 -200)/48)));
lap5:=sort(expand(simplify( \
 +c3*exp(I*(+k1+k2-k3))+c3*exp(I*(+k1-k2-k3))+c3*exp(I*(-k1+k2-k3))+c3*exp(I*(-k1-k2-k3)) \
 +c3*exp(I*(+k1+k2+k3))+c3*exp(I*(+k1-k2+k3))+c3*exp(I*(-k1+k2+k3))+c3*exp(I*(-k1-k2+k3)) \
 +c2*exp(I*(+k1+k2))+c2*exp(I*(+k1-k2))+c2*exp(I*(-k1+k2))+c2*exp(I*(-k1-k2)) \
 +c2*exp(I*(+k1+k3))+c2*exp(I*(+k1-k3))+c2*exp(I*(-k1+k3))+c2*exp(I*(-k1-k3)) \
 +c2*exp(I*(+k2+k3))+c2*exp(I*(+k2-k3))+c2*exp(I*(-k2+k3))+c2*exp(I*(-k2-k3)) \
 +c1*exp(I*(+k1))+c1*exp(I*(-k1)) \
 +c1*exp(I*(+k2))+c1*exp(I*(-k2)) \
 +c1*exp(I*(+k3))+c1*exp(I*(-k3)) \
 -6*c1-12*c2-8*c3)));
 lap5:=expand(series(subs({k1=a*p1,k2=a*p2,k3=a*p3},lap5),a,6));
 lap50:=sort(coeff(lap5,a,0),[p1,p2,p3]); # must be zero
 lap51:=sort(coeff(lap5,a,1),[p1,p2,p3]); # must be zero
 lap52:=sort(coeff(lap5,a,2),[p1,p2,p3]); # must be -p1^2-p2^2-p3^2
 lap53:=sort(coeff(lap5,a,3),[p1,p2,p3]); # must be zero
 lap54:=sort(coeff(lap5,a,4),[p1,p2,p3]); # should be isotropic
 lap50:=expand(subs(c1=1-4*c2-4*c3,lap50)); # is zero
 lap51:=expand(subs(c1=1-4*c2-4*c3,lap51)); # is zero
 lap52:=expand(subs(c1=1-4*c2-4*c3,lap52)); # is -p1^2-p2^2-p3^2
 lap53:=expand(subs(c1=1-4*c2-4*c3,lap53)); # is zero
 lap54:=expand(subs(c1=1-4*c2-4*c3,lap54)); # no way to eliminate all p^4 terms
der1:=simplify((exp(I*k1)-exp(-I*k1))/2);
der2:=factor(expand(simplify(( \
 +4*exp(+I*k1) \
 +2*exp(I*(+k1+k2))+2*exp(I*(+k1-k2)) \
 +2*exp(I*(+k1+k3))+2*exp(I*(+k1-k3)) \
 +1*exp(I*(+k1+k2+k3))+1*exp(I*(+k1+k2-k3)) \
 +1*exp(I*(+k1-k2+k3))+1*exp(I*(+k1-k2-k3)) \
 -4*exp(-I*k1) \
 -2*exp(I*(-k1+k2))-2*exp(I*(-k1-k2))-2*exp(I*(-k1+k3))-2*exp(I*(-k1-k3)) \
 -1*exp(I*(-k1+k2+k3))-1*exp(I*(-k1+k2-k3)) \
 -1*exp(I*(-k1-k2+k3))-1*exp(I*(-k1-k2-k3)) \
 )/32)));
der3:=factor(expand(simplify(( \
 +16*exp(+I*k1) \
 +4*exp(I*(+k1+k2))+4*exp(I*(+k1-k2)) \
 +4*exp(I*(+k1+k3))+4*exp(I*(+k1-k3)) \
 +1*exp(I*(+k1+k2+k3))+1*exp(I*(+k1+k2-k3)) \
 +1*exp(I*(+k1-k2+k3))+1*exp(I*(+k1-k2-k3)) \
 -16*exp(-I*k1) \
 -4*exp(I*(-k1+k2))-4*exp(I*(-k1-k2)) \
 -4*exp(I*(-k1+k3))-4*exp(I*(-k1-k3)) \
 -1*exp(I*(-k1+k2+k3))-1*exp(I*(-k1+k2-k3)) \
 -1*exp(I*(-k1-k2+k3))-1*exp(I*(-k1-k2-k3)) \
 )/72)));
\endcomment

\clearpage


\section*{App.\,C: Discrete Laplacians and derivatives in 4D}


We give four Laplace stencils in 4D along with their momentum-space
representation:
\begin{itemize}
\itemsep-1pt
\item
Standard Laplacian in 4D:\\[2mm]
\begin{tabular*}{2.1cm}{@{\extracolsep{\fill}}|ccc|}
\hline 0&0&0\\0&0&0\\0&0&0\\ \hline
\end{tabular*}
\begin{tabular*}{2.1cm}{@{\extracolsep{\fill}}|ccc|}
\hline 0&0&0\\0&1&0\\0&0&0\\ \hline
\end{tabular*}
\begin{tabular*}{2.1cm}{@{\extracolsep{\fill}}|ccc|}
\hline 0&0&0\\0&0&0\\0&0&0\\ \hline
\end{tabular*}\\[1mm]
\begin{tabular*}{2.1cm}{@{\extracolsep{\fill}}|ccc|}
\hline 0&0&0\\0&1&0\\0&0&0\\ \hline
\end{tabular*}
\begin{tabular*}{2.1cm}{@{\extracolsep{\fill}}|ccc|}
\hline 0&1&0\\1&-8&1\\0&1&0\\ \hline
\end{tabular*}
\begin{tabular*}{2.1cm}{@{\extracolsep{\fill}}|ccc|}
\hline 0&0&0\\0&1&0\\0&0&0\\ \hline
\end{tabular*}\,\,/1\\[1mm]
\begin{tabular*}{2.1cm}{@{\extracolsep{\fill}}|ccc|}
\hline 0&0&0\\0&0&0\\0&0&0\\ \hline
\end{tabular*}
\begin{tabular*}{2.1cm}{@{\extracolsep{\fill}}|ccc|}
\hline 0&0&0\\0&1&0\\0&0&0\\ \hline
\end{tabular*}
\begin{tabular*}{2.1cm}{@{\extracolsep{\fill}}|ccc|}
\hline 0&0&0\\0&0&0\\0&0&0\\ \hline
\end{tabular*}\\[1mm]
$\hat\lap=2\cos(k_1)+2\cos(k_2)+2\cos(k_3)+2\cos(k_4)-8$
\item
Tilted Laplacian in 4D:\\[2mm]
\begin{tabular*}{2.3cm}{@{\extracolsep{\fill}}|ccc|}
\hline 1&0&1\\0&0&0\\1&0&1\\ \hline
\end{tabular*}
\begin{tabular*}{2.3cm}{@{\extracolsep{\fill}}|ccc|}
\hline 0&0&0\\0&0&0\\0&0&0\\ \hline
\end{tabular*}
\begin{tabular*}{2.3cm}{@{\extracolsep{\fill}}|ccc|}
\hline 1&0&1\\0&0&0\\1&0&1\\ \hline
\end{tabular*}\\[1mm]
\begin{tabular*}{2.3cm}{@{\extracolsep{\fill}}|ccc|}
\hline 0&0&0\\0&0&0\\0&0&0\\ \hline
\end{tabular*}
\begin{tabular*}{2.3cm}{@{\extracolsep{\fill}}|ccc|}
\hline 0&0&0\\0&-16&0\\0&0&0\\ \hline
\end{tabular*}
\begin{tabular*}{2.3cm}{@{\extracolsep{\fill}}|ccc|}
\hline 0&0&0\\0&0&0\\0&0&0\\ \hline
\end{tabular*}\,\,/8\\[1mm]
\begin{tabular*}{2.3cm}{@{\extracolsep{\fill}}|ccc|}
\hline 1&0&1\\0&0&0\\1&0&1\\ \hline
\end{tabular*}
\begin{tabular*}{2.3cm}{@{\extracolsep{\fill}}|ccc|}
\hline 0&0&0\\0&0&0\\0&0&0\\ \hline
\end{tabular*}
\begin{tabular*}{2.3cm}{@{\extracolsep{\fill}}|ccc|}
\hline 1&0&1\\0&0&0\\1&0&1\\ \hline
\end{tabular*}\\[1mm]
$\hat\lap=2\cos(k_1)\cos(k_2)\cos(k_3)\cos(k_4)-2$
\hfill [mind the seven additional zeros]
\item
Brillouin Laplacian in 4D:\\[2mm]
\begin{tabular*}{2.5cm}{@{\extracolsep{\fill}}|ccc|}
\hline 1&2&1\\2&4&2\\1&2&1\\ \hline
\end{tabular*}
\begin{tabular*}{2.5cm}{@{\extracolsep{\fill}}|ccc|}
\hline 2&4&2\\4&8&4\\2&4&2\\ \hline
\end{tabular*}
\begin{tabular*}{2.5cm}{@{\extracolsep{\fill}}|ccc|}
\hline 1&2&1\\2&4&2\\1&2&1\\ \hline
\end{tabular*}\\[1mm]
\begin{tabular*}{2.5cm}{@{\extracolsep{\fill}}|ccc|}
\hline 2&4&2\\4&8&4\\2&4&2\\ \hline
\end{tabular*}
\begin{tabular*}{2.5cm}{@{\extracolsep{\fill}}|ccc|}
\hline 4&8&4\\8&-240&8\\4&8&4\\ \hline
\end{tabular*}
\begin{tabular*}{2.5cm}{@{\extracolsep{\fill}}|ccc|}
\hline 2&4&2\\4&8&4\\2&4&2\\ \hline
\end{tabular*}\,\,/64\\[1mm]
\begin{tabular*}{2.5cm}{@{\extracolsep{\fill}}|ccc|}
\hline 1&2&1\\2&4&2\\1&2&1\\ \hline
\end{tabular*}
\begin{tabular*}{2.5cm}{@{\extracolsep{\fill}}|ccc|}
\hline 2&4&2\\4&8&4\\2&4&2\\ \hline
\end{tabular*}
\begin{tabular*}{2.5cm}{@{\extracolsep{\fill}}|ccc|}
\hline 1&2&1\\2&4&2\\1&2&1\\ \hline
\end{tabular*}\\[1mm]
$\hat\lap=4\cos^2(k_1/2)\cos^2(k_2/2)\cos^2(k_3/2)\cos^2(k_4/2)-4$
\hfill [constant at boundary of BZ]
\item
Isotropic Laplacian in 4D:\\[2mm]
\begin{tabular*}{2.8cm}{@{\extracolsep{\fill}}|ccc|}
\hline 1&7&1\\7&40&7\\1&7&1\\ \hline
\end{tabular*}
\begin{tabular*}{2.8cm}{@{\extracolsep{\fill}}|ccc|}
\hline 7&40&7\\40&100&40\\7&40&7\\ \hline
\end{tabular*}
\begin{tabular*}{2.8cm}{@{\extracolsep{\fill}}|ccc|}
\hline 1&7&1\\7&40&7\\1&7&1\\ \hline
\end{tabular*}\\[1mm]
\begin{tabular*}{2.8cm}{@{\extracolsep{\fill}}|ccc|}
\hline 7&40&7\\40&100&40\\7&40&7\\ \hline
\end{tabular*}
\begin{tabular*}{2.8cm}{@{\extracolsep{\fill}}|ccc|}
\hline 40&100&40\\$\!\!$100$\!\!$&$\!\!\!$-2000$\!\!\!$&$\!\!$100$\!\!$\\40&100&40\\ \hline
\end{tabular*}
\begin{tabular*}{2.8cm}{@{\extracolsep{\fill}}|ccc|}
\hline 7&40&7\\40&100&40\\7&40&7\\ \hline
\end{tabular*}\,\,/432\\[1mm]
\begin{tabular*}{2.8cm}{@{\extracolsep{\fill}}|ccc|}
\hline 1&7&1\\7&40&7\\1&7&1\\ \hline
\end{tabular*}
\begin{tabular*}{2.8cm}{@{\extracolsep{\fill}}|ccc|}
\hline 7&40&7\\40&100&40\\7&40&7\\ \hline
\end{tabular*}
\begin{tabular*}{2.8cm}{@{\extracolsep{\fill}}|ccc|}
\hline 1&7&1\\7&40&7\\1&7&1\\ \hline
\end{tabular*}\\[1mm]
$\hat\lap=[2c_1c_2c_3c_4+7c_1c_2c_3+...+20c_1c_2+...+25c_1+...-250]/54$
\hfill [with $c_1\!=\!\cos(k_1)$ etc.]
\end{itemize}
We give three $x$-derivative stencils in 4D along with their momentum-space
representation:
\begin{itemize}
\itemsep-1pt
\item
Standard $x$-derivative in 4D:\\[2mm]
\begin{tabular*}{2.1cm}{@{\extracolsep{\fill}}|ccc|}
\hline 0&0&0\\0&0&0\\0&0&0\\ \hline
\end{tabular*}
\begin{tabular*}{2.1cm}{@{\extracolsep{\fill}}|ccc|}
\hline 0&0&0\\0&0&0\\0&0&0\\ \hline
\end{tabular*}
\begin{tabular*}{2.1cm}{@{\extracolsep{\fill}}|ccc|}
\hline 0&0&0\\0&0&0\\0&0&0\\ \hline
\end{tabular*}\\[1mm]
\begin{tabular*}{2.1cm}{@{\extracolsep{\fill}}|ccc|}
\hline 0&0&0\\0&0&0\\0&0&0\\ \hline
\end{tabular*}
\begin{tabular*}{2.1cm}{@{\extracolsep{\fill}}|ccc|}
\hline 0&0&0\\-1&0&1\\0&0&0\\ \hline
\end{tabular*}
\begin{tabular*}{2.1cm}{@{\extracolsep{\fill}}|ccc|}
\hline 0&0&0\\0&0&0\\0&0&0\\ \hline
\end{tabular*}\,\,/2\\[1mm]
\begin{tabular*}{2.1cm}{@{\extracolsep{\fill}}|ccc|}
\hline 0&0&0\\0&0&0\\0&0&0\\ \hline
\end{tabular*}
\begin{tabular*}{2.1cm}{@{\extracolsep{\fill}}|ccc|}
\hline 0&0&0\\0&0&0\\0&0&0\\ \hline
\end{tabular*}
\begin{tabular*}{2.1cm}{@{\extracolsep{\fill}}|ccc|}
\hline 0&0&0\\0&0&0\\0&0&0\\ \hline
\end{tabular*}\\[1mm]
$\hat\pad_x=\ri\sin(k_1)$
\item
Brillouin $x$-derivative in 4D:\\[2mm]
\begin{tabular*}{2.1cm}{@{\extracolsep{\fill}}|ccc|}
\hline -1&0&1\\-2&0&2\\-1&0&1\\ \hline
\end{tabular*}
\begin{tabular*}{2.1cm}{@{\extracolsep{\fill}}|ccc|}
\hline -2&0&2\\-4&0&4\\-2&0&2\\ \hline
\end{tabular*}
\begin{tabular*}{2.1cm}{@{\extracolsep{\fill}}|ccc|}
\hline -1&0&1\\-2&0&2\\-1&0&1\\ \hline
\end{tabular*}\\[1mm]
\begin{tabular*}{2.1cm}{@{\extracolsep{\fill}}|ccc|}
\hline -2&0&2\\-4&0&4\\-2&0&2\\ \hline
\end{tabular*}
\begin{tabular*}{2.1cm}{@{\extracolsep{\fill}}|ccc|}
\hline -4&0&4\\-8&0&8\\-4&0&4\\ \hline
\end{tabular*}
\begin{tabular*}{2.1cm}{@{\extracolsep{\fill}}|ccc|}
\hline -2&0&2\\-4&0&4\\-2&0&2\\ \hline
\end{tabular*}\,\,/128\\[1mm]
\begin{tabular*}{2.1cm}{@{\extracolsep{\fill}}|ccc|}
\hline -1&0&1\\-2&0&2\\-1&0&1\\ \hline
\end{tabular*}
\begin{tabular*}{2.1cm}{@{\extracolsep{\fill}}|ccc|}
\hline -2&0&2\\-4&0&4\\-2&0&2\\ \hline
\end{tabular*}
\begin{tabular*}{2.1cm}{@{\extracolsep{\fill}}|ccc|}
\hline -1&0&1\\-2&0&2\\-1&0&1\\ \hline
\end{tabular*}\\[1mm]
$\hat\pad_x=\ri\sin(k_1)[\cos(k_2)+1][\cos(k_3)+1][\cos(k_4)+1]/8$
\item
Isotropic $x$-derivative in 4D:\\[2mm]
\begin{tabular*}{2.5cm}{@{\extracolsep{\fill}}|ccc|}
\hline -1&0&1\\ -4&0&4\\ -1&0&1\\ \hline
\end{tabular*}
\begin{tabular*}{2.5cm}{@{\extracolsep{\fill}}|ccc|}
\hline -4&0&4\\ -16&0&16\\ -4&0&4\\ \hline
\end{tabular*}
\begin{tabular*}{2.5cm}{@{\extracolsep{\fill}}|ccc|}
\hline -1&0&1\\ -4&0&4\\ -1&0&1\\ \hline
\end{tabular*}\\[1mm]
\begin{tabular*}{2.5cm}{@{\extracolsep{\fill}}|ccc|}
\hline -4&0&4\\ -16&0&16\\ -4&0&4\\ \hline
\end{tabular*}
\begin{tabular*}{2.5cm}{@{\extracolsep{\fill}}|ccc|}
\hline -16&0&16\\ -64&0&64\\ -16&0&16\\ \hline
\end{tabular*}
\begin{tabular*}{2.5cm}{@{\extracolsep{\fill}}|ccc|}
\hline -4&0&4\\ -16&0&16\\ -4&0&4\\ \hline
\end{tabular*}\,\,/432\\[1mm]
\begin{tabular*}{2.5cm}{@{\extracolsep{\fill}}|ccc|}
\hline -1&0&1\\ -4&0&4\\ -1&0&1\\ \hline
\end{tabular*}
\begin{tabular*}{2.5cm}{@{\extracolsep{\fill}}|ccc|}
\hline -4&0&4\\ -16&0&16\\ -4&0&4\\ \hline
\end{tabular*}
\begin{tabular*}{2.5cm}{@{\extracolsep{\fill}}|ccc|}
\hline -1&0&1\\ -4&0&4\\ -1&0&1\\ \hline
\end{tabular*}\\[1mm]
$\hat\pad_x=\ri\sin(k_1)[\cos(k_2)+2][\cos(k_3)+2][\cos(k_4)+2]/27$
\end{itemize}

\begincomment
lap1:=sort(expand(simplify(( \
 +exp(+I*k1)+exp(-I*k1) \
 +exp(+I*k2)+exp(-I*k2) \
 +exp(+I*k3)+exp(-I*k3) \
 +exp(+I*k4)+exp(-I*k4) \
 -8)/1)));
lap2:=sort(expand(simplify(( \
 +exp(I*(+k1+k2+k3+k4))+exp(I*(-k1+k2+k3+k4))+exp(I*(+k1-k2+k3+k4))+exp(I*(-k1-k2+k3+k4)) \
 +exp(I*(+k1+k2-k3+k4))+exp(I*(-k1+k2-k3+k4))+exp(I*(+k1-k2-k3+k4))+exp(I*(-k1-k2-k3+k4)) \
 +exp(I*(+k1+k2+k3-k4))+exp(I*(-k1+k2+k3-k4))+exp(I*(+k1-k2+k3-k4))+exp(I*(-k1-k2+k3-k4)) \
 +exp(I*(+k1+k2-k3-k4))+exp(I*(-k1+k2-k3-k4))+exp(I*(+k1-k2-k3-k4))+exp(I*(-k1-k2-k3-k4)) \
 -16)/8)));
lap3:=sort(expand(simplify(( \
 +exp(I*(+k1+k2+k3+k4))+exp(I*(-k1+k2+k3+k4))+exp(I*(+k1-k2+k3+k4))+exp(I*(-k1-k2+k3+k4)) \
 +exp(I*(+k1+k2-k3+k4))+exp(I*(-k1+k2-k3+k4))+exp(I*(+k1-k2-k3+k4))+exp(I*(-k1-k2-k3+k4)) \
 +exp(I*(+k1+k2+k3-k4))+exp(I*(-k1+k2+k3-k4))+exp(I*(+k1-k2+k3-k4))+exp(I*(-k1-k2+k3-k4)) \
 +exp(I*(+k1+k2-k3-k4))+exp(I*(-k1+k2-k3-k4))+exp(I*(+k1-k2-k3-k4))+exp(I*(-k1-k2-k3-k4)) \
 +2*exp(I*(+k1+k2+k3))+2*exp(I*(-k1+k2+k3))+2*exp(I*(+k1-k2+k3))+2*exp(I*(-k1-k2+k3)) \
 +2*exp(I*(+k1+k2-k3))+2*exp(I*(-k1+k2-k3))+2*exp(I*(+k1-k2-k3))+2*exp(I*(-k1-k2-k3)) \
 +2*exp(I*(+k1+k2+k4))+2*exp(I*(-k1+k2+k4))+2*exp(I*(+k1-k2+k4))+2*exp(I*(-k1-k2+k4)) \
 +2*exp(I*(+k1+k2-k4))+2*exp(I*(-k1+k2-k4))+2*exp(I*(+k1-k2-k4))+2*exp(I*(-k1-k2-k4)) \
 +2*exp(I*(+k1+k3+k4))+2*exp(I*(-k1+k3+k4))+2*exp(I*(+k1-k3+k4))+2*exp(I*(-k1-k3+k4)) \
 +2*exp(I*(+k1+k3-k4))+2*exp(I*(-k1+k3-k4))+2*exp(I*(+k1-k3-k4))+2*exp(I*(-k1-k3-k4)) \
 +2*exp(I*(+k2+k3+k4))+2*exp(I*(-k2+k3+k4))+2*exp(I*(+k2-k3+k4))+2*exp(I*(-k2-k3+k4)) \
 +2*exp(I*(+k2+k3-k4))+2*exp(I*(-k2+k3-k4))+2*exp(I*(+k2-k3-k4))+2*exp(I*(-k2-k3-k4)) \
 +4*exp(I*(+k1+k2))+4*exp(I*(+k1-k2))+4*exp(I*(-k1+k2))+4*exp(I*(-k1-k2)) \
 +4*exp(I*(+k1+k3))+4*exp(I*(+k1-k3))+4*exp(I*(-k1+k3))+4*exp(I*(-k1-k3)) \
 +4*exp(I*(+k1+k4))+4*exp(I*(+k1-k4))+4*exp(I*(-k1+k4))+4*exp(I*(-k1-k4)) \
 +4*exp(I*(+k2+k3))+4*exp(I*(+k2-k3))+4*exp(I*(-k2+k3))+4*exp(I*(-k2-k3)) \
 +4*exp(I*(+k2+k4))+4*exp(I*(+k2-k4))+4*exp(I*(-k2+k4))+4*exp(I*(-k2-k4)) \
 +4*exp(I*(+k3+k4))+4*exp(I*(+k3-k4))+4*exp(I*(-k3+k4))+4*exp(I*(-k3-k4)) \
 +8*exp(I*(+k1))+8*exp(I*(-k1)) \
 +8*exp(I*(+k2))+8*exp(I*(-k2)) \
 +8*exp(I*(+k3))+8*exp(I*(-k3)) \
 +8*exp(I*(+k4))+8*exp(I*(-k4)) \
 -240)/64)));
lap4:=sort(expand(simplify(( \
 +exp(I*(+k1+k2+k3+k4))+exp(I*(-k1+k2+k3+k4))+exp(I*(+k1-k2+k3+k4))+exp(I*(-k1-k2+k3+k4)) \
 +exp(I*(+k1+k2-k3+k4))+exp(I*(-k1+k2-k3+k4))+exp(I*(+k1-k2-k3+k4))+exp(I*(-k1-k2-k3+k4)) \
 +exp(I*(+k1+k2+k3-k4))+exp(I*(-k1+k2+k3-k4))+exp(I*(+k1-k2+k3-k4))+exp(I*(-k1-k2+k3-k4)) \
 +exp(I*(+k1+k2-k3-k4))+exp(I*(-k1+k2-k3-k4))+exp(I*(+k1-k2-k3-k4))+exp(I*(-k1-k2-k3-k4)) \
 +7*exp(I*(+k1+k2+k3))+7*exp(I*(-k1+k2+k3))+7*exp(I*(+k1-k2+k3))+7*exp(I*(-k1-k2+k3)) \
 +7*exp(I*(+k1+k2-k3))+7*exp(I*(-k1+k2-k3))+7*exp(I*(+k1-k2-k3))+7*exp(I*(-k1-k2-k3)) \
 +7*exp(I*(+k1+k2+k4))+7*exp(I*(-k1+k2+k4))+7*exp(I*(+k1-k2+k4))+7*exp(I*(-k1-k2+k4)) \
 +7*exp(I*(+k1+k2-k4))+7*exp(I*(-k1+k2-k4))+7*exp(I*(+k1-k2-k4))+7*exp(I*(-k1-k2-k4)) \
 +7*exp(I*(+k1+k3+k4))+7*exp(I*(-k1+k3+k4))+7*exp(I*(+k1-k3+k4))+7*exp(I*(-k1-k3+k4)) \
 +7*exp(I*(+k1+k3-k4))+7*exp(I*(-k1+k3-k4))+7*exp(I*(+k1-k3-k4))+7*exp(I*(-k1-k3-k4)) \
 +7*exp(I*(+k2+k3+k4))+7*exp(I*(-k2+k3+k4))+7*exp(I*(+k2-k3+k4))+7*exp(I*(-k2-k3+k4)) \
 +7*exp(I*(+k2+k3-k4))+7*exp(I*(-k2+k3-k4))+7*exp(I*(+k2-k3-k4))+7*exp(I*(-k2-k3-k4)) \
 +40*exp(I*(+k1+k2))+40*exp(I*(+k1-k2))+40*exp(I*(-k1+k2))+40*exp(I*(-k1-k2)) \
 +40*exp(I*(+k1+k3))+40*exp(I*(+k1-k3))+40*exp(I*(-k1+k3))+40*exp(I*(-k1-k3)) \
 +40*exp(I*(+k1+k4))+40*exp(I*(+k1-k4))+40*exp(I*(-k1+k4))+40*exp(I*(-k1-k4)) \
 +40*exp(I*(+k2+k3))+40*exp(I*(+k2-k3))+40*exp(I*(-k2+k3))+40*exp(I*(-k2-k3)) \
 +40*exp(I*(+k2+k4))+40*exp(I*(+k2-k4))+40*exp(I*(-k2+k4))+40*exp(I*(-k2-k4)) \
 +40*exp(I*(+k3+k4))+40*exp(I*(+k3-k4))+40*exp(I*(-k3+k4))+40*exp(I*(-k3-k4)) \
 +100*exp(I*(+k1))+100*exp(I*(-k1)) \
 +100*exp(I*(+k2))+100*exp(I*(-k2)) \
 +100*exp(I*(+k3))+100*exp(I*(-k3)) \
 +100*exp(I*(+k4))+100*exp(I*(-k4)) \
 -2000)/432)));
lap5:=sort(expand(simplify( \
 +c4*exp(I*(+k1+k2+k3+k4))+c4*exp(I*(-k1+k2+k3+k4))+c4*exp(I*(+k1-k2+k3+k4))+c4*exp(I*(-k1-k2+k3+k4)) \
 +c4*exp(I*(+k1+k2-k3+k4))+c4*exp(I*(-k1+k2-k3+k4))+c4*exp(I*(+k1-k2-k3+k4))+c4*exp(I*(-k1-k2-k3+k4)) \
 +c4*exp(I*(+k1+k2+k3-k4))+c4*exp(I*(-k1+k2+k3-k4))+c4*exp(I*(+k1-k2+k3-k4))+c4*exp(I*(-k1-k2+k3-k4)) \
 +c4*exp(I*(+k1+k2-k3-k4))+c4*exp(I*(-k1+k2-k3-k4))+c4*exp(I*(+k1-k2-k3-k4))+c4*exp(I*(-k1-k2-k3-k4)) \
 +c3*exp(I*(+k1+k2+k3))+c3*exp(I*(-k1+k2+k3))+c3*exp(I*(+k1-k2+k3))+c3*exp(I*(-k1-k2+k3)) \
 +c3*exp(I*(+k1+k2-k3))+c3*exp(I*(-k1+k2-k3))+c3*exp(I*(+k1-k2-k3))+c3*exp(I*(-k1-k2-k3)) \
 +c3*exp(I*(+k1+k2+k4))+c3*exp(I*(-k1+k2+k4))+c3*exp(I*(+k1-k2+k4))+c3*exp(I*(-k1-k2+k4)) \
 +c3*exp(I*(+k1+k2-k4))+c3*exp(I*(-k1+k2-k4))+c3*exp(I*(+k1-k2-k4))+c3*exp(I*(-k1-k2-k4)) \
 +c3*exp(I*(+k1+k3+k4))+c3*exp(I*(-k1+k3+k4))+c3*exp(I*(+k1-k3+k4))+c3*exp(I*(-k1-k3+k4)) \
 +c3*exp(I*(+k1+k3-k4))+c3*exp(I*(-k1+k3-k4))+c3*exp(I*(+k1-k3-k4))+c3*exp(I*(-k1-k3-k4)) \
 +c3*exp(I*(+k2+k3+k4))+c3*exp(I*(-k2+k3+k4))+c3*exp(I*(+k2-k3+k4))+c3*exp(I*(-k2-k3+k4)) \
 +c3*exp(I*(+k2+k3-k4))+c3*exp(I*(-k2+k3-k4))+c3*exp(I*(+k2-k3-k4))+c3*exp(I*(-k2-k3-k4)) \
 +c2*exp(I*(+k1+k2))+c2*exp(I*(+k1-k2))+c2*exp(I*(-k1+k2))+c2*exp(I*(-k1-k2)) \
 +c2*exp(I*(+k1+k3))+c2*exp(I*(+k1-k3))+c2*exp(I*(-k1+k3))+c2*exp(I*(-k1-k3)) \
 +c2*exp(I*(+k1+k4))+c2*exp(I*(+k1-k4))+c2*exp(I*(-k1+k4))+c2*exp(I*(-k1-k4)) \
 +c2*exp(I*(+k2+k3))+c2*exp(I*(+k2-k3))+c2*exp(I*(-k2+k3))+c2*exp(I*(-k2-k3)) \
 +c2*exp(I*(+k2+k4))+c2*exp(I*(+k2-k4))+c2*exp(I*(-k2+k4))+c2*exp(I*(-k2-k4)) \
 +c2*exp(I*(+k3+k4))+c2*exp(I*(+k3-k4))+c2*exp(I*(-k3+k4))+c2*exp(I*(-k3-k4)) \
 +c1*exp(I*(+k1))+c1*exp(I*(-k1)) \
 +c1*exp(I*(+k2))+c1*exp(I*(-k2)) \
 +c1*exp(I*(+k3))+c1*exp(I*(-k3)) \
 +c1*exp(I*(+k4))+c1*exp(I*(-k4)) \
 -8*c1-24*c2-32*c3-16*c4)));
 lap5:=expand(series(subs({k1=a*p1,k2=a*p2,k3=a*p3,k4=a*p4},lap5),a,6));
 lap50:=sort(coeff(lap5,a,0),[p1,p2,p3,p4]); # must be zero
 lap51:=sort(coeff(lap5,a,1),[p1,p2,p3,p4]); # must be zero
 lap52:=sort(coeff(lap5,a,2),[p1,p2,p3,p4]); # must be -p1^2-p2^2-p3^2-p4^2
 lap53:=sort(coeff(lap5,a,3),[p1,p2,p3,p4]); # must be zero
 lap54:=sort(coeff(lap5,a,4),[p1,p2,p3,p4]); # should be isotropic
 lap50:=expand(subs(c1=1-6*c2-12*c3-8*c4,lap50)); # is zero
 lap51:=expand(subs(c1=1-6*c2-12*c3-8*c4,lap51)); # is zero
 lap52:=expand(subs(c1=1-6*c2-12*c3-8*c4,lap52)); # is -p1^2-p2^2-p3^2-p4^2
 lap53:=expand(subs(c1=1-6*c2-12*c3-8*c4,lap53)); # is zero
 lap54:=expand(subs(c1=1-6*c2-12*c3-8*c4,lap54)); # no way to eliminate all p^4 terms
der1:=simplify((exp(I*k1)-exp(-I*k1))/2);
der2:=factor(expand(simplify(( \
 +8*exp(+I*k1) \
 +4*exp(I*(+k1+k2))+4*exp(I*(+k1-k2)) \
 +4*exp(I*(+k1+k3))+4*exp(I*(+k1-k3)) \
 +4*exp(I*(+k1+k4))+4*exp(I*(+k1-k4)) \
 +2*exp(I*(+k1+k2+k3))+2*exp(I*(+k1+k2-k3)) \
 +2*exp(I*(+k1-k2+k3))+2*exp(I*(+k1-k2-k3)) \
 +2*exp(I*(+k1+k2+k4))+2*exp(I*(+k1+k2-k4)) \
 +2*exp(I*(+k1-k2+k4))+2*exp(I*(+k1-k2-k4)) \
 +2*exp(I*(+k1+k3+k4))+2*exp(I*(+k1+k3-k4)) \
 +2*exp(I*(+k1-k3+k4))+2*exp(I*(+k1-k3-k4)) \
 +1*exp(I*(+k1+k2+k3+k4))+1*exp(I*(+k1+k2+k3-k4)) \
 +1*exp(I*(+k1+k2-k3+k4))+1*exp(I*(+k1+k2-k3-k4)) \
 +1*exp(I*(+k1-k2+k3+k4))+1*exp(I*(+k1-k2+k3-k4)) \
 +1*exp(I*(+k1-k2-k3+k4))+1*exp(I*(+k1-k2-k3-k4)) \
 -8*exp(-I*k1) \
 -4*exp(I*(-k1+k2))-4*exp(I*(-k1-k2)) \
 -4*exp(I*(-k1+k3))-4*exp(I*(-k1-k3)) \
 -4*exp(I*(-k1+k4))-4*exp(I*(-k1-k4)) \
 -2*exp(I*(-k1+k2+k3))-2*exp(I*(-k1+k2-k3)) \
 -2*exp(I*(-k1-k2+k3))-2*exp(I*(-k1-k2-k3)) \
 -2*exp(I*(-k1+k2+k4))-2*exp(I*(-k1+k2-k4)) \
 -2*exp(I*(-k1-k2+k4))-2*exp(I*(-k1-k2-k4)) \
 -2*exp(I*(-k1+k3+k4))-2*exp(I*(-k1+k3-k4)) \
 -2*exp(I*(-k1-k3+k4))-2*exp(I*(-k1-k3-k4)) \
 -1*exp(I*(-k1+k2+k3+k4))-1*exp(I*(-k1+k2+k3-k4)) \
 -1*exp(I*(-k1+k2-k3+k4))-1*exp(I*(-k1+k2-k3-k4)) \
 -1*exp(I*(-k1-k2+k3+k4))-1*exp(I*(-k1-k2+k3-k4)) \
 -1*exp(I*(-k1-k2-k3+k4))-1*exp(I*(-k1-k2-k3-k4)) \
 )/128)));
der3:=factor(expand(simplify(( \
 +64*exp(+I*k1) \
 +16*exp(I*(+k1+k2))+16*exp(I*(+k1-k2)) \
 +16*exp(I*(+k1+k3))+16*exp(I*(+k1-k3)) \
 +16*exp(I*(+k1+k4))+16*exp(I*(+k1-k4)) \
 +4*exp(I*(+k1+k2+k3))+4*exp(I*(+k1+k2-k3)) \
 +4*exp(I*(+k1-k2+k3))+4*exp(I*(+k1-k2-k3)) \
 +4*exp(I*(+k1+k2+k4))+4*exp(I*(+k1+k2-k4)) \
 +4*exp(I*(+k1-k2+k4))+4*exp(I*(+k1-k2-k4)) \
 +4*exp(I*(+k1+k3+k4))+4*exp(I*(+k1+k3-k4)) \
 +4*exp(I*(+k1-k3+k4))+4*exp(I*(+k1-k3-k4)) \
 +1*exp(I*(+k1+k2+k3+k4))+1*exp(I*(+k1+k2+k3-k4)) \
 +1*exp(I*(+k1+k2-k3+k4))+1*exp(I*(+k1+k2-k3-k4)) \
 +1*exp(I*(+k1-k2+k3+k4))+1*exp(I*(+k1-k2+k3-k4)) \
 +1*exp(I*(+k1-k2-k3+k4))+1*exp(I*(+k1-k2-k3-k4)) \
 -64*exp(-I*k1) \
 -16*exp(I*(-k1+k2))-16*exp(I*(-k1-k2)) \
 -16*exp(I*(-k1+k3))-16*exp(I*(-k1-k3)) \
 -16*exp(I*(-k1+k4))-16*exp(I*(-k1-k4)) \
 -4*exp(I*(-k1+k2+k3))-4*exp(I*(-k1+k2-k3)) \
 -4*exp(I*(-k1-k2+k3))-4*exp(I*(-k1-k2-k3)) \
 -4*exp(I*(-k1+k2+k4))-4*exp(I*(-k1+k2-k4)) \
 -4*exp(I*(-k1-k2+k4))-4*exp(I*(-k1-k2-k4)) \
 -4*exp(I*(-k1+k3+k4))-4*exp(I*(-k1+k3-k4)) \
 -4*exp(I*(-k1-k3+k4))-4*exp(I*(-k1-k3-k4)) \
 -1*exp(I*(-k1+k2+k3+k4))-1*exp(I*(-k1+k2+k3-k4)) \
 -1*exp(I*(-k1+k2-k3+k4))-1*exp(I*(-k1+k2-k3-k4)) \
 -1*exp(I*(-k1-k2+k3+k4))-1*exp(I*(-k1-k2+k3-k4)) \
 -1*exp(I*(-k1-k2-k3+k4))-1*exp(I*(-k1-k2-k3-k4)) \
 )/432)));
\endcomment

\clearpage


\section*{App.\,D: Isotropic stencils via Kumar's trick}


A somewhat systematic overview that includes ``isotropic'' Laplacians in 2D and
3D is presented in \cite{Patra:2006}.
In this appendix we review a particularly practical approach for deriving an
isotropic stencil in any dimension, due to Kumar \cite{Kumar:2004}.

The standard discretization of the first derivative operator in two
dimensions (2D) is
\beq
(\ps_x)^\mr{std}_{i,j}={1\ovr2a}(\ps_{i+1,j}-\ps_{i-1,j})
\;.
\eeq
From a Taylor expansion [of
$\ri\sin(ak_1)\!=\!\ri ak_1(1\!-\!a^2k_1^2/6\!+\!O(a^4))$ in momentum
space or directly in position space] one finds that the standard discrete
derivative deviates from the continuum derivative through $O(a^2)$-suppressed
terms.
This may be summarized in the form
\beq
(\ps_x)^\mr{std}_{i,j}=(1+{a^2\ovr6}\pad_{xx})(\ps_x)_{i,j}
\label{kumar_summary}
\eeq
where both derivatives on the r.h.s.\ refer to the continuum.
A possible strategy to improve rotational symmetry is thus to define the
discretized first derivative such that the deviation from the continuum
behavior $\ri ak_1$ is the same in either direction, that is through
\beq
(\ps_x)^\mr{iso}_{i,j}=(1+{a^2\ovr6}\lap)(\ps_x)_{i,j}
\eeq
where again the operators on the r.h.s.\ refer to the continuum.
The idea by Kumar \cite{Kumar:2004} is to factor the bracket and to define the
discretized ``isotropic'' first derivative through
\beq
(\ps_x)^\mr{iso}_{i,j}=
(1+{a^2\ovr6}\pad_{yy})(1+{a^2\ovr6}\pad_{xx})(\ps_x)_{i,j}=
(1+{a^2\ovr6}\pad_{yy})(\ps_x)^\mr{std}_{i,j}
\label{kumar_factor}
\eeq
where (\ref{kumar_summary}) has been used in the second step.
Moreover, we may replace the second derivative in the $y$ direction by its
simplest discrete version (i.e.\ the 1/-2/1 stencil operator), since the
difference is another $a^4$ term [of which we did not keep track in
(\ref{kumar_summary}-\ref{kumar_factor}) anyway].
This gives\\[4mm]
\hspace*{2.3cm}$(\ps_x)^\mr{iso}_{i,j}\;=\;$
\begin{tabular}{|ccc|}
\hline 0&0&0\\-1/2&0&-1/2\\0&0&0\\ \hline
\end{tabular}
$+{1\ovr6}$
\begin{tabular}{|ccc|}
\hline -1/2&0&1/2\\1&0&-1\\-1/2&0&1/2\\ \hline
\end{tabular}
$\;=\;$
\begin{tabular}{|ccc|}
\hline -1&0&1\\-4&0&4\\-1&0&1\\ \hline
\end{tabular}\,\,/12
\beq
=\Big[
\ps_{i+1,j+1}+4\ps_{i+1,j}+\ps_{i+1,j-1}-
\ps_{i-1,j+1}-4\ps_{i-1,j}-\ps_{i-1,j-1}
\Big]/12
\eeq
where we have used the stencil notation.
Compared to the standard discrete derivative, there is a spreading in the
transverse direction with a factor ${1\ovr6}/{4\ovr6}/{1\ovr6}$, respectively.
It is easy to generalize this procedure to higher dimensions, and the pertinent
``isotropic'' first derivative operators have been given in App.\,A,\,B,\,C for
2D,\,3D,\,4D, respectively.

The standard discretization of the second derivative operator in two
dimensions (2D) is
\beq
(\ps_{xx})^\mr{std}_{i,j}={1\ovr a^2}(\ps_{i+1,j}-2\ps_{i,j}+\ps_{i-1,j})
\eeq
and from a Taylor expansion one finds that this is equivalent to
\beq
(\ps_{xx})^\mr{std}_{i,j}=(1+{a^2\ovr12}\pad_{xx})(\ps_{xx})_{i,j}
\eeq
where both derivatives on the r.h.s.\ refer to the continuum.
The ``isotropic'' second derivative operator follows by deliberately
introducing the same discretization error in the $y$-direction
\beq
(\ps_{xx})^\mr{iso}_{i,j}=(1+{a^2\ovr12}\lap)(\ps_{xx})_{i,j}
\eeq
and Kumar's trick \cite{Kumar:2004} of factorizing (to the order we are
interested in) the continuum expression
\bea
(\ps_{xx})^\mr{iso}_{i,j}\!&\!=\!&\!
(1\!+\!{a^2\ovr12}\pad_{yy})(1\!+\!{a^2\ovr12}\pad_{xx})(\ps_{xx})_{i,j}=
(1\!+\!{a^2\ovr12}\pad_{yy})(\ps_{xx})^\mr{std}_{i,j}
=\Big[\ps_{i+1,j+1}+10\ps_{i+1,j}+\ps_{i+1,j-1}
\nonumber
\\
&&
-2\ps_{i,j+1}-20\ps_{i,j}-2\ps_{i,j-1}
+ \ps_{i-1,j+1}+10\ps_{i-1,j}+ \ps_{i-1,j-1}
\Big]/12 
\eea
yields (the simplest) ``isotropic'' second derivative operator.
Compared to the standard discrete second derivative in 2D, there is a spreading
in the transverse direction by a factor ${1\ovr12}/{10\ovr12}/{1\ovr12}$.
It is easy to generalize this procedure to higher dimensions, and we shall
just give the result:
\begin{itemize}
\itemsep-1pt
\item
Isotropic second $x$-derivative in 2D:\\[2mm]
\begin{tabular*}{2.1cm}{@{\extracolsep{\fill}}|@{\,\,}c@{}c@{}c@{\,\,}|}
\hline 1&-2&1\\10&-20&10\\1&-2&1\\ \hline
\end{tabular*}\,\,/12\\[1mm]
$\hat\pad_x^2=[\cos(k_1)-1][\cos(k_2)+5]/3$
\item
Isotropic second $x$-derivative in 3D:\\[2mm]
\begin{tabular*}{2.4cm}{@{\extracolsep{\fill}}|@{\,\,}c@{}c@{}c@{\,\,}|}
\hline 1&-2&1\\10&-20&10\\1&-2&1\\ \hline
\end{tabular*}
\begin{tabular*}{2.4cm}{@{\extracolsep{\fill}}|@{\,\,}c@{}c@{}c@{\,\,}|}
\hline 10&-20&10\\100&-200&100\\10&-20&10\\ \hline
\end{tabular*}
\begin{tabular*}{2.4cm}{@{\extracolsep{\fill}}|@{\,\,}c@{}c@{}c@{\,\,}|}
\hline 1&-2&1\\10&-20&10\\1&-2&1\\ \hline
\end{tabular*}\,\,/144\\[1mm]
$\hat\pad_x^2=[\cos(k_1)-1][\cos(k_2)+5][\cos(k_3)+5]/18$
\item
Isotropic second $x$-derivative in 4D:\\[2mm]
\begin{tabular*}{2.7cm}{@{\extracolsep{\fill}}|@{\,\,}c@{}c@{}c@{\,\,}|}
\hline 1&-2&1\\ 10&-20&10\\ 1&-2&1\\ \hline
\end{tabular*}
\begin{tabular*}{2.7cm}{@{\extracolsep{\fill}}|@{\,\,}c@{}c@{}c@{\,\,}|}
\hline 10&-20&10\\ 100&-200&100\\ 10&-20&10\\ \hline
\end{tabular*}
\begin{tabular*}{2.7cm}{@{\extracolsep{\fill}}|@{\,\,}c@{}c@{}c@{\,\,}|}
\hline 1&-2&1\\ 10&-20&10\\ 1&-2&1\\ \hline
\end{tabular*}\\[1mm]
\begin{tabular*}{2.7cm}{@{\extracolsep{\fill}}|@{\,\,}c@{}c@{}c@{\,\,}|}
\hline 10&-20&10\\ 100&-200&100\\ 10&-20&10\\ \hline
\end{tabular*}
\begin{tabular*}{2.7cm}{@{\extracolsep{\fill}}|@{}c@{}c@{}c@{}|}
\hline 100&-200&100\\ 1000&-2000&1000\\ 100&-200&100\\ \hline
\end{tabular*}
\begin{tabular*}{2.7cm}{@{\extracolsep{\fill}}|@{\,\,}c@{}c@{}c@{\,\,}|}
\hline 10&-20&10\\ 100&-200&100\\ 10&-20&10\\ \hline
\end{tabular*}\,\,/1728\\[1mm]
\begin{tabular*}{2.7cm}{@{\extracolsep{\fill}}|@{\,\,}c@{}c@{}c@{\,\,}|}
\hline 1&-2&1\\ 10&-20&10\\ 1&-2&1\\ \hline
\end{tabular*}
\begin{tabular*}{2.7cm}{@{\extracolsep{\fill}}|@{\,\,}c@{}c@{}c@{\,\,}|}
\hline 10&-20&10\\ 100&-200&100\\ 10&-20&10\\ \hline
\end{tabular*}
\begin{tabular*}{2.7cm}{@{\extracolsep{\fill}}|@{\,\,}c@{}c@{}c@{\,\,}|}
\hline 1&-2&1\\ 10&-20&10\\ 1&-2&1\\ \hline
\end{tabular*}\\[1mm]
$\hat\pad_x^2=[\cos(k_1)-1][\cos(k_2)+5][\cos(k_3)+5][\cos(k_4)+5]/108$
\end{itemize}
Upon adding the ``isotropic'' second derivative in the $y$ (and possibly $z$,
$t$) direction, one gets the ``isotropic'' Laplacian stencil in 2D,\,3D,\,4D,
as given in previous appendices.
We emphasize that the ``isotropic'' Laplacian establishes better rotational
symmetry near the \emph{center} of the Brillouin zone.
What proves most useful in many applications (including our goal of designing
more continuum-like lattice Dirac operators), however, is isotropy at the
\emph{boundary} of the Brillouin zone, and this is achieved through the
``Brillouin'' Laplacian, as given in App.\,A,\,B,\,C.

\begincomment
sec1:=factor(expand(simplify(( \
 +1*exp(I*(-k1+k2))-2*exp(I*(+k2))+1*exp(I*(+k1+k2)) \
 +10*exp(I*(-k1))-20+10*exp(I*(+k1)) \
 +1*exp(I*(-k1-k2))-2*exp(I*(-k2))+1*exp(I*(+k1-k2)) \
 )/12)));
sec2:=factor(expand(simplify(( \
 +1*exp(I*(-k1+k2-k3))-2*exp(I*(+k2-k3))+1*exp(I*(+k1+k2-k3)) \
 +10*exp(I*(-k1-k3))-20*exp(I*(-k3))+10*exp(I*(+k1-k3)) \
 +1*exp(I*(-k1-k2-k3))-2*exp(I*(-k2-k3))+1*exp(I*(+k1-k2-k3)) \
 +10*exp(I*(-k1+k2))-20*exp(I*(+k2))+10*exp(I*(+k1+k2)) \
 +100*exp(I*(-k1))-200+100*exp(I*(+k1)) \
 +10*exp(I*(-k1-k2))-20*exp(I*(-k2))+10*exp(I*(+k1-k2)) \
 +1*exp(I*(-k1+k2+k3))-2*exp(I*(+k2+k3))+1*exp(I*(+k1+k2+k3)) \
 +10*exp(I*(-k1+k3))-20*exp(I*(+k3))+10*exp(I*(+k1+k3)) \
 +1*exp(I*(-k1-k2+k3))-2*exp(I*(-k2+k3))+1*exp(I*(+k1-k2+k3)) \
 )/144)));
sec3:=factor(expand(simplify(( \
 +1*exp(I*(-k1+k2-k3-k4))-2*exp(I*(+k2-k3-k4))+1*exp(I*(+k1+k2-k3-k4)) \
 +10*exp(I*(-k1-k3-k4))-20*exp(I*(-k3-k4))+10*exp(I*(+k1-k3-k4)) \
 +1*exp(I*(-k1-k2-k3-k4))-2*exp(I*(-k2-k3-k4))+1*exp(I*(+k1-k2-k3-k4)) \
 +10*exp(I*(-k1+k2-k4))-20*exp(I*(+k2-k4))+10*exp(I*(+k1+k2-k4)) \
 +100*exp(I*(-k1-k4))-200*exp(I*(-k4))+100*exp(I*(+k1-k4)) \
 +10*exp(I*(-k1-k2-k4))-20*exp(I*(-k2-k4))+10*exp(I*(+k1-k2-k4)) \
 +1*exp(I*(-k1+k2+k3-k4))-2*exp(I*(+k2+k3-k4))+1*exp(I*(+k1+k2+k3-k4)) \
 +10*exp(I*(-k1+k3-k4))-20*exp(I*(+k3-k4))+10*exp(I*(+k1+k3-k4)) \
 +1*exp(I*(-k1-k2+k3-k4))-2*exp(I*(-k2+k3-k4))+1*exp(I*(+k1-k2+k3-k4)) \
 +10*exp(I*(-k1+k2-k3))-20*exp(I*(+k2-k3))+10*exp(I*(+k1+k2-k3)) \
 +100*exp(I*(-k1-k3))-200*exp(I*(-k3))+100*exp(I*(+k1-k3)) \
 +10*exp(I*(-k1-k2-k3))-20*exp(I*(-k2-k3))+10*exp(I*(+k1-k2-k3)) \
 +100*exp(I*(-k1+k2))-200*exp(I*(+k2))+100*exp(I*(+k1+k2)) \
 +1000*exp(I*(-k1))-2000+1000*exp(I*(+k1)) \
 +100*exp(I*(-k1-k2))-200*exp(I*(-k2))+100*exp(I*(+k1-k2)) \
 +10*exp(I*(-k1+k2+k3))-20*exp(I*(+k2+k3))+10*exp(I*(+k1+k2+k3)) \
 +100*exp(I*(-k1+k3))-200*exp(I*(+k3))+100*exp(I*(+k1+k3)) \
 +10*exp(I*(-k1-k2+k3))-20*exp(I*(-k2+k3))+10*exp(I*(+k1-k2+k3)) \
 +1*exp(I*(-k1+k2-k3+k4))-2*exp(I*(+k2-k3+k4))+1*exp(I*(+k1+k2-k3+k4)) \
 +10*exp(I*(-k1-k3+k4))-20*exp(I*(-k3+k4))+10*exp(I*(+k1-k3+k4)) \
 +1*exp(I*(-k1-k2-k3+k4))-2*exp(I*(-k2-k3+k4))+1*exp(I*(+k1-k2-k3+k4)) \
 +10*exp(I*(-k1+k2+k4))-20*exp(I*(+k2+k4))+10*exp(I*(+k1+k2+k4)) \
 +100*exp(I*(-k1+k4))-200*exp(I*(+k4))+100*exp(I*(+k1+k4)) \
 +10*exp(I*(-k1-k2+k4))-20*exp(I*(-k2+k4))+10*exp(I*(+k1-k2+k4)) \
 +1*exp(I*(-k1+k2+k3+k4))-2*exp(I*(+k2+k3+k4))+1*exp(I*(+k1+k2+k3+k4)) \
 +10*exp(I*(-k1+k3+k4))-20*exp(I*(+k3+k4))+10*exp(I*(+k1+k3+k4)) \
 +1*exp(I*(-k1-k2+k3+k4))-2*exp(I*(-k2+k3+k4))+1*exp(I*(+k1-k2+k3+k4)) \
 )/1728)));
iso1:=simplify(expand(sec1+ \
 subs({k1=k2,k2=k1},sec1) \
 ));
iso2:=simplify(expand(sec2+ \
 subs({k1=k2,k2=k1},sec2)+subs({k1=k3,k3=k1},sec2) \
 ));
iso3:=simplify(expand(sec3+ \
 subs({k1=k2,k2=k1},sec3)+subs({k1=k3,k3=k1},sec3)+subs({k1=k4,k4=k1},sec3) \
 ));
ser1:=factor(coeff( \
 series(subs({k1=a*k1,k2=a*k2},iso1),a,6) \
 ,a,4));
ser2:=factor(coeff( \
 series(subs({k1=a*k1,k2=a*k2,k3=a*k3},iso2),a,6) \
 ,a,4));
ser3:=factor(coeff( \
 series(subs({k1=a*k1,k2=a*k2,k3=a*k3,k4=a*k4},iso3),a,6) \
 ,a,4));
\endcomment


\begincomment
\section*{App.\,E: Dispersion relation details}
\endcomment


\begincomment


\bdm
O_1=-\sum\nab_{\!\mu}^2=
\left\{
\begin{array}{lc}
\sin^2(k_1)+...                        & \mbox{(std)} \\
\sin^2(k_1)[\cos(k_2)\!+\!1]^2/2^2+... & \mbox{(bri)} \\
\sin^2(k_1)[\cos(k_2)\!+\!2]^2/3^2+... & \mbox{(iso)} \\
\end{array}
\right.
\edm
\bdm
O_2=[\frac{1}{2}\lap-m]^2=
\left\{
\begin{array}{lc}
{[\cos(k_1)+...-2-m]^2}                          & \mbox{(std)} \\
{[\cos(k_1)\cos(k_2)-1-m]^2}                     & \mbox{(til)} \\
{[\cos(k_1)\cos(k_2)+\cos(k_1)+...-3-2m]^2/2^2}  & \mbox{(bri)} \\
{[\cos(k_1)\cos(k_2)+2\cos(k_1)+...-5-3m]^2/3^2} & \mbox{(iso)} 
\end{array}
\right.
\edm
\bdm
O_2=[\frac{1}{2}\lap-m]^2=
\left\{
\begin{array}{lc}
{[2\sin^2(k_1/2)+...+m]^2}                                    & \mbox{(std)} \\
{[4\cos^2(k_1/2)\cos^2(k_2/2)-2\cos^2(k_1/2)-...-m]^2}        & \mbox{(til)} \\
{[2\cos^2(k_1/2)\cos^2(k_2/2)-2-m]^2}                         & \mbox{(bri)} \\
{[4\cos^2(k_1/2)\cos^2(k_2/2)+2\cos^2(k_1/2)+...-8-3m]^2/3^2} & \mbox{(iso)} 
\end{array}
\right.
\edm



\bdm
O_1=
\left\{
\begin{array}{lc}
\sin^2(k_1)+...                                           & \mbox{(std)} \\
\sin^2(k_1)[\cos(k_2)\!+\!1]^2[\cos(k_3)\!+\!1]^2/2^4+... & \mbox{(bri)} \\
\sin^2(k_1)[\cos(k_2)\!+\!2]^2[\cos(k_3)\!+\!2]^2/3^4+... & \mbox{(iso)}
\end{array}
\right.
\edm
\bdm
O_2=
\left\{
\begin{array}{lc}
{[\cos(k_1)+...-3-m]^2}                                                          & \mbox{(std)} \\
{[\cos(k_1)\cos(k_2)\cos(k_3)-1-m]^2}                                            & \mbox{(til)} \\
{[\cos(k_1)\cos(k_2)\cos(k_3)+\cos(k_1)\cos(k_2)+...+\cos(k_1)+...-7-4m]^2/4^2}  & \mbox{(bri)} \\
{[\cos(k_1)\cos(k_2)\cos(k_3)+3\cos(k_1)\cos(k_2)+...+5\cos(k_1)-25-12m]^2/12^2} & \mbox{(iso)} 
\end{array}
\right.
\edm
\bdm
O_2=
\left\{
\begin{array}{lc}
{[2\sin^2(k_1/2)+...+m]^2}                                                              & \!\mbox{(std)} \\
{[8\cos^2(k_1/2)\cos^2(k_2/2)\cos^2(k_3/2)-4\cos^2(k_1/2)\cos^2(k_2/2)-...}             &                \\
 {\quad+2\cos^2(k_1/2)+...-2-m]^2}                                                        & \!\mbox{(til)} \\
{[2\cos^2(k_1/2)\cos^2(k_2/2)\cos^2(k_3/2)-2-m]^2}                                      & \!\mbox{(bri)} \\
{[2\cos^2(k_1/2)\cos^2(k_2/2)\cos^2(k_3/2)+2\cos^2(k_1/2)\cos^2(k_2/2)+...-8-3m]^2/3^2} & \!\mbox{(iso)} 
\end{array}
\right.
\edm



\bdm
O_1=
\left\{
\begin{array}{lc}
\sin^2(k_1)+...                                                              & \mbox{(std)} \\
\sin^2(k_1)[\cos(k_2)\!+\!1]^2[\cos(k_3)\!+\!1]^2[\cos(k_4)\!+\!1]^2/2^6+... & \mbox{(bri)} \\
\sin^2(k_1)[\cos(k_2)\!+\!2]^2[\cos(k_3)\!+\!2]^2[\cos(k_4)\!+\!2]^2/3^6+... & \mbox{(iso)}
\end{array}
\right.
\edm
\bdm
O_2=
\left\{
\begin{array}{lc}
{[\cos(k_1)+...-4-m]^2}                                                   & \mbox{(std)} \\
{[\cos(k_1)\cos(k_2)\cos(k_3)\cos(k_4)-1-m]^2}                            & \mbox{(til)} \\
{[\cos(k_1)\cos(k_2)\cos(k_3)\cos(k_4)+\cos(k_1)\cos(k_2)\cos(k_3)+...}   &              \\
 {\quad+\cos(k_1)\cos(k_2)+...+\cos(k_1)-15-8m]^2/8^2}                    & \mbox{(bri)} \\
{[2\cos(k_1)\cos(k_2)\cos(k_3)\cos(k_4)+7\cos(k_1)\cos(k_2)\cos(k_3)+...} &              \\
 {\quad+20\cos(k_1)\cos(k_2)+...+25\cos(k_1)+...-250-108m]^2/108^2}       & \mbox{(iso)} 
\end{array}
\right.
\edm
\bdm
O_2=
\left\{
\begin{array}{lc}
{[2\sin^2(k_1/2)+...+m]^2}                                                 & \mbox{(std)} \\
{[16\cos^2(k_1/2)\cos^2(k_2/2)\cos^2(k_3/2)\sin^2(k_4/2)}                  &              \\
 {\quad-8\cos^2(k_1/2)\cos^2(k_2/2)\cos^2(k_3/2)-...}                      &              \\
 {\quad+4\cos^2(k_1/2)\cos^2(k_2/2)+...-2\cos^2(k_1/2)-...-m]^2}           & \mbox{(til)} \\
{[2\cos^2(k_1/2)\cos^2(k_2/2)\cos^2(k_3/2)\sin^2(k_4/2)-2-m]^2}            & \mbox{(bri)} \\
{[8\cos^2(k_1/2)\cos^2(k_2/2)\cos^2(k_3/2)\sin^2(k_4/2)}                   &              \\
 {\quad+10\cos^2(k_1/2)\cos^2(k_2/2)\cos^2(k_3/2)+...}                     &              \\
 {\quad+8\cos^2(k_1/2)\cos^2(k_2/2)+...-8\cos^2(k_1/2)-...-64-27m]^2/27^2} & \mbox{(iso)}
\end{array}
\right.
\edm

\endcomment



\end{document}